\documentclass[aps,prx,reprint,floatfix]{revtex4-2}

\usepackage{amsmath,amssymb,amsfonts,mathrsfs}
\usepackage{graphicx}
\usepackage{epsfig}
\usepackage{color}
\usepackage{esint}

\usepackage{empheq}

\newcommand{\ie}{\textit{i.e.\ }}%
\newcommand{\eg}{\textit{e.g.\ }}%

\newcommand{\bPhi}{\boldsymbol{\Phi}}%
\newcommand{\bigwp}{\mathpalette\bigwpaux\relax}
\newcommand{\bigwpaux}[2]{%
  \mathord{\scalebox{1.20}{$#1\wp$}}%
}

\newcommand{\bigvarsigma}{\mathpalette\bigvarsigmaaux\relax}
\newcommand{\bigvarsigmaaux}[2]{%
  \mathord{\scalebox{1.15}{$#1\varsigma$}}%
}
\newcommand{\bnabla}{\boldsymbol{\nabla}}%
\newcommand{\Div}{\text{div}\,}%
\newcommand{\Tr}{\text{tr}\,}%
\newcommand*{\cddot}{\mathrel{\rotatebox{90}{.{}.}}}%
\newcommand*{\itbf}[1]{\boldsymbol{#1}}
\newcommand*{\eu}[1]{\hspace{0.2pt}{\boldsymbol{e}}^{#1}}
\newcommand*{\ed}[1]{\hspace{0.2pt}{\boldsymbol{e}}_{#1}}
\newcommand*{\nv}{\hspace{0.2pt}{\boldsymbol{n}}}
\newcommand*{\rv}{\boldsymbol{\hat{r}}}
\newcommand*{\thetav}{\boldsymbol{\hat{\theta}}}
\newcommand*{\phiv}{\boldsymbol{\hat{\varphi}}}

\newcommand*{\CircleArrow}{\scalebox{0.9}{$\circlearrowleft$}}
\newcommand*{\bMk}{\mathbb{M}^{\hspace{0.5pt}k}_{\CircleArrow}}
\newcommand*{\bMkp}{\mathbb{M}^{\hspace{0.5pt}k'}_{\CircleArrow}}
\newcommand*{\bVk}{\mathbb{V}^{\hspace{0.5pt}k}_{\hspace{-0.85pt}\CircleArrow}}

\newcommand*{\bfk}{\mathbb{F}^{\hspace{0.5pt}k}_{\hspace{-0.35pt}\CircleArrow}}
\newcommand*{\bPk}{\mathbb{P}^{\hspace{0.5pt}k}_{\hspace{-0.35pt}\CircleArrow}}
\newcommand*{\bQk}{\mathbb{Q}^{\hspace{0.5pt}k}_{\hspace{-0pt}\CircleArrow}}
\newcommand*{\bJk}{\mathbb{J}^{\hspace{0.5pt}k}_{\CircleArrow}}
\newcommand*{\bRk}{\mathbb{R}^{\hspace{0.0pt}k}_{\CircleArrow}}

\newcommand*{\bMone}{\mathbb{M}^{\hspace{0.5pt}1}_{\CircleArrow}}
\newcommand*{\bMtwo}{\mathbb{M}^{\hspace{0.5pt}2}_{\CircleArrow}}

\newcommand{\cc}[1]{\textcolor{black}{#1}}

\begin{document}
	
	\title{Active morphodynamics of intracellular organelles in the trafficking pathway
	}
	
	
	\author{S.\ Alex Rautu$^{1,2}$}
	\author{Richard G. Morris$^{3,4,5}$}
	\author{Madan Rao$^{2}$}
	\affiliation{
		$^1$Center for Computational Biology, Flatiron Institute, New York, NY 10010, USA\\
		$^2$Simons center for the Study of Living Machines, National center for Biological Sciences (TIFR), Bangalore 560065, India\\
		$^3$School of Physics, UNSW, Sydney, NSW 2052, Australia.\\
		$^4$ARC center of Excellence for the Mathematical Analysis of Cellular Systems, UNSW Node, Sydney, NSW 2052, Australia.\\
		$^5$EMBL Australia Node in Single Molecule Science, School of Biomedical Sciences, UNSW, Sydney 2052, Australia.
	}
	
	\begin{abstract}	
		From the Golgi apparatus to endosomes, organelles in the endomembrane system exhibit complex and varied morphologies that are often related to their function. These membrane-bound organelles operate far from equilibrium due to directed fluxes of trafficking vesicles, yet the physical principles governing the emergence and maintenance of these morphologies remain elusive. By coarse-graining the active mechanochemical cycles associated with individual fission and fusion events, we show how vesicular trafficking manifests at the hydrodynamic scale, generating not only fluxes of membrane area and encapsulated volume but also active stresses that mediate momentum transfer between an organelle and its cytosolic environment. The fluid and deformable nature of the bounding membrane gives rise to novel physics by coupling these nonequilibrium forces to organelle composition, morphology, and hydrodynamic flows. We demonstrate how both stable compartment drift and ramified sac-like morphologies, reminiscent of Golgi cisternal motion and shape, emerge from the same nonequilibrium dynamics of fission and fusion.
	\end{abstract}
	
	\maketitle
    
	\section{Introduction}
    
	Nonequilibrium self-assembly is thought to play a crucial role in the emergence and  {spatiotemporal}  patterning of intracellular structures \cite{Marshall2020rev, Marshall2020jcub, Misteli2001, Hara2009, Rafelski2008}, such as organelles, 
	mitotic spindle \cite{Oriola2018, Hara2009}, centrioles \cite{Banterle2017},  filopodia, flagella and cilia~\cite{Gilpin2020}.
	A key feature of such nonequilibrium structure formation and maintenance is the ATP/GTP-driven addition-removal of its subunits; for instance, actin or microtubule based structures form and adapt dynamically via the polymerization-depolymerization of monomers maintained out of equilibrium~\cite{Mohapatra2016, Banerjee2022}. 
	\\
	\indent
	In this work, we explore such ideas in the context of the organelles in the endomembrane system \cite{Phuyal2020}, such as the Golgi apparatus, lysosomes, endosomes, and vacuoles. Such organelles are often identified by their lipid and protein composition, and governed by local material fluxes and synthesis \cite{Glick1998, Papanikou2014, Bonifacino2004, Glick2009, Glick2019}. They are also identified by their morphology, which remarkably is maintained in the face of continual nonequilibrium driving from the flux of transport vesicles \cite{Malhotra1988,Lippincott-Schwartz2000, Ward2001, Glick2009, Dunlop2017, Glick2019}. 
	Indeed, perturbations to vesicular flux have been observed to dramatically alter the shape of Golgi cisternae \cite{Lippincott1989, Lippincott1990, Lippincott1992rev, Dinter1998, Martinez-Alonso2005, Singh2018}, highlighting the role of nonequilibrium driving in determining the steady-state shape.
	\\
	\indent
	Due to the complexity and intricate nature of the relevant soft-matter physics, a principled physical description of the nonequilibrium dynamics of such micron-size membrane compartments embedded in the cytosol and subject to vesicular trafficking has been a challenge. Although a few studies have treated the out-of-equilibrium shape changes induced by the incorporation of additional membrane material \cite{Rao2001, Girard2004, Morris2011, Sens2013, Ramakrishnan2015, Sachdeva2016, Tachikawa2017, Vagne2018, Ruiz-Herrero2019}, developing a principled physical description of organelle morphodynamics under sustained vesicular trafficking remains a challenge.
	\\
    \indent
    To this end, we incorporate three fundamental features that are common to the dynamics of membrane-bound organelles in the trafficking pathway. First, fission and fusion of transport vesicles are active processes \cite{Rao2001, Ramaswamy2001} that require specific energy-consuming macromolecules associated with {\it nonequilibrium mechanochemical cycles}~\cite{Alberts2008}. \cc{Second, each completed fission or fusion cycle has two distinct coarse-grained effects: it exchanges membrane area and lumenal volume with the organelle, and it exerts a localized distribution of active forces on the adjacent cytosolic fluid \cite{BatchelorBook, Siggia1979, Rao2001, Ramaswamy2001}.} Third, membrane shape deformations couple hydrodynamically to the surrounding medium, inducing both membrane and fluid flows, which dynamically change the local membrane composition. \cc{Taken together, these ingredients yield a coarse-grained hydrodynamic framework for membrane-bound organelles subject to active trafficking. By representing fission and fusion through generic material fluxes and active force distributions rather than organelle-specific molecular details, the framework applies to a broad class of organelles whose morphology and motion are regulated by trafficking. Organelle morphodynamics therefore provides a rich context for the physics of nonequilibrium soft matter, involving an interplay between fluid dynamics, deformable interfaces, composition, and active stresses arising from the microscopic transduction of chemical energy~\cite{Maitra2014, Salbreux2017, Morris2019, Marchetti2013}.}

	\section{Results}
    
\begin{figure*}[t]\includegraphics[width=\textwidth]{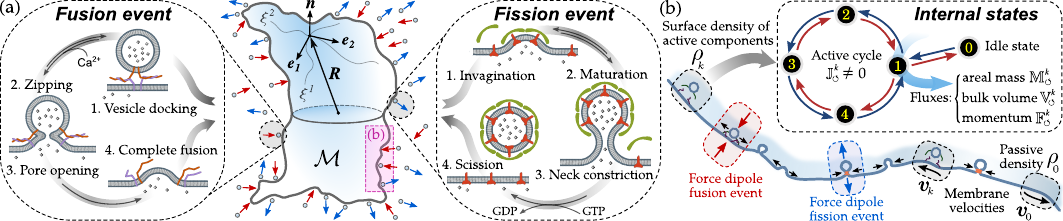}
		\caption{\label{fig:fig1} {\bf Organelle subject to active fission and fusion of transport vesicles.} (a) A membrane organelle that is continually subject to the fission and fusion of transport vesicles (small circles). The blue arrows indicate outgoing vesicles (fission), while the red arrows label incoming vesicles (fusion). We model fission and fusion as separate processes, each associated with a biochemical reaction cycle that violates detailed balance (insets depict four states). (b) Cross-section of a patch of the deformable membrane $\mathcal{M}$, showing the thin active layer adjacent to the membrane in which numerous fission and fusion events may occur. On large length and time scales, these events can be described by local densities $\rho_k$, $\rho_0$ and velocities $\itbf{v}_k$, $\itbf{v}_0$ of active (fissogens and fusogens) and passive components. \cc{Each completed cycle produces two distinct coarse-grained effects: exchange of membrane area and lumenal volume with the organelle, and localized forces exerted on the adjacent cytosolic fluid. The opposing arrows schematically depict these forces as a force dipole for fission and fusion events. More generally, the forcing is represented by a force-free distribution acting across the active layer, characterized by its first and second moments.} 
	}\end{figure*}

    We derive active hydrodynamic equations for a closed membrane compartment embedded in a viscous solvent subject to active fission and fusion of transport vesicles, coupling membrane shape and composition to the surrounding fluid flow (see Appendix A). \cc{We find that each trafficking cycle is associated with a cycle-averaged force profile, characterized by its first and second moments (Fig.~\ref{fig:fig1}). These active force moments dynamically renormalize membrane tension and spontaneous curvature: the former can drive shape instabilities, while the latter can promote segregation of fission and fusion components}---our first main result. Linear stability analysis of the derived hydrodynamic equations (Appendix B) about a uniform spherical membrane compartment in the presence of isotropic vesicular flux leads to our second result: a spontaneous drift instability of the compartment beyond a threshold activity (Fig.~\ref{fig:fig2}). Such motion can be either anterograde or retrograde relative to the incident vesicular flux, depending on the magnitude of the active force moment and the relative fission and fusion rates (Fig.~\ref{fig:fig3}). {Indeed the active stresses control a combined drift-shape instability, leading to moving spheres, prolates, and oblates. Importantly, the drift instability does not require a fore-aft symmetry breaking in shape and can be driven by a \cc{dipolar membrane-area flux}.} This has important implications for cisternal progression \cite{Glick2009, Nakano2006, Losev2006, Glick2019}. In addition, the compartment exhibits higher-order shape instabilities in the form of flattened sacs or ramified tubular structures (Fig.~\ref{fig:fig2} and \ref{fig:fig3}). Such cisternal morphologies, commonly exhibited in a variety of cell types \cite{Engel2015, Klumperman2011, Glick2019}, {constitute} our third result. \cc{To investigate whether the growing compositional perturbations saturate at finite amplitude, we extend the perturbative analysis to weakly nonlinear order, retaining composition dynamics through cubic order while keeping the mechanical shape response linear (Appendix~C). The resulting reduced amplitude description reveals stable stationary and steadily drifting states, constituting our fourth main result; see Fig.~\ref{fig:fig4}.} We emphasize that all of these results are contingent on a hydrodynamic description of nonequilibrium fission and fusion processes and will not arise from a purely kinetic treatment. 

	\subsection{Fission and fusion processes are active mechanochemical pumps}
	
    \cc{Our coarse-grained hydrodynamic equations for membrane composition and shape are derived by treating fission and fusion as spatially local surface processes and averaging their effects over repeated trafficking events (see Appendix~A). Individual trafficking events act over vesicle-sized membrane patches with a characteristic lateral size of approximately {$50$--$100$\,nm}, much smaller than the characteristic length $L\sim1\,\mu\mathrm{m}$ over which the membrane fields vary. Their effects can be represented as spatially local contributions relative to the membrane-scale variations. The estimated interval between vesicle-sized trafficking events, $T_a\sim0.01\text{--}0.6$\,s, is much shorter than the lateral diffusion time, $T_d\sim2\text{--}100$\,s, and shorter than the shape-relaxation time, $T_s\sim2.5\text{--}50$\,s (Table~\ref{tab:scaled-eqs}, Appendix~A). Repeated fission and fusion can therefore maintain the membrane
    away from equilibrium. This is the regime considered here. If instead successive trafficking events were separated by times much longer than the membrane-relaxation time, the membrane would relax between events toward a quasi-equilibrium shape at the updated area and volume.}

    The composition fields on the membrane consist of passive membrane components (lipids and proteins) and the protein machinery associated with fission and fusion, referred to throughout as {\it fissogens} and {\it fusogens}. Locally, these compositional fields evolve via recruitment, active mechanochemistry, and dynamics along the membrane. \cc{We describe the mechanochemical kinetics of fission and fusion using Markov cycles with multiple intermediate internal states. We adopt an adiabatic approximation in which the populations of these internal states relax rapidly relative to the membrane composition and shape dynamics. They can therefore be treated as remaining in an instantaneous stationary nonequilibrium distribution, with their relative populations determined by the {transition rates} and the instantaneous total fraction of fissogens or fusogens; as derived in Appendices~A.1 and A.5 and illustrated in Fig.~\ref{fig:fig1}.} 
    
    The transitions between these states do not satisfy detailed balance because the cycle is driven by energy-consuming reactions. For simplicity, we model the biochemical cycles \cc{using first-order kinetics with constant transition rates. This neglects possible feedback of membrane elasticity, such as surface tension, on the {transition rates}} \cite{Foret2008, Vagne2018, Thottacherry2018}. \cc{We assume that the molecular machinery required for fission and fusion, together with incoming vesicles for fusion, is locally available and that their transport does not limit the prescribed cycle kinetics. Consequently, the nonequilibrium kinetics generate a local loop current for each trafficking cycle, denoted by $\bJk$, where $k=1,2$ labels fission and fusion, respectively. The loop current measures the net rate at which the corresponding mechanochemical cycle is completed locally on the membrane {(see Appendix~A.5)}. For the prescribed kinetics, $\bJk$ is proportional to the local abundance of the corresponding fission or fusion machinery, and we denote the proportionality coefficient by the cycle rate $\bRk$.}
    
    \cc{This cycle rate is determined entirely by the internal transition rates and is proportional to the difference between the products of the transition rates around the cycle in the forward and reverse directions. At thermodynamic equilibrium, detailed balance is restored and the Kolmogorov cycle condition requires these two products to be equal \cite{Schnakenberg1976}. Consequently, $\bRk$ and the loop current vanish. A nonzero loop current therefore reflects the nonequilibrium driving of the mechanochemical cycle. This nonequilibrium cycling also has direct physical consequences: each completed fission (fusion) cycle removes (adds) membrane area and lumenal volume, including solvent and solute cargo, with the corresponding local transfer rates determined by $\bJk$ and the amounts transferred per completed cycle. Fissogens and fusogens therefore act as {\it active pumps} \cite{Ramaswamy2000, Ramaswamy2001, Chen2010} for membrane area and lumenal volume.}

    \cc{In addition to exchanging membrane area and lumenal volume, each trafficking cycle also exerts a distribution of forces on the surrounding cytosolic fluid. We represent these forces by a body-force density localized within a thin layer of thickness {$\Lambda\sim50$--$100$\,nm} on the cytosolic side of the membrane, extending along the membrane normal \cite{Siggia1979, BatchelorBook, Anderson1989}. The resulting stresses and flows within this layer transmit the active forcing to the membrane through the hydrodynamic traction exerted by the adjacent fluid. Since $\Lambda/L\ll1$, the finite active layer can be replaced, to leading order, by a sharp-interface description at the membrane. In this reduction, the moments of the force distribution are retained through effective interfacial stresses, bending moments, and a modified interfacial velocity condition \cite{Lomholt2006, Lomholt2006a, Lomholt2006b}. These effective contributions are described in Section~II.B; see Appendices~A.7 and A.9 for details.}

	\subsection{Active hydrodynamics of membranes subject to fission and fusion}
    
    The density of active membrane components is given by $\rho_k$, where $k=1$, $2$ \cc{labels} fissogens and fusogens, respectively, whilst the density of passive membrane components is $\rho_0$ (see Fig.~\ref{fig:fig1}). In our coarse-grained description, the membrane is incompressible, with a constant total density $\rho=\rho_0+\sum_{k}\rho_k$. By defining the local \cc{areal fractions} $\Phi_0=\rho_0/\rho$ and $\Phi_k=\rho_k/\rho$ and assuming \cc{that the active components are dilute} ($\Phi_k\ll\Phi_0$), the \cc{free energy of the closed organelle membrane} $\mathcal{M}$ is given by
    $\mathcal{F}=\int_{\mathcal{M}}F\,\mathrm{d}S$ (see Appendix~A.4), with
    \begin{equation}
        F=\Sigma+2\kappa H^2+2\kappa H C_\mathrm{m}+V_\mathrm{m},
    \end{equation}
    where $\Sigma$ is the surface tension, $\kappa$ is the bending rigidity, and $H$ is the mean curvature \cc{of the membrane. The adiabatic approximation allows the fractions of the internal mechanochemical states to be eliminated in favor of the total areal fractions $\Phi_k$. The spontaneous curvature and composition free-energy density consequently take the reduced forms} $C_\mathrm{m}=\mathcal{C}_0\Phi_0+\sum_k\mathcal{C}_k\Phi_k$ \cc{and} $V_\mathrm{m}=\frac{k_B T}{b_0}\big[\Phi_0(\ln\Phi_0+\mathcal{E}_0)+\sum_k\Phi_k(\ln\Phi_k+\mathcal{E}_k)\big]$. \cc{Here, $\mathcal{C}_0$ is the curvature-coupling strength of the passive components, while $\mathcal{C}_k$ is the effective curvature-coupling strength obtained by averaging over the internal states of cycle $k$. Similarly, $\mathcal{E}_0$ is the dimensionless chemical-potential parameter of the passive components, while $\mathcal{E}_k$ is the corresponding effective parameter for cycle $k$, and $b_0$ is a coarse-graining area scale on the membrane. The explicit expressions for $\mathcal{C}_k$ and $\mathcal{E}_k$ are derived in Appendix~A.5.}

    \cc{The evolution of the membrane {areal} fractions is governed by passive {diffusive transport} derived from $\mathcal{F}$ and active {membrane-area exchange} associated with the mechanochemical cycles} (Fig.~\ref{fig:fig1}):
    \begin{equation}
        \label{eqn:phi-k-dynamics-text}
        \frac{\partial{\Phi}_k}{\partial t}
        +v^\alpha{\nabla_{\hspace{-1pt}\alpha}\Phi}_{k}
        =
        -\Phi_k\mathbb{M}
        +2\hspace{1pt}\Omega_k\Delta\hspace{-0.25pt}H
        +\gamma_k\hspace{1pt}\Delta\hspace{-0.5pt}
        \ln\frac{\Phi_k}{\Phi_0},
    \end{equation}
    where $\gamma_k$ is \cc{the} surface diffusivity, $\Omega_k$ is \cc{the curvature-coupling coefficient}, $\Delta$ is \cc{the} Laplace--Beltrami operator, and $\nabla_{\hspace{-1pt}\alpha}$ is the covariant derivative, \cc{with summation over repeated Greek indices implied}. \cc{The active membrane-area flux is $\mathbb{M}=\sum_k\bMk\Phi_k$, where $\bMk$ is the membrane-area transfer rate for cycle $k$, determined by $\bRk$ and the net area change accumulated across its internal transitions. The corresponding local rate of cycle completion is the loop current $\bJk=\Phi_k\bRk$, obtained under adiabatic approximation (see Appendix~A.5). At equilibrium, $\bJk=0$ and hence $\mathbb{M}=0$. Away from equilibrium, $\mathbb{M}$ must be accommodated by the local membrane motion. For an incompressible membrane, this requires}
    \begin{equation}
        \label{eqn:membrane-area-balance}
        \mathbb{M}=\nabla_{\!\alpha}v^\alpha-2Hv.
    \end{equation} \cc{where the right-hand side is the surface divergence of the membrane velocity} $\boldsymbol{v}=v^\alpha\ed{\alpha}+v\hspace{0.2pt}\nv$, \cc{with tangential components $v^\alpha$ along the surface basis vectors $\ed{\alpha}$ and normal component $v$ along the outward unit normal $\nv$.}

    \cc{The evolution of membrane shape is determined by passive elastic forces, the hydrodynamic traction exerted by the ambient fluids, and effective active stresses associated with fission and fusion. We write the membrane force balance in the sharp-interface description, with the active contributions defined below.} By conservation of linear and angular momentum, the force balance at the membrane is given by (Appendix~A.8):
	\begin{equation}
		\label{eqn:force-balance-membrane}
		\boldsymbol{f} + \nabla_{\alpha}\!\left[\left(\sigma^{\alpha\beta}+b^\beta_\gamma M^{\gamma\alpha}\right)\ed{\beta}-\cc{\nabla_\beta} M^{\alpha\beta}\nv\right]\!  = \boldsymbol{0}, 
	\end{equation} where \cc{$\itbf{f}$ is the external hydrodynamic force per unit area exerted by the ambient fluids}, $\sigma^{\alpha\beta}=\sigma^{\alpha\beta}_{\hspace{-1.5pt}\scriptstyle A}+\sigma^{\alpha\beta}_0$ is the total in-plane stress tensor, and $M^{\alpha\beta}=M^{\alpha\beta}_{\hspace{-1.5pt}\scriptstyle A}+M^{\alpha\beta}_0$ is the total bending moment tensor. \cc{The subscripts $0$ and $A$ denote the passive and  active contributions, respectively.} The passive elastic stresses are given by (Appendix~A.3):
	\begin{equation}
		\sigma_0^{\alpha\beta} = \frac{2}{\sqrt{g}}\frac{\partial(\sqrt{g}\hspace{1pt}F)}{\partial g_{\alpha\beta}},\quad M^{\alpha\beta}_{0} = \frac{\partial F}{\partial b_{\alpha\beta}},
	\end{equation} where $g_{\alpha\beta}$ and $b_{\alpha\beta}$ are the metric and curvature tensors, respectively, and $g = \det[g_{\alpha\beta}]$. 

    \cc{To obtain the effective active contributions to the membrane force balance, we assume that the machinery driving fission and fusion generates localized body forces within a finite active layer on the cytosolic side of the membrane. We assume that this {force profile} has no preferred tangential direction and retain only its component along the outward normal $\nv$. We denote the normal distance from the membrane by height $h$, with $0\leq h\leq\Lambda$, and define the cycle-averaged normal force profile associated with cycle $k$ by $\bfk(h)$, such that $\bfk(h)\,\mathrm{d}h$ is the normal force per unit membrane area contributed by a slice between $h$ and $h+\mathrm{d}h$ (see Appendix~A.7). We assume that each cycle-specific profile is locally force-free, so that its net force per unit membrane area vanishes:}
    \begin{equation}
    	\int_0^\Lambda \bfk(h)\,\mathrm{d}h=0.
    \end{equation}
    \cc{This force-free assumption amounts to representing the molecular machinery driving fission and fusion as generating internal force pairs within the layer, with oppositely directed forces acting at different normal heights. It also assumes that this machinery is not mechanically coupled to a larger structure, such as the cytoskeleton, that could support a nonzero force monopole.}

    \cc{Since $\Lambda/L\ll1$, the fluid equations within the finite active layer can be expanded in this small ratio. A slice at height $h$ occupies the physical volume $J(h)\,\mathrm{d}S\,\mathrm{d}h$, where $J(h)=1-2hH+h^2K$ and $K$ is the Gaussian curvature. Since $\bfk(h)\,\mathrm{d}S\,\mathrm{d}h$ is the force acting within this slice, the corresponding body-force density is $\bfk(h)/J(h)$. At leading order, this body-force density is balanced by the normal gradient of the fluid pressure $p_k$ within the finite active layer, $-\partial_h p_k+{\bfk(h)}/{J(h)}=0$. Integrating this balance with respect to $h$ yields the pressure profile within the finite active layer. Since pressure is isotropic, it also contributes to the in-plane stress, while tangential viscous stresses are negligible at leading order. Integrating across the layer gives the total in-plane force per unit length. For zero net force, its leading contribution is set by the first moment of the force profile,}
    \begin{equation}
    	\bPk=\int_0^\Lambda h\,\bfk(h)\,\mathrm{d}h.
    \end{equation}
    \cc{Matching the force transmitted in the finite-layer and sharp-interface descriptions identifies this contribution as an isotropic active membrane stress.} 
    
    \cc{Moreover, because the in-plane stress is distributed over finite distances from the membrane, it also exerts a torque about $h=0$. Integrating this torque across the layer gives the bending moment, which is set by the second moment of the force profile,}
    \begin{equation}
    	\bQk=\int_0^\Lambda h^2\bfk(h)\,\mathrm{d}h.
    \end{equation}
    \cc{For a curved interface, the curvature dependence of $J(h)$ causes $\bQk$ to contribute also to the curvature-dependent part of the active in-plane stress. Collecting these contributions gives the effective active in-plane stress and bending moment (Appendix~A.7):}
    \begin{equation}
    	\sigma^{\alpha\beta}_{\hspace{-1.5pt}\scriptstyle A}\! = \mathbb{P}\hspace{1.5pt}g^{\alpha\beta} + \mathbb{Q}\,\left(\hspace{-0.6pt} H g^{\alpha\beta}\!+b^{\alpha\beta}\right),\;\; M^{\alpha\beta}_{\hspace{-1pt}\scriptstyle A}\! =\! - \frac{1}{2}\hspace{1pt}\mathbb{Q}\,g^{\alpha\beta},
    \end{equation}
    where $\mathbb{P}=\sum_k\bPk\Phi_k$ and $\mathbb{Q}=\sum_k\bQk\Phi_k$ are the total first and second moments of the active force profile.

    \cc{In the sharp-interface description, the active layer is no longer resolved explicitly, and} the inner ($-$) and outer ($+$) ambient fluids satisfy the \cc{unforced} incompressible Stokes equations~\cite{Happel1981}: $\bnabla\cdot\boldsymbol{\mathsf{\Omega}}_{\boldsymbol{\pm}}=\boldsymbol{0}$ and $\bnabla\cdot\itbf{V}_{\!{\boldsymbol{\pm}}}=0$, where \cc{the fluid stress tensors are}
    \begin{equation}
        \boldsymbol{\mathsf{\Omega}}_{\boldsymbol{\pm}}
        =
        -p_{\hspace{0.5pt}{\boldsymbol{\pm}}}\itbf{I}
        +
        \eta\big[
        \left(\bnabla\hspace{-1pt}\itbf{V}_{\!{\boldsymbol{\pm}}}\right)
        +
        \left(\bnabla\hspace{-1pt}\itbf{V}_{\!{\boldsymbol{\pm}}}\right)^{\!\mathsf{T}}
        \big],
    \end{equation}
    \cc{and $\itbf{I}$ is the identity tensor, while $p_{\hspace{0.5pt}{\boldsymbol{\pm}}}$ and $\itbf{V}_{\!{\boldsymbol{\pm}}}$ are the corresponding fluid pressures and velocities.} For simplicity, we choose the inner and outer fluids to have the same shear viscosity $\eta$. \cc{The hydrodynamic force per unit area $\boldsymbol{f}$ in Eq.~(\ref{eqn:force-balance-membrane}) is given by the jump in fluid traction across the sharp interface:}
    \begin{equation}
        \itbf{f} = \left(\boldsymbol{\mathsf{\Omega}}_{\boldsymbol{+}}-\boldsymbol{\mathsf{\Omega}}_{\boldsymbol{-}}\right)\cdot\nv\big|_{\,\mathcal{M}}
    \end{equation}  \cc{Accordingly, $\itbf{f}$ is computed solely from the stresses of the unforced bulk fluids. The force and torque transmitted through the active layer are accounted for separately through the effective membrane stress and bending moment. The finite active layer also modifies the kinematic coupling between the membrane and the bulk fluid. Although no slip holds at the physical membrane, flow across the layer can cause the fluid velocity at its outer edge to differ from the membrane velocity. Together with volume transport across the membrane, this gives the following interfacial velocity condition:}
    \begin{equation}
    	\label{eqn:boundary-condintions}
    	\itbf{v}+\itbf{v}_{\!\scriptscriptstyle S}-{j}_{\scriptscriptstyle V}\hspace{1pt}\nv
    	=
    	\itbf{V}_{\!{\boldsymbol{\pm}}}\big|_{\,\mathcal{M}}.
    \end{equation}
    
    \cc{The tangential term $\itbf{v}_{\!\scriptscriptstyle S}$ accounts for the velocity offset due to flow within the finite active layer. Variations in the active force profile along the membrane generate pressure gradients and shear within the layer. When the layer is reduced to a sharp interface, the resulting offset appears as the slip velocity (see Appendix~A.7),}
    \begin{equation}
    	\itbf{v}_{\!\scriptscriptstyle S}
    	=
    	\frac{1}{2\eta}\eu{\alpha}\hspace{1pt}\nabla_{\!\alpha}\mathbb{Q}.
    \end{equation}
    
    \cc{The normal term, $-{j}_{\scriptscriptstyle V}\nv$, instead represents the velocity difference associated with volume transport across the membrane. The scalar flux ${j}_{\scriptscriptstyle V}$ gives the local rate of volume transfer per unit area (see Appendix~A.10),}
    \begin{equation}
    	{{j}_{\scriptscriptstyle V}
    	=
    	\mathbb{V}
    	-
    	w_0\left(\Pi_{\scriptscriptstyle\Delta}-P_{\hspace{-0.25pt}\scriptscriptstyle\Delta}\right),}
    \end{equation}
    \cc{where $\mathbb{V}=\sum_k\Phi_k\bVk$ is the active volume flux generated by fission and fusion, with $\bVk$ the volume-transfer rate for cycle $k$. The second term describes passive osmotic permeation, with $w_0>0$ the solvent-permeation coefficient, and $\Pi_{\scriptscriptstyle\Delta}$ and $P_{\hspace{-0.25pt}\scriptscriptstyle\Delta}$ the osmotic and hydrostatic pressure differences between the outside and inside of the organelle, respectively. Positive values of ${j}_{\scriptscriptstyle V}$ increase the enclosed volume, so that $\mathrm{d}V/\mathrm{d}t=\int_{\mathcal M}{j}_{\scriptscriptstyle V}\,\mathrm{d}S$.}

    The composition dynamics~(\ref{eqn:phi-k-dynamics-text}), \cc{local membrane-area balance~(\ref{eqn:membrane-area-balance}), membrane} force--balance~(\ref{eqn:force-balance-membrane}), incompressible Stokes equations \cc{for the inner and outer bulk fluids}, and \cc{interfacial velocity} boundary condition~(\ref{eqn:boundary-condintions}) form a closed set of equations that describe the organelle morphodynamics, see the flowchart in Fig.~\ref{fig:flowchart}. \cc{For a given interfacial traction jump $\boldsymbol{f}$, the inner and outer fluid flows can be obtained from a boundary-integral representation over the membrane, as derived in Appendix~A.11.} 
    
    Even before solving the governing equations, \cc{the force balance already reveals how activity modifies the effective membrane mechanics. In its normal component, the terms linear in mean curvature define an} \cc{activity-renormalized surface tension  whose direct mechanical contribution is set by the total first moment $\mathbb{P}$, see Eq.~(\ref{eqn:renormalized-surface-tension}). Likewise, the total second moment $\mathbb{Q}$ shifts the effective spontaneous curvature, giving the activity-renormalized form in Eq.~(\ref{eqn:renormalized-spontaneous-curvature}).}

    \cc{To illustrate the possible signs of these active moments, consider a two-point force distribution, with a distal force associated with the bud or transport vesicle at $h=\Lambda$ and an equal and opposite force acting closer to the mother membrane. During fission, coat-driven budding and scission can generate forces that separate the nascent vesicle from the mother membrane~\cite{Bonifacino2004,Bykov2017}, whereas during fusion, docking and SNARE-mediated zippering pull the incoming vesicle towards the membrane~\cite{Bonifacino2004,Mayer2017}. We therefore represent fission by an extensile force dipole and fusion by a contractile one. The same force polarity can arise when the distal force is dominated by Stokesian drag, since the force exerted by the bud or vesicle on the fluid follows its normal motion, outward during fission and inward during fusion. Since the compensating force acts closer to the membrane, both $\bPk$ and $\bQk$ have the sign of the distal force, giving $\mathrm{sgn}(\bPk)=\mathrm{sgn}(\bQk)=(-1)^{k+1}$. More generally, this sign relation need not hold, since the two moments weight the force profile $\bfk(h)$ differently and may have either sign independently of one another.}
  
    \cc{For the particular case described above, these force moment signs imply that the first-moment contribution raises the activity-renormalized surface tension during fission and lowers it during fusion, whereas the second-moment contribution shifts the activity-renormalized spontaneous curvature in opposite directions for the two processes. Sufficiently strong fusion activity can reduce the effective tension and potentially drive it negative. This tendency is qualitatively consistent with theoretical studies of membranes exchanging material~\cite{Rao2001,Girard2004} and with {\it in vitro} observations of lipid uptake and membrane-area regulation~\cite{Solon2006,Staykova2011}. The associated curvature preference can also favor enrichment of membrane components in regions of matching curvature, consistent with curvature-dependent partitioning at highly curved Golgi cisternal rims~\cite{Ernst2019}.}

    \begin{figure*}[t]\includegraphics[width=\textwidth]{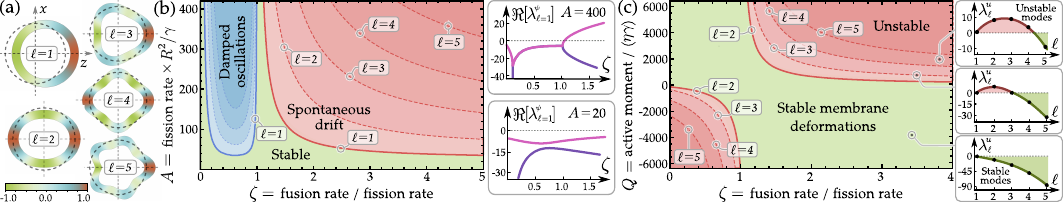}
        \caption{\label{fig:fig2} {\bf Generic instabilities in the composition and shape of a membrane organelle.} (a) \cc{Normalized axisymmetric spherical harmonics $Y_{\ell,0}(\theta)$}, illustrating the spherical decomposition of \cc{membrane composition fields} (color map) and shape distortions from a homogeneous sphere (dashed circle). (b) Stability diagram corresponding to composition perturbations for each spherical harmonic, in terms of the fission and fusion rates, made dimensionless using the characteristic diffusion time $R^2/\gamma$. The onset of drift instability ($\ell=1$) is marked by the solid red line, while the boundaries of the \cc{higher-order} ($\ell>1$) shape instabilities are shown as red dashed lines. Both green and blue regions correspond to the linearly stable regime, with the latter showing stable damped oscillations (each mode is shown by the dotted lines in the blue region). Insets show the growth rates $\lambda^{\psi}_{\ell=1}$ of composition at two fixed values of $A$ as a function of $\zeta$. (c) Stability diagram corresponding to shape perturbations. \cc{The fission and fusion active moments are taken to have equal magnitudes and opposite signs and are parameterized by the dimensionless quantity $Q$.} The insets show the growth rates $\lambda_{\ell}^{u}$ as a function of $\ell$ at fixed values of $\zeta$, $Q$, and dimensionless bending rigidity $\mathcal{K}=10$. Red lines show the onset of shape instabilities for the different $\ell$ modes. \cc{The composition stability diagram in (b) is independent of the active moment $Q$, whereas the shape stability diagram in (c) is independent of the dimensionless fission rate $A$. In both panels (b) and (c), the homogeneous base-state areal fraction is fixed at $\bar{\Phi}_0=3/5$.}}
    \end{figure*}
 
    \cc{We emphasize that activity affects the effective surface tension and spontaneous curvature in two distinct ways. The active force profile contributes directly through its first and second moments, $\bPk$ and $\bQk$, respectively. Material exchange associated with fission and fusion provides an additional, indirect contribution by changing the local fractions $\Phi_k$, whose dynamics depend on the membrane-area transfer rates $\bMk$.}

\begin{figure*}[t]
    \includegraphics[width=\textwidth]{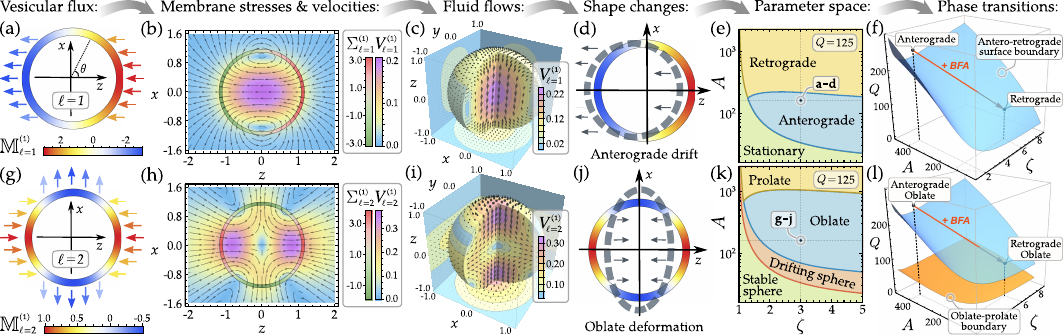}
    \caption{\label{fig:fig3}
    {\bf Vesicular flux drives shape changes by inducing membrane stresses and fluid flows.}
    \cc{An axisymmetric composition perturbation produces a total membrane-area flux profile $\mathbb{M}^{\mathsf{(1)}}_{\ell}(\theta)=\sum_k\bMk\Psi_{k,\ell,0}Y_{\ell,0}(\theta)$, which generates membrane stresses and fluid flows. We choose $\zeta=3$, $A=180$, $Q=125$, and $\mathcal{K}=10$, for which the $\ell=1$ and $\ell=2$ composition modes are unstable; see Fig.~\ref{fig:fig2}(b). Panels (a)--(d) and (g)--(j) are evaluated at $t=0.1\,R^2/\gamma$, with initial amplitudes $\Psi_{k,\ell,0}(0)=0.01$ and $u_{\ell,0}(0)=0$. For the $\ell=1$ mode, (a) shows the membrane-area flux profile, (b) the membrane and bulk fluid flows in the $y=0$ cross-section together with the activity-renormalized surface tension $\Sigma^{(\mathtt{1})}_{\ell=1}$, and (c) the three-dimensional bulk flow, which decays as $1/r^3$ in the far field. Panel (d) illustrates the resulting anterograde drift away from the region of net membrane addition. Panel (e) shows the stationary, anterograde, and retrograde regimes in the $A$--$\zeta$ plane at fixed $Q$, with the marked point corresponding to (a)--(d), while (f) gives the anterograde--retrograde boundary surface in $A$--$\zeta$--$Q$. The red arrow in (f) shows how a threefold reduction in the fission cycle rate shifts the system across the displayed boundary, from anterograde to retrograde motion, consistent with the retrograde motion observed following Brefeldin A (BFA) treatment.
    For the $\ell=2$ mode, (g) shows the membrane-area flux profile, (h) the associated fluid flows and activity-renormalized surface tension in the $y=0$ cross-section, and (i) the three-dimensional bulk flow, which decays as $1/r^2$ in the far field and produces the oblate deformation in (j). Panel (k) shows the spherical, prolate, and oblate regimes in the $A$--$\zeta$ plane at fixed $Q$, with the red line marking the drift-instability boundary and the marked point corresponding to (g)--(j). Panel (l) shows the anterograde--retrograde and oblate--prolate boundary surfaces in $A$--$\zeta$--$Q$, in blue and orange, respectively. The prolate and oblate phases in (k) and (l) correspond to the asymptotic linear response to the chosen initial perturbations, which produce a positive amplitude of the quadrupolar membrane-area flux, as illustrated in (g). The red arrow in (l) shows the corresponding shift into the retrograde-oblate region and, for stronger inhibition, towards the retrograde-prolate region.}
    }
\end{figure*}

	\subsection{Spontaneous drift and shape instabilities}
    
    The set of nonlinear equations can be solved using a suitable numerical scheme (see Fig.~\ref{fig:flowchart}), such as boundary integral methods~\cite{Hou2001}. Instead, we find it instructive to study the dynamics perturbatively about the homogeneous spherical nonequilibrium steady-state (see Appendix B). We parameterize the membrane shape as $\itbf{R} = R\left[1+u\hspace{-1pt}\left(\theta,\varphi\right)\right]\rv$, where $\rv$ is the radial unit vector, and $u(\theta,\varphi)$ is a small radial distortion about the undeformed sphere of radius $R$, with $\theta$ and $\varphi$ as the inclination and azimuthal angles. We perturb the composition fields $\Phi_k = \bar{\Phi}_k + \Psi_k(\theta,\varphi)$ about the homogeneous value $\bar{\Phi}_k$. 
    The velocity fields of the fluid and membrane are expanded to linear order, with Eq.~(\ref{eqn:boundary-condintions}) appearing as a boundary condition on the undeformed sphere. Lamb's solution to the Stokes equations \cite{BatchelorBook} allows us to \cc{express the leading-order inner and outer fluid flows, $\!\itbf{V}_{\!{\boldsymbol{\pm}}}^{(\mathtt{1})}$, in terms of the membrane dynamics}. \cc{Here, the superscript ${\scriptstyle(\mathtt{1})}$ denotes the first-order contribution in the perturbative expansion about the homogeneous spherical base state.} By expanding Eqs.~(\ref{eqn:phi-k-dynamics-text}) and (\ref{eqn:force-balance-membrane}) in spherical harmonics \cc{$Y_{\ell,m}(\theta,\varphi)$}, we \cc{obtain the linear equations for the individual shape and composition amplitudes $u_{\ell,m}$ and $\Psi_{k,\ell,m}$, together with the corresponding surface-tension amplitudes $\Sigma_{\ell,m}$}; see Appendix~B for details. \cc{At linear order, the azimuthal modes $m$ are degenerate and obey the same equations for a given $\ell$. For compactness, we use $u_\ell$ and $\Psi_{k,\ell}$ to denote the complete contributions of degree $\ell$, constructed by summing over all $m$ as defined in Eqs.~(\ref{eqn:surface-harmonic-def}) and (\ref{eqn:psi-harmonic-def}) of Appendix~B.}

    This approach allows us to study the instabilities of membrane composition and shape (Appendix B.7), arising from spatial modulations in composition and coupling to hydrodynamic flows induced by membrane stresses (Fig.~\ref{fig:fig2} and \ref{fig:fig3}). \cc{The far-field decay of the surrounding fluid flow depends on the mode: the force-free $\ell=1$ flow decays as $1/r^3$, whereas the $\ell=2$ flow decays as $1/r^2$} (Appendix B.9)\cc{; see Fig.~\ref{fig:fig3}(c) and (i)}. Our key result is the spontaneous drift instability of the organelle ($\ell=1$), along with higher--order shape instabilities ($\ell\geq2$) in the form of flattened sacs and tubular ramified structures.

    We construct stability diagrams, \cc{shown in Fig.~\ref{fig:fig2}(b) and (c)}, in terms of the dimensionless fission parameter $A=-\mathbb{M}^{\hspace{0.5pt}1}_{\CircleArrow}R^2/\gamma$, the ratio $\zeta=-{\mathbb{M}^{\hspace{0.5pt}2}_{\CircleArrow}}/{\mathbb{M}^{\hspace{0.5pt}1}_{\CircleArrow}}$ between fusion and fission rates, \cc{and the dimensionless active second moments $Q_k=\bQk/(\gamma\eta)$}, with both the dimensionless bending rigidity $\mathcal{K}=\kappa/(R\gamma\eta)$ \cc{and homogeneous base-state areal fraction $\bar{\Phi}_0$ of passive membrane components held fixed}. \cc{Although the two second moments are independent in general, we restrict to the one-parameter case $Q_2=-Q_1=Q$, corresponding to equal magnitudes and opposite signs, with $Q$ allowed to take either sign \footnote{\cc{For the illustrative two-point force model, $Q<0$ corresponds to an extensile force dipole during fission and a contractile one during fusion, whereas $Q>0$ reverses this arrangement.}}}. \cc{We also take the surface diffusivities to be equal, $\gamma_k=\gamma$,} and neglect spontaneous curvature terms and \cc{active volume fluxes associated with fission and fusion, which are subdominant for transport vesicles much smaller than the organelle}. \cc{We also assume a uniform, isotonic base state, with vanishing osmotic pressure $\Pi_{\scriptscriptstyle\Delta}^{(\mathtt{0})}=0$, and set the isotropic $\ell=0$ perturbations to zero, excluding uniform dilation or shrinkage at linear order; see Appendix~B.5 and B.6. We thus consider only the nonuniform modes, with the $\ell=1$ mode corresponding to organelle translation, while modes with $\ell\geq2$ describe shape deformations of the membrane.}

    The spontaneous drift instability \cc{occurs when}
    \begin{equation}
        A\geq A_\star=\frac{2(1+\zeta)^2}{\zeta(\zeta-1)(1\hspace{-1pt}-\hspace{-1pt}\bar{\Phi}_0)^2}\quad\text{and}\quad\zeta>1, 
    \end{equation} where $A_\star$ is the critical dimensionless fission parameter, corresponding to a threshold organelle radius. In smaller compartments, lateral diffusion rapidly homogenizes composition differences, whereas in larger compartments these differences can build up and drive the instability. To understand the spontaneous organelle drift, we solve the linearized equations for the \cc{degree-$\ell$ components $u_\ell$ and $\Psi_{k,\ell}$}, obtaining the membrane stresses and fluid flows to linear order. Fig.~\ref{fig:fig3} highlights this interplay between vesicular flux and shape deformations through the induced fluid flows and membrane stresses.

    \cc{To make the physical origin and direction of the drift explicit, we take the perturbation in membrane-area exchange to have a dipolar profile aligned with the $z$-axis, as in Fig.~\ref{fig:fig3}. For the translational $\ell=1$ mode, this corresponds to retaining only the axisymmetric $m=0$ component; since the $\ell=1$ modes are rotationally degenerate, this choice is without loss of generality. The resulting imbalance between the rates at which membrane area is added on one side of the organelle and removed on the opposite side is} \cc{described by the membrane-area flux profile $\mathbb{M}^{\mathsf{(1)}}_{\ell=1}(\theta)=\sum_k\bMk\Psi_{k,1,0}Y_{1,0}(\theta)$.} \cc{The underlying compositional perturbation enters the linearized shape equation and generates the corresponding} \cc{translational displacement. We define $V^{\mathsf{(1)}}_{\mathrm{cm}}\equiv\itbf{V}^{\mathsf{(1)}}_{\mathrm{cm}}\cdot\boldsymbol{e}_z$ as the center-of-mass velocity along the chosen $z$-axis, which at linear order is}
    \begin{equation}
        V^{\mathsf{(1)}}_{\mathrm{cm}}
        =
        {\frac{1}{\sqrt{12\pi}}}
        \sum_k
        \left(
            R\,\bMk-\frac{1}{\eta R}\bQk
        \right)
        {\Psi_{k,1,0}}.
    \end{equation}
    \cc{The two terms in this expression have distinct physical origins. The term proportional to $\bMk$ is kinematic: local membrane-area addition and removal drive a tangential membrane flow from the region of net addition towards the region of net removal. This membrane motion exerts viscous traction on the surrounding fluid, causing the force-free organelle to translate in the opposite direction, towards the region of net addition, as in treadmilling. The term proportional to $\bQk$ is mechanical: spatial gradients of the active second moment drive shear within the {finite active layer}, producing a {Marangoni-like slip velocity} along the membrane. These two contributions can reinforce or oppose one another. The direction of motion relative to the incident membrane-area (vesicular) flux is determined by the competition between these kinematic and mechanical contributions. We call the motion {\it retrograde} when} \cc{it follows the treadmilling direction, and {\it anterograde} when it is reversed.} \cc{For the parameters used in Fig.~\ref{fig:fig3}(d), the $\bQk$-dependent contribution opposes and dominates the treadmilling contribution, reversing the direction expected from membrane-area exchange alone.}

    \begin{figure*}[t!]
        \includegraphics[width=\textwidth]{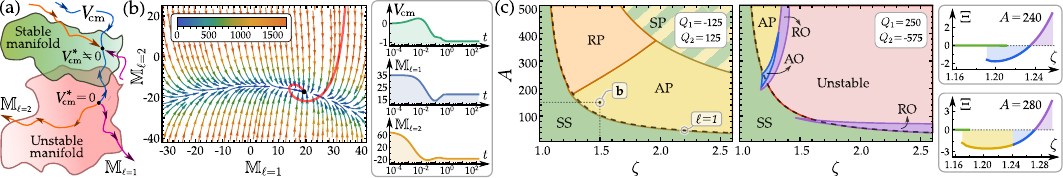}
        \caption{\label{fig:fig4}
        \cc{\textbf{Weakly nonlinear analysis reveals stable
        organelle motion and shape.}}
        \cc{(a) Schematic phase portrait in the space of center-of-mass velocity $V_{\mathrm{cm}}$ and membrane-area flux amplitudes $\mathbb{M}_{\ell=1}$ and $\mathbb{M}_{\ell=2}$. Trajectories along the unstable manifold (pink) move away from the stationary fixed point ($V_{\mathrm{cm}}^\ast=0$), while those along the stable manifold (green) approach the drifting fixed point ($V_{\mathrm{cm}}^\ast\neq0$).}
        \cc{(b) Streamlines of the nonlinear dynamics in the $\mathbb{M}_{\ell=1}$--$\mathbb{M}_{\ell=2}$ plane, with the remaining dynamical variables held at their values at the stable fixed point. Color indicates the magnitude of the vector field governing the evolution of these two amplitudes. In contrast, the red curve shows the projection of a trajectory of the full reduced system, with all five dynamical amplitudes evolving subject to the fixed-volume and active-component conservation constraints. Starting from an initial state with $V_{\mathrm{cm}}(0)=0$ within the basin of attraction of the stable drifting fixed point, the system relaxes to steady anterograde prolate motion (black point). The right-hand insets show the time evolution of $V_{\mathrm{cm}}$, $\mathbb{M}_{\ell=1}$, and $\mathbb{M}_{\ell=2}$ along the red trajectory. Here, $\zeta=1.5$, $A=150$, $Q_1=-125$, and $Q_2=125$.}
        \cc{(c) Phase diagrams in the $\zeta$--$A$ plane for the two choices of active second moments $(Q_1,Q_2)$. The phases are stationary sphere (SS), stationary prolate (SP), anterograde prolate (AP), anterograde oblate (AO), retrograde oblate (RO), and retrograde prolate (RP), with SP describing a stationary state ($V_{\mathrm{cm}}=0$) with finite prolate deformation. The dashed curve marks the linear $\ell=1$ instability threshold of the homogeneous sphere, and the marked point in the left-hand diagram corresponds to the parameters used in (b). A region where no stable fixed points were found is also indicated (red). The insets beside the right-hand phase diagram show $\Xi=\operatorname{sgn}(\mathbb{M}_{\ell=1\!})V_{\mathrm{cm}}$ along cuts at $A=240$ (top) and $A=280$ (bottom). In both (b) and (c), $\mathcal{K}=10$ and $\bar{\Phi}_0=3/5$. Here, time, velocity, and membrane-area flux amplitudes are nondimensionalized by $R^2/\gamma$, $\gamma/R$, and $\gamma/R^2$, respectively.}
        }
    \end{figure*}

    Fig.~\ref{fig:fig3}(e) shows transitions from stationary $\to$ anterograde $\to$ retrograde as we increase $A$, keeping $\zeta$ and $Q$ fixed. \cc{Along this path, the stationary state first becomes unstable, while at larger $A$ the competition between the kinematic and mechanical contributions described above can reverse the drift direction.} Similarly, the higher-order ($\ell\geq2$) shape instabilities are driven by the imbalance $\zeta$ and active moment $Q$, \cc{with their onset marked by the critical lines in Fig.~\ref{fig:fig2}(c). The lowest shape mode, $\ell=2$, produces prolate or oblate deformations.}

    To find the shape corresponding to the $\ell=2$ mode, we solve the linear equations for \cc{$u_{\ell=2}$ and $\Psi_{k,\ell=2}$}. \cc{We consider axisymmetric perturbations about the same $z$-axis used for drift, taking the initial quadrupolar membrane-area flux amplitude to be positive, with membrane addition near the poles and removal near the equator. Such a pattern may arise from vesicle exchange with neighboring compartments on opposite sides of the organelle. The deformation is prolate for $u_2>0$, elongated along the $z$-axis, and oblate for $u_2<0$, flattened along that axis. Reversing the initial $\ell=2$ perturbation reverses both the flux and $u_2$.} Fig.~\ref{fig:fig3}(k) shows transitions from stationary sphere $\to$ moving sphere $\to$ moving oblate $\to$ moving prolate as we increase $A$ while holding $\zeta$ and $Q$ fixed. \cc{Therefore, combining Figs.~\ref{fig:fig3}(e) and (k), we find that, within the linear theory, spherical and oblate shapes can accompany either anterograde or retrograde motion, whereas prolate shapes occur only with retrograde motion. Figs.~\ref{fig:fig3}(f) and (l) show that this ordering persists throughout the $A$--$\zeta$--$Q$ parameter space.}
    \\
    \indent
    \cc{The combined drift and shape phase diagrams reveal two features of particular biological relevance. First, they permit anterograde motion accompanied by an oblate linear response, reminiscent of Golgi cisternal progression \cite{Glick2009, Glick2019}. Second, the phase diagrams suggest how selective inhibition of fission can reverse organelle motion. We model this by reducing the fission cycle rate, namely, $\mathbb{R}^{\hspace{0.5pt}1}_{\CircleArrow}\to\chi\mathbb{R}^{\hspace{0.5pt}1}_{\CircleArrow}$, where $0<\chi\leq1$, while leaving the fusion cycle rate unchanged. Hence, $A\to\chi A$ and $\zeta\to\zeta/\chi$. Within the reduced parameter setting used to construct these phase diagrams, this inhibition defines a one-parameter trajectory through the four-dimensional space spanned by $A$, $\zeta$, and the two independent active second moments. 
    The arrows in Fig.~\ref{fig:fig3}(f) and (l) show the projection of this trajectory onto the reduced $A$--$\zeta$--$Q$ space, with $Q=-Q_1$ in the plotted parameterization, so that $Q\to\chi Q$. The displayed boundary surfaces correspond to the equal-magnitude, opposite-sign choice defined above, whereas the arrows show the projection of the full trajectory onto this reduced phase space. This projection indicates how fission inhibition can drive an anterograde organelle into the retrograde regime.} The same reduction in fission relative to fusion may also lower the activity--renormalized surface tension and, if sufficiently strong, drive it negative, potentially leading to a break-up of the cisterna~\cite{Staykova2011, Solon2006, Rao2001}. \cc{The predicted retrograde motion and cisternal break-up are} consistent with Brefeldin A {treatment} \cite{Lippincott-Schwartz2000,Helms1992, Niu2005}, \cc{which inhibits} vesicle coat formation and hence fission, causing retrograde redistribution and fragmentation of Golgi membranes, followed by their fusion with the endoplasmic reticulum (ER). This is also consistent with the observed fragmentation of Golgi cisternae in fibroblast cells upon loss of adhesion from a substrate~\cite{Singh2018}. To our knowledge, this is the first theoretical justification from first principles that may explain these experimental observations.
    
    \subsection{Weakly Nonlinear Analysis}

    \cc{The spontaneous drift instability identified above originates in the membrane composition. In the unstable regime, the growing compositional asymmetry generates membrane stresses and hydrodynamic flows that move the organelle. The preceding phase diagrams classify motion and shape according to the dominant modes of the linearized dynamics, but these perturbations eventually grow beyond the validity of the linear approximation. We thus extend the composition dynamics beyond linear order to investigate whether nonlinear interactions can arrest their growth and which states ultimately persist; see Fig.~\ref{fig:fig4}(a).}
    \\
    \indent    
    \cc{To study steadily drifting states, we work in a frame comoving with the organelle. In the laboratory frame, the $\ell=1$ mode represents rigid translation only at linear order; at higher orders, translation mixes this mode with higher spherical harmonics. The comoving frame allows us to separate the motion of the organelle as a whole from changes in its intrinsic shape and membrane composition. As in the linear analysis, we neglect active volume fluxes and consider an isotonic reference state. Since passive solvent permeation is slow on the timescales of interest, we fix the enclosed volume at its homogeneous reference value. We also conserve the total numbers of fissogens and fusogens on the membrane, as required by the general model (Appendix~C). Here, the excess membrane area is determined by the deformed shape rather than imposed as an additional constraint.}
    \\
    \indent    
    \cc{We perform a Galerkin expansion of the composition dynamics in axisymmetric spherical harmonics, retaining modes up to $\ell=2$ and nonlinear terms through cubic order in the composition dynamics and global constraints, while keeping the mechanical shape response linear (Appendix~C). The resulting system has five evolving amplitudes: the dipolar ($\Psi_{k,\ell=1}$) and quadrupolar ($\Psi_{k,\ell=2}$) composition amplitudes for $k=1,2$, together with the quadrupolar shape amplitude $u_2$. The fixed-volume constraint and conservation of fissogen and fusogen numbers determine the three isotropic amplitudes $\Psi_{1,0}$, $\Psi_{2,0}$, and $u_0$ algebraically in terms of these evolving modes. Nonlinear coupling between harmonics introduces feedback absent from the linear theory. For example, a growing dipolar compositional asymmetry induces isotropic and quadrupolar responses, which in turn influence the dipolar dynamics. This feedback can arrest the growth of the instability at finite amplitude.}
    \\
    \indent    
    \cc{By numerically determining the fixed points of the reduced system and assessing their linear stability, we identify stable stationary and steadily drifting states across parameter space. Fig.~\ref{fig:fig4}(b) illustrates relaxation towards one such state: the organelle develops a finite drift velocity and settles into stable anterograde prolate motion, starting from an initial condition with $V_{\mathrm{cm}}(0)=0$ within its basin of attraction. We classify the direction of motion by the signed order parameter $\Xi=V_{\mathrm{cm}}\operatorname{sgn}(\mathbb{M}_{\ell=1})$, with $\Xi<0$ for anterograde and $\Xi>0$ for retrograde motion, while $u_2>0$ and $u_2<0$ identify prolate and oblate shapes. By reflection symmetry, each stable drifting solution has an equally stable counterpart with opposite $V_{\mathrm{cm}}$ and $\mathbb{M}_{\ell=1}$ but unchanged $u_2$, so that both have the same $\Xi$ and belong to the same phase.}
    \\
    \indent    
    \cc{Allowing $Q_1$ and $Q_2$ to vary independently, we extend the fixed-point analysis across parameter space to obtain the phase diagrams in Fig.~\ref{fig:fig4}(c) for two representative choices of the active second moments. The left-hand diagram, with $Q_1=-125$ and $Q_2=125$, corresponding to $Q=125$ in Fig.~\ref{fig:fig3}, contains four phases: stationary sphere (SS), anterograde prolate (AP), retrograde prolate (RP), and stationary prolate (SP). In the SP phase, the drift velocity and dipolar flux vanish, while the quadrupolar composition and flux sustain a finite prolate deformation. The right-hand diagram, with $Q_1=250$ and $Q_2=-575$, contains four phases: stationary sphere (SS), anterograde prolate (AP), anterograde oblate (AO), and retrograde oblate (RO). Of particular significance are the stable AO states, which combine anterograde motion with flattening along the direction of motion, consistent with Golgi cisternal progression. Unlike the left-hand diagram, the right-hand diagram also contains a region where no stable fixed points were found despite the nonlinear coupling.}
    \\
    \indent
    \cc{To examine the transitions between these phases, we follow the nonlinear fixed points as $\zeta$ varies at fixed $A$, $Q_1$, and $Q_2$. The insets beside the right-hand phase diagram in Fig.~\ref{fig:fig4}(c) show $\Xi$ along cuts at $A=240$ (top) and $A=280$ (bottom), revealing a region where both stationary and finite-amplitude drifting states are stable near the linear instability threshold, consistent with a subcritical onset of organelle drift. Along the $A=280$ cut, the AP--AO shape transition precedes the AO--RO drift reversal, as the organelle changes from prolate to oblate while moving anterogradely before reversing its direction of motion. The corresponding changes in drift velocity, shape deformation, and membrane-area flux amplitudes are detailed in Appendix~C.4 (Fig.~\ref{fig:C-bifurcation-A280}). For the left-hand diagram in Fig.~\ref{fig:fig4}(c), the bifurcation curves along the $A=300$ cut also reveal a discontinuous onset of drift from SS to RP, followed by an RP--AP drift reversal and a continuous AP--SP transition, in which the drift velocity and dipolar flux vanish smoothly while the prolate deformation and quadrupolar flux remain nonzero (Fig.~\ref{fig:C-bifurcation-A300}).}
    \\
    \indent
    \cc{The phase regions in Fig.~\ref{fig:fig4}(c) are two-dimensional cross sections of a larger four-dimensional parameter space $(\zeta,A,Q_1,Q_2)$. To explore the broader phase structure, we determine, at fixed $\zeta$ and $A$, the critical pair $(Q_1^\ast,Q_2^\ast)$ at which the drift-reversal ($V_{\mathrm{cm}}=0$) and prolate--oblate ($u_2=0$) boundaries intersect. These conditions yield two equations linear in the active second moments, allowing $Q_1^\ast$ and $Q_2^\ast$ to be expressed in terms of the composition amplitudes. Substituting these into the remaining nonlinear amplitude equations, we solve for the fixed points and retain only stable nontrivial solutions. The intersecting boundaries locally divide parameter space into four regimes combining anterograde or retrograde motion with prolate or oblate morphology. Fig.~\ref{fig:C-critical-moments} in Appendix~C.4 shows how the critical moments vary across the $\zeta$--$A$ plane. We find $Q_2^\ast<0$ throughout the region of stable critical solutions, while $Q_1^\ast$ is predominantly positive at moderate $\zeta$ and becomes negative at larger $\zeta$. Stable AO states occur predominantly for $Q_1>0$ and $Q_2<0$, as illustrated by the right-hand phase diagram in Fig.~\ref{fig:fig4}(c). The narrow AO region in this cross section does not necessarily reflect its extent in the full parameter space. These signs are consistent with the illustrative two-point force model discussed above, in which fission is associated with an extensile force pair and fusion with a contractile one.}
    \\
    \indent
    \cc{The center-of-mass speeds predicted in the AO region of Fig.~\ref{fig:fig4}(c) reach up to approximately $250~\mathrm{nm/min}$ and average around $130~\mathrm{nm/min}$ across AO states, using representative physical parameters from Tables I and II (see Appendix~C.4). For comparison, the effective cis-to-trans cargo progression through the Golgi is $\sim10~\mathrm{nm/min}$~\cite{Tie2025}. Since cargo progression does not directly measure physical cisternal motion, a quantitative comparison remains difficult. We expect interactions with the cytoskeleton and neighbouring cisternae, which are not included in our model, to slow cisternal motion. Such a comparison would require extending our model to account for these mechanical constraints and for how cargo exchange and biochemical maturation govern cargo progression.}

    \section{Discussion, caveats and outlook}
    
	In this paper, we provide a systematic analysis of organelle shapes subjected to nonequilibrium forces and hydrodynamic flows driven by energy-consuming active fission and fusion. Our work represents \cc{a step forward} towards a quantitative study of the nonequilibrium mechanics of membrane-bound organelles, such as the Golgi apparatus, within trafficking pathways. \cc{Here, we focus on the maintenance of cisternal shape under continuous trafficking and on the instabilities that drive shape changes and organelle motion.} While our approach centers on the dynamics of existing organelle structures, the question of {\it de novo} nonequilibrium assembly remains an intriguing direction for extending these concepts.  We have demonstrated that the fluid, deformable nature of the membrane provides a coupling between composition and shape, which drives spontaneous drift and shape instabilities; nonlinearities in the fields result in stable organelle motion that depends on the levels of activity and {membrane-area fluxes}.

	Over the years, there have been several proposed  models of Golgi trafficking, such as vesicle transport, cisternal progression and their variants, often taken to be ``diametrically" opposed \cite{Rothman2010, Glick2009}. Recent studies have provided strong evidence for cisternal progression in the budding yeast \cite{Nakano2006, Losev2006}. An outstanding question in the field has been: {\it  ``What drives cisternal progression?"}~\cite{Glick2009}; our work may provide an answer to this, and identifies, for the first time, nonequilibrium fission-fusion as a driving force for cisternal movement. 
	An important result of our study, Fig.~\ref{fig:fig4}(c), with far-reaching implications for intracellular	patterning~\cite{Marshall2020jcub}, is that cisternal progression and vesicle transport can be realised as the moving and stationary phases, respectively, arising from the same underlying nonequilibrium dynamics of fission and fusion. 

	Another unresolved question has been~{\it ``What processes define cisternal morphology?"}~\cite{Glick2009}, particularly, how Golgi cisternae maintain their flattened shapes. Earlier theoretical studies \cite{Svetina2006, Tachikawa2017} proposed membrane adhesion in stacked cisternae as a cause for the flattened cisternal shapes. However, this is in conflict with two lines of evidence: (i) the observation of flattened shapes in unstacked cisternae in {\it S.~cerevisiae} \cite{Preuss1992}, and (ii) isolated Golgi cisternae in mammalian cells that have been unstacked by protease treatment \cite{Cluett1992} maintain their flattened shapes. This  suggests that stacking is not likely the cause of flattened cisternae.		
	\cc{Our results show that fission–fusion stresses can produce oblate deformations and sustain stable oblate states in the weakly nonlinear regime, suggesting a mechanism for maintaining flattened Golgi cisternae. These stresses can also induce higher-order shape instabilities, which may contribute to the ramified morphology of cisternae within Golgi stacks.} Indeed, an implication of our study is that the Golgi morphology and dynamics is intimately associated with its primary function, namely glycosylation, which is dependent on the nonequilibrium trafficking of vesicles \cite{Glick1998,Yadav2022}. We emphasise that the key results highlighted here could not have been derived from a purely kinetic approach and require a careful treatment of active mechanics and hydrodynamics, as attempted here. \cc{In particular, active force moments control both anterograde--retrograde drift reversal and oblate--prolate shape transitions.} 
 
	We now list the caveats regarding the present work and suggest ways of addressing them.  One major shortcoming is that the {transition rates} in the fission--fusion cycles have been taken to be independent of {the state} of the organelle, such as surface tension and local composition or the availability of fissogens and fusogens. Such mechanochemical feedback may play an important role in the homeostatic regulation of organelle size.

	Another important cellular aspect that we have ignored is the influence of the cytoskeleton. The presence of the cytoskeletal meshwork can result in a screening of hydrodynamic flows in the exterior fluid generated by active membrane stresses on the organelle. Direct interactions between the cytoskeleton and the membrane could suppress higher-order shape instabilities. In addition, in cells with stacked Golgi cisternae, the cytoskeletal scaffolding could lead to a suppression of the trafficking-induced cisternal drift instability.

	We hope the ideas presented here will stimulate further theoretical studies and careful biophysical experiments on a system of organelles, such as Golgi and endosomes, in the trafficking pathway. An exciting challenge is to construct synthetic realisations of nonequilibrium  soft fluidics capable of controlled fission and fusion.

	\section*{Acknowledgements}

	AR acknowledges support from the Simons Foundation, and stimulating discussions with M.\ Shelley and B.\ Chakrabarti. MR acknowledges support and funding  from the Department of Atomic Energy (India), under project no.~RTI4006, the Simons Foundation (Grant No.~287975), and the JC Bose Fellowship from DST-SERB (India). RGM acknowledges the EMBL Australia program and support from the Australian Research Council Center of Excellence for Mathematical Analysis of Cellular Systems (CE230100001). 
	
	\section*{Appendix}\appendix
	
	\section{Nonequilibrium Model and Active Hydrodynamic Equations}
	
    \begin{figure*}\includegraphics[width=\textwidth]{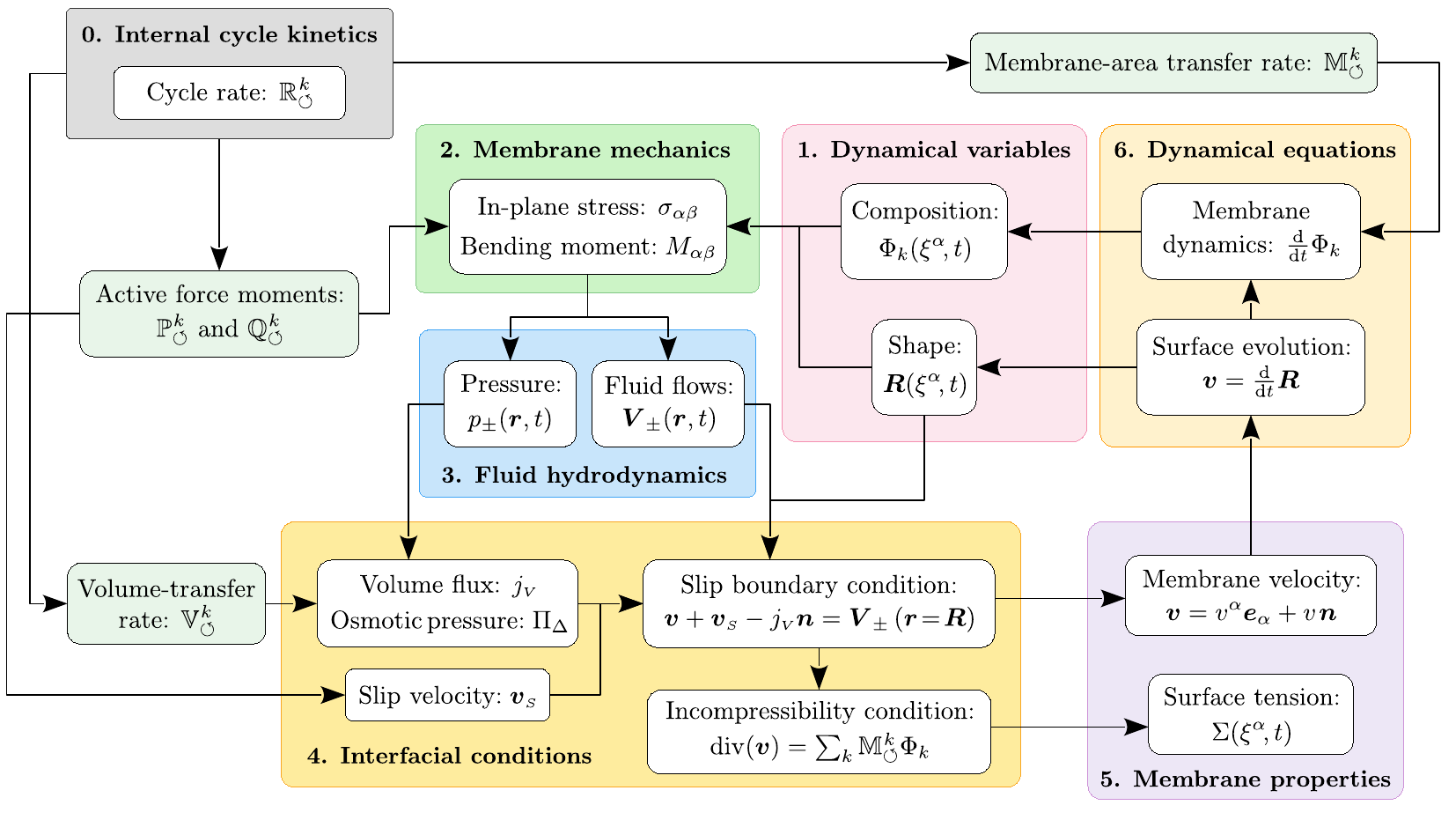}
    	\caption{\label{fig:flowchart} Flow chart summarizing the model. Starting at box $0$, each fission or fusion cycle is characterized by an \cc{cycle rate $\bRk$, determined by the underlying biochemical transition kinetics}. \cc{This converts the net area and volume changes accumulated over one completed cycle into the transfer rates $\bMk$ and $\bVk$, and similarly sets the cycle-averaged active force profile.}  The latter is described in terms of the lowest nontrivial moments of the \cc{force profile normal to the membrane}; namely, its first and second moments $\bPk$ and $\bQk$, respectively. The dynamical variables are the {areal fractions} $\Phi_k(\xi^\alpha,t)$ of active components (fissogens and fusogens) that reside on the membrane surface, which itself evolves, being described by the shape $\boldsymbol{R}(\xi^\alpha,t)$ embedded in a fluid with shear viscosity $\eta$. Here, $\xi^\alpha$ are curvilinear coordinates that parameterize the two-dimensional surface of the compartment. Initial data {for} the composition and shape {allow} us to compute the in-plane stress tensor $\sigma^{\alpha\beta}$ and the bending moment tensor $M^{\alpha\beta}$, which include passive contributions (due to the membrane elasticity and entropy-of-mixing terms), as well as active stresses that depend on $\bPk$ and $\bQk$. The \cc{membrane traction balance couples these stresses to the \cc{hydrodynamic tractions exerted by}} the inner and outer ambient fluids, which satisfy the incompressible Stokes equations. \cc{Together with the membrane boundary conditions, the traction balance allows us to determine} the fluid pressures \cc{$p_\pm(\boldsymbol{r},t)$} and the fluid flows $\boldsymbol{V}_\pm(\boldsymbol{r},t)$ at each point $\boldsymbol{r}$, using a boundary integral representation. All of the terms in $\sigma^{\alpha\beta}$ and $M^{\alpha\beta}$ are known, apart from the local surface tension $\Sigma(\xi^\alpha,t)$, which acts as a Lagrange multiplier for the \cc{local membrane-area balance}; the surface divergence of the membrane velocity $\boldsymbol{v}$ must equal the \cc{total active membrane-area flux}, $\mathbb{M} = \sum_k\bMk\Phi_k$. \cc{The traction balance, fluid equations, slip boundary condition, and local membrane-area balance are solved self-consistently to determine the fluid fields, surface tension, and membrane velocity.} The \cc{slip velocity} $\itbf{v}_{\!\scriptscriptstyle S} = \frac{1}{2\eta}\eu{\alpha}\hspace{1pt}\nabla_{\!\alpha}\mathbb{Q}$, where the \cc{total second moment} $\mathbb{Q} = \sum_k\bQk\Phi_k$, \cc{arises from flow within the active layer, whereas the volume flux per unit area ${j}_{\scriptscriptstyle V}$ includes passive leakage of solvent through the membrane and the active volume flux $\mathbb{V}=\sum_k\bVk\Phi_k$}. The passive volume flux is driven by the difference between the osmotic pressure $\Pi_\Delta$ and the hydrostatic pressure across the interface. \cc{Once the coupled mechanical and hydrodynamic problem is solved,} the membrane velocity $\boldsymbol{v}$ \cc{is obtained together with the surface tension $\Sigma$}. This velocity can be decomposed \cc{into its tangential and normal components, $v^\alpha\ed{\alpha}$ and $v\nv$, respectively}. The {tangential component} tells us how the membrane constituents are advected on the deformed surface. The material derivative of the compositional fields is thus given by $\frac{\mathrm{d}}{\mathrm{d}t} = \frac{\partial}{\partial t} + v^\alpha\nabla_\alpha$, where $\nabla_\alpha$ denotes the covariant derivative. Here, the evolution of {the areal fractions} $\frac{\mathrm{d}}{\mathrm{d}t}\Phi_k$ is governed by the surface diffusion of membrane components and the active exchange of membrane {area}, with the rates given by $\bMk$. From an algorithmic viewpoint, this allows us to compute the {areal fractions} $\Phi_k$ at the next time-step. Similarly, by using the normal velocity $v\nv$, we can evolve the surface to determine its shape at the subsequent time point. This method of time-marching describes a {scheme for numerically solving} the system of equations, starting from some initial data for the composition fields and the shape. Note that we have assumed that the rates entering the internal cycle kinetics are {\it a priori} given and not regulated by feedback from the membrane mechanics, allowing us to operate in a regime where the transitions can be modeled {by} first-order kinetics with constant {transition rates}.
    }\end{figure*}

    \begin{table}[b!]
        \begin{ruledtabular}
            \begin{tabular}{lr}
            {Effective organelle size} &
            {$R\sim1~\mu\mathrm{m}$}~\cite{Ladinsky1999,Svetina2006}\\[5pt]\hline

            Transport vesicle radius & $r=30$--$40~\mathrm{nm}$~\cite{Orci1986,Bykov2017,Wieland2019}\\[5pt]\hline

            Total area of Golgi & $\mathcal{A}_0\sim(1\text{--}3)\times10^3~\mu\mathrm{m}^2$~\cite{Griffiths1989}\\[5pt]\hline

            Turnover time of Golgi & $\tau=10$--$25~\mathrm{min}$~\cite{Wieland1987,Patterson2008}\\[5pt]\hline
            
            {Membrane bending rigidity} &
            {$\kappa\sim(2\text{--}4)\times10^{-20}~\mathrm{J}$}~\cite{Rawicz2000}\\[5pt]\hline
            
            {Effective cytoplasmic viscosity} &
            {$\eta\sim0.1\text{--}1~\mathrm{Pa}\cdot\mathrm{s}$}~\cite{Chevry2013}\\[5pt]\hline
            
            {Effective surface diffusivity} &
            {$D\sim0.01\text{--}0.5~\mu\mathrm{m}^2/\mathrm{s}$}~\cite{Cole1996,Manley2008}\\[5pt]\hline
            
            {Trafficking event interval} &
            {$T_a\sim0.01\text{--}0.6~\mathrm{s}$}\\[5pt]\hline
            
            {Surface diffusion time} &
            {$T_d\sim R^2/D\sim2\text{--}100~\mathrm{s}$}\\[5pt]\hline
            
            {Shape relaxation time} &
            {$T_s\sim\eta R^3/\kappa\sim2.5\text{--}50~\mathrm{s}$}\\[5pt]
            \end{tabular}
        \end{ruledtabular}        
        \caption{\label{tab:scaled-eqs}
        Length scales, membrane mechanical and transport properties, and estimated time scales relevant to Golgi trafficking and cisternal dynamics.
        \cc{The effective surface diffusivity $D$ characterizes the relaxation of membrane composition. The trafficking event interval $T_a$ is estimated from Golgi-wide turnover, with $f_0\sim0.05$--$0.5$ denoting the fraction of Golgi membrane area participating in vesicular turnover, corresponding to an effective turnover area $f_0\mathcal{A}_0\sim50$--$1500~\mu\mathrm{m}^2$.}}
    \end{table}

    We study the role of membrane trafficking \cc{in} the morphology of subcellular compartments, such as those encountered within the eukaryotic cell. These continuously interchange material amongst themselves via fission and fusion of small transport vesicles. The fission and fusion are orchestrated by a highly complex biochemical network of proteins driven by energy-consuming processes. 
    
    Table I contains a list of {representative physiological quantities and ranges}, which {help} us to determine the appropriate length and time scales in our system. This includes the time-scale associated with the lateral redistribution of membrane composition ($T_d$); the time-scale corresponding to the relaxation of membrane shape deformations ($T_s$); and lastly the time-scale associated with {fission and fusion events of transport vesicles} ($T_a$). \cc{Here, $R$ is the radius of the spherical reference state used in the model and provides an effective measure of the organelle size. We estimate the surface-diffusion and shape-relaxation time scales from the corresponding scaling relations, $T_d\sim R^2/D$ and $T_s\sim\eta R^3/\kappa$, respectively, where $D$ is the effective surface diffusivity, $\eta$ is the effective viscosity of the cytoplasmic fluid, and $\kappa$ is the membrane bending rigidity.}
    
    \cc{The estimate of $T_a$ requires an additional assumption. The measured Golgi turnover time $\tau$ reflects cargo trafficking rather than membrane turnover directly, and we use it here as a proxy for the latter. We introduce an effective turnover fraction $f_0$ to account for the fraction of the total Golgi membrane area participating in vesicular turnover over this time. We take $f_0\sim0.05$--$0.5$ as a broad estimate, reflecting both that cargo turnover does not correspond directly to membrane turnover and that vesicular trafficking is spatially heterogeneous across the Golgi membrane (for example, trafficking is enriched at the cisternal rims~\cite{Ladinsky1999,Rothman2010}). If an area $f_0\mathcal{A}_0$ is turned over during a time $\tau$, and each transport vesicle of radius $r$ carries a membrane area $4\pi r^2$, the characteristic interval between trafficking events is $T_a\sim4\pi r^2\tau/(f_0\mathcal{A}_0)$, giving $T_a\sim0.01$--$0.6$\,s. These estimates indicate that vesicular trafficking across the Golgi occurs on a shorter time scale than surface diffusion and membrane shape relaxation. This implies that repeated fission and fusion can maintain the cisternal membrane out of equilibrium.}
    
    {To describe this nonequilibrium regime, we develop a coarse-grained hydrodynamic theory coupling fission and fusion, membrane morphology, and the ambient fluid.} {Fig.~\ref{fig:flowchart} provides a flow chart} that outlines the model with its key elements and the connections between them. This can be viewed as an {algorithmic scheme for computing and time-marching} the {coupled dynamics}, including membrane surface velocities, membrane stresses and surface tension, bulk fluid flows and pressures, as well as membrane composition and shape. {The principal equations are boxed for ease of reference.}

	\subsection{Mass balance and kinematics}
    
	We consider a closed membrane that is described by $\boldsymbol{R}(\xi^\alpha,t)$ at time $t$, where $\xi^\alpha$ are the local surface coordinates parametrizing the membrane. The membrane comprises three distinct components: {active components associated with membrane fission and fusion}, denoted by the superscript $k=1$ (fissogens) and $k=2$ (fusogens), respectively; and the passive membrane background component, indicated by the subscript `0'. Fission and fusion involve highly specific proteins and lipids, which undergo a number of biochemical reactions that primarily require the hydrolysis of either adenosine triphosphate~(ATP) or guanosine triphosphate~(GTP). 

\begin{figure}[t]\includegraphics[width=0.85\columnwidth]{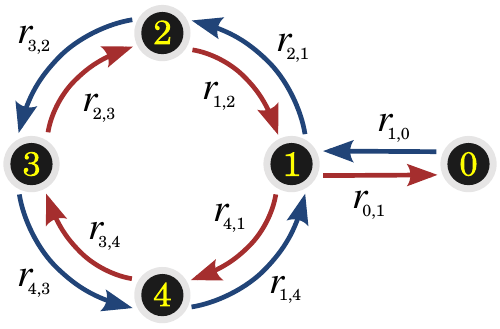}
    	\caption{\label{fig:cycles} Kinetic diagram corresponding to the biochemical cycles of fission or fusion, where the number of internal active states is $\hspace{0.5pt}\aleph_k=4$, in addition to the idle state $n=0$. The diagram therefore contains 10 possible directed transitions. Here, the forward (backward) transitions are indicated by the blue (red) arrows, and each has a first-order transition rate associated with it. We denote the latter by $r_{nm}$ for $m\!\rightarrow n$ (superscript $k$ is omitted here).
    }\end{figure}

    We describe the internal mechanochemical processes of fission and fusion using a nonequilibrium Markov chain model, in which the different intermediary stages of each process are represented by a set of states connected by transitions that form at least one closed loop, or cycle, in state space, as shown in Fig.~\ref{fig:cycles}. The completion of one such cycle corresponds to a single fission or fusion event. Starting from an idle state denoted by $n=0$, the membrane constituents {associated with each cycle $k$ can enter} a sequence of intermediate states indexed by $n\in\{1,2,3,\dots,\aleph_k\}$, where $\aleph_k$ is {the} number of internal states associated with that cycle. {The transition from the final state back to the first closes the loop and completes the cycle.} The system may then proceed directly into another cycle or first return to the idle state $n=0$, where it remains for a dwell time before a new cycle begins, as in Fig.~\ref{fig:cycles}. This dwell time therefore represents the waiting time between two successive fission (or fusion) events.
    
	The {density of active membrane component $k$ in the $n$-th internal state} at time $t$ is denoted by $\rho^{k}_{n}\left(\xi^{\alpha}\!,\hspace{0.5pt}t\right)$, whereas the density of the {passive background component} is given by $\rho_0\left(\xi^{\alpha}\!,\hspace{0.5pt}t\right)$. The total {membrane density} is written as
	\begin{equation}
		\label{eqn:total-density}
		\rho\left(\xi^{\alpha}\!,\hspace{0.5pt}t\right)=\rho_0\!\left(\xi^{\alpha}\!,\hspace{0.5pt}t\right) + \sum_{k,n}\,\rho^{k}_{n}\!\left(\xi^{\alpha}\!,\hspace{0.5pt}t\right)\!,
	\end{equation} where we use the notation that $\sum_{k,n}\left(\boldsymbol{\cdot}\right)=\sum^{}_{k}\sum^{\aleph_k}_{n=0}\left(\boldsymbol{\cdot}\right)$. Moreover, the barycentric velocity of the membrane in the local surface basis is given by $\itbf{v}=v^\alpha\ed{\alpha} +v\hspace{0.2pt}\nv$, with $\nv$ and $\ed{\alpha}$ as the normal and in-plane tangent vectors to the membrane, respectively (see Appendix D.1 for definitions). This {barycentric velocity} is defined by 
	\begin{equation}
		\label{eqn:barycentric-velocity}
		\rho\itbf{v}=\rho_0\itbf{v}_0 + \sum_{k,n}\rho^k_n\itbf{v}^k_n,
	\end{equation} where the background membrane velocity can be decomposed as $\itbf{v}_0 = v^\alpha_0\ed{\alpha} + v\hspace{0.2pt}\nv$, and the velocity of component $k$ in state $n$ is given by $\itbf{v}_n^k = v^{k\hspace{0.5pt}\alpha}_n\ed{\alpha} + v\hspace{0.2pt}\nv$. Notice that the normal velocities of $\itbf{v}$, $\itbf{v}_0$, and $\itbf{v}^k_n$ must all be the same for all species and their associated internal states. 
	
	Furthermore, we denote the in-plane diffusive flux of component $k$ in state $n$, by $\itbf{j}^k_n$, defined as follows:
	\begin{equation}
		\label{eqn:diffusion-flux-species}
		\itbf{j}^k_n = \rho^k_n\hspace{-1pt}\left(\hspace{-0.5pt}\itbf{v}^k_n-\!\itbf{v}\right) = \rho^k_n\hspace{-1pt}\left(v^{k\hspace{0.5pt}\alpha}_n - v^\alpha\right)\!\ed{\alpha} = j^{k\hspace{0.5pt}\alpha}_n\ed{\alpha},
	\end{equation} which shows that this flux lies solely within the tangent plane of the membrane, with components $j^{k\hspace{0.5pt}\alpha}_n$. Similarly, the diffusive flux corresponding to $\rho_0$ is defined by
	\begin{equation}
		\label{eqn:diffusion-flux-background}
		\itbf{j}_0 = \rho_0\hspace{-1pt}\left(\itbf{v}_0-\itbf{v}\right)\!.
	\end{equation}
	As a corollary, it can be shown that the summation over all diffusive fluxes in Eqs.~(\ref{eqn:diffusion-flux-species}) and (\ref{eqn:diffusion-flux-background}) gives
	\begin{equation}
		\label{eqn:diffusion-current-sum}
		\itbf{j}_0 + \sum_{k,n}\,\itbf{j}^k_n=\mathbf{0}.
	\end{equation}
	
	The {amount of {component $k$ in internal state $n$}} over a membrane patch $\mathcal{S}$ changes due to the diffusive {flux} at its boundary $\partial\mathcal{S}$, as well as due to the {internal transitions between states $n$ and $m$ in cycle $k$. Here, $r^k_{nm}$ denotes the transition rate from state $m$ to state $n$ in cycle $k$} (see Fig.~\ref{fig:cycles}). Thus,
	\begin{equation}
		\label{eqn:mass-change}
		\frac{\mathrm{d}}{\mathrm{d}t}\int\limits_{\mathcal{S}\;\;}\!\rho^k_n\,\mathrm{d}S = -\!\!\int\limits_{\partial\mathcal{S}\;\;}\!\!\itbf{j}^k_n\cdot\,\boldsymbol{\nu}\;\mathrm{d}l\,+\!\int\limits_{\mathcal{S}\;\;}\!\sum_{m\hspace{0.5pt}\in\hspace{1pt}\mathfrak{I}_{\hspace{-1pt}n}^{\hspace{-0.15pt}k}}\rho\,\mathbb{J}^{k}_{nm}\,\mathrm{d}S,
	\end{equation} where $\boldsymbol{\nu}$ is the {outward unit vector tangent to the membrane and normal to the boundary $\partial\mathcal{S}$}, with $\boldsymbol{\nu}\cdot\itbf{n}=0$. The index set $\mathfrak{I}_{\hspace{-0.5pt}n}^{\hspace{-0.1pt}k}$ gives all the internal states of {cycle $k$} which are directly connected to state $n$; for example, the states one and zero in Fig.~\ref{fig:cycles} have $\mathfrak{I}_{\hspace{-0.5pt}1}^{\hspace{-0.15pt}k} = \left\{0,2,4\right\}$ and $\mathfrak{I}_{\hspace{-0.5pt}0}^{\hspace{-0.15pt}k} = \left\{1\right\}$, respectively. Here, we define the material derivative to be $\frac{\mathrm{d}}{\mathrm{d}t}(\boldsymbol{\cdot})=\frac{\partial}{\partial t}+v^\alpha\cc{\partial_{\alpha}}(\boldsymbol{\cdot}).$ In the last term of Eq.~(\ref{eqn:mass-change}), $\mathbb{J}^k_{nm}$ denotes the {internal transition current}:
	\begin{equation}
		\mathbb{J}^k_{nm} = r^k_{nm}\phi^k_m - r^k_{mn}\phi^k_n,
	\end{equation} where the {internal-state fractions are defined by $\phi^k_n=\rho^k_n/\rho$, while the areal fraction of the passive background component is $\Phi_0=\rho_0/\rho$.} Note that $\mathbb{J}^k_{nm}$ is an antisymmetric object with respect to the lower indices, that is, we have $\mathbb{J}^k_{nm}\!=\!-\mathbb{J}^k_{mn}$. By using the Reynolds transport theorem on the left-hand-side of (\ref{eqn:mass-change}), namely 
    \begin{equation}
    	\frac{\mathrm{d}}{\mathrm{d}t}\int_\mathcal{S}{\rho_n^k}\,\mathrm{d}S = \int_\mathcal{S}\left(\frac{\mathrm{d}{\rho_n^k}}{\mathrm{d}t} + {\rho_n^k}\,\Div\itbf{v}\right)\!\mathrm{d}S,
    \end{equation} and then by applying the divergence theorem to the first term on the right-hand-side of Eq.~(\ref{eqn:mass-change}), the {local density balance is}
	\begin{equation}
		\label{eqn:mass-balance}
		\boxed{\dot{\rho}^k_n + \rho^k_n\,\Div\itbf{v}= - \Div\hspace{-1pt}\big(\hspace{0.5pt}\itbf{j}^k_n\big) + \rho\sum_{m\hspace{0.5pt}\in\hspace{1pt}\mathfrak{I}_{\hspace{-1pt}n}^{\hspace{-0.15pt}k}}\mathbb{J}^k_{nm},}
	\end{equation} where the dot notation denotes the material derivative. Here, the divergence of the diffusive current can be also written as $\Div\hspace{-1pt}\big(\hspace{0.5pt}\itbf{j}^k_n\big) = \cc{\nabla_{\!\alpha}j^{k\hspace{0.5pt}\alpha}_{n}}$. At equilibrium, \cc{detailed balance requires} $\mathbb{J}^k_{mn}=0$, \cc{whereas in a nonequilibrium steady state these currents may be nonzero}.
	
	Similarly, the {rate of change in the amount of the passive background component} is given by
    \begin{equation}
    	\label{eqn:mass-change-0}
    	\frac{\mathrm{d}}{\mathrm{d}t}\int\limits_{\mathcal{S}\;\;}\!\rho_0\,\mathrm{d}S = -\!\!\int\limits_{\partial\mathcal{S}\;\;}\!\!\itbf{j}_0\cdot\,\boldsymbol{\nu}\;\mathrm{d}l\,+\!\int\limits_{\mathcal{S}\;\;}\!\rho\,\mathbb{M}\;\mathrm{d}S,
    \end{equation}
    where the last term \cc{accounts for membrane material removed from ($k\!=\!1$) or added to ($k\!=\!2$) the surface $\mathcal{S}$ as a result of the internal chemical reactions. The corresponding \cc{total active membrane-area flux is}}
    \begin{equation}
    	\label{eqn:rate-mass-flux}
    	\mathbb{M}\!=\!\sum_{k,n,m}\mathbb{M}^k_{nm}r^k_{nm}\phi^k_m,
    \end{equation}
    with $\sum_{k,n,m}\left(\boldsymbol{\cdot}\right)=\sum_k\sum^{\aleph_k}_{n=0}\sum_{m\hspace{0.5pt}\in\hspace{1pt}\mathfrak{I}_{\hspace{-1pt}n}^{\hspace{-0.15pt}k}}\left(\boldsymbol{\cdot}\right)$, and $\mathbb{M}^k_{nm}$ \cc{characterizes the signed membrane-area transfer associated with a transition from state $m$ to $n$ in cycle $k$}. Again, by means of the Reynolds transport theorem and the divergence theorem, this leads to the following local balance for the background density:
	\begin{equation}
		\label{eqn:mass-balance-0}
		\boxed{\dot{\rho}_0 + \rho_0\,\Div\itbf{v}= -\Div\hspace{-1pt}\big(\hspace{0.5pt}\itbf{j}_0\big) + \rho\,\mathbb{M},}
	\end{equation} with $\Div\hspace{-1pt}\big(\hspace{0.5pt}\itbf{j}_0\big)=\cc{\nabla_{\!\alpha}j^\alpha_0}$. Thus, by summing over all densities from Eqs.~(\ref{eqn:mass-balance}) and (\ref{eqn:mass-balance-0}), the local balance of the total membrane density can be written as follows:
	\begin{equation}
		\label{eqn:mass-balance-total}
		\boxed{\dot{\rho} + \rho\;\Div\itbf{v} = \rho\,\mathbb{M}.}
	\end{equation} Notice that the diffusive currents vanish due to Eq.~(\ref{eqn:diffusion-current-sum}), and also $\sum_n\sum_{m\hspace{0.5pt}\in\hspace{1pt}\mathfrak{I}_{\hspace{-1pt}n}^{\hspace{-0.15pt}k}}\,\mathbb{J}^k_{nm} = 0$ as $\mathbb{J}^k_{nm}=-\mathbb{J}^k_{mn}$. 
	
	Using the total-density balance in Eq.~(\ref{eqn:mass-balance-total}), the dynamics of the areal fraction $\Phi_0$ is governed by
	\begin{equation}
		\label{eqn:mass-fraction-0}
		\dot{\Phi}_0 =\left(1-\Phi_0\right)\mathbb{M}+\frac{1}{\rho}\sum_{k,n}\hspace{1pt}\Div\big(\hspace{0.5pt}\itbf{j}^k_n\big),
	\end{equation} where Eq.~(\ref{eqn:diffusion-current-sum}) and the chain rule, $\dot{\rho}_0 = \rho\hspace{1.5pt}\dot{\Phi}_0 + \dot{\rho}\hspace{1.5pt}\Phi_0$, are used to derive the above expression. Note that $\Phi_0$ is linked  to the other areal fractions through Eq.~(\ref{eqn:total-density}), 
	\begin{equation}
		\label{eqn:mass-fraction-condition}
		1-\Phi_0 = \sum_{k,n}\phi^k_n,
	\end{equation} where $k=1,2$. Thus there are only $\sum_k (1+\aleph_k)$ linearly independent fractions. Analogously to Eq.~(\ref{eqn:mass-fraction-0}), a dynamical equation for the internal-state fractions $\phi_n^k$ can be derived, that is,
	\begin{equation}
		\label{eqn:mass-fractions}
		\dot{\phi}^k_n =-\phi^k_n\,\mathbb{M}\,+\!\sum_{m\hspace{0.5pt}\in\hspace{1pt}\mathfrak{I}_{\hspace{-1pt}n}^{\hspace{-0.15pt}k}}\mathbb{J}^k_{nm}-\frac{1}{\rho}\hspace{1pt}\Div\big(\hspace{0.5pt}\itbf{j}^k_n\big).
	\end{equation}

	\subsection{Force balance and kinetics}
	
    The membrane patch $\mathcal{S}$ {is} subjected to a traction $\itbf{T}$ at the boundary $\partial\mathcal{S}$ of the domain. As a consequence, the total linear momentum over all membrane constituents, that is, $\int_\mathcal{S}\big(\rho_0\itbf{v}_0+\sum_{k,n}\rho^k_n\itbf{v}^k_n\big)\hspace{1pt}\mathrm{d}S$, changes as follows:
    \begin{equation}
    	\label{eqn:momentum-change}
    	\frac{\mathrm{d}}{\mathrm{d}t}\int\limits_{\mathcal{S}\;\;}\!\rho\itbf{v}\,\mathrm{d}S = \!\!\int\limits_{\mathcal{S}\;\;}\!\itbf{f}\,\mathrm{d}S\,+\!\!\!\int\limits_{\partial\mathcal{S}\;\;}\!\!\!\itbf{T}^\alpha\nu_\alpha\,\mathrm{d}l\,+\!\int\limits_{\mathcal{S}\;\;}\!\rho\itbf{v}\,\mathbb{M}\,\mathrm{d}S,
    \end{equation}
    where $\boldsymbol{\nu}=\nu_\alpha\eu{\alpha}$ is the \cc{outward unit vector tangent to the surface patch $\mathcal{S}$ and normal to its boundary $\partial\mathcal{S}$}, and $\itbf{f}$ is the \cc{external force per unit area acting on the membrane}.
    The last term in Eq.~(\ref{eqn:momentum-change}) gives the \cc{momentum carried by membrane material added to or removed from the membrane through active membrane-area exchange}. This assumes that the added or removed material moves with a velocity given by the barycentric velocity $\itbf{v}(t)$. By means of {the} Reynolds transport theorem, $\frac{\mathrm{d}}{\mathrm{d}t}\int_\mathcal{S}(\rho\itbf{v})\,\mathrm{d}S = \!\int_\mathcal{S}\left[\frac{\mathrm{d}}{\mathrm{d}t}(\rho\itbf{v}) + \rho\itbf{v}\,\Div\itbf{v}\right]\mathrm{d}S$, together with the mass balance relation in Eq.~(\ref{eqn:mass-balance-total}), we obtain that
    \begin{equation}
    	\frac{\mathrm{d}}{\mathrm{d}t}\int\limits_{\mathcal{S}\;\;}\!\rho\itbf{v}\,\mathrm{d}S = \!\int\limits_{\mathcal{S}\;\;}\!\left(\rho\dot{\itbf{v}}+\rho\itbf{v}\,\mathbb{M}\right)\mathrm{d}S.
    \end{equation}
    Moreover, by applying the divergence theorem {to} the boundary term of Eq.~(\ref{eqn:momentum-change}), this results in the following local form of linear momentum balance:
    \begin{equation}
    	\label{eqn:momentum-local}
    	\rho\dot{\itbf{v}} = \itbf{f} + \cc{\nabla_{\!\alpha}\itbf{T}^{\alpha}}.
    \end{equation}
    The stress vectors $\itbf{T}^\alpha$ are the projections of the tractions onto the constant surface lines, parametrized by~$\xi^\alpha$. This can be further decomposed into the basis of the surface:
    \begin{equation}
    	\label{eqn:decomposition-T}
    	\itbf{T}^\alpha = T^{\alpha\beta}\ed{\beta}+N^\alpha\nv,
    \end{equation}
    where $N^\alpha$ describes the out-of-plane shear stresses, and $T^{\alpha\beta}$ are the components of the in-plane stress tensor. {Decomposing the force} $\itbf{f}$ in the basis $\{\ed{\alpha},\nv\}$, {that is,} $\itbf{f}= f^\alpha\ed{\alpha}+f\nv$, {the normal component} of the momentum balance relation in Eq.~(\ref{eqn:momentum-local}) is found to be
    \begin{equation}
    	\rho\dot{\itbf{v}}\cdot\nv = f+\,T^{\alpha\beta}b_{\alpha\beta}+\cc{\nabla_{\!\alpha}N^\alpha},
    \end{equation}
    whereas the in-plane component is given by
    \begin{equation}
    	\rho\dot{\itbf{v}}\cdot\eu{\beta} =f^\beta+\,\cc{\nabla_{\!\alpha}T^{\alpha\beta}}\!-N^\alpha b_\alpha^\beta,
    \end{equation}
    by using {\it Gauss-Weingarten} equations, cf.~Eq.~(\ref{eqn:gauss-weingarten}).
    	
    In addition, the rate of change of angular momentum is given by the sum of the torques that arise from the body forces and tractions, as well as a director traction $\itbf{M}$ that acts on the boundary of membrane patch $\mathcal{S}$. In our case, the coarse-grained membrane is assumed to be an object of zero thickness, and its associated director field is chosen to be the same as the unit vector normal $\nv$ to the surface, which is known as the Kirchhoff--Love assumption \cite{Sahu2017}. At each point along the boundary $\partial\mathcal{S}$, the director traction $\itbf{M}$ produces a moment per unit length $\itbf{m}=\nv\times\itbf{M}$, giving rise to an overall boundary torque $\int_{\partial\mathcal{S}}\itbf{m}\,\mathrm{d}l$. This implies that $\itbf{m}$ has only tangential components, even though the director traction may have in general both normal and in-plane components. Hence, the corresponding stress vectors $\itbf{M}^\alpha$, which are obtained by projecting $\itbf{M}$ along the curves of constant $\xi^\alpha$ (that is, $\itbf{M} = \itbf{M}^\alpha\nu_\alpha$), can be chosen without any loss of generality to be in the tangent plane of the membrane; namely, we define
    \begin{equation}
    	\label{eqn:moment-bending-tensor}
    	\itbf{M}^\alpha= -M^{\alpha\beta}\ed{\beta},
    \end{equation}
    where $M^{\alpha\beta}$ is known as the bending moment tensor. As a result, the conservation of angular momentum yields
    \begin{align}
    	\label{eqn:angular-momentum-change}
    	\frac{\mathrm{d}}{\mathrm{d}t}\int\limits_{\mathcal{S}\;\;}\!\itbf{R}\times\rho\itbf{v}\,\mathrm{d}S\;=& \int\limits_{\mathcal{S}\;\;}\!\itbf{R}\times\left(\itbf{f}+\rho\itbf{v}\,\mathbb{M}\right)\mathrm{d}S\notag\\[-3pt]
    	&\;\;\;+{\int\limits_{\partial\mathcal{S}\;\;}\!\!\!\left(\itbf{m}+\itbf{R}\times\itbf{T}\right)\mathrm{d}l},
    \end{align}
    which, by applying Reynolds transport theorem, becomes
    \begin{equation}
    	\int\limits_{\mathcal{S}\;\;}\!\!\itbf{R}\times\rho\dot{\itbf{v}}\,\mathrm{d}S= \!\!\int\limits_{\mathcal{S}\;\;}\!\!\itbf{R}\times\itbf{f}\,\mathrm{d}S\, +{\!\!\int\limits_{\partial\mathcal{S}\;\;}\!\!\!\left(\itbf{m}+\itbf{R}\times\itbf{T}\right)\mathrm{d}l}.
    \end{equation}
    By means of divergence theorem, a local form of the angular momentum balance can be written as
    \begin{align}
    	\itbf{R}\times\rho\dot{\itbf{v}} =& \;\itbf{R}\times\itbf{f}+\ed{\alpha}\times\itbf{T}^\alpha+\itbf{R}\times\cc{\nabla_{\!\alpha}\itbf{T}^\alpha}\notag\\
    	&\qquad\quad+\nv\times\cc{\nabla_{\!\alpha}\itbf{M}^{\alpha}} - b^{\beta}_\alpha\ed{\beta}\times\itbf{M}^\alpha,
    \end{align}
    where the covariant derivatives have been expanded out and the Gauss-Weingarten equations have been used to derive the above expression. By subtracting out the cross product of $\itbf{R}$ with the momentum balance relation in Eq.~(\ref{eqn:momentum-local}), then $\ed{\alpha}\times\itbf{T}^\alpha +\nv\times\cc{\nabla_{\!\alpha}\itbf{M}^{\alpha}} - b^{\beta}_\alpha\ed{\beta}\times\itbf{M}^\alpha = \boldsymbol{0}$.
    Therefore, by using the definition of the bending moment tensor in Eq.~(\ref{eqn:moment-bending-tensor}), the latter equation leads to
    \begin{equation}
    	\Big[\!\left(T^{\alpha\beta}\!-b^{\beta}_{\gamma}M^{\gamma\alpha}\right)\!\ed{\beta} + \left(N^\alpha+\cc{\nabla_{\!\beta}M^{\beta\alpha}}\right)\!\nv\,\Big]\!\times\ed{\alpha}=\boldsymbol{0},
    \end{equation}
    which readily implies that
    \begin{equation}
    	\label{eqn:def-sigma}
    	\sigma^{\alpha\beta}= T^{\alpha\beta}\!-b^{\beta}_{\gamma}M^{\gamma\alpha}
    \end{equation}
    must be a symmetric tensor, and
    \begin{equation}
    	\label{eqn:shear-bending-relation}
    	N^\alpha = -\cc{\nabla_{\!\beta}M^{\beta\alpha}},
    \end{equation}
    so that the linear and angular momentum are conserved. Notice that in a system where the boundary moment $\itbf{m}$ vanishes identically, the in-plane stress tensor $T^{\alpha\beta}$ must be symmetric, and the out-of-plane shear stress $N^\alpha=0$.
    	
    As a result, the equation of motion of the membrane along the normal direction is found to be
    \begin{equation}
    	\label{eqn:eqaution-motion-n}
    	\boxed{\rho\dot{\itbf{v}}\cdot\nv = f+\sigma^{\alpha\beta}b_{\alpha\beta}+M^{\alpha\beta}b_{\alpha\gamma}b_\beta^\gamma-\cc{\nabla_{\!\alpha}\!\nabla_{\!\beta}M^{\alpha\beta}}}
    \end{equation}
    which is known as {\it the shape equation}. The identity $b_{\alpha\gamma}b_\beta^\gamma = 2H\hspace{1pt}b_{\alpha\beta}-K\hspace{0.5pt}g_{\alpha\beta}$ can be used to further simplify this expression. The force balance along the surface reads
    \begin{equation}
    	\label{eqn:eqaution-motion-t}
    	\boxed{\rho\dot{\itbf{v}}\cdot\eu{\alpha} =f^\alpha+\cc{\nabla_{\!\beta}\sigma^{\alpha\beta}}+\cc{\nabla_{\!\gamma}\!\left(b^{\alpha}_{\beta}M^{\beta\gamma}\right)}+ b^\alpha_\beta\hspace{0.5pt}\cc{\nabla_{\!\gamma}M^{\gamma\beta}}.}
    \end{equation}
    To solve these equations of motion, the symmetrized form of the in-plane stress tensor $\sigma^{\alpha\beta}$ and the bending moment tensor $M^{\alpha\beta}$ must be first prescribed. In general, these stress tensors include both passive (\eg elasticity and viscous dissipation) and active contributions.
    	
    Hereinafter, the passive contributions to the traction and the director traction are denoted by $\itbf{T}_{\hspace{-1pt}0}$ and $\itbf{M}_{\hspace{-1pt}0}$, respectively. \cc{These contributions follow from the elastic free-energy of the membrane and are derived in the next section. The active contributions are denoted by the subscript~$A$, namely the traction $\itbf{T}_{\!\scriptstyle A}$ and director traction $\itbf{M}_{\!\scriptstyle A}$, and arise from forces acting within the finite active layer on the cytosolic side of the membrane.} Thus, the net traction is $\itbf{T}=\itbf{T}_{\hspace{-1pt}0} + \itbf{T}_{\!\scriptstyle A}$, while the total director traction is given by $\itbf{M}=\itbf{M}_{\hspace{-1pt}0} + \itbf{M}_{\!\scriptstyle A}$.

	\subsection{Surface variation and membrane stresses}
	
    We consider the geometric dependence of the elastic free-energy $\mathcal{F}$ of a membrane patch $\mathcal{S}$ at fixed membrane composition. Accordingly, $\mathcal{F}$ is treated as a functional of the metric tensor $g_{\alpha\beta}$ and the curvature tensor $b_{\alpha\beta}$. For this geometric dependence,
    \begin{equation}
    	\mathcal{F}(g_{\alpha\beta},b_{\alpha\beta})=\int\limits_{\mathcal{S}\;\;}\!{F}(g_{\alpha\beta},b_{\alpha\beta})\;\mathrm{d}S,
    \end{equation}
    with $F$ as the surface free-energy density, we seek to obtain the mechanical response of the membrane, namely its free-energy change $\delta\mathcal{F}$, due to a shape deformation $\itbf{R}\mapsto\itbf{R}+\delta\hspace{-1pt}\itbf{R}$. This requires us to compute the corresponding changes in the basis vectors, $\delta\ed{\alpha}$, and the normal, $\delta\nv$, as well as the corresponding variations in both the metric tensor, $\delta g_{\alpha\beta}$, and the curvature tensor, $\delta b_{\alpha\beta}$. However, computing these objects is typically cumbersome \cite{Deserno2015}, so we adopt a Lagrange multiplier technique \cite{Guven2004} which circumvents this difficulty by treating the variables $\itbf{R}$, $\nv$, $\ed{\alpha}$, $g_{\alpha\beta}$, and $b_{\alpha\beta}$ as independent, including the known geometrical relationships between these variables as constraints. {Therefore, we construct the constrained functional} $\mathcal{F}_{\scriptscriptstyle C}[\itbf{R}, \nv, \ed{\alpha}, g_{\alpha\beta}, b_{\alpha\beta}]$, defined by
    \begin{align}
    	\label{eqn:free-energy-Lagrange}
    	\mathcal{F}_{\scriptscriptstyle C} = \mathcal{F} &\;+\; \int_{\mathcal{S}}\!\mathrm{d}S\;\big[\hspace{1pt}\boldsymbol{\Upsilon}^\alpha\cdot\left(\ed{\alpha}-\cc{\partial_{\alpha}\itbf{R}}\right)+\lambda^{\alpha}\left(\ed{\alpha}\cdot\nv\right)\big]\notag\\[5pt]
    	&\;+\;\int_{\mathcal{S}}\!\mathrm{d}S\,\left[\frac{\lambda}{2}\left(\nv^2-1\right) + \frac{\lambda^{\alpha\beta}}{2}\left(g_{\alpha\beta}-\ed{\alpha}\cdot\ed{\beta}\right)\right]\notag\\[5pt]
    	&\;+\;\int_{\mathcal{S}}\!\mathrm{d}S\;\left[\Lambda^{\alpha\beta}\left(b_{\alpha\beta}+\ed{\alpha}\cdot\cc{\partial_{\beta}\nv}\right)\right]\!
    \end{align}
    where the Lagrange multiplier $\boldsymbol{\Upsilon}^\alpha$ enforces the relation between the basis $\ed{\alpha}$ and the embedding function~$\itbf{R}$, $\lambda^\alpha$~imposes the orthogonality condition between $\ed{\alpha}$ and $\nv$, and $\lambda$ enforces the normalization of the unit vector $\nv$. Similarly, the multipliers $\lambda^{\alpha\beta}$ and $\Lambda^{\alpha\beta}$ ensure the definitions of the metric tensor $g_{\alpha\beta}$ and the curvature tensor $b_{\alpha\beta}$, respectively, in terms of the basis vectors $\ed{\alpha}$ and the normal vector $\nv$. Note that both these surface tensors are required to be symmetric, namely we have $\lambda^{\alpha\beta} = \lambda^{\beta\alpha}$ and $\Lambda^{\alpha\beta} = \Lambda^{\beta\alpha}$. The advantage of this approach is that we do not need to explicitly consider the changes in $g_{\alpha\beta}$ and $b_{\alpha\beta}$ due to a variation in the embedding $\itbf{R}$, as these tensors are now treated as independent variables~\cite{Guven2004}.
    	
    To determine $\boldsymbol{\Upsilon}^{\alpha}\!$, the functional derivative of $\mathcal{F}_{\scriptscriptstyle C}$ with respect to $\ed{\alpha}$ must vanish ($\delta\mathcal{F}_{\scriptscriptstyle C}/\delta\ed{\alpha} = 0$), which yields
    \begin{equation}
    	\label{eqn:Upsilon-stress}
    	\boldsymbol{\Upsilon}^\alpha = \left(\lambda^{\alpha\beta}+\Lambda^{\alpha\gamma}\hspace{1pt}b^{\beta}_{\gamma}\hspace{1pt}\right)\!\ed{\beta}-\lambda^\alpha\nv,
    \end{equation}
    where the {\it Gauss--Weingarten} equations have been used, that is, \cc{$\nabla_{\!\beta}\ed{\alpha}=b_{\alpha\beta}\nv$}, and \cc{$\partial_{\alpha}\nv = -b^{\hspace{0.5pt}\beta}_{\alpha}\,\ed{\beta}$}. These relations follow from the geometrical constraints imposed in Eq.~(\ref{eqn:free-energy-Lagrange}). Likewise, $\lambda^{\alpha\beta}$ and $\Lambda^{\alpha\beta}$ are given by the Euler-Lagrange equations associated {with} $g_{\alpha\beta}$ and $b_{\alpha\beta}$, namely,
    \begin{equation}
    	\lambda^{\alpha\beta}=-\frac{2}{\sqrt{g}}\frac{\partial(\sqrt{g}\hspace{1pt}F)}{\partial g_{\alpha\beta}},\quad\text{and}\quad\Lambda^{\alpha\beta} = -\frac{\partial F}{\partial b_{\alpha\beta}},
    \end{equation}
    respectively. Since the variable $\itbf{R}$ appears only in the second term of Eq.~(\ref{eqn:free-energy-Lagrange}), the infinitesimal variation of the functional $\mathcal{F}_{\scriptscriptstyle C}$ due to a deformation $\delta\hspace{-1pt}\itbf{R}$ is written as
    \begin{equation}
    	\label{eqn:variation-R}
    	\delta_{\hspace{-0.5pt}\scriptstyle\itbf{R}}\!\left(\mathcal{F}_{\scriptscriptstyle C}\right) = \!\int_{\mathcal{S}\;\;}\!\mathrm{d}S\left(\cc{\nabla_{\!\alpha}\boldsymbol{\Upsilon}^{\alpha}}\hspace{-1pt}\cdot\delta\hspace{-1pt}\itbf{R}\right) - \!\oint_{\partial\mathcal{S}\;}\!\mathrm{d}l\;\nu_\alpha\left(\boldsymbol{\Upsilon}^\alpha\hspace{-1pt}\cdot\delta\hspace{-1pt}\itbf{R}\right)\!,
    \end{equation}
    with $\boldsymbol{\nu}=\nu_\alpha\eu{\alpha}$ being the outward unit vector tangent to the membrane and normal to the boundary curve $\partial\mathcal{S}$. If the shape variation $\delta\hspace{-1pt}\itbf{R}$ vanishes on $\partial\mathcal{S}$, that is, the position vector $\itbf{R}$ is fixed at the boundary, then the last term in Eq.~(\ref{eqn:variation-R}) is identically zero, and thus we have
    \begin{equation}
    	\label{eqn:EL-R}
    	\frac{\delta\mathcal{F}_{\scriptscriptstyle C}}{\delta\hspace{-1pt}\itbf{R}} = \cc{\nabla_{\!\alpha}\boldsymbol{\Upsilon}^{\alpha}}=0,
    \end{equation}
    which states that at equilibrium the stress vector $\boldsymbol{\Upsilon}^\alpha$ has vanishing surface divergence.
    	
    Similarly, the local variation of $\mathcal{F}_{\scriptscriptstyle C}$ due to \cc{an} infinitesimal change $\delta\nv$ in the normal unit vector is given by
    \begin{align}
    	\delta_{\nv}\!\left(\mathcal{F}_{\scriptscriptstyle C}\right) &= \!\int\limits_{\mathcal{S}\;\;}\!\mathrm{d}S\left[\left(\lambda+\Lambda^{\alpha\beta}b_{\alpha\beta}\right)\nv+\Big(\hspace{-1pt}\lambda^\alpha - \cc{\nabla_{\!\beta}\Lambda^{\alpha\beta}}\Big)\ed{\alpha}\right]\!\cdot\delta\hspace{-0.5pt}\nv\notag\\[-3pt]
    	&\;\qquad+\oint\limits_{\partial\mathcal{S}\;}\!\mathrm{d}l\;\nu_\beta\left(\Lambda^{\alpha\beta}\ed{\alpha}\cdot\delta\hspace{-0.5pt}\nv\right)\!.
    \end{align}
    If the normal unit vector is fixed at the boundary, that is, $\delta\hspace{-0.5pt}\nv=0$ on $\partial\mathcal{S}$, then the associated Euler-Lagrange equation with respect to $\nv$ is found to be
    \begin{equation}
    	\label{eqn:EL-n}
    	\frac{\delta\mathcal{F}_{\scriptscriptstyle C}}{\delta\hspace{-0.5pt}\nv} = \left(\lambda+\Lambda^{\alpha\beta}b_{\alpha\beta}\right)\nv+\Big(\hspace{-1pt}\lambda^\alpha - \cc{\nabla_{\!\beta}\Lambda^{\alpha\beta}}\Big)\ed{\alpha} =0,
    \end{equation}
    and therefore we have that
    \begin{equation}
    	\label{eqn:lambda-alpha}
    	\lambda=-\Lambda^{\alpha\beta}b_{\alpha\beta}\quad\text{and}\quad\lambda^\alpha= \cc{\nabla_{\!\beta}\Lambda^{\alpha\beta}}.
    \end{equation}
    	
    However, in general, neither $\nv$ nor $\itbf{R}$ is fixed on the boundary of the patch $\mathcal{S}$, and both are free to vary. Consequently, the boundary terms give rise to the traction and director traction. Using the Euler-Lagrange equations in (\ref{eqn:EL-R}) and (\ref{eqn:EL-n}), the total infinitesimal variation of the free-energy $\mathcal{F}_{\scriptscriptstyle C}$ is given by
    \begin{equation}
    	\label{eqn:total-variation-free-energy}
    	\delta\mathcal{F}_{\scriptscriptstyle C} = \oint\limits_{\partial\mathcal{S}\;}\!\mathrm{d}l\;\nu_\beta\left(\Lambda^{\alpha\beta}\ed{\alpha}\cdot\delta\hspace{-0.5pt}\nv - \boldsymbol{\Upsilon}^\beta\cdot\delta\hspace{-1pt}\itbf{R}\right)\!.
    \end{equation}
    This must be balanced by the traction $\itbf{T}_{\hspace{-1pt}0}$ and the director traction $\itbf{M}_{\hspace{-1pt}0}$ at the boundary $\partial\mathcal{S}$, which are (by definition) the passive contributions to the traction $\itbf{T}$ and director traction $\itbf{M}$, respectively. Thus, $\delta\mathcal{F}_{\scriptscriptstyle C}$ equals the virtual work,
    \begin{equation}
    	\label{eqn:virtual-work}
    	\delta\mathcal{W} = \oint\limits_{\partial\mathcal{S}\;}\!\mathrm{d}l\;\nu_\alpha\left(\itbf{T}_{\hspace{-1pt}0}^\alpha\cdot\delta\hspace{-1pt}\itbf{R} + \itbf{M}_{\hspace{-1pt}0}^\alpha\cdot\delta\hspace{0.5pt}\nv\right)\!,
    \end{equation}
    where $\hspace{-1pt}\itbf{T}^\alpha_{\hspace{-1pt}0}$ and $\hspace{-1pt}\itbf{M}^\alpha_{\hspace{-1pt}0}$ are the projections of the tractions onto the surface lines of constant $\xi^\alpha$. From Eq.~(\ref{eqn:total-variation-free-energy}) and (\ref{eqn:virtual-work}), we find that $\itbf{T}^\alpha_{\hspace{-1pt}0} = -\boldsymbol{\Upsilon}^\alpha$ and $\itbf{M}^\alpha_{\hspace{-1pt}0} = \Lambda^{\alpha\beta}\ed{\beta}$. By decomposing these traction vectors into the surface basis, analogously to Eqs.~(\ref{eqn:moment-bending-tensor}) and (\ref{eqn:decomposition-T}),  $\itbf{M}^\alpha_{\hspace{-1pt}0} = -M^{\alpha\beta}_{0}\ed{\beta}$ and $\itbf{T}^\alpha_{\hspace{-1pt}0} = T^{\alpha\beta}_{0}\ed{\beta}+N_{0}^\alpha\nv$, then we have that the bending moment tensor $M^{\alpha\beta}_{0} = -\Lambda^{\alpha\beta}$, the in-plane stress tensor $\sigma^{\alpha\beta}_{0}= T_0^{\alpha\beta}\!-b^{\beta}_{\gamma}M_0^{\gamma\alpha}=-\lambda^{\alpha\beta}\!$, cf.~Eq.~(\ref{eqn:def-sigma}), and the out-of-plane shear stress $N^\alpha_0 = \lambda^\alpha$. The latter, together with the second equation in (\ref{eqn:lambda-alpha}), gives
    \begin{equation}
    	\cc{N^\alpha_0=-\nabla_{\!\beta}M^{\beta\alpha}_0,}
    \end{equation}
    as found previously in Eq.~(\ref{eqn:shear-bending-relation}) for the overall shear stress $N^\alpha$. Since $\sigma_0^{\alpha\beta}$ is symmetric, this shows that the passive contributions to the tractions, which arise from the free-energy $\mathcal{F}$, respect angular-momentum conservation. Thus, the passive stresses are given by
    \begin{equation}
    	\label{eqn:membrane-stresses}
    	\boxed{\sigma_0^{\alpha\beta} = \frac{2}{\sqrt{g}}\frac{\partial(\sqrt{g}\hspace{1pt}F)}{\partial g_{\alpha\beta}},\quad\text{and}\quad M^{\alpha\beta}_{0} = \frac{\partial F}{\partial b_{\alpha\beta}},}
    \end{equation}
    and the stress vector reduces to
    \begin{equation}
    	{\boldsymbol{\Upsilon}^\alpha = -T^{\alpha\beta}_0\hspace{-1pt}\ed{\beta} + \nabla_{\!\beta}M^{\alpha\beta}_{0}\hspace{1pt}\nv.}
    \end{equation}

	\subsection{Free-energy and constitutive relations}
    
	At a coarse-grained level, the Helmholtz free-energy $\mathcal{F}$ of a multi-component biomembrane (described by three distinct membrane components and their internal states) includes an elastic bending energy together with the surface-tension contribution (\ie the Helfrich energy density) \cite{Helfrich1973}, an entropy of mixing of the various species, {and a phenomenological coupling} between the mean curvature $H$ of the surface and the local density of membrane components~\cite{Leibler1986}. We consider that the membrane forms a closed surface, a membrane compartment, and we denote its surface by $\mathcal{M}$ (formally, we say that $\mathcal{M}$ is homeomorphic to the Euclidean 2-sphere). The overall free-energy of the compartment is therefore given by $\mathcal{F} = \int_{\mathcal{M}}F\,\mathrm{d}S$, where the free-energy density
	\begin{equation}
		\label{eqn:total-free-energy}
		\boxed{F\hspace{-1.5pt}=\hspace{-1pt}\Sigma+2\kappa\hspace{0.25pt} H^2 + 2\kappa\hspace{0.25pt} H\hspace{0.5pt}C_{\mathrm{m}}+ V_{\mathrm{m}},}
	\end{equation} with $\kappa\hspace{0.25pt}$ and $\Sigma$ as the bending rigidity and surface tension of the membrane, respectively \cite{Deserno2015}. The latter can be in general a function of the local coordinates $\xi^\alpha$. Here, $V_{\mathrm{m}}$ is the mean-field free-energy associated with membrane components in the dilute limit (expanded to lowest order in the areal fractions of each membrane component):
	\begin{equation}
		\label{eqn:meanfield-energy}
		V_{\mathrm{m}} = \frac{k_B T}{b_0\,}\hspace{1pt}\Big[\Phi_0\big(\!\ln\Phi_0+\mathcal{E}_0\big)+\sum_{k,n}\phi^k_n\big(\!\ln\phi^k_n+\mathcal{E}^k_n\big)\Big]\hspace{-1pt},
	\end{equation} which contains the entropy of mixing terms, as well as the linear-order energetic penalties for each membrane component. Herein, $\mathcal{E}_0$ and $\mathcal{E}^k_n$ are dimensionless constants, representing the standard chemical potentials, measured in units of $k_B T$ (thermal energy). \cc{The parameter} $b_0$ is \cc{a coarse-graining area scale on the membrane, with an associated length scale $\sqrt{b_0}\sim50$--$100$~nm}. The local spontaneous curvature $C_{\mathrm{m}}$ gives the interaction term via the coupling to curvature, namely
	\begin{equation}
		\label{eqn:coupling-mean-curvature-energy}
		C_{\mathrm{m}}  = \Phi_0\hspace{1pt}\mathcal{C}_0 + \sum_{k,n}\phi^k_n\hspace{1pt}\mathcal{C}^k_n = \,\mathcal{C}_0 + \sum_{k,n}\phi^k_n\hspace{-1pt}\left(\mathcal{C}^k_n-\mathcal{C}_0\right)\!,
	\end{equation} where $\mathcal{C}_0$ \cc{specifies the curvature-coupling strength of the background membrane component, while $\mathcal{C}^k_n$ specifies that of internal state $n$ in cycle $k$.} The second equality in (\ref{eqn:coupling-mean-curvature-energy}) follows from Eq.~(\ref{eqn:mass-fraction-condition}), and~$\mathcal{C}_0$ can be interpreted as a global spontaneous curvature of the membrane, often included in the Helfrich theory~\cite{Helfrich1973}. Strictly positive values of $\mathcal{C}_0$ and $\mathcal{C}^k_n$ indicate a tendency to enrich regions of negative curvature with the corresponding membrane components. Note that the standard Helfrich free-energy functional~\cite{Helfrich1973} can also contain a Gaussian curvature term of the form $\int_\mathcal{M}\bar{\kappa}\hspace{1pt}K\,\mathrm{d}S$~\cite{Deserno2015}.
    \cc{For the closed surface considered here, the Gauss--Bonnet theorem fixes this contribution to the constant $4\pi\bar{\kappa}$~\cite{Kreyszig1991}, so it does not affect the membrane dynamics and can be omitted.}
	
	Using Eq.~(\ref{eqn:membrane-stresses}), the membrane surface stresses can thus be calculated from the free-energy density $F$, which is a function of the metric $g_{\alpha\beta}$ and the curvature tensor $b_{\alpha\beta}$. The free-energy density $F$ depends on the tensors $g_{\alpha\beta}$ and $b_{\alpha\beta}$ only via the mean curvature $H$ for the constitutive form in Eq.~(\ref{eqn:total-free-energy}). Hence, by the chain rule,
	\begin{equation}
		\frac{\partial F}{\partial g_{\alpha\beta}}=-\frac{1}{2}\,b^{\alpha\beta}\frac{\partial F}{\partial H}\quad\text{and}\quad\frac{\partial F}{\partial b_{\alpha\beta}}=\frac{1}{2}\,g^{\alpha\beta}\frac{\partial F}{\partial H},
	\end{equation} which follow from the derivative identities $\frac{\partial H}{\partial g_{\alpha\beta}} = -\frac{1}{2}\hspace{0.5pt}b^{\alpha\beta}$ and $\frac{\partial H}{\partial b_{\alpha\beta}} = \frac{1}{2}\hspace{0.5pt}g^{\alpha\beta}$. The derivative of $F$ with respect to the mean curvature $H$ is obtained by using Eq.~(\ref{eqn:total-free-energy}),
	\begin{equation}
		\frac{\partial F}{\partial H} = {4\kappa\hspace{0.25pt}H+2\kappa\hspace{0.25pt}C_{\mathrm{m}}},
	\end{equation} then the in-plane stress tensor $\sigma^{\alpha\beta}_0$ readily follows:
	\begin{equation}
		\label{eqn:sigma-0-F}
		\sigma^{\alpha\beta}_0= F g^{\alpha\beta} -\frac{\partial F}{\partial H}\,b^{\alpha\beta},
	\end{equation} by applying the identity $\partial g/\partial g_{\alpha\beta} = g\hspace{1pt}g^{\alpha\beta}$. Similarly, the tensor $M^{\alpha\beta}_0\!$, as defined in Eq.~(\ref{eqn:membrane-stresses}), is given by
	\begin{equation}
		\label{eqn:M-0-F}
		M^{\alpha\beta}_0 = \frac{1}{2}\frac{\partial F}{\partial H}\,g^{\alpha\beta}.
	\end{equation} {Hence, the passive in-plane stress $T^{\alpha\beta}_0$ and out-of-plane shear stress $N_0^{\alpha\phantom{\beta}}\!$} are found to be
    \begin{equation}
    	T^{\alpha\beta}_0 = F g^{\alpha\beta}-\frac{1}{2}\frac{\partial F}{\partial H}\,b^{\alpha\beta},
    \end{equation}
    and
    \begin{equation}
    	N_0^{\alpha\phantom{\beta}}=-\frac{1}{2}\cc{\nabla^{\alpha}}\!\left(\frac{\partial F}{\partial H}\right).
    \end{equation}
	
    In addition to \cc{the passive mechanical stresses}, the free-energy density $F$ also \cc{determines the chemical potentials of the membrane species}. In general, \cc{these are defined by functional derivatives} of $\mathcal{F}$ with respect to the \cc{areal fractions}. \cc{Since $F$ is local in composition and contains no composition-gradient terms, these reduce here to partial derivatives of $F$.} \cc{Because} $\phi_n^k$ and $\Phi_0$ are \cc{constrained by} Eq.~(\ref{eqn:mass-fraction-condition}), \cc{the relevant quantity is the chemical-potential difference relative to the background component,}
    \begin{equation}
        \label{eqn:chemical-potential}
        \boxed{\mu_n^k=\frac{\partial F}{\partial\phi_n^k}-\frac{\partial F}{\partial\Phi_0},}
    \end{equation}
    \cc{which gives the local free-energy cost per unit area associated with replacing background membrane by component $k$ in internal state $n$.}
	
	A local chemical potential imbalance {generates gradients in} $\mu_n^k$, which {drive} the migration of membrane components {and decrease} the total free-energy of the system. Thus, the diffusive currents $j^{k\hspace{0.5pt}\alpha}_n$ must be directly related to the gradients \cc{$\nabla^{\alpha}\mu_n^k$}. For general irreversible processes in the linear regime (\ie close to equilibrium), the Onsager's principle holds~\cite{Onsager1931a}, which establishes a linear relationship between the generalized forces and their corresponding thermodynamic fluxes \cite{Sahu2017}. For isothermal systems, this principle is shown to reduce to the principle of least energy dissipation \cite{Onsager1931b}; namely, the system evolves dynamically so that it minimizes the competition between dissipation and energy release. 
	
    Therefore, this can be cast as a variational principle problem, where we seek to minimize a functional that is known as the {\it Rayleighian}, and which is defined by
    \begin{equation}
    	\mathfrak{R}=\mathfrak{D}+\dot{\mathcal{F}},
    \end{equation}
    with $\mathfrak{D}$ as the dissipation potential \cite{Arroyo2018}. $\mathcal{F}$ is typically a function of the state variables, such as the metric tensor $g_{\alpha\beta}$, the curvature tensor $b_{\alpha\beta}$, and the {internal-state fractions} $\phi_n^k$, describing the state of the system \cite{Arroyo2018}. On the other hand, $\mathfrak{D}$ is a function of the {\it process variables}, such as the diffusive fluxes $\itbf{j}^k_n$, which characterize how the system changes its state and generates dissipation~\cite{Arroyo2018}. \cc{The Rayleighian is minimized with respect to the process variables rather than the state variables}~\cite{Arroyo2018}. In general, the Rayleighian $\mathfrak{R}$ may include terms that correspond to external power sources, as well as any other constraints that the process variables must additionally satisfy~\cite{Arroyo2018}. \cc{In Onsager's variational approach, the dissipation potential $\mathfrak{D}$ is quadratic in the process variables}~\cite{Doi2011}. \cc{This can be generalized to nonlinear convex functions satisfying} $\frac{\mathrm{d}}{\mathrm{d}\hspace{0.5pt}t}\mathcal{F}\leq0$~\cite{Arroyo2018}. \cc{Such a dissipation potential is positive and vanishes only when all process variables vanish}~\cite{Arroyo2018}.
	
	Since the free-energy density $F\equiv F\!\left[g_{\alpha\beta},b_{\alpha\beta},\phi_n^k\right]$, we find the rate of change of the total free-energy to be
	\begin{equation}
		\dot{\mathcal{F}} = \int\limits_{\mathcal{M}\;}\!\mathrm{d}S\,\Big(\,\frac{1}{2}\sigma_0^{\alpha\beta}\,\dot{g}_{\alpha\beta}+M_0^{\alpha\beta}\,\dot{b}_{\alpha\beta}+\sum_{k,n}\mu_n^k\,\dot{\phi}_n^k\,\Big),
	\end{equation} which follows \cc{from the Reynolds transport theorem for the closed surface $\mathcal{M}$, together with} the results in Eq.~(\ref{eqn:membrane-stresses}) and the definition of $\mu_n^k$ in Eq.~(\ref{eqn:chemical-potential}). 
	
    In general, the in-plane membrane flow and the out-of-plane bending may also result in energy dissipation, \eg the in-plane dissipation due to the two-dimensional shear and bulk viscosities of the membrane \cite{Arroyo2009}. Here we neglect their contributions, since the viscosity of the ambient fluid adjacent to the membrane provides the dominant mechanism of dissipation in the system.

    \cc{By contrast, dissipation associated with the diffusive motion of the membrane species relative to the background component must be retained, since it determines their finite mobility and hence the diffusive fluxes entering the composition dynamics.} Consequently, in order to construct the dissipation potential associated with this mechanism, we assume that the drag force $\itbf{d}_n^k$ (a force per unit area) on {component $k$ in internal state $n$} within a {coarse-grained membrane patch of area $b_0$} is given by $\itbf{d}^k_n = -\zeta^k_n\hspace{1pt}\phi^k_n\!\left(\hspace{-0.5pt}\itbf{v}^k_n-\!\itbf{v}\right)$, where $\zeta^k_n$ are the drag coefficients. Here, $\phi^k_n\!\left(\hspace{-0.5pt}\itbf{v}^k_n-\!\itbf{v}\right)$ \cc{is the relative flux of this membrane component with respect to the barycentric motion}. Thus the rate of energy loss per unit membrane area can be written as follows: $\mathfrak{D}^k_n=\frac{1}{2}\zeta^k_n\hspace{1pt}\phi^k_n\left|\hspace{-0.5pt}\itbf{v}^k_n-\!\itbf{v}\right|^2\hspace{-1pt}$. Since the concentration of membrane species is assumed to follow a dilute distribution of non-interacting particles, \cc{the drag on each component is unaffected by the others, so the local dissipation is additive over all components}. As a result, we write the total dissipation potential as follows:
	\begin{equation}
		\mathfrak{D}=\int\limits_{\mathcal{M}\;}\!\mathrm{d}S\;\frac{1}{2\rho}\sum_{k,n}\zeta^k_n\,\big|\itbf{j}_n^k\big|^2,
	\end{equation} using Eq.~(\ref{eqn:diffusion-flux-species}). The minimization of the Rayleighian $\mathfrak{R}$ with respect to each diffusive flux $\itbf{j}_n^k$ gives us that
	\begin{equation}
		-\zeta^k_n\,\itbf{j}^k_n  = \rho\,\frac{\delta\dot{\mathcal{F}}}{\delta\itbf{j}^k_n} = \rho\,\frac{\delta}{\delta\itbf{j}^k_n}\left(\int_{\mathcal{M}\;}\!\mathrm{d}S\,\sum_{k,n}\mu_{n}^{k}\hspace{1pt}\dot{\phi}_{n}^{k}\right)\!. %
	\end{equation} As the dynamics of $\phi_n^k$ depends on their diffusive fluxes, via Eq.~(\ref{eqn:mass-fractions}), we find, by integration by parts, that
    \begin{equation}
        \label{eqn:current-chemical-potential}
        \boxed{\cc{\itbf{j}^k_n=-\frac{\rho}{\zeta^k_n}\,\eu{\alpha}\hspace{1pt}\nabla_{\!\alpha}\!\left(\frac{\mu_n^k}{\rho}\right)\!,}}
    \end{equation} which linearly relates the diffusive flux with the gradient in the chemical potential if the total density $\rho\left(\xi^{\alpha}\!,\hspace{0.5pt}t\right)$ is \cc{spatially uniform along the membrane, namely $\nabla_{\!\alpha}\rho=0$}.

	\subsection{Adiabatic approximation}
	
	Henceforth, only cyclic transitions are considered for the biochemical cycles associated with fusion and fission events, as depicted in Fig.~\ref{fig:cycles}. In other words, the index set $\mathfrak{I}_{\hspace{-1pt}n}^{\hspace{-0.15pt}k}$, whose elements are the states $m$ which directly connect to state $n$ for a given $k$-th component (fissogens or fusogens), is defined to be
	\begin{equation}
		\label{eqn:index-set}
		\mathfrak{I}_{\hspace{-1pt}n}^{\hspace{-0.15pt}k} = 
		\begin{cases}
			\hspace{0.5pt}\left\{1\right\}                  
			& \quad\mathrm{if}~n=0,\\[1.5pt]
			\hspace{0.5pt}\left\{0,2,\aleph_k\hspace{-0.5pt}\right\}         
			& \quad\mathrm{if}~n=1,\\[1pt]
			\hspace{0.5pt}\{n-1,n+1\}                         
			& \quad\mathrm{otherwise},
		\end{cases}
	\end{equation} with the cyclic condition that state $1+\aleph_k$ is the same as, or maps back to, state one. Here, the state zero plays the role of an idle or inactivated state. \cc{The label $n=0$ refers to the biochemical state of the membrane-associated machinery, rather than to a state carried by the transport vesicle.} In the case of a fusion cycle, the transition out of this inactive state physically corresponds to the arrival of a small transport vesicle, \cc{such as a COPII-derived vesicle, near the membrane of the organelle, \eg a Golgi cisterna, and the activation of the local fusion machinery}; while the transition into the idle state signifies the inactivation of the membrane components that were previously recruited at that site, \eg the SNARE-complex \mbox{disassembly}. \mbox{Similarly}, in the case of a fission cycle, the transition out of the idle state \cc{corresponds to the recruitment and activation of proteins that initiate fission at a local site on the membrane}, such as coatomers; whereas a transition into the idle state represents the scission of a newly formed vesicle from the \cc{mother membrane and the return of the associated machinery to its inactive state}.

    In the physiological regime, the characteristic times associated with the mechanochemical kinetics of the internal, cyclic transitions are much smaller (on the order of \mbox{milliseconds}) than the relaxation times of the membrane \mbox{dynamics} on length scales larger than the coarse-graining length scale (\cc{$\sim50$--$100$~nm}). Typically, we find the relaxation times of the latter to be in the range of $1$--$100$~s (see Table~I). This separation of scales allows us to consider the transitional dynamics of the internal states as an adiabatic process with respect to the  shape deformations of the membrane, and the diffusion of its components within the membrane. 
	
	By using this adiabatic approximation, the densities~$\rho_n^k$ can be treated as \cc{rapidly relaxing fields that adjust to an instantaneous nonequilibrium steady state}, which must satisfy the following flux relations:
	\begin{equation}
		\label{eqn:adiabatic-condition-general}
		\sum_{m\hspace{0.5pt}\in\hspace{1pt}\mathfrak{I}_{\hspace{-1pt}n}^{\hspace{-0.15pt}k}}(r^k_{nm}\rho^k_m - r^k_{mn}\rho^k_n)=0,\;\,\mathrm{or}\;\sum_{m\hspace{0.5pt}\in\hspace{1pt}\mathfrak{I}_{\hspace{-1pt}n}^{\hspace{-0.15pt}k}}\mathbb{J}^{k}_{nm} = 0,
	\end{equation} for all states $n$ in cycle $k$. This follows from Eq.~(\ref{eqn:mass-balance}) in which the \cc{material-derivative and local surface-dilation terms}, together with the diffusive terms, are neglected \cc{relative to the internal kinetics}. Note that the internal currents $\mathbb{J}^{k}_{nm}\hspace{-1pt}= r^k_{nm}\phi^k_m-r^k_{mn}\phi^k_n$ vanish identically at thermodynamic equilibrium due to the detailed balance condition. In a {nonequilibrium steady state, detailed balance is no longer satisfied, and the currents} $\mathbb{J}^{k}_{nm}\hspace{-1pt}\neq 0$ in general. Since the transitions are cyclic (see Fig.~\ref{fig:cycles}), {Eq.~(\ref{eqn:adiabatic-condition-general}) reduces, using Eq.~(\ref{eqn:index-set}), to}
	\begin{equation}
		\label{eqn:adiabatic-condition-general-new}
		\begin{cases}
			\,\mathbb{J}^{k}_{0,1} = 0,               
			& \quad\mathrm{if}~n=0,\\[2pt]
			\,\mathbb{J}^{k}_{1,0} + \mathbb{J}^{k}_{1,2} + \mathbb{J}^{k}_{1,\aleph_k} = 0,       
			& \quad\mathrm{if}~n=1,\\[2pt]
			\,\mathbb{J}^{k}_{n,n-1} + \mathbb{J}^{k}_{n,n+1} = 0,                    
			& \quad\mathrm{otherwise}.
		\end{cases}
	\end{equation} Due to the antisymmetric property, that is, $\mathbb{J}^k_{nm}=-\mathbb{J}^k_{mn}$, we can easily see that the only non-trivial solution to this recurrence equation is a constant current solution,
	\begin{equation}
		\label{eqn:Jnm-currents}
		\boxed{\mathbb{J}^{k}_{n+1,n} = \bJk\;\;\textnormal{and}\;\;\mathbb{J}^{k}_{n,n+1} = -\bJk,}
	\end{equation} for all $k$, and $n\in\left\{1,2,\,\dots,\aleph_k\right\}$, with the property that the state $1+\aleph_k$ is the same as state $1$. Here, the {loop current} $\bJk\geq0$, such that the internal currents $\mathbb{J}^k_{nm}$ along a forward cyclic direction (\mbox{$1\rightarrow2\rightarrow3\rightarrow\dots\rightarrow1$}) are strictly positive, and negative otherwise. In addition to this, we must have the {transition currents}
	\begin{equation}
		\label{eqn:J01-currents}
		\mathbb{J}^{k}_{0,1} = -\mathbb{J}^{k}_{1,0} = 0,
	\end{equation} for each {species $k$}. \cc{Note that if the cyclic part of cycle $k$ contains only two internal states, excluding the idle state $n=0$, then $\bJk=0$. With only a single pair of transitions between these states, stationarity requires the forward and reverse fluxes to balance. Therefore, a nonzero loop current requires at least three internal states in the cycle} ($\aleph_k>2$). 
    The nonzero current $\bJk$ is a nonequilibrium feature of the system \cc{that describes the steady circulation through the internal mechanochemical cycle} (vanishing identically at equilibrium {by detailed balance}).
	
	The {internal-state fractions} $\phi_n^k$ are determined by solving the simultaneous equations in (\ref{eqn:Jnm-currents}) and (\ref{eqn:J01-currents}), which can be rewritten in the matrix form
	\begin{equation*}  
		\underbrace{\begin{bmatrix}
				-r^k_{1,0} & r^k_{0,1} & 0 & 0 & \cdots & 0 \\[4pt]
				0 & \!-r^k_{2,1} & r^k_{1,2} & 0 & \cdots & 0 \\[4pt]
				0 & 0 & \!-r^k_{3,2} & r^k_{2,3} & \cdots & 0 \\[1pt]
				\vdots  & \vdots  & \vdots & \vdots & \ddots & \vdots  \\[4pt]
				0 & r^k_{\aleph_k,1} & 0 & 0 & \cdots & \!-r^k_{1, \aleph_k}\,
		\end{bmatrix}}_{=\,\mathsf{R}_k\,\textnormal{(a square matrix of order $1+\aleph_k$)}}
		\!\overbrace{\begin{bmatrix}
				\phi_0^k \\[4pt]
				\phi_1^k \\[4pt]
				\phi_2^k \\[1pt]
				\vdots\\[4pt]
				\phi_{\aleph_k}^k
		\end{bmatrix}}^{=\,\mathsf{f}_k}
		=\overbrace{\begin{bmatrix}
				0 \\[4pt]
				\,\bJk \\[4pt]
				\,\bJk \\[1pt]
				\vdots\\[4pt]
				\bJk
		\end{bmatrix}}^{=\,\mathsf{j}_k}
	\end{equation*}
	and thus succinctly expressed as follows: $\mathsf{R}_k\hspace{0.75pt}\mathsf{f}_k\hspace{-0.5pt}=\mathsf{j}_k$. This can be easily solved by a matrix inversion, provided that $\det\mathsf{R}_k\neq0$; namely, $\mathsf{f}_k = \mathsf{R}_k^{-1}\mathsf{j}_k$, where $\mathsf{R}_k^{-1}$ is the inverse matrix of $\mathsf{R}_k$. This yields the solution of the mass fractions $\phi^k_n$ in terms of the loop current~$\bJk$; namely,
	\begin{equation}
		\label{eqn:mass-fracs-after-oneloop}
		\phi^k_n = {\bJk}\,/\,{\mathbb{R}^k_n},
	\end{equation} where $\mathbb{R}^k_n$ is a coefficient that depends solely on the transition rates $r^k_{nm}$. 
    The loop current $\bJk$ is not an independent parameter, and its explicit form can be obtained in terms of the {total areal fraction associated with cycle $k$}, which we define as follows:
	\begin{equation}
		\boxed{\Phi_k\!=\!\sum_{n=0}^{\aleph_k}\hspace{1pt}\phi_n^k.}
    \end{equation} Thus, the internal loop current can be rewritten as
    \begin{equation}
    	\label{eqn:bJK-equation}
    	\boxed{\bJk=\Phi_k\hspace{1pt}\bRk\,}
    \end{equation}
    with the parameter $\bRk$ satisfying ${1}/{\bRk}=\sum_{n=0}^{\aleph_k}{1}/{\mathbb{R}_n^k}$. \cc{A more explicit expression for $\bRk$ in terms of the underlying transition rates follows from Eq.~(\ref{eqn:Jnm-currents}). Along each cyclic transition, the constant-current condition can be rearranged as $\phi^k_{n+1}=(r^k_{n+1,n}/r^k_{n,n+1})\phi^k_n-\bJk/r^k_{n,n+1}$. Iterating this relation once around the cycle and imposing $\phi^k_{\aleph_k+1}=\phi^k_1$ shows that}
    \begin{equation}
    	\cc{\bRk\propto
    	\prod_{n=1}^{\aleph_k}r^k_{n+1,n}
    	-
    	\prod_{n=1}^{\aleph_k}r^k_{n,n+1}.}
    \end{equation}
    \cc{Thus, at thermodynamic equilibrium, the {\it Kolmogorov cycle condition} requires the products of the forward and reverse transition rates to be equal \cite{Schnakenberg1976}, giving $\bRk=0$ and hence $\bJk=0$. Using Eq.~(\ref{eqn:bJK-equation}) in Eq.~(\ref{eqn:mass-fracs-after-oneloop}), the internal-state fractions can then be written as}
    \begin{equation}
    	\label{eqn:phikn}
    	\boxed{\phi^k_n = \Phi_k\hspace{1pt}{\bRk}/{\mathbb{R}^k_n},}
    \end{equation}
    which tells us that each of the {relative internal-state fractions} $\phi^k_n/\Phi_k$ is set entirely by the biochemical transition rates. Nonetheless, the {total areal fractions} $\Phi_k$ themselves are not determined by {these transition rates, but remain} slow dynamical variables, {whose evolution is discussed} in the next section.

    \cc{Under the adiabatic approximation, the mean-field free-energy density in Eq.~(\ref{eqn:meanfield-energy}) can be written in terms of the total component fractions $\Phi_0$ and $\Phi_k$. This follows by eliminating the internal states $n$ using Eq.~(\ref{eqn:phikn}), which expresses $\phi_n^k$ in terms of $\Phi_k$ and the transition rates of the $k$th cycle; namely,}
    \begin{equation}
        \label{eqn:meanfield-energy-reduced}
        \cc{V_{\mathrm{m}} = \frac{k_B T}{b_0\,}\Big[\Phi_0\big(\!\ln\Phi_0\!+\mathcal{E}_0\big)+\sum_{k}\Phi_k\big(\!\ln\Phi_k\!+\mathcal{E}_k\big)\Big]\hspace{-1pt},}
    \end{equation}
    \cc{where}
    \begin{equation}
        \label{eqn:effective-energy-k}
        \cc{\mathcal{E}_k =
        \sum\limits_{n=0}^{\aleph_k}
        \frac{\bRk}{\mathbb{R}_n^k}
        \Bigg[
        \mathcal{E}^k_n+
        \ln\frac{\bRk}{\mathbb{R}_n^k}
        \Bigg],}
    \end{equation}
    \cc{using
    ${1}/{\bRk}=\sum_{n=0}^{\aleph_k}{1}/{\mathbb{R}_n^k}$.
    Similarly, the local spontaneous curvature $C_\mathrm{m}$ in Eq.~(\ref{eqn:coupling-mean-curvature-energy}) can be written in terms of the total component fractions:}
    \begin{equation}
        \cc{C_\mathrm{m} = \mathcal{C}_0\Phi_0 + \sum_k\mathcal{C}_k\Phi_k},
    \end{equation}
    \cc{where the effective curvature-coupling coefficient}
    \begin{equation}
        \label{eqn:effective-curvature-k}
        \cc{\mathcal{C}_k=
        \sum\limits_{n=0}^{\aleph_k}
        \mathcal{C}^k_n\hspace{1pt}
        \frac{\bRk}{\mathbb{R}^k_n}.}
    \end{equation}

	\subsection{Dynamics of the areal fractions}
    
    In the adiabatic approximation, the total {membrane-area flux}, as given by Eq.~(\ref{eqn:rate-mass-flux}), can be expressed in terms of the loop currents as follows:
	\begin{equation}
		\label{eqn:net-rate-mass-flux}
		{\mathbb{M} = \sum_{k}\,\bJk\,\sum_{n=1}^{\aleph_k}\,\mathbb{M}^k_{n+1,n},}
	\end{equation} {where $\aleph_k+1\equiv1$}. This has been obtained by using the fact that the transition rates are cyclic, together with {the antisymmetry} $\mathbb{M}^k_{nm}=-\mathbb{M}^k_{mn}$; namely, the membrane area transferred during the biochemical transition $m\rightarrow n$ must have the same magnitude as that transferred during the reverse transition $n\rightarrow m$, but with opposite sign. Notice that the {membrane-area flux} $\mathbb{M}$ vanishes, as expected, if the system is at equilibrium. Thus, by using Eq.~(\ref{eqn:bJK-equation}), $\mathbb{M}$ can be rewritten as
	\begin{equation}
		\label{eqn:total-mass-flux-rate}
		\boxed{\mathbb{M} = \sum_k\,\Phi_k\,\bMk\,,\vspace{-1pt}}
	\end{equation} where $\bMk$ is the {membrane-area transfer rate associated with the $k$-th trafficking cycle}, namely
	\begin{equation}
		\label{eqn:mass-flux-rate}
		\bMk = \bRk\,\sum_{n=1}^{\aleph_k}\,\mathbb{M}^k_{n+1,n},
	\end{equation} whose sign is negative for $k=1$ (fissogens) and positive for $k=2$ (fusogens); specifically, we say that
	\begin{equation}
		\mathrm{sgn}\hspace{-1pt}\left(\bMk\right)  =(-1)^k,
	\end{equation} with $\mathrm{sgn}\hspace{-1pt}\left(\boldsymbol{\cdot}\right)$ denoting the {\it signum} function.
	
	Moreover, we assume that the {total} membrane density $\rho$ is a constant (say $\bar{\rho}$) {on length scales larger than} the coarse-graining length scale in our model (\cc{$\sim50$--$100$~nm}). In other words, the membrane is assumed to be locally incompressible, and thus gradients in the total density $\rho$ are not supported at that scale. As a corollary, Eq.~(\ref{eqn:mass-balance-total}) leads to the following condition:
    \begin{equation}
	\label{eqn:divergence-law-eq}
	\boxed{\Div\itbf{v} = \cc{\nabla_{\!\alpha}v^\alpha} - 2Hv =\mathbb{M},}
    \end{equation} where the identity (\ref{eqn:div-surface-3D}) is used. This equation acts as an imposed local constraint for the membrane area, with the surface tension $\Sigma$ being the Lagrange multiplier. 
	
	As a result, the kinematic equation for the mass fraction $\Phi_0$ in Eq.~(\ref{eqn:mass-fraction-0}), together with Eq.~(\ref{eqn:current-chemical-potential}), yields
	\begin{equation}
		\label{eqn:phi-0-dynamics}
		\dot{\Phi}_0 =\left(1-\Phi_0\right)\mathbb{M}-\sum_{k,n}\hspace{1pt}\frac{\Delta\mu_n^k}{\bar{\rho}\hspace{1.2pt}\zeta^k_n},
	\end{equation} where $\Delta$ is the two-dimensional Laplace--Beltrami operator.
	On the other hand, the dynamics of the {areal fractions} $\Phi_k$ is obtained from Eq.~(\ref{eqn:mass-fractions}), and given by
	\begin{equation}
		\label{eqn:phi-k-dynamics}
		\dot{\Phi}_k = -\Phi_k\hspace{1pt}\mathbb{M}+\sum_{n=0}^{\aleph_k}\hspace{1pt}\frac{\Delta\mu_n^k}{\bar{\rho}\hspace{1.2pt}\zeta^k_n}.
	\end{equation} Notice that by summing this equation with respect to $k$, and using that $1-\Phi_0 = \Phi_1 + \Phi_2$, we retrieve Eq.~(\ref{eqn:phi-0-dynamics}). 
	
	From the free-energy density~$F$ in Eq.~(\ref{eqn:total-free-energy}), the local chemical potential difference is found to be
	\begin{equation}
		\label{eqn:chemical-potential-n-k}
		\mu_n^k\hspace{-1pt} = 2\kappa\hspace{0.25pt} H\hspace{-1pt}\left(\mathcal{C}_n^k-\mathcal{C}_0\right)+\frac{k_B T}{b_0\,}\!\hspace{-1pt}\left[\hspace{-1pt}\left(\mathcal{E}_n^k-\mathcal{E}_0\right)\!+\ln\frac{\phi_n^k}{\Phi_0}\right]\!\hspace{-1pt},
	\end{equation} which must vanish identically at equilibrium. By using the form of the chemical potential in Eq.~(\ref{eqn:chemical-potential-n-k}), this allows us to write the last term in Eq.~(\ref{eqn:phi-k-dynamics}) as follows:
	\begin{equation}
		\sum_{n=0}^{\aleph_k}\frac{\Delta\mu_n^k}{\bar{\rho}\hspace{1.2pt}\zeta^k_n} = 2\hspace{1pt}\Omega_k\hspace{1pt}\Delta H + \gamma_k\,\Delta\Big(\hspace{-1pt}\ln\frac{\Phi_k}{\Phi_0}\hspace{1pt}\Big),
	\end{equation} where the {effective surface diffusivities} $\gamma_k$ are defined by
	\begin{equation}
		\gamma_k = \frac{k_B T}{\bar{\rho}\hspace{0.5pt}b_0\,}\,\sum_{n=0}^{\aleph_k}\hspace{1pt}\frac{1}{\zeta_n^k},
	\end{equation} and the effective constants $\Omega_k$ are given by
	\begin{equation}
		\Omega_k = \frac{\kappa\hspace{0.25pt}}{\bar{\rho}}\sum\limits_{n=0}^{\aleph_k}\frac{\mathcal{C}_n^k-\mathcal{C}_0}{\zeta_n^k}.
	\end{equation} \cc{As in Eqs.~(\ref{eqn:effective-energy-k}) and (\ref{eqn:effective-curvature-k}), these sums include the idle state $n=0$, since $\Phi_k$ includes all internal states. The membrane-area transfer sums in Eqs.~(\ref{eqn:net-rate-mass-flux}) and (\ref{eqn:mass-flux-rate}) instead begin at $n=1$, because transitions to and from the idle state carry no loop current, cf.~Eq.~(\ref{eqn:J01-currents}).}
    
    Accordingly, Eq.~(\ref{eqn:phi-k-dynamics}) can be rewritten as follows:
	\begin{equation}
		\label{eqn:phi-k-dynamics-expanded}
		\boxed{\dot{\Phi}_k = -\Phi_k\hspace{1pt}\mathbb{M}+2\hspace{1pt}\Omega_k\hspace{1pt}\Delta H + \gamma_k\,\Delta\Big(\hspace{-1pt}\ln\frac{\Phi_k}{\Phi_0}\hspace{1pt}\Big),}
	\end{equation}  where the Laplacian term, by chain rule, expands to
	\begin{align}
	\Delta\Big(\hspace{-1.5pt}\ln\hspace{-1pt}\frac{\Phi_k}{\Phi_0}\Big)
	&=\frac{\Delta\Phi_k}{\Phi_k}
	-\frac{\cc{\nabla_{\!\alpha}\Phi_k\,\nabla^{\alpha}\Phi_k}}{(\Phi_k)^2}\notag\\[4pt]
	&\quad\cc{-\frac{\Delta\Phi_0}{\Phi_0}}
	+\frac{\cc{\nabla_{\!\alpha}\Phi_0\,\nabla^{\alpha}\Phi_0}}{(\Phi_0)^2}.
    \end{align} Thus, Eq.~(\ref{eqn:phi-k-dynamics-expanded}) represents the governing equation (at large scales) of the areal concentration of membrane components, \ie fissogens ($k=1$) and fusogens ($k=2$).

    \subsection{Active membrane stresses}
    
	\cc{The internal biochemical transitions associated with fission and fusion generate localized forces in the neighboring cytosolic fluid. These forces act within a finite region on the cytosolic side of the membrane, which we refer to as the active layer. Its characteristic thickness is of order the coarse-graining length \cc{($\sim\!50$--$100$~nm)}.}
	
	\cc{Because the forces are distributed over a finite normal distance, the fluid transmits both force and torque to the membrane. In the sharp-interface description derived below, these effects are retained through the active contributions $\itbf{T}_{\!\scriptstyle A}$ and $\itbf{M}_{\!\scriptstyle A}$ to the membrane traction and director traction, respectively, together with the {slip velocity} $\itbf{v}_{\!\scriptscriptstyle S}$. We first describe the force distribution within the finite layer and then derive this sharp-interface reduction for a layer whose thickness is small compared with the tangential length scales of the membrane~\cite{Lomholt2006,Lomholt2006a,Lomholt2006b,Salbreux2017}.}

	\subsubsection{Force distribution within the active layer}
	
    \cc{We denote by $\itbf{F}_{\hspace{-2.5pt}\scriptstyle A}(\itbf{r})$ the three-dimensional active body-force density exerted on the cytosolic fluid by the molecular machinery associated with the fission and fusion cycles. Using the surface coordinates $\xi^\alpha$ of the membrane compartment $\mathcal{M}$ and the normal distance $h$ from the membrane, this force density can be represented as}
    \begin{equation}
    	\label{eqn:F_A_def}
    	\itbf{F}_{\hspace{-2.5pt}\scriptstyle A}\!\left(\itbf{r}\right) = \int_{\mathcal{M}\;}\!\mathrm{d}S\int_{0}^{\infty}\!\mathrm{d}h\;\boldsymbol{\mathfrak{F}}_{\hspace{-1pt}\scriptstyle A}(h)\,\delta\big(\itbf{r}-\tilde{\!\itbf{R}}\big),
    \end{equation}
    where the position vector $\tilde{\!\itbf{R}}$ is defined by
    \begin{equation}
    	\label{eqn:active-layer-position}
    	\tilde{\!\itbf{R}}(\xi^1\hspace{-1pt},\hspace{1pt}\xi^2\hspace{-1pt},\hspace{0.5pt}h)= h\nv(\xi^1\hspace{-1pt},\hspace{1pt}\xi^2)+\itbf{R}(\xi^1\hspace{-1pt},\hspace{1pt}\xi^2).
    \end{equation}
    \cc{Here and below, the dependence of local surface fields on the membrane coordinates $\xi^\alpha$ is left implicit unless it is required explicitly. The function $\boldsymbol{\mathfrak{F}}_{\hspace{-1pt}\scriptstyle A}(h)$ is defined in these membrane-based coordinates such that $\boldsymbol{\mathfrak{F}}_{\hspace{-1pt}\scriptstyle A}\,\mathrm{d}S\,\mathrm{d}h$ gives the force associated with the normal interval between $h$ and $h+\mathrm{d}h$ above a membrane area element $\mathrm{d}S$.} \cc{The force profile is obtained by coarse-graining the microscopic forces associated with the fission and fusion machinery over a thin active layer of thickness $\Lambda$, whose scale is set by the size of transport vesicles. Accordingly, $\bfk(h)$ has support only over $0\leq h\leq\Lambda$ and vanishes for $h>\Lambda$.} \cc{As the area of a surface displaced by a normal distance $h$ differs from that of the membrane itself, the corresponding physical volume element is (see Fig.~\ref{fig:S2}):}
    \begin{equation}
    	\label{eqn:active-layer-Jacobian}
    	\cc{\mathrm{d}\tilde V=J(h)\,\mathrm{d}S\,\mathrm{d}h,\quad J(h)=1-2hH+h^2K.}
    \end{equation}
    \cc{For a point within the active layer located at $\tilde{\!\itbf{R}}(\xi^\alpha,h)$, the three-dimensional body-force density is therefore}
    \begin{equation}
    	\label{eqn:physical-active-force-density}
    	\cc{\itbf{F}_{\hspace{-2.5pt}\scriptstyle A}\big(\tilde{\itbf{R}}\,\big)={\boldsymbol{\mathfrak{F}}_{\hspace{-1pt}\scriptstyle A}(h)}/{J(h)}.}
    \end{equation}
    \cc{Hence, $\itbf{F}_{\hspace{-2.5pt}\scriptstyle A}$ represents the force per unit physical volume entering the bulk momentum balance, while $\boldsymbol{\mathfrak{F}}_{\hspace{-1pt}\scriptstyle A}$ gives the same force distribution per unit membrane area and unit normal distance. The normal moments introduced below are defined in terms of $\boldsymbol{\mathfrak{F}}_{\hspace{-1pt}\scriptstyle A}$.}
	
    \cc{We assume that, at the coarse-grained level considered here, the distribution of active forces has no preferred direction within the local tangent plane. The tangential components therefore average to zero, and we retain only the force component along the outward normal, writing}
	\begin{equation}
		\label{eqn:F_A}
		\boldsymbol{\mathfrak{F}}_{\hspace{-1pt}\scriptstyle A}(h) = \mathfrak{F}_{\hspace{-1pt}\scriptstyle A}(h)\nv,
	\end{equation}
	\cc{where $\mathfrak{F}_{\hspace{-1pt}\scriptstyle A}$ is the scalar normal force profile. It is obtained by summing the cycle-specific contributions,}
	\begin{equation}
		\label{eqn:F_A_scalar}
		\mathfrak{F}_{\hspace{-1pt}\scriptstyle A}(h) = \sum_k\Phi_k\,\bfk(h),
	\end{equation}
	which includes the corresponding active force distributions $\bfk$ due to fissogens ($k=1$) and fusogens ($k=2$) over their entire cycle. \cc{Analogous to the membrane-area flux in Eq.~(\ref{eqn:net-rate-mass-flux}), the force profile associated with cycle $k$ is obtained by summing the mechanical contributions from its internal transitions over one complete cycle. Within the adiabatic approximation, this gives}
    \begin{equation}
    	\bfk(h) = \bRk\,\sum_{n=1}^{\aleph_k}\mathbb{F}^k_{n+1,n}(h),
    \end{equation}
    \cc{where $\mathbb{F}^{k}_{nm}(h)$ denotes the momentum transferred during the active transition from the internal state $m$ to $n$ at normal height $h$. As for the membrane-area exchange, we assume that the transitions form a cyclic process and that $\mathbb{F}^k_{nm}=-\mathbb{F}^k_{mn}$. Using the loop current $\bJk=\Phi_k\bRk$, the total active force profile can thus be written as}
    \begin{equation}
    	\mathfrak{F}_{\hspace{-1pt}\scriptstyle A}(h)
    	=
    	\sum_k\bJk\sum_{n=1}^{\aleph_k}\mathbb{F}^k_{n+1,n}(h).
    \end{equation}
    
    \cc{The active force $\itbf{F}_{\hspace{-2.5pt}\scriptstyle A}$ is exerted on the neighboring cytosolic fluid and is localized within $0<h<\Lambda$. We denote by $\boldsymbol{\mathsf{\Omega}}^{(\Lambda)}_{\boldsymbol{+}}$ the stress in the outer cytosolic fluid when this finite active layer is retained explicitly. The momentum balance for $h>0$ is then}
    \begin{equation}
    	\label{eqn:external_fluid_and active_force} \cc{\bnabla\cdot\boldsymbol{\mathsf{\Omega}}^{(\Lambda)}_{\boldsymbol{+}}+\hspace{0.5pt}\itbf{F}_{\hspace{-2.5pt}\scriptstyle A}=\boldsymbol{0},}
    \end{equation}
    \cc{where $\itbf{F}_{\hspace{-2.5pt}\scriptstyle A}=\boldsymbol{0}$ for $h>\Lambda$, so that Eq.~(\ref{eqn:external_fluid_and active_force}) reduces to the unforced Stokes equation outside the active layer.}
    
    \cc{Since $\Lambda$ is much smaller than the characteristic tangential variation length $L$ of the membrane geometry, composition, and outer fluid fields, we introduce the small parameter $\epsilon_\Lambda=\Lambda/L\ll1$ and take the corresponding sharp-interface limit. In this description, the finite active layer is no longer resolved explicitly, and the outer-fluid stress $\boldsymbol{\mathsf{\Omega}}_{\boldsymbol{+}}$ satisfies}
    \begin{equation}
    	\label{eqn:external_fluid_without_active_force}
    	\cc{\bnabla\cdot\boldsymbol{\mathsf{\Omega}}_{\boldsymbol{+}}=\boldsymbol{0},\qquad h>0.}
    \end{equation}
    \cc{The effects of the active layer are instead retained through effective active membrane stresses and a modified interfacial velocity condition at $h=0$ \cite{Lomholt2006, Lomholt2006a, Lomholt2006b}.}

    As written in Eq.~(\ref{eqn:external_fluid_and active_force}), $\itbf{F}_{\hspace{-2.5pt}\scriptstyle A}$ enters the bulk momentum balance as a body-force density. \cc{We assume that the net force associated with each coarse-grained trafficking cycle vanishes locally. Since $\bfk(h)\,\mathrm{d}S\,\mathrm{d}h$ gives the force associated with the normal interval $\mathrm{d}h$ above a membrane area element $\mathrm{d}S$, this requires, at each position along the membrane,}
    \begin{equation}
    	\label{eqn:monopole}
    	\int_0^{\Lambda}\!\mathrm{d}h\;\bfk(h) = 0.
    \end{equation}
    \cc{The zeroth normal moment therefore vanishes. The first and second moments are consequently the first nonzero moments of the force distribution and, as shown below, provide the leading contributions retained in the sharp-interface description.}
	
	\cc{The first and second normal moments associated with the $k$-th trafficking cycle are defined by}
    \begin{equation}
    	\label{eqn:cycle-active-force-dipole}
    	\boxed{\bPk=\int_0^\Lambda\!\mathrm{d}h\;h\,\bfk(h),}
    \end{equation}
    \cc{and}
    \begin{equation}
    	\label{eqn:cycle-active-force-quadrupole}
    	\boxed{\bQk=\int_0^\Lambda\!\mathrm{d}h\;h^2\,\bfk(h),}
    \end{equation}
    \cc{respectively. Using Eq.~(\ref{eqn:F_A_scalar}), the corresponding total first and second moments of the active force profile are}
    \begin{equation}
    	\label{eqn:active-force-dipole}
    	\mathbb{P}=\sum_k\Phi_k\bPk,
    	\qquad
    	\mathbb{Q}=\sum_k\Phi_k\bQk.
    \end{equation}  
    \cc{The moments $\bPk$ and $\bQk$ are determined by the corresponding cycle-specific force profile and may independently take either sign. Their signs are not fixed by the type of trafficking event alone, although particular force profiles may favor specific sign choices.}
    
    For illustrative purposes, we consider a simple point-force dipolar model:
    \begin{equation}
    	\label{eqn:point-force-moment-model}
    	\bfk(h)\!=f^k_0\left[\delta\!\left(h-\Lambda\right)-\delta(h)\right]\!,
    \end{equation}
    \cc{where $f_0^k$ sets the magnitude and sign of the cycle-averaged force profile for cycle $k$.} Note that the monopole of this force distribution is identically zero, by construction. This is a simple model that captures the magnitude of the momentum transfer and its characteristic separation along the membrane normal. In this simple model, the first and second moments of the force {profile} are
    \begin{equation}
    	\label{eqn:point-force-moments}
    	\bPk=f^k_0\Lambda,\qquad\bQk=f^k_0\Lambda^2,
    \end{equation}
    respectively. \cc{Hence, for this particular force profile, $\bPk$ and $\bQk$ necessarily have the same sign, which is set by $f_0^k$. The sign of $f_0^k$ is not determined simply by whether the event is fission or fusion.}

    \cc{One possible physical interpretation of this sign follows if the distal force is dominated by Stokesian drag on the bud or transport vesicle. The drag exerted by the fluid opposes its motion, so the equal and opposite force exerted by the bud or vesicle on the fluid is directed along its motion. Taking the outward normal as positive, this gives $f_0^k>0$ for fission ($k=1$), when the bud moves outward, and $f_0^k<0$ for fusion ($k=2$), when the vesicle moves towards the membrane. For this drag-dominated realisation, Eq.~(\ref{eqn:point-force-moments}) therefore gives $\mathrm{sgn}(\bPk)=\mathrm{sgn}(\bQk)=(-1)^{k+1}$. The corresponding force dipoles are therefore extensile for fission and contractile for fusion. This sign relation is specific to this illustrative force profile and is not imposed in the general model.}
	
    As another illustrative example, \cc{we consider a different force profile to show that the first and second moments need not have the same sign. Specifically,} we consider the force \cc{profile} $\bfk$ to be \cc{quadratic in the dimensionless normal coordinate $h/\Lambda$}, that is,
    \begin{equation}
    	\bfk(h) = f^k_0+f^k_1\left({h}/{\Lambda}\right)+f^k_2\left({h}/{\Lambda}\right)^2,
    \end{equation}
    where $f^k_0$, $f^k_1$ and $f^k_2$ are some constants. By imposing that force monopoles vanish, from Eq.~(\ref{eqn:monopole}), we find
    \begin{equation}
    	f_0^k = -\frac{1}{6}\left(3f_1^k+2f_2^k\right),
    \end{equation}
    which yields the first moment
    \begin{equation}
    	\label{eqn:bPk-toy-example}
    	\bPk = \frac{1}{12}(f_1^k+f_2^k)\Lambda^2,
    \end{equation}
    while the second moment is
    \begin{equation}
    	\label{eqn:bQk-toy-example}
    	\bQk = \frac{1}{180}(15f_1^k+16f_2^k)\Lambda^3.
    \end{equation}
    In this case, the sign of $\bPk$ can be different from the sign of $\bQk$, provided that $-1<f_2^k/f_1^k<-15/16$. \cc{Thus, unlike the two-point force profile above, this quadratic force profile can produce first and second moments with opposite signs.}

    \subsubsection{Locally planar sharp-interface reduction}

    \cc{To relate the explicit finite-thickness description to its sharp-interface reduction, we compare the corresponding stresses in the outer cytosolic fluid. In the finite-thickness formulation, the active layer of thickness $\Lambda$ is retained explicitly; for brevity, we refer to this below as the {\it finite-$\Lambda$ description}. In the sharp-interface description, the active layer is instead collapsed onto the membrane at $h=0$. The boundary conditions in the sharp-interface description are chosen so that its fluid solution coincides with that of the finite-$\Lambda$ description outside the active layer, $h>\Lambda$. We define the resulting excess bulk stress by}
    \begin{equation}
        \label{eqn:active-excess-stress-definition}
        \cc{\boldsymbol{\mathsf{\Omega}}_{\scriptstyle A}=\boldsymbol{\mathsf{\Omega}}^{(\Lambda)}_{\boldsymbol{+}}-\boldsymbol{\mathsf{\Omega}}_{\boldsymbol{+}}.}
    \end{equation}
    \cc{Here and below, the superscript ${\scriptstyle(\Lambda)}$ identifies quantities in the finite-$\Lambda$ description, while the corresponding quantities without this superscript are those of the sharp-interface description. Accordingly, the excess stress vanishes outside the active layer, $h>\Lambda$. Defining similarly the excess pressure and velocity by}
    \begin{equation}
        \label{eqn:active-excess-pressure-velocity}
        \cc{p_A=p^{(\Lambda)}_{\boldsymbol{+}}-p_{\boldsymbol{+}},\qquad
        \itbf{V}_{\!A}=\itbf{V}^{(\Lambda)}_{\boldsymbol{+}}-\itbf{V}_{\!{\boldsymbol{+}}},}
    \end{equation}
    \cc{the excess stress for a Newtonian fluid is}
    \begin{equation}
        \label{eqn:active-excess-Newtonian-stress}
        \cc{\boldsymbol{\mathsf{\Omega}}_{\scriptstyle A}=-p_A\itbf{I}+\eta\left[\bnabla\itbf{V}_{\!A}+\left(\bnabla\itbf{V}_{\!A}\right)^{\!\mathsf{T}}\right].}
    \end{equation}

    \cc{We first determine these excess fields locally for a planar membrane, for which $H=K=0$ and $J=1$, while allowing variations along both tangential directions. We denote the tangential coordinates by $x^\alpha$, with $\alpha=1,2$, and the corresponding gradient by $\bnabla_{\!\parallel}$. Since $\epsilon_\Lambda=\Lambda/L\ll1$, normal variations across the active layer occur over the scale $\Lambda$, whereas tangential variations occur over the much larger scale $L$, so that $\partial_h\sim\Lambda^{-1}$ and $\bnabla_{\!\parallel}\sim L^{-1}$. Although $h/\Lambda$ is generally of order unity within $0\leq h\leq\Lambda$, the ratio $h/L\leq\epsilon_\Lambda$ remains small throughout the active layer. We decompose the excess velocity into tangential and normal components as}
    \begin{equation}
    	\label{eqn:flat-active-velocity-decomposition}
    	\cc{\itbf{V}_{\!A}(\itbf{x}_{\parallel},h)=\itbf{u}_{\!A}(\itbf{x}_{\parallel},h)+w_A(\itbf{x}_{\parallel},h)\nv,}
    \end{equation}
    \cc{where $\itbf{x}_{\parallel}=(x^1,x^2)$ and $\itbf{u}_{\!A}$ contains both tangential components. Subtracting the homogeneous Stokes equations of the sharp-interface description from those of the finite-$\Lambda$ description gives, within $0<h<\Lambda$,}
    \begin{equation}
    	\label{eqn:flat-active-tangential-momentum}
    	\cc{-\bnabla_{\!\parallel}p_A+\eta\left(\partial_h^2\itbf{u}_{\!A}+\nabla_{\!\parallel}^2\itbf{u}_{\!A}\right)=\boldsymbol{0},}
    \end{equation}
    \begin{equation}
    	\label{eqn:flat-active-normal-momentum}
    	\cc{-\partial_hp_A+\eta\left(\partial_h^2w_A+\nabla_{\!\parallel}^2w_A\right)+\mathfrak{F}_{\hspace{-1pt}\scriptstyle A}=0,}
    \end{equation}
    \cc{together with the incompressibility condition}
    \begin{equation}
    	\label{eqn:flat-active-incompressibility}
    	\cc{\bnabla_{\!\parallel}\cdot\itbf{u}_{\!A}+\partial_hw_A=0.}
    \end{equation}
    
    \cc{If $\itbf{u}_{\!A}$ has characteristic magnitude $U$, the incompressibility condition gives $w_A\sim\epsilon_\Lambda U$. In Eq.~(\ref{eqn:flat-active-tangential-momentum}), the viscous term involving tangential derivatives, $\nabla_{\!\parallel}^2\itbf{u}_{\!A}$, is smaller than $\partial_h^2\itbf{u}_{\!A}$ by a factor $\epsilon_\Lambda^2$. Similarly, the viscous terms in the normal momentum balance are higher order relative to the normal pressure gradient. To leading order in $\epsilon_\Lambda$, Eqs.~(\ref{eqn:flat-active-tangential-momentum}) and (\ref{eqn:flat-active-normal-momentum}) reduce to}
    \begin{equation}
    	\label{eqn:flat-active-pressure-balance}
    	\cc{-\partial_hp_A+\mathfrak{F}_{\hspace{-1pt}\scriptstyle A}+\mathcal{O}(\epsilon_\Lambda^2) = 0,}
    \end{equation}
    \cc{and}
    \begin{equation}
    	\label{eqn:flat-active-tangential-balance}
    	\cc{-\bnabla_{\!\parallel}p_A + \eta\partial_h^2\itbf{u}_{\!A} + \mathcal{O}(\epsilon_\Lambda^2) = 0.}
    \end{equation}
    
    \cc{The tangential--tangential components of the excess viscous stress are smaller than the pressure contribution by the same order. In the locally planar Cartesian coordinates, $g^{\alpha\beta}$ reduces to $\delta^{\alpha\beta}$, with $\delta^{\alpha\beta}$ the Kronecker delta, so that}
    \begin{equation}
    	\label{eqn:flat-active-inplane-stress}
    	\cc{\Omega^{\alpha\beta}_{\scriptstyle A}=-p_A\delta^{\alpha\beta}+\mathcal{O}(\epsilon_\Lambda^2).}
    \end{equation}
    \cc{The tangential--normal component of the excess stress must nevertheless be retained. We define the corresponding tangential traction across a surface of constant $h$ by}
    \begin{equation}
    	\label{eqn:flat-active-shear-definition}
    	\cc{\itbf{\tau}_{\!A}=(\itbf{I}-\nv\nv)\cdot\boldsymbol{\mathsf{\Omega}}_{\scriptstyle A}\cdot\nv.}
    \end{equation}
    \cc{To the same order,}
    \begin{equation}
    	\label{eqn:flat-active-shear-velocity}
    	\cc{\itbf{\tau}_{\!A}=\eta\partial_h\itbf{u}_{\!A}+\mathcal{O}(\epsilon_\Lambda^2).}
    \end{equation}
    \cc{and Eq.~(\ref{eqn:flat-active-tangential-balance}) therefore gives}
    \begin{equation}
    	\label{eqn:flat-active-shear-balance}
    	\cc{\partial_h\itbf{\tau}_{\!A}=\bnabla_{\!\parallel}p_A+\mathcal{O}(\epsilon_\Lambda^2).}
    \end{equation}
    
    \cc{Since the two descriptions coincide outside the active layer, the excess pressure vanishes at its outer edge, $p_A(\Lambda)=0$. Integrating Eq.~(\ref{eqn:flat-active-pressure-balance}) with respect to the normal coordinate from $h$ to $\Lambda$ then gives}
    \begin{equation}
    	\label{eqn:flat-active-pressure-profile}
    	\cc{p_A(\itbf{x}_{\parallel},h)=-\int_h^\Lambda\!\mathrm{d}s\;\mathfrak{F}_{\hspace{-1pt}\scriptstyle A}(\itbf{x}_{\parallel},s)+\mathcal{O}(\epsilon_\Lambda^2).}
    \end{equation}
    \cc{Using the force-free condition in Eq.~(\ref{eqn:monopole}), Eq.~(\ref{eqn:flat-active-pressure-profile}) also shows that the leading-order excess pressure vanishes at the membrane, $p_A(\itbf{x}_{\parallel},h=0)=0$. Integrating the pressure profile across the active layer and interchanging the order of integration then gives}
    \begin{equation}
    	\label{eqn:flat-pressure-first-identity}
    	\cc{-\int_0^\Lambda\!\mathrm{d}h\;p_A(h)=\int_0^\Lambda\!\mathrm{d}h\;h\,\mathfrak{F}_{\hspace{-1pt}\scriptstyle A}(h)=\mathbb{P}.}
    \end{equation}  \cc{Also, its first moment about the membrane gives}
    \begin{equation}
    	\label{eqn:flat-pressure-second-identity}
    	\cc{\int_0^\Lambda\!\mathrm{d}h\;h\,p_A(h)=-\frac{1}{2}\int_0^\Lambda\!\mathrm{d}h\;h^2\,\mathfrak{F}_{\hspace{-1pt}\scriptstyle A}(h)=-\frac{1}{2}\mathbb{Q}.}
    \end{equation}

    \cc{In the sharp-interface limit, the detailed stress distribution across the active layer is no longer resolved, but the net force and torque transmitted by the layer must be preserved. The effective in-plane stress is therefore obtained by integrating the tangential excess stress across the layer. In the locally planar limit, $b^\beta_{\gamma}=0$, so Eq.~(\ref{eqn:def-sigma}) reduces to $T^{\alpha\beta}_{\hspace{-1.5pt}\scriptstyle A}=\sigma^{\alpha\beta}_{\hspace{-1.5pt}\scriptstyle A}$, and we obtain}
    \begin{equation}
    	\label{eqn:flat-active-surface-stress}
    	\cc{\sigma^{\alpha\beta}_{\hspace{-1.5pt}\scriptstyle A}
    	=
    	-\int_0^\Lambda\!\mathrm{d}h\;p_A(h)\delta^{\alpha\beta}
    	=
    	\mathbb{P}\,\delta^{\alpha\beta}.}
    \end{equation}
    
    \cc{Because the excess stress is distributed at different normal distances from the membrane, it also transmits a torque about the surface $h=0$. Weighting the stress by the corresponding lever arm $h$ therefore gives the effective bending moment. Using the definition of the bending moment tensor introduced in Eq.~(\ref{eqn:moment-bending-tensor}), we obtain}
    \begin{equation}
    	\label{eqn:flat-active-bending-moment}
    	\cc{M^{\alpha\beta}_{\hspace{-1pt}\scriptstyle A}
    	=
    	\int_0^\Lambda\!\mathrm{d}h\;h\,p_A(h)\delta^{\alpha\beta}
    	=
    	-\frac{1}{2}\mathbb{Q}\,\delta^{\alpha\beta}.}
    \end{equation}
    
    \cc{The pressure profile also determines the tangential traction transmitted through the active layer. Since the two descriptions coincide at the outer edge of the layer, the excess tangential traction vanishes, $\itbf{\tau}_{\!A}(h=\Lambda)=\boldsymbol{0}$. Integrating Eq.~(\ref{eqn:flat-active-shear-balance}) with respect to the normal coordinate from $h$ to $\Lambda$ gives}
    \begin{equation}
    	\label{eqn:flat-active-shear-profile}
    	\cc{\itbf{\tau}_{\!A}(h)
    	=
    	-\bnabla_{\!\parallel}\int_h^\Lambda\!\mathrm{d}s\;p_A(s).}
    \end{equation}
    \cc{Evaluating Eq.~(\ref{eqn:flat-active-shear-profile}) at the membrane, $h=0$, and using Eq.~(\ref{eqn:flat-pressure-first-identity}) gives the active-layer contribution to the tangential traction,}
    \begin{equation}
    	\label{eqn:flat-active-excess-traction}
    	\cc{\itbf{\tau}_{\!A}(h=0)=\bnabla_{\!\parallel}\mathbb{P}.}
    \end{equation}
    \cc{Using the definition of the excess stress in Eq.~(\ref{eqn:active-excess-stress-definition}), together with $\partial_\beta \sigma^{\alpha\beta}_{\hspace{-1.5pt}\scriptstyle A}=\partial^\alpha\mathbb{P}$ from Eq.~(\ref{eqn:flat-active-surface-stress}), we obtain}
    \begin{equation}
    	\label{eqn:flat-active-traction-matching}
    	\cc{\itbf{\tau}^{(\Lambda)}_{\boldsymbol{+}}(h=0)
    	=
    	\itbf{\tau}_{\boldsymbol{+}}(h=0)
    	+
    	\partial_\beta \sigma^{\alpha\beta}_{\hspace{-1.5pt}\scriptstyle A}\,\ed{\alpha}.}
    \end{equation}
    \cc{Equation~(\ref{eqn:flat-active-traction-matching}) makes the correspondence between the two descriptions explicit: the difference between the fluid traction at the physical membrane in the finite-$\Lambda$ description and the fluid traction in the sharp-interface description is precisely the surface force generated by the active membrane surface stress $\sigma^{\alpha\beta}_{\hspace{-1.5pt}\scriptstyle A}$. Thus, when the finite active layer is collapsed onto the membrane, the effect of the stresses distributed across that layer is retained through this active surface stress, while $\itbf{\tau}_{\boldsymbol{+}}(h=0)$ remains the fluid traction acting directly on the membrane. The same tangential force balance is therefore preserved.}
    	
    \cc{Finally, the finite active layer also modifies the relation between the tangential fluid and membrane velocities in the sharp-interface description. In the finite-$\Lambda$ description, the cytosolic fluid obeys no slip at the physical membrane, whereas in the sharp-interface description its tangential velocity at the membrane differs from the membrane velocity by $\itbf{v}_{\!\scriptscriptstyle S}$. Since $\itbf{V}_{\!A}=\itbf{V}^{(\Lambda)}_{\boldsymbol{+}}-\itbf{V}_{\!{\boldsymbol{+}}}$, and the two fluid solutions coincide at the outer edge of the active layer, the tangential excess velocity satisfies}
    \begin{equation}
    	\label{eqn:flat-active-slip-boundaries}
    	\cc{\itbf{u}_{\!A}(h=0)=-\itbf{v}_{\!\scriptscriptstyle S},\qquad\itbf{u}_{\!A}(h=\Lambda)=\boldsymbol{0}.}
    \end{equation}
    \cc{The difference between these two boundary values can be written as the integral of the normal derivative of the excess tangential velocity across the active layer,}
    \begin{equation}
    	\label{eqn:flat-active-slip-integral}
    	\cc{\itbf{v}_{\!\scriptscriptstyle S}
    	=
    	\int_0^\Lambda\!\mathrm{d}h\;\partial_h\itbf{u}_{\!A}
    	=
    	\frac{1}{\eta}\int_0^\Lambda\!\mathrm{d}h\;\itbf{\tau}_{\!A}(h),}
    \end{equation}
    \cc{where the last equality follows from Eq.~(\ref{eqn:flat-active-shear-velocity}). Substituting the shear profile from Eq.~(\ref{eqn:flat-active-shear-profile}) and interchanging the order of integration gives}
    \begin{equation}
    	\label{eqn:flat-active-slip-pressure}
    	\cc{\itbf{v}_{\!\scriptscriptstyle S}
    	=
    	-\frac{1}{\eta}\bnabla_{\!\parallel}\int_0^\Lambda\!\mathrm{d}h\;h\,p_A(h).}
    \end{equation}
    \cc{Using Eq.~(\ref{eqn:flat-pressure-second-identity}), which relates this integral to the second moment $\mathbb{Q}$ of the active force profile, we obtain}
    \begin{equation}
    	\label{eqn:active-slip-flat}
    	\cc{\boxed{\itbf{v}_{\!\scriptscriptstyle S}=\frac{1}{2\eta}\bnabla_{\!\parallel}\mathbb{Q}.}}
    \end{equation}

	\subsubsection{Sharp-interface reduction for a curved membrane}
    
    \begin{figure}[t]\includegraphics[width=0.95\columnwidth]{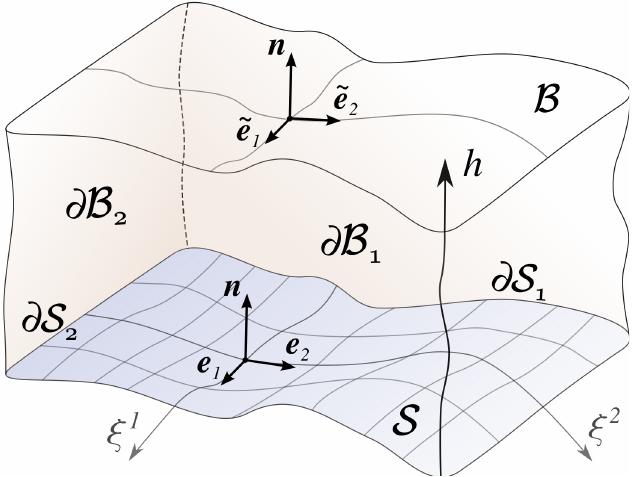}
		\caption{\label{fig:S1} Diagram of a three-dimensional cell that includes the ambient fluid above a membrane surface patch $\mathcal{S}$, where the side boundaries are labeled by $\partial\mathcal{S}_\alpha$ with $\alpha\in\{1,2\}$. Each point above the membrane is parameterized by the internal curvilinear coordinates $\xi^\alpha$ and the height $h$ along the surface normal $\nv$. Also, the side faces of the volume cell~$\mathcal{B}$, that are associated with the boundaries $\partial\mathcal{S}_\alpha$, are indicated by $\partial\mathcal{B}_\alpha$. Note that local tangent basis at height $h$ above the membrane surface are slightly different than those on $\mathcal{S}$, due to the local change in curvature $b_{\alpha\beta}$; namely, $\tilde{\itbf{e}}_\alpha = \ed{\beta}\hspace{0.5pt}\big(\delta^\beta_\alpha-h\hspace{1pt}b^\beta_\alpha\hspace{1pt}\big)$.
	}\end{figure}
    
    \cc{The locally planar calculation above shows how the force and torque transmitted through the finite active layer are retained, in the sharp-interface limit, through an effective membrane stress, bending moment, and \cc{slip velocity}. We now extend the same construction to a curved membrane. The finite-$\Lambda$ and sharp-interface descriptions must transmit the same net force across any surface cutting through the active layer. In particular, the force carried by the excess bulk stress across a side face $\partial\mathcal{B}_\alpha$ must be equal to that transmitted by the active membrane traction across the corresponding boundary curve $\partial\mathcal{S}_\alpha$.} Thus, we must impose that
    \begin{equation}
    	\label{eqn:excess-stress-condition}
    	\int_{\partial\mathcal{B}_\alpha}\!\!\mathrm{d}\tilde{S}\,\;\tilde{\boldsymbol{\nu}}\cdot\boldsymbol{\mathsf{\Omega}}_{\scriptstyle A} = \int_{\partial\mathcal{S}_\alpha}\!\!\mathrm{d}l\;\nu_\alpha\,\itbf{T}^\alpha_{\hspace{-2.5pt}\scriptstyle A}.
    \end{equation}  
    \cc{Here, $\partial\mathcal{B}_\alpha$ denotes the side face obtained by extending the corresponding boundary curve $\partial\mathcal{S}_\alpha$ of the membrane patch $\mathcal{S}$ along the surface normal, as illustrated in Fig.~\ref{fig:S1}; its intersection with the membrane at $h=0$ is $\partial\mathcal{S}_\alpha$. Parameterizing this boundary curve by $\lambda$, such that $\itbf{R}=\itbf{R}[\xi^1(\lambda),\xi^2(\lambda)]$, the unit vector $\tilde{\boldsymbol{\nu}}$ normal to the side face $\partial\mathcal{B}_\alpha$ is given by}
    \begin{equation}
    	\tilde{\boldsymbol{\nu}}  = \textstyle\left(\tilde{\itbf{e}}_\beta \frac{\mathrm{d}\xi^\beta}{\mathrm{d}\lambda}\times\nv\right)\left|\tilde{\itbf{e}}_\beta \frac{\mathrm{d}\xi^\beta}{\mathrm{d}\lambda}\right|^{-1}\!\!,
    \end{equation}
    whereas the tangent basis are found to be
    \begin{equation}
    	\tilde{\itbf{e}}_\alpha=\cc{\partial_{\alpha}\,\tilde{\!\itbf{R}}} = \ed{\beta}\hspace{0.5pt}\big(\delta^\beta_\alpha-h\hspace{1pt}b^\beta_\alpha\hspace{1pt}\big).
    \end{equation}
    Furthermore, the area element $\mathrm{d}\tilde{S}$ can be expressed in this parametrization as $\mathrm{d}\tilde{S}\! = \!\mathrm{d}h\hspace{1pt}\mathrm{d}\lambda\left|\tilde{\itbf{e}}_\alpha \frac{\mathrm{d}\xi^\alpha}{\mathrm{d}\lambda}\right|$, leading to
    \begin{equation}
    	\mathrm{d}\tilde{S}\;\tilde{\boldsymbol{\nu}} = \mathrm{d}h\hspace{1pt}\mathrm{d}\lambda\;\varepsilon_{\alpha\gamma}\eu{\alpha}\big(\delta_\beta^\gamma-h\hspace{1pt}b_\beta^\gamma\hspace{1pt}\big)\frac{\mathrm{d}\xi^\beta}{\mathrm{d}\lambda},
    \end{equation}
    where the identity $\varepsilon_{\alpha\gamma}\eu{\alpha} = \ed{\gamma}\times\nv$ is used, with $\varepsilon_{\alpha\gamma}$ as the Levi-Civita tensor. Similarly, the line element is
    \begin{equation}
    	\mathrm{d}l\,\nu_\alpha  = \varepsilon_{\alpha\beta}\frac{\mathrm{d}\xi^\beta}{\mathrm{d}\lambda}.
    \end{equation}
    \cc{Substituting these geometrical relations into Eq.~(\ref{eqn:excess-stress-condition}), and requiring the force matching to hold for an arbitrary boundary curve, gives the active stress vectors}~\cite{Lomholt2006a,Lomholt2006b}:
    \begin{equation}
    	\label{eqn:linear-stress-vector-A}
    	\boxed{\itbf{T}^\alpha_{\hspace{-2.5pt}\scriptstyle A} = \!\int_{0\,}^{\,\infty}\!\mathrm{d}h\;\ed{\beta}\cdot\boldsymbol{\mathsf{\Omega}}_{\scriptstyle A}\left(\hspace{1pt} g^{\alpha\beta}\!-h\,\bar{b}^{\alpha\beta}\right)\!.}
    \end{equation} \cc{where $\bar{b}^{\alpha\beta} = 2H g^{\alpha\beta}-b^{\alpha\beta}$ is cofactor of the curvature tensor, cf.~Eq.~(\ref{eqn:cofactor-curvature}). Thus, Eq.~(\ref{eqn:linear-stress-vector-A}) is the curved-surface counterpart of the thickness-integrated excess stress used in the locally planar calculation above. It follows from the force-matching condition and the geometry of the surfaces displaced along the membrane normal, without yet making an approximation for the form of $\boldsymbol{\mathsf{\Omega}}_{\scriptstyle A}$. We next evaluate the excess stress in Eq.~(\ref{eqn:linear-stress-vector-A}) using the same separation between the normal scale $\Lambda$ and the tangential scale $L$ introduced in the locally planar calculation, with ratio $\epsilon_\Lambda=\Lambda/L\ll1$.}
    
    \cc{The physical body-force density entering the bulk momentum balance is $\mathfrak{F}_{\hspace{-1pt}\scriptstyle A}(h)\nv/J(h)$, cf.~Eq.~(\ref{eqn:physical-active-force-density}). The leading normal momentum balance within the curved active layer is therefore}
    \begin{equation}
    	\label{eqn:curved-active-pressure-balance}
    	\cc{-\partial_h p_A+\frac{\mathfrak{F}_{\hspace{-1pt}\scriptstyle A}(h)}{J(h)}=0,}
    \end{equation}
    \cc{where the neglected viscous terms are smaller by a relative factor of order $\epsilon_\Lambda^2$, as in the locally planar calculation. Since the finite-$\Lambda$ and sharp-interface descriptions coincide at the outer edge of the active layer, $p_A(\Lambda)=0$, and Eq.~(\ref{eqn:curved-active-pressure-balance}) gives}
    \begin{equation}
    	\label{eqn:curved-active-pressure-profile}
    	\cc{p_A(h)=-\int_h^\Lambda\!\mathrm{d}s\;\frac{\mathfrak{F}_{\hspace{-1pt}\scriptstyle A}(s)}{J(s)}.}
    \end{equation}
    \cc{The tangential--tangential components of the excess viscous stress are again smaller than the pressure contribution by a relative factor of order $\epsilon_\Lambda^2$. Hence, to the order required here, $\ed{\beta}\cdot\boldsymbol{\mathsf{\Omega}}_{\scriptstyle A}=-p_A\ed{\beta}$. Substituting this result into Eq.~(\ref{eqn:linear-stress-vector-A}) gives the in-plane part of the active stress vector as}
    \begin{equation}
    	\label{eqn:curved-active-stress-pressure}
    	\cc{T^{\alpha\beta}_{\hspace{-1.5pt}\scriptstyle A}
    	=
    	-\int_0^\Lambda\!\mathrm{d}h\;p_A(h)\left(g^{\alpha\beta}-h\bar{b}^{\alpha\beta}\right).}
    \end{equation}
    \cc{Using Eq.~(\ref{eqn:curved-active-pressure-profile}) and interchanging the order of integration, this becomes}
    \begin{equation}
    	\label{eqn:curved-active-stress-kernel}
    	\cc{T^{\alpha\beta}_{\hspace{-1.5pt}\scriptstyle A}
    	=
    	\int_0^\Lambda\!\mathrm{d}h\;\frac{\mathfrak{F}_{\hspace{-1pt}\scriptstyle A}(h)}{J(h)}
    	\left[h\,g^{\alpha\beta}-\frac{h^2}{2}\bar{b}^{\alpha\beta}\right].}
    \end{equation}
    
    \cc{The remaining expansion is purely geometrical. Since $h/L\leq\epsilon_\Lambda\ll1$ throughout the active layer, we use $J^{-1}(h)\simeq1+2hH$ together with $\bar{b}^{\alpha\beta}=2Hg^{\alpha\beta}-b^{\alpha\beta}$. The geometrical factor in Eq.~(\ref{eqn:curved-active-stress-kernel}) then becomes}
    \begin{equation}
    	\label{eqn:curved-active-stress-kernel-expansion}
    	\cc{\frac{h\,g^{\alpha\beta}-{h^2}\bar{b}^{\alpha\beta}/2}{J(h)}
    	\simeq
    	h\,g^{\alpha\beta}
    	+h^2(Hg^{\alpha\beta}+b^{\alpha\beta}/2).}
    \end{equation}
    \cc{The neglected terms are of order $(h/L)^2$ and involve the third and higher moments of the active force profile. Retaining only the first and second moments defined in Eq.~(\ref{eqn:active-force-dipole}), Eq.~(\ref{eqn:curved-active-stress-kernel}) therefore reduces to}
    \begin{equation}
    	\label{eqn:TAintegral}
    	T^{\alpha\beta}_{\hspace{-1.5pt}\scriptstyle A}\hspace{-1pt}
    	=
    	\int_0^\Lambda\!\mathrm{d}h\,\mathfrak{F}_{\hspace{-1pt}\scriptstyle A}(h)
    	\left[h\,g^{\alpha\beta}
    	+h^2\!\left(Hg^{\alpha\beta}+b^{\alpha\beta}/2\right)\right].
    \end{equation}
    \cc{Thus, the active in-plane stress tensor is}
    \begin{equation}
    	\label{eqn:active-inplane-traction-stress}
    	\boxed{T^{\alpha\beta}_{\hspace{-1.5pt}\scriptstyle A}
    	=
    	\mathbb{P}\,g^{\alpha\beta}
    	+\frac{1}{2}\mathbb{Q}\left(2Hg^{\alpha\beta}+b^{\alpha\beta}\right).}
    \end{equation}
    \cc{For a locally flat membrane, where the curvature tensor vanishes, $b^{\alpha\beta}=0$ and hence $H=0$, Eq.~(\ref{eqn:active-inplane-traction-stress}) reduces to the locally planar result in Eq.~(\ref{eqn:flat-active-surface-stress}). The additional term proportional to $\mathbb{Q}$ therefore arises from the local membrane curvature.}

    \begin{figure}[t]\includegraphics[width=0.97\columnwidth]{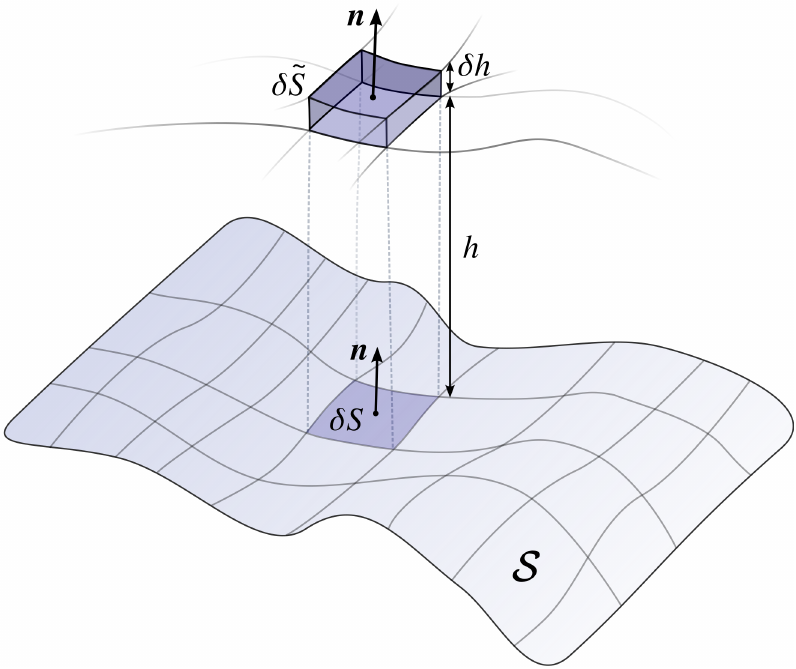}
		\caption{\label{fig:S2} Schematic of an infinitesimal volume $\delta\tilde{V}=\delta h\,\delta\tilde{S}$ of~the ambient fluid, which is positioned away from the membrane between the heights $h$ and $h+\delta h$, along the unit vector $\nv$ normal to the membrane $\mathcal{S}$. The area element $\delta\tilde S$ shown at height $h$ differs from its associated membrane area $\delta S$ because of the local curvature. \cc{They are related by $\delta\tilde{S}=J(h)\delta S$, where $J(h)=1-2hH+h^2K$, so that the physical volume of the segment is $\delta\tilde{V}=J(h)\delta S\,\delta h$. The force associated with this segment is $\mathfrak{F}_{\hspace{-1pt}\scriptstyle A}(h)\delta S\,\delta h\,\nv$, and the corresponding three-dimensional body-force density is therefore $\mathfrak{F}_{\hspace{-1pt}\scriptstyle A}(h)\nv/J(h)$.}}\end{figure}
    
    \cc{The same matching argument can be applied to the torque transmitted through the active layer. Requiring the integrated torque to be identical in the finite-$\Lambda$ and sharp-interface descriptions gives}
    \begin{equation}
    	\label{eqn:active-torque-matching}
        \int_{\partial\mathcal{B}_\alpha}\!\tilde{\!\itbf{R}}\times\left(\tilde{\boldsymbol{\nu}}\cdot\boldsymbol{\mathsf{\Omega}}_{\scriptstyle A}\right)\mathrm{d}\tilde{S}
    	=
    	\int_{\partial\mathcal{S}_\alpha}\!\!\nu_\alpha
    	\left(\nv\times\itbf{M}_{\hspace{-2pt}\scriptstyle A}^{\alpha}
    	+\itbf{R}\times\itbf{T}^\alpha_{\hspace{-2.5pt}\scriptstyle A}\right)\mathrm{d}l.
    \end{equation}
    \cc{By the same geometrical construction as for Eq.~(\ref{eqn:linear-stress-vector-A}), the corresponding active director traction is~\cite{Lomholt2006a,Lomholt2006b}:}
    \begin{equation}
    	\label{eqn:director-stress-vector-A}
    	\itbf{M}_{\hspace{-2pt}\scriptstyle A}^{\alpha}
    	=
    	\int_0^\infty\!\mathrm{d}h\;h
    	\left(\ed{\beta}\cdot\boldsymbol{\mathsf{\Omega}}_{\scriptstyle A}\right)
    	\left(g^{\alpha\beta}-h\bar{b}^{\alpha\beta}\right).
    \end{equation}
    \cc{Using the pressure profile in Eq.~(\ref{eqn:curved-active-pressure-profile}) and the definition $M^{\alpha\beta}_{\hspace{-1pt}\scriptstyle A}=-\itbf{M}^{\alpha}_{\hspace{-2pt}\scriptstyle A}\cdot\eu{\beta}$, we obtain}
    \begin{equation}
    	\label{eqn:curved-active-bending-kernel}
    	\cc{M^{\alpha\beta}_{\hspace{-1pt}\scriptstyle A}
    	=
    	-\int_0^\Lambda\!\mathrm{d}h\;
    	\frac{\mathfrak{F}_{\hspace{-1pt}\scriptstyle A}(h)}{J(h)}
    	\left[\frac{h^2}{2}g^{\alpha\beta}
    	-\frac{h^3}{3}\bar{b}^{\alpha\beta}\right].}
    \end{equation}
    \cc{The leading contribution in Eq.~(\ref{eqn:curved-active-bending-kernel}) is proportional to $h^2\mathfrak{F}_{\hspace{-1pt}\scriptstyle A}$, whose integral over the active layer gives $\mathbb{Q}$. The curvature-dependent corrections carry one additional power of $h$ and contribute through the third and higher normal moments. Truncating consistently at the second moment therefore gives}
    \begin{equation}
    	\label{eqn:active-bending-moment}
    	\boxed{M^{\alpha\beta}_{\hspace{-1pt}\scriptstyle A}
    	=
    	-\frac{1}{2}\mathbb{Q}\,g^{\alpha\beta}.}
    \end{equation}
    
    In addition, due to the conservation of angular momentum, cf.~Eq.~(\ref{eqn:shear-bending-relation}), the active out-of-plane shear stress is thus found to be
    \begin{equation}
    	\label{eqn:active-out-of-plane-shear}
    	N^\alpha_{\hspace{-2pt}\scriptstyle A}
    	=
    	-\nabla_{\!\beta}M^{\beta\alpha}_{\hspace{-2pt}\scriptstyle A}
    	=
    	\frac{1}{2}g^{\alpha\beta}\nabla_{\!\beta}\mathbb{Q}.
    \end{equation}
    The symmetric in-plane stress tensor due to activity, which is defined by $\sigma^{\alpha\beta}_{\hspace{-1.5pt}\scriptstyle A}=T^{\alpha\beta}_{\hspace{-1pt}\scriptstyle A}-b^{\beta}_{\gamma}M^{\gamma\alpha}_{\hspace{-1pt}\scriptstyle A}$,  reduces to
    \begin{equation}
    	\label{eqn:active-symmetric-inplane-stress}
    	\boxed{\sigma^{\alpha\beta}_{\hspace{-1.5pt}\scriptstyle A}
    	=
    	\mathbb{P}\,g^{\alpha\beta}
    	+\mathbb{Q}\left(Hg^{\alpha\beta}+b^{\alpha\beta}\right).}
    \end{equation}
    This readily shows how the active force distribution can contribute to the total in-plane stress of the membrane, through its nonzero first and second moments.
    
    \cc{It remains to determine the corresponding \cc{slip velocity} on the curved membrane. To the order at which the first two normal moments are retained, the tangential momentum balance within the thin active layer has the same local form as in Eq.~(\ref{eqn:flat-active-shear-balance}), with the planar gradient replaced by the surface covariant gradient. Curvature corrections to this relation contain an additional factor of $h/L$ and contribute only through higher moments. Thus, the tangential excess traction satisfies}
    \begin{equation}
    	\label{eqn:curved-active-shear-balance}
    	\cc{\partial_h\itbf{\tau}_{\!A}
    	=
    	\eu{\alpha}\nabla_{\!\alpha}p_A.}
    \end{equation}
    \cc{Since $\itbf{\tau}_{\!A}(\Lambda)=\boldsymbol{0}$, integration across the active layer gives}
    \begin{equation}
    	\label{eqn:curved-active-shear-profile}
    	\cc{\itbf{\tau}_{\!A}(h)
    	=
    	-\eu{\alpha}\nabla_{\!\alpha}
    	\int_h^\Lambda\!\mathrm{d}s\;p_A(s).}
    \end{equation}
    \cc{Likewise, the leading relation between tangential traction and the normal gradient of the excess tangential velocity remains $\itbf{\tau}_{\!A}=\eta\partial_h\itbf{u}_{\!A}$. Hence, using the same boundary matching as in Eq.~(\ref{eqn:flat-active-slip-boundaries}), the {slip velocity} is}
    \begin{equation}
    	\label{eqn:curved-active-slip-pressure}
    	\cc{\itbf{v}_{\!\scriptscriptstyle S}
    	=
    	-\frac{1}{\eta}\eu{\alpha}\nabla_{\!\alpha}
    	\int_0^\Lambda\!\mathrm{d}h\;h\,p_A(h).}
    \end{equation}
    \cc{Using Eq.~(\ref{eqn:curved-active-pressure-profile}), the pressure moment appearing here can be written as}
    \begin{equation}
    	\label{eqn:curved-pressure-second-identity}
    	\cc{\int_0^\Lambda\!\mathrm{d}h\;h\,p_A(h)
    	=
    	-\frac{1}{2}\int_0^\Lambda\!\mathrm{d}h\;
    	\frac{h^2\,\mathfrak{F}_{\hspace{-1pt}\scriptstyle A}(h)}{J(h)}.}
    \end{equation}
    \cc{Expanding $J^{-1}(h)$ in $h/L$, the leading term is $-\mathbb{Q}/2$, while the first curvature-dependent correction involves the third moment and is beyond the order retained here. Hence we recover the sharp-interface slip velocity}
    \begin{equation}
    	\label{eqn:active-slip-curved}
    	\cc{\boxed{\itbf{v}_{\!\scriptscriptstyle S}
    	=
    	\frac{1}{2\eta}\eu{\alpha}\nabla_{\!\alpha}\mathbb{Q}.}}
    \end{equation}
    \cc{Thus, the locally planar and curved calculations give the same leading {slip velocity}, with the ordinary tangential gradient replaced by the covariant surface gradient. Together, Eqs.~(\ref{eqn:active-symmetric-inplane-stress}), (\ref{eqn:active-bending-moment}), and (\ref{eqn:active-slip-curved}) give the active contributions retained in the sharp-interface description through the first two moments of the active force profile.}

    \subsection{Membrane equations of motion}
    
    The free-energy density $F$ in Eq.~(\ref{eqn:total-free-energy}) \cc{can be written using the adiabatically reduced quantities $V_{\mathrm{m}}$ and $C_{\mathrm{m}}$ derived in Appendix~A.5 as}
    \begin{equation}
        \label{eqn:free-energy-renormalized-1}
        F = \cc{\Sigma_{\mathrm{m}}} + 2\kappa H\cc{C_{\mathrm{m}}} + 2\kappa\hspace{0.25pt} H^2.
    \end{equation}
    Here, \cc{$\Sigma_{\mathrm{m}}$ represents the composition-renormalized surface tension, defined by}
    \begin{equation}
        \label{eqn:new-Sigma}
        \Sigma_{\mathrm{m}}=\Sigma+V_{\mathrm{m}},
    \end{equation}
    \cc{where $V_{\mathrm{m}}$ is given in Eq.~(\ref{eqn:meanfield-energy-reduced}). We also define the difference in the dimensionless chemical-potential of component $k$ relative to the passive membrane background,}
    \begin{equation}
        \label{eqn:new-E}
        \cc{E_k\equiv\mathcal{E}_k-\mathcal{E}_0}
        =\sum_{n=0}^{\aleph_k}\frac{\bRk}{\mathbb{R}^k_n}\left[\mathcal{E}^k_n-\mathcal{E}_0+\ln\frac{\bRk}{\mathbb{R}^k_n}\right].
    \end{equation}
    
    Using $1-\Phi_0=\sum_k\Phi_k$, the \cc{composition-dependent spontaneous curvature can be equivalently written as} $\cc{C_{\mathrm{m}}}\!=\mathcal{C}_0+\sum_k\Phi_k\hspace{1pt}C_k$,
    where \cc{$C_k$ denotes the difference in curvature coupling of component $k$ relative to the passive membrane background,}
    \begin{equation}
        \label{eqn:new-C}
        C_k \cc{\equiv \mathcal{C}_k-\mathcal{C}_0} = \sum_{n=0}^{\aleph_k}\frac{\bRk\hspace{-1pt}\left(\mathcal{C}^k_n\hspace{-1pt}-\mathcal{C}_0\right)}{\mathbb{R}^k_n}.
    \end{equation}
    
    By using the free-energy in Eq.~(\ref{eqn:free-energy-renormalized-1}), the in-plane stress tensor $\sigma^{\alpha\beta}_0$ and the bending moment tensor $M^{\alpha\beta}_0$ are determined from Eq.~(\ref{eqn:sigma-0-F}) and (\ref{eqn:M-0-F}), respectively. Furthermore, the overall in-plane (symmetrized) stress tensor $\sigma^{\alpha\beta} = \sigma_0^{\alpha\beta} + \sigma^{\alpha\beta}_{\hspace{-1.5pt}\scriptstyle A}$ is found to be
    \begin{equation}
        \label{eqn:sigma-final}
        \sigma^{\alpha\beta}\hspace{-1.5pt} = \hspace{-0.5pt}\left(F+\mathbb{P} + H\mathbb{Q}\right)\hspace{-0.5pt}g^{\alpha\beta}\hspace{-1pt} +\left(\hspace{-1.5pt}\mathbb{Q}-\frac{\partial F}{\partial H}\right)b^{\alpha\beta}\hspace{-1pt},
    \end{equation}
    whereas the bending moment tensor $M^{\alpha\beta}$ is given by
    \begin{equation}
        \label{eqn:M-final}
        M^{\alpha\beta} = -\frac{1}{2}\left(\hspace{-1pt}\mathbb{Q}-\frac{\partial F}{\partial H}\right)g^{\alpha\beta}.
    \end{equation}
    This allows us to obtain the equations of motion, as derived in Eq.~(\ref{eqn:eqaution-motion-n}) and (\ref{eqn:eqaution-motion-t}), along the normal and tangential direction of the membrane, respectively.
    
    As a result, the force balance equation along tangential direction is found to be
    \begin{align}
        \quad\rho\dot{\itbf{v}}\cdot\ed{\alpha} = f_\alpha &+\cc{\nabla_{\!\alpha}}\cc{\Sigma_{\mathrm{m}}}+2\hspace{0.5pt}\kappa\hspace{0.25pt} H\hspace{0.85pt}\cc{\nabla_{\!\alpha}}\cc{C_{\mathrm{m}}}\notag\\[5pt]
        &+\cc{\nabla_{\!\alpha}}\mathbb{P}+H\,\cc{\nabla_{\!\alpha}}\mathbb{Q}+2\,\mathbb{Q}\hspace{1pt}\cc{\nabla_{\!\alpha}}H\,,
    \end{align}
    by substituting Eq.~(\ref{eqn:free-energy-renormalized-1}) into (\ref{eqn:sigma-final}) and (\ref{eqn:M-final}), and then expanding out the terms in Eq.~(\ref{eqn:eqaution-motion-t}). Moreover, by using Eq.~(\ref{eqn:meanfield-energy-reduced}), the definitions in Eq.~(\ref{eqn:active-force-dipole}), as well as Eqs.~(\ref{eqn:new-Sigma}--\ref{eqn:new-C}), this can be further rewritten as
    \begin{align}
        \label{eqn:membrane-equation-tangential}
        &\rho\dot{\itbf{v}}\cdot\ed{\alpha}=f_\alpha+\cc{\nabla_{\!\alpha}}\Sigma+\sum_k\!\left[\bPk+H\!\left(2\kappa C_k+\bQk\right)\right]\cc{\nabla_{\!\alpha}}\Phi_k\notag\\[4pt]
        &+\frac{k_BT}{b_0}\sum_k\!\left(E_k+\ln\frac{\Phi_k}{\Phi_0}\right)\cc{\nabla_{\!\alpha}}\Phi_k+2\sum_k\Phi_k\bQk\,\cc{\nabla_{\!\alpha}}H.
    \end{align}
    
    Similarly, the force-balance along the normal direction to the membrane can be found via Eq.~(\ref{eqn:eqaution-motion-n}) as follows:
    \begin{align}
        \label{eqn:membrane-equation-normal-c}
        &\rho\dot{\itbf{v}}\cdot\nv = f+2H\!\left(\cc{\Sigma_{\mathrm{m}}}+\mathbb{P}\right)-2\hspace{0.5pt}\kappa\hspace{0.25pt}\!\left[\Delta H+2H\!\left(H^2\hspace{-1pt}-K\right)\hspace{-1pt}\right]\notag\\[1pt]
        &-\hspace{-1pt}\kappa\hspace{-1pt}\left(\Delta\cc{C_{\mathrm{m}}}\hspace{-1pt}-\hspace{-1pt}2K\cc{C_{\mathrm{m}}}\right)+\frac{\Delta\mathbb{Q}}{2}+\mathbb{Q}\left(4H^2\hspace{-2.5pt}-\hspace{-1pt}K\right)\!,
    \end{align}
    which can be subsequently expanded as
    \begin{align}
        \label{eqn:membrane-equation-normal}
        \!\!\rho\dot{\itbf{v}}\cdot\nv &= f-2\hspace{0.5pt}\kappa\hspace{0.25pt}\!\left[\Delta H+2H\left(H^2\hspace{-1pt}-K\right)\right]+\hspace{-1pt}2\hspace{0.5pt}\kappa\hspace{0.25pt}K\hspace{0.5pt}\mathcal{C}_0\,\notag\\[3pt]
        &+\hspace{-1pt}2H{\left[\Sigma+\frac{k_BT}{b_0}\left(\mathcal{E}_0+\ln\Phi_0\right)\right]}\notag\\[2pt]
        &+{2H}\sum_{k}\Phi_k\!\left[\bPk+\frac{k_BT}{b_0}\!\left(E_k+\ln\frac{\Phi_k}{\Phi_0}\right)\right]\notag\\[2pt]
        &+\,\sum_k\Phi_k\!\left[2\hspace{0.5pt}\kappa\hspace{0.25pt}K\hspace{0.5pt}C_k+\left(4H^2\!-\!K\right)\bQk\right]\notag\\[2pt]
        &-\sum_k\Delta\Phi_k\!\left(\kappa\hspace{0.25pt}C_k\hspace{-1pt}-\hspace{-1pt}{\bQk}/{2}\hspace{0.75pt}\right)\hspace{-1pt}\!,
    \end{align}
    using Eq.~(\ref{eqn:meanfield-energy-reduced}) and Eqs.~(\ref{eqn:new-Sigma}--\ref{eqn:new-C}). 
    \cc{The terms proportional to $2H$ in Eqs.~(\ref{eqn:membrane-equation-normal-c}) and (\ref{eqn:membrane-equation-normal}) identify the activity-renormalized surface tension $\Sigma_A$:}
    \begin{equation}
        \label{eqn:renormalized-surface-tension}
        \boxed{{\Sigma_A=\Sigma_{\mathrm{m}}+\sum_k\Phi_k\bPk=\Sigma_{\mathrm{m}}+\mathbb{P}.}}
    \end{equation}
    \cc{The direct mechanical contribution to $\Sigma_A$ is the total first moment $\mathbb{P}$. For the simple point-force dipolar model in Eq.~(\ref{eqn:point-force-moment-model}), with the sign set by viscous drag, $\mathbb{P}^1_{\CircleArrow}>0$ for fission and $\mathbb{P}^2_{\CircleArrow}<0$ for fusion. The fission first-moment contribution therefore raises the activity-renormalized surface tension, whereas the fusion contribution lowers it. More generally, these signs depend on the cycle-averaged force profiles and are not fixed by whether the underlying process is fission or fusion.}
    
    Similarly, we identify an activity-renormalized spontaneous curvature \cc{that includes both the composition-dependent contribution} \cc{$C_{\mathrm{m}}$} \cc{and the direct mechanical contributions from the second force moments $\bQk$}, namely
    \begin{equation}
        \label{eqn:renormalized-spontaneous-curvature}
        \boxed{C_A={C_{\mathrm{m}}-\frac{\mathbb{Q}}{2\kappa}}=\mathcal{C}_0+\!\sum_{k}\hspace{-1pt}\Phi_k\!\left(C_k-\frac{\bQk}{2\kappa}\right).}
    \end{equation}
    \cc{For the same simple point-force dipolar model, with forcing set by viscous drag, $\mathbb{Q}_{\CircleArrow}^1>0$ for fission and $\mathbb{Q}_{\CircleArrow}^2<0$ for fusion. The second-moment contribution therefore decreases $C_A$ for fission and increases it for fusion. More generally, the direction of this shift is determined by the sign of the corresponding active second moment.}

	\subsection{Coupling to the ambient fluid}
    
	On either {side} of the organelle, its closed membrane surface $\mathcal{M}$ is in contact with an ambient fluid. In general, this surrounding fluid provides the dominant dissipative mechanism {for the relaxation of shape deformations on large length scales} \cite{Cai1994}, {corresponding here to length scales larger than} the coarse-graining length in our model of the membrane system (\cc{that is, $\sim50$--$100\;$nm}). \cc{Accordingly, on the length scales resolved by our coarse-grained model,} the membrane is hydrodynamically coupled to the bulk fluid through interfacial conditions on both the interior and exterior faces of the membrane~\cite{Seifert1999}. \cc{The normal component of the fluid velocity couples to shape deformations of the membrane, whereas tangential fluid motion drives in-plane membrane flows.}
	
	Here, we assume that the {inner and outer bulk fluids are} locally incompressible and obey the Stokes {equations}. This is motivated by the small Reynolds number in our system ($\mathrm{Re}\lessapprox10^{-4}$), with the advective inertial terms being negligible, compared with the viscous forces. Thus, the fluid motion is governed by
	\begin{equation}
		\label{eqn:Stokes-Eq}
		\boxed{\eta\hspace{1pt}\bnabla^2\hspace{-1pt}\itbf{V}_{\!{\boldsymbol{\pm}}}=\bnabla p_{\hspace{0.5pt}{\boldsymbol{\pm}}},\quad\text{and}\quad\bnabla\hspace{-1pt}\cdot\hspace{-1pt}\itbf{V}_{\!{\boldsymbol{\pm}}}=0,}
	\end{equation} where the Laplacian $\bnabla^2\!=\!\bnabla\cdot\bnabla$, with $\bnabla$ as the usual three-dimensional gradient operator, and $\eta$ is the shear viscosity of the bulk fluid, {taken to be the same inside and outside the organelle}. Here, $\itbf{V}_{\!{\boldsymbol{\pm}}}(\itbf{r})$ and $p_{\hspace{0.5pt}{\boldsymbol{\pm}}}(\itbf{r})$ are the fluid velocities and pressures at position $\itbf{r}$, with the subscripts ``$\scriptstyle\boldsymbol{-}$" and ``$\scriptstyle\boldsymbol{+}$" indicating respectively whether the fluid is located inside and outside of the membrane compartment. The last relation in Eq.~(\ref{eqn:Stokes-Eq}) is the incompressibility condition of the bulk fluid, whereas the former equation is the momentum conservation relationship, which follows from the stress balance condition:
	\begin{equation}
		\label{eqn:Stokes-stress-equation}
		\bnabla\cdot\boldsymbol{\mathsf{\Omega}}_{\boldsymbol{\pm}} = \hspace{0.5pt}\boldsymbol{0},
	\end{equation} where the Stokesian stresses, which are denoted by $\boldsymbol{\mathsf{\Omega}}_{\boldsymbol{\pm}}$ for the respective inner and outer flows,  are given by
	\begin{equation}
		\label{eqn:Stokes-stress}
		\boxed{\boldsymbol{\mathsf{\Omega}}_{\boldsymbol{\pm}} = -p_{\hspace{0.5pt}{\boldsymbol{\pm}}} \,\itbf{I} + \eta\left[\left(\bnabla\hspace{-1pt}\itbf{V}_{\!{\boldsymbol{\pm}}}\right)+\left(\bnabla\hspace{-1pt}\itbf{V}_{\!{\boldsymbol{\pm}}}\right)^{\!\mathsf{T}}\hspace{0.5pt}\right]\hspace{-0.5pt}\!.}
	\end{equation} 

    \cc{As discussed in Appendix~A.7, the active body-force density acts within a finite layer on the cytosolic side of the membrane. In the sharp-interface description used here, this finite active layer is no longer resolved explicitly, and the inner and outer bulk fluids therefore obey the homogeneous Stokes equations in Eq.~(\ref{eqn:Stokes-stress-equation}). The force and torque transmitted through the active layer are instead retained through the active in-plane stress $\sigma^{\alpha\beta}_{\!\scriptstyle A}$ and bending moment $M^{\alpha\beta}_{\!\scriptstyle A}$, while flow within the layer produces the {slip velocity} $\itbf{v}_{\!\scriptscriptstyle S}$. The in-plane stress depends on both the first and second moments of the active force profile, whereas the bending moment and {slip velocity} depend only on the second moment.}
	
	\cc{The hydrodynamic force per unit area $\itbf{f}$ exerted by the ambient fluids on the membrane is given by the jump in bulk-fluid traction}, cf.~Eq.~(\ref{eqn:momentum-local}). Hence,
	\begin{equation}
		\label{eqn:stress-boundary-condition}
		\boxed{\itbf{f} = \left(\hspace{1pt}\boldsymbol{\mathsf{\Omega}}_{\boldsymbol{+}}-\boldsymbol{\mathsf{\Omega}}_{\boldsymbol{-}}\right)\cdot\nv\,\equiv\,\boldsymbol{\pi}_{\boldsymbol{+}}-\boldsymbol{\pi}_{\boldsymbol{-}},\;\mathrm{at}\;\,\itbf{r}=\!\itbf{R},}
	\end{equation} where $\itbf{R}$ is the position vector of the membrane surface, $\nv$ is the normal unit vector to the membrane, and we define the {traction vector} acting across the surface as 
	\begin{equation}
		\boldsymbol{\pi}_{\boldsymbol{\pm}}=\boldsymbol{\mathsf{\Omega}}_{\boldsymbol{\pm}}\cdot\nv.
	\end{equation} In the local coordinate system of the membrane, we can decompose Eq.~(\ref{eqn:stress-boundary-condition}) into the normal component 
	\begin{equation}
		f = \nv\cdot\Big[\hspace{1pt}\boldsymbol{\pi}_{\boldsymbol{+}}\!\left(\itbf{r}=\!\itbf{R}\right)-\,\boldsymbol{\pi}_{\boldsymbol{-}}\!\left(\itbf{r}=\!\itbf{R}\right)\Big],
	\end{equation} as well as along the in-plane membrane components
	\begin{equation}
		f^\alpha = \eu{\alpha}\cdot\Big[\hspace{1pt}\boldsymbol{\pi}_{\boldsymbol{+}}\!\left(\itbf{r}=\!\itbf{R}\right)-\,\boldsymbol{\pi}_{\boldsymbol{-}}\!\left(\itbf{r}=\!\itbf{R}\right)\Big].
	\end{equation}
	
	\cc{The coupling between the membrane and the bulk fluids is specified by the interfacial velocity condition,}
    \begin{equation}
        \label{eqn:boundary-cond-fluid}
        \boxed{\itbf{v}\, +\,\itbf{v}_{\!\scriptscriptstyle S}\, {- \,{j}_{\scriptscriptstyle V}\nv}\, =\, \itbf{V}_{\!{\boldsymbol{\pm}}}\hspace{-1pt}\left(\itbf{r}=\itbf{R}\right)\!,}
    \end{equation}
    where $\itbf{v}$ is the membrane velocity. \cc{The tangential correction $\itbf{v}_{\!\scriptscriptstyle S}$ is the slip velocity generated by flow within the finite active layer and is derived in Appendix~A.7. The normal correction $-{j}_{\scriptscriptstyle V}\nv$ arises from volume exchange at the membrane, where ${j}_{\scriptscriptstyle V}$ is the volume flux per unit area associated with passive osmosis and active volume exchange due to fission and fusion, as discussed in Appendix~A.10.}
	
	Additionally, we consider that the fluid flow is at rest at infinity and bounded at the origin of the compartment:
    \begin{equation}
    	\label{eqn:condition-infinity}
    	\begin{cases}
    		\cc{\displaystyle \lim_{|\itbf r|\to0}\left|\itbf V_{\!{\boldsymbol{-}}}(\itbf r)\right|<\infty},&
    		\cc{\displaystyle \lim_{|\itbf r|\to0}p_{\hspace{0.5pt}{\boldsymbol{-}}}(\itbf r)=P_-},\\[5pt]
    		\cc{\displaystyle \lim_{|\itbf r|\to\infty}\left|\itbf V_{\!{\boldsymbol{+}}}(\itbf r)\right|=0},&
    		\cc{\displaystyle \lim_{|\itbf r|\to\infty}p_{\hspace{0.5pt}{\boldsymbol{+}}}(\itbf r)=P_+}.
    	\end{cases}
    \end{equation} where $P_\pm$ are scalar pressure constants. \cc{The exterior pressure $P_+$ sets the overall pressure reference, while the pressure difference $P_{\scriptscriptstyle\Delta}\equiv P_+-P_-$ is determined by the global volume constraint in Eq.~(\ref{eqn:volume-integral-condition}).}
	
    \cc{These boundary conditions, together with the interfacial velocity condition in Eq.~(\ref{eqn:boundary-cond-fluid}) and the traction balance in Eq.~(\ref{eqn:stress-boundary-condition}), complete the specification of the inner and outer Stokes flows. On the length and time scales considered here, the membrane dynamics is overdamped by the viscous traction exerted by the ambient fluids. We therefore neglect the inertial term $\rho\dot{\itbf{v}}$ in Eqs.~(\ref{eqn:membrane-equation-normal}) and (\ref{eqn:membrane-equation-tangential}) relative to the hydrodynamic traction $\itbf{f}$, so that}
    \begin{equation}
    	\label{eqn:low-Reynolds-no}
    	\rho\dot{\itbf{v}}=\boldsymbol{0}.
    \end{equation}
    \cc{The membrane force balance thus reduces to a balance between the hydrodynamic traction and the passive and active membrane stresses.}
    
	\subsection{Osmosis and volume dynamics}
    
    \cc{Organelle membranes are semipermeable, allowing solvent to cross the membrane more readily than solute, so that differences in solute concentration can drive osmotic transport \cite{Doi2013}. For dilute solutions, where solute--solute interactions can be neglected, the solvent chemical potential is given by \cite{LandauLifshitz1980}:}
    \begin{equation}
    	\mu^{\scriptstyle\hspace{-1pt}\mathfrak{B}}\!\left(P_{\boldsymbol{\pm}}\right) = \mu^{\scriptstyle\hspace{-1pt}\mathfrak{B}}_0\!\left(P_{\boldsymbol{\pm}}\right) - k_B T \hspace{1pt}\nu_0\hspace{1pt}c_{\boldsymbol{\pm}},
    \end{equation} where $c_{\boldsymbol{\pm}}$ are the solute number densities, while $\mu^{\scriptstyle\hspace{-1pt}\mathfrak{B}}_0$ is the bulk chemical potential associated with the pure solvent at a uniform pressure $P_{\boldsymbol{\pm}}$ and temperature $T$, \cc{with the subscripts ``$\scriptstyle\boldsymbol{+}$'' and ``$\scriptstyle\boldsymbol{-}$'' referring to quantities outside and inside of the organelle, respectively.} \cc{The scalars $P_\pm$ are the reference pressures defined by the boundary conditions in Eq.~(\ref{eqn:condition-infinity}), while $\nu_0=\partial\mu^{\scriptstyle\hspace{-1pt}\mathfrak{B}}_0/\partial P$ is the molecular volume of the pure solvent at fixed $T$.}
    
    \cc{Osmotic equilibrium corresponds to a vanishing solvent chemical-potential difference across the membrane,}
    \begin{equation}
    	\label{eqn:chemical-equilibrium}
    	\mu^{\scriptstyle\hspace{-1pt}\mathfrak{B}}\!\left(P_{\boldsymbol{+}}\right)=\mu^{\scriptstyle\hspace{-1pt}\mathfrak{B}}\!\left(P_{\boldsymbol{-}}\right)\!.
    \end{equation}
    \cc{Defining the hydrostatic pressure difference across the membrane as $P_{\hspace{-0.5pt}\scriptscriptstyle\Delta}\equiv P_{\boldsymbol{+}}-P_{\boldsymbol{-}}$, and taking $\nu_0$ to be approximately constant over this pressure difference, Eq.~(\ref{eqn:chemical-equilibrium}) gives the dilute-solution {\it van't Hoff} relation \cite{LandauLifshitz1980},}
    \begin{equation}
    	\label{eqn:vantHoff}
    	\Pi_{\scriptscriptstyle\Delta}
    	=
    	k_B T\left(c_{\boldsymbol{+}}-c_{\boldsymbol{-}}\right),
    	\qquad
    	P_{\hspace{-0.5pt}\scriptscriptstyle\Delta}
    	=
    	\Pi_{\scriptscriptstyle\Delta}.
    \end{equation}
    
    \cc{Away from osmotic equilibrium, the solvent chemical potentials on the two sides of the membrane differ, with}
    \begin{equation}
    	{
    	\delta\hspace{-0.5pt}\mu^{\scriptstyle\hspace{-1pt}\mathfrak{B}}
    	\equiv
    	\mu^{\scriptstyle\hspace{-1pt}\mathfrak{B}}\!\left(P_{\boldsymbol{+}}\right)
    	-
    	\mu^{\scriptstyle\hspace{-1pt}\mathfrak{B}}\!\left(P_{\boldsymbol{-}}\right)
    	=
    	\nu_0\left(P_{\hspace{-0.25pt}\scriptscriptstyle\Delta}-\Pi_{\scriptscriptstyle\Delta}\right).
    	}
    \end{equation}
    \cc{We model passive solvent permeation by taking the corresponding flux to depend linearly on this chemical-potential difference. Since solvent permeation through the membrane typically occurs on timescales of order minutes \cite{Staykova2013}, the passive osmotic contribution to the volume dynamics is correspondingly slow.}

    \cc{Fission and fusion provide an active contribution to the exchange of lumenal volume. The corresponding local volume flux generated by the trafficking cycles is}
    \begin{equation}
    	\mathbb{V}=\!\sum_{k,n,m}\!\mathbb{V}^k_{nm}r^k_{nm}\phi^k_m
    	=
    	\sum_k\,\bJk\sum_{n=1}^{\aleph_k}\mathbb{V}^k_{n+1,n},
    \end{equation}\cc{where the second equality uses the adiabatic approximation of Appendix~A.5.} \cc{Using} Eq.~(\ref{eqn:bJK-equation}), we can write
    \begin{equation}
    	\label{eqn:total-volume-flux-rate}
    	\boxed{\mathbb{V} = \sum_k\,\Phi_k\bVk,}
    \end{equation} where $\bVk$ is the {volume transfer rate associated with cycle $k$}, defined by
    \begin{equation}
    	\label{eqn:volume-flux-rate}
    	\bVk=\bRk{\sum_{n=1}^{\aleph_k}\mathbb{V}^k_{n+1,n}}.
    \end{equation}
    Its sign is negative for the fissogens ($k=1$) and positive for the fusogens ($k=2$):
    \begin{equation}
    	\mathrm{sgn}\hspace{-1pt}\left(\bVk\right)=(-1)^k.
    \end{equation}

    \cc{The total volume flux per unit membrane area combines the active trafficking contribution with passive osmotic permeation. Denoting by $j_{\scriptscriptstyle V}$ the scalar volume flux into the organelle and by $\itbf{j}_{\!\scriptscriptstyle V}$ the corresponding vector volume flux, we write}
    \begin{equation}
        \boxed{{{j}_{\scriptscriptstyle V} = \mathbb{V}-w_0\hspace{-0.5pt}\left(\hspace{1pt}\Pi_{\scriptscriptstyle\Delta}-P_{\hspace{-0.25pt}\scriptscriptstyle\Delta}\right),\quad \itbf{j}_{\!\scriptscriptstyle V}=-{j}_{\scriptscriptstyle V}\nv.}}
    \end{equation}
    \cc{Since $\nv$ is the outward membrane normal, $j_{\scriptscriptstyle V}>0$ corresponds to volume transport into the organelle and thus increases the enclosed volume. The corresponding change in the enclosed volume is therefore}
    \begin{equation}
        \label{eqn:volume-dynamics}
        \frac{\mathrm{d}V}{\mathrm{d}t} = \int_\mathcal{M}{j}_{\scriptscriptstyle V}\,\mathrm{d}S.
    \end{equation}
    
    \cc{A nonzero volume flux also modifies the kinematic coupling between the membrane and the surrounding fluids. In the absence of volume exchange, the normal fluid and membrane velocities coincide; however, with volume exchange, this instead becomes}
    \begin{equation}
        {\left(\itbf{V}_{\!{\boldsymbol{\pm}}}\big|_{\mathcal M}-\itbf{v}\right)\cdot\nv=-{j}_{\scriptscriptstyle V}.}
    \end{equation}
    \cc{Thus, the interfacial fluid velocity acquires the normal correction $-{j}_{\scriptscriptstyle V}\nv$. Including the tangential slip velocity $\itbf{v}_{\!\scriptscriptstyle S}$ generated by the active layer then we recover the full interfacial condition in Eq.~(\ref{eqn:boundary-cond-fluid}).}

    \cc{Let $\mathfrak{n}_{\scriptscriptstyle\mathcal{S}}$ denote the number of solute molecules enclosed by the organelle. Assuming that the interior solution is well mixed, its concentration is $c_{\boldsymbol{-}}=\mathfrak{n}_{\scriptscriptstyle\mathcal{S}}/V$, and hence}
    \begin{equation}
    	\label{eqn:osmotic-pressure-concentrations}
    	\Pi_{\scriptscriptstyle\Delta}
    	=
    	k_B T\,\bigg[
    	c_{\boldsymbol{+}}
    	-
    	\frac{\mathfrak{n}_{\scriptscriptstyle\mathcal{S}}}{V}
    	\bigg].
    \end{equation}
    \cc{In general, both $V$ and $\mathfrak{n}_{\scriptscriptstyle\mathcal{S}}$ may change through vesicular trafficking or lumenal cargo processing. Fission and fusion can exchange solvent and solute cargo, while cargo maturation may also alter the enclosed solute content. A more detailed description would require additional balance equations for the lumenal solute species, including both trafficking fluxes and internal reactions, which we do not consider here.}

    \cc{The hydrostatic pressure difference $P_{\scriptscriptstyle\Delta}$ is determined by requiring consistency between volume exchange across the membrane and the resulting motion of the membrane surface. Equation~(\ref{eqn:volume-dynamics}) gives the rate at which volume is added to or removed from the organelle by transport, whereas the same change in enclosed volume can be expressed geometrically in terms of the normal membrane velocity. As derived in Appendix~D.3,}
    \begin{equation}
        \frac{\mathrm{d}V}{\mathrm{d}t}=\int_{\mathcal M}v\,\mathrm{d}S,
    \end{equation}
    \cc{where $v$ is the normal membrane velocity. Since both expressions represent the same change in the enclosed volume, equating them gives the volume constraint}
    \begin{equation}
        \label{eqn:volume-integral-condition}
        \int_\mathcal{M}\!\mathrm{d}S\,\bigg[
        v+w_0\left(\Pi_{\scriptscriptstyle\Delta}-P_{\hspace{-0.25pt}\scriptscriptstyle\Delta}\right)
        -\sum_k\Phi_k\bVk
        \bigg]=0.
    \end{equation}
    \cc{This gives us $P_{\scriptscriptstyle\Delta}$ for a given membrane velocity, osmotic pressure, and active volume exchange. This is a global condition, as its integrand need not vanish pointwise.}

    \subsection{Boundary integral formulation}
        
    A more general approach {to solving} the {incompressible} Stokes equations in Eq.~(\ref{eqn:Stokes-Eq}) in the presence of an interface with a surface force $\boldsymbol{f}\neq\boldsymbol{0}$, cf.~Eq.~(\ref{eqn:stress-boundary-condition}), involves a boundary integral representation~\cite{Pozrikidis1992}. A starting point is the use of {the} Lorentz reciprocal identity \cite{Kim1991}, which states that any two flows $\boldsymbol{V}$ and $\boldsymbol{V}'$ with their corresponding stress tensors $\boldsymbol{\Omega}$ and $\boldsymbol{\Omega}'$ satisfy:
    \begin{equation}
    	\frac{\partial}{\partial x_k}\left[{V}'_i\,\Omega_{ik}-{V}_i\,\Omega'_{ik}\right] = 0.
    \end{equation} \cc{Here, $\boldsymbol{x}=(x_1,x_2,x_3)$ denotes the spatial position in the fluid domain.}
    \cc{We choose $\boldsymbol{V}'$ to be the flow generated by a point force $\boldsymbol{g}=g_i\hat{\boldsymbol{e}}_i$ located at $\boldsymbol{x}_0$, with $\hat{\boldsymbol{e}}_i$ denoting the Cartesian basis vectors.} \cc{We write the velocity field as $V'_i(\boldsymbol{x})=G_{ij}(\boldsymbol{x},\boldsymbol{x}_0)g_j/(8\pi\eta)$, where $G_{ij}$ is the velocity Green's function for a point force at $\boldsymbol{x}_0$. The corresponding stress is $\Omega'_{ik}(\boldsymbol{x})=T_{ijk}(\boldsymbol{x},\boldsymbol{x}_0)g_j/(8\pi)$, where $T_{ijk}$ is the associated stress Green's function~\cite{Pozrikidis1992}. Since the point force $\boldsymbol{g}$ is arbitrary, we obtain}
    \begin{equation}
    	\label{eqn:Lorentz-id-1}
    	\frac{\partial}{\partial x_k}\big[G_{ij}(\boldsymbol{x},\boldsymbol{x}_0)\Omega_{ik}(\boldsymbol{x})-\eta{V}_i(\boldsymbol{x})T_{ijk}(\boldsymbol{x},\boldsymbol{x}_0)\big]\!=0.
    \end{equation}

    \cc{Integrating Eq.~(\ref{eqn:Lorentz-id-1}) over a control volume bounded by a closed surface $\mathcal{S}_{\scriptscriptstyle C}$ and applying the divergence theorem, with} $\mathcal{L}_{jk}=G_{ij}\Omega_{ik}-\eta V_iT_{ijk}$, \cc{gives, when the source point $\boldsymbol{x}_0$ lies outside the control volume,}
    \begin{equation}
        \label{eqn:boundary-integral-out}
        \int_{\mathcal{S}_{\scriptscriptstyle C}} \mathcal{L}_{jk}(\boldsymbol{x},\boldsymbol{x}_0)\,n_k(\boldsymbol{x})\,\mathrm{d}S(\boldsymbol{x})=0.
    \end{equation}
    \cc{When $\boldsymbol{x}_0$ lies inside the control volume, the point-force Green's functions $G_{ij}$ and $T_{ijk}$ are singular at $\boldsymbol{x}=\boldsymbol{x}_0$. We first exclude a small spherical neighborhood around $\boldsymbol{x}_0$ before applying the divergence theorem and then take its radius to zero, yielding \cite{Kim1991,Pozrikidis1992}}
    \begin{equation}
        \label{eqn:boundary-integral-in}
        \int_{\mathcal{S}_{\scriptscriptstyle C}} \mathcal{L}_{jk}(\boldsymbol{x},\boldsymbol{x}_0)\,n_k(\boldsymbol{x})\,\mathrm{d}S(\boldsymbol{x})=-8\pi\eta V_j(\boldsymbol{x}_0).
    \end{equation}   \cc{As the source point approaches the boundary, the two terms in the boundary-integral representation behave differently. The term involving $G_{ij}$ acts on the surface traction and is referred to as {\it the single-layer potential}; it remains continuous across the interface. The term involving $T_{ijk}$ acts on the surface velocity and is referred to as {\it the double-layer potential}; its limiting value has a finite jump across the interface, with the remaining singular integral understood in the principal-value sense \cite{Pozrikidis1992}.}
    
    \cc{We now apply these identities to the exterior and interior flows $\boldsymbol{V}_{\!\pm}$ separated by the closed membrane interface $\mathcal{M}$. For the exterior flow, together with its decay at infinity, we obtain}
    \begin{align}
    	\label{eqn:boundary-eq-1}
    	\boldsymbol{V_+}(\boldsymbol{r}) = & \frac{1}{8\pi}\int_\mathcal{M\;}\boldsymbol{V}^{\mathsf{T}}\hspace{-1pt}(\boldsymbol{x})\hspace{1.5pt}\boldsymbol{T}(\boldsymbol{x},\boldsymbol{r})\hspace{1pt}\boldsymbol{n}(\boldsymbol{x}) \,\mathrm{d}S(\boldsymbol{x})\notag\\[4pt]
    	& - \frac{1}{8\pi\eta}\int_\mathcal{M\;}\!\boldsymbol{G}(\boldsymbol{r},\boldsymbol{x})\,\boldsymbol{\Omega}_{+}(\boldsymbol{x})\nv(\boldsymbol{x})\,\mathrm{d}S(\boldsymbol{x}),
    \end{align}
    where the symmetry property $G_{ij}(\boldsymbol{x},\boldsymbol{x}_0)=G_{ji}(\boldsymbol{x}_0,\boldsymbol{x})$ is used to switch the order of indices \cite{Pozrikidis1992}. 
    
    \cc{The corresponding interior representation contains the same two types of terms. Since the fluid velocity at the membrane is the same when approached from either side, and the inner and outer viscosities are equal, the double-layer potentials associated with the stresslet cancel when the two representations are combined. The remaining single-layer contribution depends only on the traction jump $\boldsymbol{f}=(\boldsymbol{\Omega}_+-\boldsymbol{\Omega}_-)\nv$ defined in Eq.~(\ref{eqn:stress-boundary-condition}), giving}
    \begin{equation}
    	\boxed{\boldsymbol{V}_{\!\pm}(\boldsymbol{r})
    	=
    	-\frac{1}{8\pi\eta}
    	\int_\mathcal{M}
    	\boldsymbol{G}(\boldsymbol{r},\boldsymbol{x})\,\boldsymbol{f}(\boldsymbol{x})\,\mathrm{d}S(\boldsymbol{x}).}
    \end{equation}
    \cc{Thus, the fluid velocity throughout both domains, including its limiting value at the membrane, is determined directly by the interfacial traction jump $\boldsymbol{f}$.}
    
    \cc{For the equal-viscosity case considered here, $G_{ij}$ is the same in both fluid domains and is given by the free-space Green's function, the Oseen tensor $\mathscr{S}_{ij}$~\cite{Pozrikidis1992}. Thus, $G_{ij}(\boldsymbol{x},\boldsymbol{x}_0)=\mathscr{S}_{ij}(\hat{\boldsymbol{x}})$, where $\hat{\boldsymbol{x}}=\boldsymbol{x}-\boldsymbol{x}_0$ is the separation vector from the source point $\boldsymbol{x}_0$ to the field point $\boldsymbol{x}$, with}
    \begin{equation}
        {\mathscr{S}_{ij}(\hat{\boldsymbol{x}}) = \frac{\delta_{ij}}{\,\left|\hat{\boldsymbol{x}}\right|^{\phantom{1}}\!}+\frac{\hat{x}_i\hat{x}_j}{\left|\hat{\boldsymbol{x}}\right|^3}.}
    \end{equation}

    \cc{The pressure fields follow from the corresponding free-space pressure Green's function $\mathscr{P}(\hat{\boldsymbol{y}})=2\hat{\boldsymbol{y}}/|\hat{\boldsymbol{y}}|^3$~\cite{Pozrikidis1992}, where $\hat{\boldsymbol{y}}=\boldsymbol{r}-\boldsymbol{x}$ is the separation vector from the source point $\boldsymbol{x}$ on the membrane to the field point $\boldsymbol{r}$. Taking the exterior pressure $P_+$ as the reference, this representation gives the pressure on either side of the membrane,}
    \begin{equation}
        \cc{
        p_{\hspace{0.5pt}{\boldsymbol{\pm}}}(\boldsymbol{r})
        =
        P_+
        -\frac{1}{8\pi}
        \int_\mathcal{M}
        \mathscr{P}(\boldsymbol{r}-\boldsymbol{x})\cdot\boldsymbol{f}(\boldsymbol{x})\,\mathrm{d}S(\boldsymbol{x}).
        }
    \end{equation}
    \cc{Here, $p_+$ and $p_-$ correspond to $\boldsymbol{r}$ lying outside and inside the membrane, respectively. Although both are written relative to $P_+$, the pressure difference $P_{\scriptscriptstyle\Delta}=P_+-P_-$ enters through the hydrostatic contribution $-P_{\scriptscriptstyle\Delta}\nv$ to the traction jump $\boldsymbol{f}$. For a closed membrane, the integral $\int_{\mathcal M}\mathscr{P}(\boldsymbol{r}-\boldsymbol{x})\cdot\nv(\boldsymbol{x})\,\mathrm{d}S$ vanishes outside the membrane and equals $-8\pi$ inside. This contribution leaves the exterior pressure unchanged and shifts the interior pressure by $-P_{\scriptscriptstyle\Delta}$, thereby recovering the interior reference pressure $P_-=P_+-P_{\scriptscriptstyle\Delta}$. Thus, while $P_{\scriptscriptstyle\Delta}$ shifts the interior pressure, it does not generate a fluid velocity for a closed membrane; its value is fixed through the global volume constraint in Eq.~(\ref{eqn:volume-integral-condition}).}

	\section{Quasi-spherical Perturbation Analysis}

    \subsection{Surface parameterization}
    
    The results derived so far are coordinate-invariant and valid for any closed surface $\mathcal{M}$, {described by} an embedding function $\itbf{R}$ (its position vector). We now assume that the {membrane surface can be described as a small deformation of a sphere}, which is given by
    \begin{equation}
    	\label{eqn:shape-parameterization}
    	\boxed{\itbf{R} = R\left[1+\varepsilon\,u\hspace{-1pt}\left(\theta,\varphi\right)\right]\rv,}
    \end{equation}
    in which we employ the spherical coordinates $(r,\hspace{1pt}\theta,\hspace{1pt}\varphi)$ {centered on} the fixed, undeformed sphere of radius {$R$}. Here, $\rv$ is the {radial unit vector}, which is given by
    \begin{equation}
    	\rv = \left(\sin\theta\cos\varphi,\,\sin\theta\sin\varphi,\,\cos\theta\hspace{0.5pt}\right)^{\textsf{T}}\!\hspace{-1pt},
    \end{equation}
    where the inclination angle $\theta\in{[0,\pi]}$, and the azimuth angle $\varphi\in[0,2\pi)$. Furthermore, the unit tangent vectors along the spherical angular directions $\theta$ and $\varphi$ are denoted by $\thetav$ and $\phiv$, respectively. By using Eq.~(\ref{eqn:tangent-vector}), the explicit {forms} of these two unit vectors are found to be
    \begin{align}
    	\,\phiv &= \left(-\sin\varphi,\,\cos\varphi,\,0\hspace{1pt}\right)^{\textsf{T}}\!,\qquad\mathrm{and}\\[4pt]
    	\thetav\, &= \left(\cos\theta\cos\varphi,\,\cos\theta\sin\varphi,\,-\sin\theta\hspace{0.5pt}\right)^{\textsf{T}}\!\hspace{-1pt}.
    \end{align}
    Herein, $\varepsilon\ll1$ is a small dimensionless parameter which can ultimately be absorbed into the arbitrary angular function $u\left(\theta,\varphi\right)$. \cc{It is introduced only to keep track of the perturbative order of the quantities subsequently computed from} the shape parameterization in Eq.~(\ref{eqn:shape-parameterization}).
    
    We assume that $u(\theta,\varphi)$ is a well-behaved function that can be further expanded in a series of surface spherical harmonics, denoted by the subscript $\ell$; namely, we have
    \begin{equation}
    	\itbf{R} = R\hspace{1pt}\Big[1+\varepsilon\sum_{\ell=0}^{\infty}u_\ell\hspace{-1pt}\left(\theta,\varphi\right)\hspace{-1pt}\Big]\rv,
    \end{equation}
    where the surface harmonics $u_\ell\hspace{-1pt}\left(\theta,\varphi\right)$ can be written as
    \begin{equation}
    	\label{eqn:surface-harmonic-def}
    	u_\ell\hspace{-1pt}\left(\theta,\varphi\right) = \sum^{\ell}\limits_{m=-\ell}u_{\ell,m}\;{Y}_{\ell,m}\!\left(\theta,\varphi\right)\!,
    \end{equation}
    \cc{where $u_\ell(\theta,\varphi)$ denotes the complete spherical-harmonic component of degree $\ell$, while $u_{\ell,m}$ is the amplitude associated with the individual spherical harmonic} ${Y}_{\ell,m}$~\cite{Abramowitz1965}. Note that the zeroth harmonic $u_0$ results in a {uniform} rescaling of the radius over which we perturb. The first harmonics correspond to a translation of the whole sphere {at linear order}, while the second harmonics describe an ellipsoidal {deformation} of the sphere.

	\subsection{Lamb's general solution of Stokes equations}
    
	In spherical coordinates, H.\ Lamb \cite{Happel1981} {gave} a general solution for the Stokes equations in the absence of body forces within the bulk fluid. {To derive this solution}, we first employ the vector identity of the Laplacian $\bnabla^2\hspace{-1pt}\itbf{V} = \bnabla\left(\bnabla\cdot\!\itbf{V}\right) - \bnabla\times\left(\bnabla\times\!\itbf{V}\right)$, and then {take} the divergence of the first vector equation in Eq.~(\ref{eqn:Stokes-Eq}), {which gives}
	\begin{equation}
		\bnabla\cdot\left(\bnabla^2\hspace{-1pt}\itbf{V}\right) = \bnabla^2\!\left(\bnabla\cdot\!\itbf{V}\right) = \frac{1}{\eta}\hspace{1pt}\bnabla^2 p,
	\end{equation} omitting for brevity the subscript ``$\scriptstyle\boldsymbol{\pm}$". This implies that the pressure field $p$ must be a harmonic function,
	\begin{equation}
		\bnabla^2 p = 0,
	\end{equation} {by the} incompressibility condition $\bnabla\cdot\!\itbf{V}=0$. We expand the pressure in a series of solid spherical harmonics:
	\begin{equation}
	p(r,\theta,\varphi) =\sum_{\ell=-\infty}^{\infty}\, p_{\ell}(r,\theta,\varphi) ,
    \end{equation}
    where {for $\ell\geq0$, the regular solid spherical harmonic $p_\ell$ of degree $\ell$} can be explicitly written as
    \begin{equation}
    	p_{\hspace{0.45pt}\ell}(r,\theta,\varphi) = \sum^{\ell}\limits_{m=-\ell}p_{\hspace{0.45pt}\ell,m}\,r^\ell\hspace{1pt}{Y}_{\ell,m}\!\left(\theta,\varphi\right)\!,
    \end{equation}
    where $p_{\hspace{0.45pt}\ell,m}$ is the spherical harmonic amplitude corresponding to the angular mode numbers $m$ and $\ell$.
	
	This expansion forms the basis of Lamb's method \cite{Happel1981}, which allows us to find a general solution for the bulk fluid velocity $\itbf{V}$ in terms of the pressure harmonics (as the inhomogeneous part), and two other solid spherical harmonics \cc{representing the homogeneous solutions}, which we denote by $\Upsilon_\ell$ and $\Xi_{\,\ell}$. Thus, \cc{the solution reads}
    \begin{align}
    	\label{eqn:Lamb-solution}
    	\itbf{V}(r,\theta,\varphi) &=\!\sum_{\substack{\ell\hspace{0.5pt}=\hspace{0.25pt}-\infty \\ \ell\hspace{0.75pt}\neq\hspace{0.5pt}-1} }^{\infty}\!\!\frac{\left(\ell+3\right)r^2\hspace{1pt}\bnabla p_\ell-2\ell\,r\hspace{1pt}\rv\,p_\ell}{2\eta\left(\ell+1\right)\!\left(2\ell+3\right)}\notag\\[-2pt]
    	&\qquad+\sum_{\ell=-\infty}^{\infty}\!\left[\bnabla\Upsilon_\ell + \bnabla\!\times\!\left(r\hspace{1pt}\rv\,\Xi_{\,\ell}\right)\right]\!.
    \end{align}These three harmonic functions can be {determined once} the velocity field is prescribed on a spherical surface, say, at $r=R$. \cc{For membrane vesicles or droplets, however, the coupling between membrane deformations and surrounding viscous flows requires boundary conditions on a deformable, time-dependent surface~\cite{Milner1987,Schneider1984}. A common approximation is to impose these boundary conditions on a sphere of an appropriately chosen mean radius. Here, we develop a perturbation scheme to determine} the solid harmonics $p_\ell$, $\Upsilon_\ell$ and $\Xi_{\,\ell}$ in Eq.~(\ref{eqn:Lamb-solution}) when the \cc{boundary conditions are instead} prescribed on the deformed spherical surface {parameterized by} Eq.~(\ref{eqn:shape-parameterization}).
	
	Instead of directly working with the individual components of the velocity field $\hspace{-1pt}\itbf{V}\hspace{-0.5pt}$, the boundary-value problem is more suitably addressed by \cc{three scalar quantities}: the radial velocity, $\rv\cdot\itbf{V}$, the radial component of the fluid vorticity, $(r\hspace{1pt}\rv)\hspace{-0.5pt}\cdot\bnabla\times\itbf{V}$, and \cc{the quantity that we refer to as the Lamb term}, $\mathfrak{L}=(r\hspace{1pt}\rv)\cdot\bnabla(\rv\cdot\itbf{V})-r\hspace{1pt}\bnabla\cdot\itbf{V}$.
	
	The radial component of the velocity can be readily calculated from Eq.~(\ref{eqn:Lamb-solution}); namely,
	\begin{equation}
		\label{eqn:radial-velocity-Lamb}
		\rv\cdot\itbf{V} = \sum_{\ell=-\infty}^{\infty}\left[\frac{\ell\,r\hspace{1pt}p_\ell}{2\eta\left(2\ell+3\right)} + \frac{\ell\,\Upsilon_\ell}{r}\right]\!\hspace{-1pt},
	\end{equation} where the following identity has been used~\cite{Happel1981}:
	\begin{equation}
		\label{eqn:Euler-formula}
		r\frac{\partial h_\ell}{\partial r} = \ell\hspace{1pt}h_\ell,
	\end{equation} with $h_\ell$ being any solid spherical harmonic of order $\ell$. Similarly, the radial component of the curl of the velocity field $\itbf{V}$ can be determined as follows:
	\begin{equation}
		\label{eqn:radial-vorticity-Lamb}
		\rv\cdot\bnabla\times\itbf{V} = \sum_{\ell=-\infty}^{\infty}\frac{\ell\left(\ell+1\right)\Xi_{\,\ell}}{r}.
	\end{equation} 
	
	{To obtain an expression for the Lamb term $\mathfrak{L}$,} we begin by differentiating Eq.~(\ref{eqn:radial-velocity-Lamb}) with respect to $r$, and then by using again the identity in Eq.~(\ref{eqn:Euler-formula}), we find that
	\begin{equation}
		\frac{\partial\hspace{-1pt}\left(\rv\hspace{-0.5pt}\cdot\!\itbf{V}\right)}{\partial r} = \sum_{\ell=-\infty}^{\infty}\left[\frac{\ell\left(\ell+1\right) p_\ell}{2\eta\left(2\ell+3\right)} + \frac{\ell\left(\ell-1\right)\hspace{-1pt}\Upsilon_\ell}{r^2}\right]\!\hspace{-1pt}.
	\end{equation} Due to the incompressibility condition, the above equation also allows us to compute the Lamb term $\mathfrak{L}$ in terms of the solid harmonic functions, that is,
	\begin{align}
		\label{eqn:Lamb-divergence-term}
		\mathfrak{L} \,=\sum_{\ell=-\infty}^{\infty}\left[\frac{\ell\left(\ell+1\right) r\hspace{1pt}p_\ell}{2\eta\left(2\ell+3\right)} + \frac{\ell\left(\ell-1\right)\hspace{-1pt}\Upsilon_\ell}{r}\right]\!\hspace{-1pt}.
	\end{align}
	
	By evaluating {Eqs.}~(\ref{eqn:radial-velocity-Lamb}), (\ref{eqn:radial-vorticity-Lamb}), and (\ref{eqn:Lamb-divergence-term}) at $r=R$, and by employing the identity $
    \left[\hspace{0.5pt} h_\ell\hspace{1pt}\right]_{r=R}=h_\ell\left({r}/{R}\hspace{0.5pt}\right)^{-\ell}$, for {any solid spherical harmonic} of order $\ell$, {we obtain}
	\begin{equation}
		\label{eqn:radial-velocity-Lamb-R}
		\left[\rv\cdot\itbf{V}\right]_{r=R}\, =\! \sum_{\ell=-\infty}^{\infty}\hspace{-0.5pt}\left[\frac{\ell\,p_\ell\hspace{1pt} R^{\ell+1}}{2\eta\hspace{-1pt}\left(2\ell\hspace{-1pt}+\hspace{-1pt}3\right)\hspace{-0.5pt}r^{\hspace{0.5pt}\ell}} + \frac{\ell\,\Upsilon_\ell}{R^{1-\ell}\,r^{\hspace{0.5pt}\ell}}\right]\!\hspace{-1pt},
	\end{equation} the radial part of the vorticity
	\begin{equation}
		\label{eqn:radial-vorticity-Lamb-R}
		\left[\rv\cdot\bnabla\times\itbf{V}\right]_{r=R}\, =\! \sum_{\ell=-\infty}^{\infty}\frac{\ell\left(\ell+1\right)\hspace{-0.5pt} R^{\ell-1}\,\Xi_{\,\ell}}{r^{\hspace{0.5pt}\ell}},
	\end{equation} and, lastly, the Lamb divergence term
	\begin{equation}
		\label{eqn:Lamb-divergence-term-R}
		\left[\mathfrak{L}\hspace{0.25pt}\right]_{r=R} =\! \sum_{\ell=-\infty}^{\infty}\!\left[\frac{\ell\hspace{-1pt}\left(\ell+1\right)\hspace{-1pt} p_\ell\hspace{1pt} R^{\ell+1}}{2\eta\left(2\ell+3\right)\hspace{-0.5pt}r^{\hspace{0.5pt}\ell}} \hspace{-1pt}+\hspace{-1pt}\frac{\ell\hspace{-0.5pt}\left(\ell-1\right)\hspace{-1pt}\Upsilon_\ell}{R^{1-\ell}\,r^{\hspace{0.5pt}\ell}}\right]\!\hspace{-1pt}.
	\end{equation}
	
	{We denote the prescribed velocity field on the spherical surface $r=R$ by} $\hspace{-0.5pt}\itbf{v}_{\mathrm{m}}\hspace{-1pt}\left(R,\hspace{0.5pt}\theta,\varphi\right)$. {The radial velocity and radial vorticity at this surface are then}
    \begin{align}
    	\left[\rv\cdot\itbf{V}\right]_{r=R}
    	&=\rv\cdot\itbf{v}_{\mathrm{m}}\hspace{-1pt}\left(R,\hspace{0.5pt}\theta,\varphi\right),\\[3pt]
    	\left[(r\hspace{1pt}\rv)\cdot\bnabla\times\itbf{V}\right]_{r=R}
    	&=(R\hspace{1pt}\rv)\cdot\bnabla\times\itbf{v}_{\mathrm{m}}\hspace{-1pt}\left(R,\hspace{0.5pt}\theta,\varphi\right).
    \end{align}
    {Similarly, the Lamb term evaluated at $r=R$ reduces to}
    \begin{equation}
    	\left[\mathfrak{L}\hspace{0.25pt}\right]_{r=R}
    	=-R\hspace{1pt}\bnabla\cdot\itbf{v}_{\mathrm{m}}\hspace{-1pt}\left(R,\hspace{0.5pt}\theta,\varphi\right)\hspace{-1pt}.
    \end{equation}
	The velocity $\!\itbf{v}_{\mathrm{m}}(R,\hspace{0.5pt}\theta,\varphi)$ is a vector field {defined on} a spherical surface of radius $R$. Hence, {following} Lamb's approach, this vector field may be decomposed in a series of surface spherical harmonics; specifically,
	\begin{align}
		\label{eqn:X-ell}
		\rv\cdot\itbf{v}_{\mathrm{m}}\hspace{-1pt}\left(R,\hspace{0.5pt}\theta,\varphi\right) =\,\sum_{\ell=1}^{\infty}\,\mathcal{X}_\ell\left(\theta,\varphi\right)\hspace{-1pt}, \\
		\label{eqn:Y-ell}
		-r\hspace{1pt}\bnabla\cdot\itbf{v}_{\mathrm{m}}\hspace{-1pt}\left(R,\hspace{0.5pt}\theta,\varphi\right)=\,\sum_{\ell=1}^{\infty}\,\mathcal{Y}_\ell\left(\theta,\varphi\right)\hspace{-1pt},\\
		\label{eqn:Z-ell}
		(r\hspace{1pt}\rv)\hspace{-0.5pt}\cdot\bnabla\times\itbf{v}_{\mathrm{m}}\hspace{-1pt}\left(R,\hspace{0.5pt}\theta,\varphi\right)=\,\sum_{\ell=1}^{\infty}\,\mathcal{Z}_\ell\left(\theta,\varphi\right)\hspace{-1pt},
	\end{align} where $\mathcal{X}_\ell$, $\mathcal{Y}_\ell$ and $\mathcal{Z}_\ell$ are {the complete surface spherical-harmonic components of degree $\ell$, defined analogously to} Eq.~(\ref{eqn:surface-harmonic-def}). Note that the summation runs only over the positive integers, as $\mathcal{X}_0\hspace{-1pt}=\hspace{-1pt}\mathcal{Y}_0\hspace{-1pt}=\hspace{-1pt}\mathcal{Z}_0\hspace{-1pt}=0$, identically. {Accordingly, the remaining relations can be indexed by $\ell\geq1$}~\cite{Happel1981}. By means of the orthogonality condition of the solid spherical harmonics, we find
	\begin{align}
		\mathcal{X}_\ell=\;& p_{-(\ell+1)}\frac{\left(\ell+1\right)r^{\ell+1}}{2\eta\hspace{-1pt}\left(2\ell\hspace{-1pt}-\hspace{-1pt}1\right)\hspace{-0.75pt}R^{\hspace{0.5pt}\ell}}-\Upsilon_{-(\ell+1)}\frac{\left(\ell+1\right)r^{\ell+1}}{R^{\hspace{0.5pt}\ell+2}}\notag\\
		&+\, p_\ell\,\frac{\ell\hspace{1pt}R^{\ell+1}}{2\eta\hspace{-1pt}\left(2\ell\hspace{-1pt}+\hspace{-1pt}3\right)\hspace{-0.75pt}r^{\hspace{0.5pt}\ell}}\,+\;\Upsilon_\ell\,\frac{\ell\hspace{1pt}R^{\ell-1}}{r^{\hspace{0.5pt}\ell}},\\[5pt]
		\mathcal{Y}_\ell=\;& p_{-(\ell+1)}\frac{\ell\left(\ell+1\right)r^{\ell+1}}{2\eta\hspace{-1pt}\left(2\ell\hspace{-1pt}-\hspace{-1pt}1\right)\hspace{-0.5pt}R^{\hspace{0.5pt}\ell}}-\Upsilon_{-(\ell+1)}\frac{\left(\ell+1\right)\!\left(\ell+2\right)r^{\ell+1}}{R^{\hspace{0.75pt}\ell+2}}\notag\\
		&+\, p_\ell\,\frac{\ell\left(\ell+1\right)R^{\ell+1}}{2\eta\hspace{-1pt}\left(2\ell\hspace{-1pt}+\hspace{-1pt}3\right)\hspace{-0.75pt}r^{\hspace{0.5pt}\ell}}\,+\;\Upsilon_\ell\,\frac{\ell\left(\ell-1\right)R^{\ell-1}}{r^{\hspace{0.5pt}\ell}},\\[6pt]
		\mathcal{Z}_\ell=\;& \hspace{1pt}\Xi_{-(\ell+1)}\frac{\ell\left(\ell+1\right)\hspace{-0.75pt}r^{\hspace{0.5pt}\ell+1}}{R^{\hspace{0.5pt}\ell+1}} + \hspace{1pt}\Xi_{\,\ell}\,\frac{\ell\left(\ell+1\right)\hspace{-0.75pt} R^{\ell}}{r^{\hspace{0.5pt}\ell}},
	\end{align} where {$\ell\geq1$. The negative-index harmonics are obtained formally by replacing $\ell$ with $-\ell-1$.}

    Since the fluid is required to be at rest at infinity, {for the outer flow} (\ie $\!\itbf{V}\!=\!\itbf{V}_{\!\boldsymbol{+}}$), we must exclude the strictly positive harmonics,
	\begin{equation}
		p_\ell\, =\, \Upsilon_\ell \,= \,\Xi_\ell \,= 0,\quad\mathrm{for}\quad \ell\geq1,
	\end{equation} as the corresponding terms diverge as $r\to\infty$. Therefore, the resulting equations can be simultaneously solved for the three unknown harmonic functions: $p_{-(\ell+1)}$, $\Upsilon_{-(\ell+1)}$, and $\Xi_{-(\ell+1)}$. {This gives}
	\begin{align}
		\label{eqn:p-outer}
		p_{-(\ell+1)} &= \frac{\eta\left(2\ell-1\right)\hspace{-1pt}R^{\hspace{0.25pt}\ell}}{\left(\ell+1\right) r^{\hspace{0.5pt}\ell+1}}\left[\hspace{1pt}\mathcal{Y}_\ell+\left(\ell+2\right)\mathcal{X}_\ell\right]\hspace{-1pt},\\[4pt]
		\label{eqn:upsilon-outer}
		\Upsilon_{-(\ell+1)} &= \frac{R^{\hspace{0.25pt}\ell+2}}{2\hspace{-0.75pt}\left(\ell+1\right)r^{\hspace{0.5pt}\ell+1}}\left[\hspace{1pt}\mathcal{Y}_\ell+\ell\hspace{0.25pt}\mathcal{X}_\ell\hspace{1pt}\right]\hspace{-1pt},\\[4pt]
		\label{eqn:xi-outer}
		\Xi_{-(\ell+1)} &= \frac{R^{\hspace{0.25pt}\ell+1}}{\ell\left(\ell+1\right)\hspace{-0.5pt}r^{\hspace{0.5pt}\ell+1}}\,\mathcal{Z}_\ell,
	\end{align} for $\ell\geq1$, and we also set $p_{-1}\!=\!\Upsilon_{-1}\!=\Xi_{-1}\!=0$, as there are no explicit sources or sinks for the ambient fluid. This completely determines both the velocity and pressure fields outside the closed membrane.

    Similarly, for the inner flow (\ie $\!\itbf{V}\!=\!\itbf{V}_{\!\boldsymbol{-}}$), bounded by the membrane, the negative harmonics must be set to zero, as these terms diverge at the origin ($r\to0$):
	\begin{equation}
		p_{-(\ell+1)} =\Upsilon_{-(\ell+1)}= \,\Xi_{-(\ell+1)}\,= 0,\;\;\mathrm{for}\;\;\ell\geq0.
	\end{equation} Thus, we find that
	\begin{align}
		\label{eqn:p-inner}
		p_{\ell} &= \frac{\eta\left(2\ell+3\right)\hspace{-0.75pt}r^{\hspace{0.25pt}\ell}}{\ell\hspace{0.5pt}R^{\hspace{0.5pt}\ell+1}}
		\left[\hspace{1pt}\mathcal{Y}_\ell - \left(\ell-1\right)\mathcal{X}_\ell\hspace{1pt}\right]\hspace{-1pt},\\[4pt]
		\label{eqn:upsilon-inner}
		\Upsilon_{\ell} &= \frac{r^{\hspace{0.25pt}\ell}}{2\hspace{0.5pt}\ell\hspace{0.5pt}R^{\hspace{0.5pt}\ell-1}}\left[\left(\ell+1\right)\mathcal{X}_\ell-\mathcal{Y}_\ell\hspace{1pt}\right]\hspace{-1pt},\\[3pt]
		\label{eqn:xi-inner}
		\Xi_{\hspace{1pt}\ell} &= \frac{r^{\hspace{0.25pt}\ell}}{\ell\left(\ell+1\right)\hspace{-0.5pt}R^{\hspace{0.25pt}\ell}}\,\mathcal{Z}_\ell,
	\end{align} for integer modes $\ell\geq1$. {These relations determine the velocity and pressure fields of the inner fluid.}

	\subsection{Flows around a slightly deformed sphere}
    
	{Here,} we assume that {the} velocity and pressure fields of the ambient fluid can be written as {power series in} the perturbation parameter $\varepsilon$, that is,
	\begin{equation}
		\label{eqn:velocity-pressure-expansion}
		\itbf{V}_{\!{\boldsymbol{\pm}}}  = \sum_{\mathtt{i}=\hspace{0.65pt}0}^{\infty}\,\varepsilon^{\mathtt{i}}\hspace{-1pt}\itbf{V}^{(\mathtt{i})}_{\!{\boldsymbol{\pm}}},\quad\mathrm{and}\quad p_{\hspace{0.5pt}{\boldsymbol{\pm}}} = \sum_{\mathtt{i}=\hspace{0.65pt}0}^{\infty}\,\varepsilon^{\mathtt{i}}\hspace{0.5pt} p^{(\mathtt{i})}_{\hspace{0.5pt}{\boldsymbol{\pm}}},\vspace{-1pt}
	\end{equation} respectively, where the subscripts ``$\scriptstyle\boldsymbol{+}$" and ``$\scriptstyle\boldsymbol{-}$" denote {the regions outside and inside} the closed membrane, {respectively}. Due to the linearity of the Stokes equations, {substituting} the above expansions into Eq.~(\ref{eqn:Stokes-Eq}) {shows} that each {perturbation field} must also satisfy the Stokes equations,
	\begin{equation}
		\label{eqn:Stokes-Eq-perturbation}
		\eta\hspace{1pt}\bnabla^2\hspace{-1pt}\itbf{V}^{(\mathtt{i})}_{\!{\boldsymbol{\pm}}}=\bnabla p^{(\mathtt{i})}_{\hspace{0.5pt}{\boldsymbol{\pm}}},\quad\text{and}\quad\bnabla\hspace{-1pt}\cdot\hspace{-1pt}\itbf{V}^{(\mathtt{i})}_{\!{\boldsymbol{\pm}}}=0.
	\end{equation}
	
	Similarly, by equating the terms corresponding to each order in $\varepsilon$, the boundary conditions in Eq.~(\ref{eqn:condition-infinity}) yield
    \begin{equation}
    	\label{eqn:condition-infinity-perturbation}
    	\begin{cases}
    		\cc{\displaystyle \lim_{|\itbf r|\to0}\left|\itbf V_{\!{\boldsymbol{-}}}^{(\mathtt{i})}(\itbf r)\right|<\infty},&
    		\cc{\displaystyle \lim_{|\itbf r|\to0}p_{\hspace{0.5pt}{\boldsymbol{-}}}^{(\mathtt{i})}(\itbf r)=P_-^{(\mathtt{i})}},\\[5pt]
    		\cc{\displaystyle \lim_{|\itbf r|\to\infty}\left|\itbf V_{\!{\boldsymbol{+}}}^{(\mathtt{i})}(\itbf r)\right|=0},&
    		\cc{\displaystyle \lim_{|\itbf r|\to\infty}p_{\hspace{0.5pt}{\boldsymbol{+}}}^{(\mathtt{i})}(\itbf r)=P_+^{(\mathtt{i})}}.
    	\end{cases}
    \end{equation}
    Likewise, the boundary condition in Eq.~(\ref{eqn:boundary-cond-fluid}) at $\itbf{r}=\itbf{R}$ can be expressed as follows:
    \begin{equation}
    	\label{eqn:boundary-cond-fluid-perturbation}
    	\itbf{v}_{\mathfrak{m}}
    	=
    	\itbf{v}+\itbf{v}_{\!\scriptscriptstyle S}
    	\cc{-j_{\scriptscriptstyle V}\nv}
    	=
    	\sum_{\mathtt{i}=0}^{\infty}
    	\varepsilon^{\mathtt{i}}
    	\itbf{V}_{\!{\boldsymbol{\pm}}}^{(\mathtt{i})}
    	\hspace{-2pt}\left(\itbf{r}=\itbf{R}\right)\!,
    \end{equation}
    where we define the composite velocity $\itbf{v}_{\mathfrak{m}}$ as \cc{the prescribed interfacial velocity that matches the surrounding fluid flow}. In addition, the membrane velocity $\itbf{v}=\frac{\mathrm{d}}{\mathrm{d}t}\itbf{R}$, the \cc{slip velocity} $\itbf{v}_{\!\scriptscriptstyle S}$, \cc{the scalar volume flux $j_{\scriptscriptstyle V}$, and the surface normal $\nv$ may each be expanded} in the perturbation parameter; namely,
    \begin{align}
    	\itbf{v}
    	&=
    	\sum_{\mathtt{i}=0}^{\infty}\varepsilon^{\mathtt{i}}\itbf{v}^{(\mathtt{i})}\!,
    	\qquad
    	\itbf{v}_{\!\scriptscriptstyle S}
    	=
    	\sum_{\mathtt{i}=0}^{\infty}\varepsilon^{\mathtt{i}}\itbf{v}^{(\mathtt{i})}_{\!\scriptscriptstyle S}\!,
    	\notag\\[5pt]
    	\cc{j_{\scriptscriptstyle V}}
    	&\cc{
    	=
    	\sum_{\mathtt{i}=0}^{\infty}\varepsilon^{\mathtt{i}}j_{\scriptscriptstyle V}^{(\mathtt{i})},
    	\qquad
    	\nv
    	=
    	\sum_{\mathtt{i}=0}^{\infty}\varepsilon^{\mathtt{i}}\nv^{(\mathtt{i})}.
    	}
    \end{align}
    Hence, \cc{at each order in the perturbation parameter,}
    \begin{equation}
    	\cc{
    	\itbf{v}^{(\mathtt{i})}_{\mathfrak{m}}
    	=
    	\itbf{v}^{(\mathtt{i})}
    	+\itbf{v}^{(\mathtt{i})}_{\!\scriptscriptstyle S}
    	-\sum_{\mathtt{q}=0}^{\mathtt{i}}
    	j_{\scriptscriptstyle V}^{(\mathtt{q})}
    	\nv^{(\mathtt{i}-\mathtt{q})}.
    	}
    \end{equation}
    By~Taylor expanding {the bulk-fluid perturbation fields in Eq.~(\ref{eqn:boundary-cond-fluid-perturbation})} about the fixed radius $r=R$, we~find that
    \begin{equation}
    	\label{eqn:perturbation-velocity-taylor}
    	\itbf{V}_{\!{\boldsymbol{\pm}}}^{(\mathtt{i})}\hspace{-2pt}\left(\itbf{r}=\itbf{R}\right)
    	=
    	\sum^{\infty}_{\mathtt{j}=\hspace{0.65pt}0}
    	\frac{\left(\hspace{0.5pt}r-R\hspace{0.5pt}\right)^\mathtt{j}}{\mathtt{j}!}
    	\left[
    	\frac{\partial^{(\mathtt{j})}}{\partial r^{(\mathtt{j})}}
    	\itbf{V}_{\!{\boldsymbol{\pm}}}^{(\mathtt{i})}
    	\right]_{\hspace{-1pt}r=R}.
    \end{equation}
    {where, on the membrane,} $r-R=\varepsilon R\,u(\theta,\varphi)$. By substituting (\ref{eqn:perturbation-velocity-taylor}) into Eq.~(\ref{eqn:boundary-cond-fluid-perturbation}), we have that
    \begin{equation}
    	\label{eqn:expansion-boundary-condition-velocities}
    	\cc{\itbf{v}_{\mathfrak{m}}^{(\mathtt{i})}
    	=}
    	\sum_{\mathtt{j}=0}^{\mathtt{i}}
    	\frac{\left(R\,u\right)^{\mathtt{j}}}{\mathtt{j}!}
    	\left[
    	\frac{\partial^{(\mathtt{j})}}{\partial r^{(\mathtt{j})}}
    	\itbf{V}_{\!{\boldsymbol{\pm}}}^{(\mathtt{i}-\mathtt{j})}
    	\right]_{r=R},
    \end{equation}
    which is derived by collecting and equating terms of the same order in $\varepsilon$. Satisfying the {velocity} boundary condition on the deformed {membrane} requires matching {the} perturbation fields on the fixed {reference} sphere $r=R$. As each {bulk-fluid perturbation} also satisfies the Stokes equations, Lamb's \cc{method can be applied order by order}. The {zeroth- and first-order} terms in Eq.~(\ref{eqn:expansion-boundary-condition-velocities}) {are}
    \begin{align}
    	\cc{\itbf{v}_{\mathfrak{m}}^{(\mathtt{0})}}
    	&=
    	\left[\itbf{V}_{\!{\boldsymbol{\pm}}}^{(\mathtt{0})}\right]_{r=R},
    	\\[6pt]
    	\label{eqn:velocity-1-BC}
    	\cc{\itbf{v}_{\mathfrak{m}}^{(\mathtt{1})}}
    	&=
    	\left[\itbf{V}_{\!{\boldsymbol{\pm}}}^{(\mathtt{1})}\right]_{r=R}
    	+\left(Ru\right)
    	\left[
    	\frac{\partial\itbf{V}_{\!{\boldsymbol{\pm}}}^{(\mathtt{0})}}{\partial r}
    	\right]_{r=R}.
    \end{align}
    \cc{Higher-order boundary conditions contain terms with progressively higher powers of} $u(\theta,\varphi)$.

    From Eq.~(\ref{eqn:divergence-law-eq}), the surface divergence of the membrane velocity field is related to the {active membrane-area flux} $\mathbb{M}$, which may {likewise} be expanded as $\mathbb{M} = \sum_{\mathtt{i}=\hspace{0.65pt}0}^{\infty}\,\varepsilon^{\mathtt{i}}\hspace{0.5pt}\mathbb{M}^{(\mathtt{i})}$, {so that} Eq.~(\ref{eqn:divergence-law-eq}) {becomes}
    \begin{equation}
    	\label{eqn:divergence-law-perturbation}
    	\mathbb{M}^{(\mathtt{i})} = \left(\Div\itbf{v}\right)^{(\mathtt{i})}\!.
    \end{equation}
    
    Moreover, the perturbation expansion in Eq.~(\ref{eqn:velocity-pressure-expansion}) {gives} the following stress tensors:
    \begin{equation}
    	\label{eqn:Stokes-stress-perturbation}
    	\boldsymbol{\mathsf{\Omega}}^{(\mathtt{i})}_{\boldsymbol{\pm}} = -p^{(\mathtt{i})}_{\hspace{0.5pt}{\boldsymbol{\pm}}}\hspace{1pt}\itbf{I} + \eta\left\{\hspace{-0.5pt}\left[\bnabla\hspace{-1pt}\itbf{V}^{(\mathtt{i})}_{\!{\boldsymbol{\pm}}}\right]+\left[\bnabla\hspace{-1pt}\itbf{V}^{(\mathtt{i})}_{\!{\boldsymbol{\pm}}}\right]^{\hspace{-0.5pt}\mathsf{T}}\right\}\hspace{-0.75pt}\hspace{-1pt},
    \end{equation}
    {and the corresponding} stress vectors $\boldsymbol{\pi}^{(\mathtt{i})}_{\boldsymbol{\pm}}$ acting across the surface of the membrane {are}
    \begin{equation}
    	\label{eqn:stress-vector-perturbation}
    	\boldsymbol{\pi}^{(\mathtt{i})}_{\boldsymbol{\pm}} = \sum^{\mathtt{i}}_{\mathtt{j}=\hspace{0.65pt}0} \,\boldsymbol{\mathsf{\Omega}}^{(\mathtt{j})}_{\boldsymbol{\pm}}\!\cdot\nv^{(\mathtt{i}-\mathtt{j})},
    \end{equation}
    which follows from the series {expansion} in $\varepsilon$ of the stress vector $\boldsymbol{\pi}_{\boldsymbol{\pm}}=\sum_{\mathtt{i}=\hspace{0.65pt}0}^{\infty}\big[\varepsilon^{\mathtt{i}}\hspace{1.5pt}\boldsymbol{\pi}^{(\mathtt{i})}_{{\boldsymbol{\pm}}}\big]$, \cc{together with the expansion of the surface normal introduced above}. In particular, the \cc{zeroth- and first-order} terms of Eq.~(\ref{eqn:stress-vector-perturbation}) are
    \begin{align}
    	\boldsymbol{\pi}^{(\mathtt{0})}_{\boldsymbol{\pm}} &=\, \boldsymbol{\mathsf{\Omega}}^{(\mathtt{0})}_{\boldsymbol{\pm}}\!\cdot\nv^{(\mathtt{0})},\quad\mathrm{and}\\[10pt]
    	\boldsymbol{\pi}^{(\mathtt{1})}_{\boldsymbol{\pm}} &=\,
    	\boldsymbol{\mathsf{\Omega}}^{(\mathtt{0})}_{\boldsymbol{\pm}}\!\cdot\nv^{(\mathtt{1})}+\,
    	\boldsymbol{\mathsf{\Omega}}^{(\mathtt{1})}_{\boldsymbol{\pm}}\!\cdot\nv^{(\mathtt{0})}.
    \end{align}
    
    Similarly, the body {force} $\itbf{f}$ on the membrane can be expanded in a power series {in} $\varepsilon$, that is, $\itbf{f}=\sum^{\infty}_{\mathtt{i}=\hspace{0.65pt}0}\,\varepsilon^{\mathtt{i}}\hspace{0.25pt}\itbf{f}^{(\mathtt{i})}$. Hence, {Taylor expanding} the stress boundary condition in Eq.~(\ref{eqn:stress-boundary-condition}) {about the fixed sphere gives}
    \begin{equation}
    	\label{eqn:force-stress-vector-eq}
    	\itbf{f}^{(\mathtt{i})}\hspace{-1pt} = \sum^{\mathtt{i}}_{\mathtt{j}=\hspace{0.65pt}0}\,\frac{\left(R\,u\hspace{1pt}\right)^\mathtt{j}}{\mathtt{j}!}\!\left[\frac{\partial^{(\mathtt{j})}}{\partial r^{(\mathtt{j})}}\,\boldsymbol{\pi}^{(\mathtt{i}-\mathtt{j})}_{\scriptscriptstyle\Delta}\hspace{-0.5pt}\right]_{\hspace{-1.0pt} r=R},
    \end{equation}
    where we define {the stress-vector jump}
    \begin{equation}
    	\boldsymbol{\pi}_{\scriptscriptstyle\Delta}^{(\mathtt{i})}=\,\boldsymbol{\pi}^{(\mathtt{i})}_{\boldsymbol{+}}\! - \boldsymbol{\pi}_{\boldsymbol{-}}^{(\mathtt{i})}.
    \end{equation}
    Eq.~(\ref{eqn:force-stress-vector-eq}) is {obtained analogously to} Eq.~(\ref{eqn:perturbation-velocity-taylor}), with the stress vectors $\boldsymbol{\pi}^{(\mathtt{i})}_{{\boldsymbol{\pm}}}(\itbf{r}=\!\itbf{R})$ {expanded} about a fixed sphere of radius $R$. This gives {the} explicit forms of the {zeroth- and first-order} terms of Eq.~(\ref{eqn:force-stress-vector-eq}):
    \begin{align}
    	\itbf{f}^{(\mathtt{0})}\! &= \left[\boldsymbol{\pi}_{\scriptscriptstyle\Delta}^{(\mathtt{0})}\hspace{-0.5pt}\right]_{\hspace{-1pt} r=R},\quad\mathrm{and}\\[3pt]
    	\label{eqn:stress-1-BC}
    	\itbf{f}^{(\mathtt{1})} &= \left[\boldsymbol{\pi}_{\scriptscriptstyle\Delta}^{(\mathtt{1})}\hspace{-0.5pt}\right]_{\hspace{-1pt} r=R} + \left(R\,u\right)\!\left[\frac{\partial}{\partial r}\,\boldsymbol{\pi}_{\scriptscriptstyle\Delta}^{(\mathtt{0})}\right]_{\hspace{-1pt} r=R}.
    \end{align}
	
	\cc{Starting from the leading-order perturbation field, the higher-order fields can then be solved successively by matching the corresponding boundary conditions.}
	
	\subsubsection{Zeroth order perturbation}
    
    At {leading order} in $\varepsilon$, the position vector $\itbf{R}$ of the membrane is given by $\itbf{R}^{(\mathtt{0})}\hspace{-1pt}=R\hspace{1pt}\rv$. Thus, the membrane velocity $\itbf{v}^{(\mathtt{0})}=\frac{\mathrm{d}}{\mathrm{d}\hspace{0.25pt}t}\itbf{R}^{(\mathtt{0})}$ may {be written as}
    \begin{align}
    	\itbf{v}^{(\mathtt{0})}\!=\!\frac{\partial\hspace{-1pt}\itbf{R}^{(\mathtt{0})}\!\!}{\partial t}\!+
    	\left[v^{\alpha}\right]^{(\mathtt{0})}\!\hspace{-1pt}\itbf{R}^{(\mathtt{0})}_{\hspace{0.25pt},\alpha}\!=\! \,v^{(\mathtt{0})}\hspace{-1pt}\nv^{(\mathtt{0})}\! + \left[v^{\alpha}\right]^{(\mathtt{0})}\!\hspace{-2pt}\ed{\alpha}^{(\mathtt{0})}\!,
    \end{align}
    where the first equality follows from \cc{the material derivative}, \cc{while the second gives the decomposition into normal and tangential components}. Since $\nv^{(\mathtt{0})}\hspace{-1pt}=\rv$, $\itbf{e}^{(\mathtt{0})}_{\theta}\!=R\,\thetav$, and $\itbf{e}^{(\mathtt{0})}_{\varphi}\!=R\sin\theta\,\phiv$, {we have} $v^{(\mathtt{0})}=0$, and thus
    \begin{equation}
    	\itbf{v}^{(\mathtt{0})} = \hat{v}^{(\mathtt{0})}_\theta\hspace{0.6pt}\thetav\, + \,\hat{v}^{(\mathtt{0})}_\varphi\hspace{0.6pt}\phiv,
    \end{equation}
    where we additionally define
    \begin{equation}
    	\hat{v}^{(\mathtt{0})}_\theta= R\left[v^{\theta}\hspace{0.5pt}\right]^{\!(\mathtt{0})}\!\!, \quad\mathrm{and}\quad
    	\hat{v}^{(\mathtt{0})}_\varphi= R\sin\theta\left[v^{\varphi}\hspace{0.5pt}\right]^{(\mathtt{0})}\!\!.
    \end{equation}
    \cc{Thus, the zeroth-order membrane velocity is purely tangential to the fixed sphere of radius $R$.}
    
    \cc{In the homogeneous base state, the zeroth-order second active force moment $\mathbb{Q}^{(\mathtt{0})}$ is spatially uniform, so the corresponding slip velocity vanishes, $\itbf{v}_{\scriptscriptstyle S}^{(\mathtt{0})}=\boldsymbol{0}$. The normal component of the zeroth-order interfacial velocity defined in Eq.~(\ref{eqn:boundary-cond-fluid-perturbation}) is therefore}
    \begin{equation}
    	{
    	\rv\cdot\itbf{v}_{\mathfrak{m}}^{(\mathtt{0})}
    	=-j_{\scriptscriptstyle V}^{(\mathtt{0})}.
    	}
    \end{equation}
    \cc{Since the zeroth-order inner flow is incompressible, its net normal flux through the closed membrane must vanish. Moreover, $j_{\scriptscriptstyle V}^{(\mathtt{0})}$ is spatially uniform in the homogeneous base state, and hence}
    \begin{equation}
    	{
    	0=
    	\int_{\mathcal{M}}\mathrm{d}S^{(\mathtt{0})}\,
    	\rv\cdot\itbf{V}_{\!{\boldsymbol{-}}}^{(\mathtt{0})}
    	=
    	-4\pi R^2j_{\scriptscriptstyle V}^{(\mathtt{0})}.
    	}
    \end{equation}
    \cc{Thus, the zeroth-order volume flux vanishes,}
    \begin{equation}
    	\label{eqn:volume-flux-zeroth-order}
    	j_{\scriptscriptstyle V}^{(\mathtt{0})}\!=\mathbb{V}^{(\mathtt{0})}+w_0\hspace{-1pt}\left[P^{(\mathtt{0})}_{\hspace{-0.25pt}\scriptscriptstyle\Delta}-\Pi^{(\mathtt{0})}_{\scriptscriptstyle\Delta}\right]\! = 0,
    \end{equation}
    where $\mathbb{V}^{(\mathtt{0})}$ is the {zeroth-order active contribution to the volume flux}, while $\Pi^{(\mathtt{0})}_{\scriptscriptstyle\Delta}$ and $P^{(\mathtt{0})}_{\hspace{-0.25pt}\scriptscriptstyle\Delta}$ are the {zeroth-order osmotic pressure and hydrostatic pressure difference}, respectively.
    
    \cc{It follows that the zeroth-order interfacial velocity has no radial component.} Using Lamb's method as described in {the} previous section, we therefore have
    \begin{equation}
    	\rv\cdot\itbf{v}^{(\mathtt{0})}_{\mathfrak{m}}
    	=
    	{\sum_{\ell=1}^{\infty}}\,\mathcal{X}^{(\mathtt{0})}_\ell
    	=0,
    \end{equation}
    \cc{and the orthogonality of the surface spherical harmonics implies $\mathcal{X}^{(\mathtt{0})}_\ell=0$ for every $\ell\geq1$. The zeroth-order Lamb term evaluated on the reference sphere is}
    \begin{equation}
    	\mathfrak{L}^{(\mathtt{0})}
    	=
    	-R\,\Div\itbf{v}_{\mathfrak{m}}^{(\mathtt{0})}
    	=
    	-R\,\Div\itbf{v}^{(\mathtt{0})}_{\hspace{-1pt}\scriptscriptstyle S}
    	+2\hspace{0.5pt}j_{\hspace{-1pt}\scriptscriptstyle V}^{(\mathtt{0})}
    	-R\hspace{1pt}\mathbb{M}^{(\mathtt{0})}.
    \end{equation}
    Here, $\mathbb{M}^{(\mathtt{0})}$ is the \cc{zeroth-order active membrane-area flux, which vanishes in the homogeneous steady state, as shown in Appendix~B.4}. \cc{Since $\itbf{v}_{\scriptscriptstyle S}^{(\mathtt{0})}=\boldsymbol{0}$ and $j_{\scriptscriptstyle V}^{(\mathtt{0})}=0$, it follows that}
    \begin{equation}
    	\mathfrak{L}^{(\mathtt{0})}
    	=
    	{\sum_{\ell=1}^{\infty}}\,\mathcal{Y}^{(\mathtt{0})}_\ell
    	=0,
    \end{equation}
    which implies that each of the {zeroth-order} surface harmonics $\mathcal{Y}^{(\mathtt{0})}_\ell=0$ for all $\ell\geq1$.
    
    \cc{Consequently, together with $p_{-1}^{(\mathtt{0})}=\Upsilon_{-1}^{(\mathtt{0})}=0$, the corresponding pressure and potential harmonics for the inner and outer flows vanish,}
    \begin{equation}
    	\begin{cases}
    		p^{(\mathtt{0})}_\ell=\Upsilon^{(\mathtt{0})}_\ell=0,
    		& \ell\geq1,\\[4pt]
    		p^{(\mathtt{0})}_{-(\ell+1)}=\Upsilon^{(\mathtt{0})}_{-(\ell+1)}=0,
    		& \ell\geq0.
    	\end{cases}
    \end{equation}
    \cc{The first and second lines correspond to the inner and outer flows, respectively.}
    
    As the normal $\nv^{(\mathtt{0})}\hspace{-1pt}=\rv$, the {zeroth-order} {traction} vectors $\boldsymbol{\pi}^{(\mathtt{0})}_{\boldsymbol{\pm}}$ acting across the membrane are given by
    \begin{equation}
    	\label{eqn:radial-stress-vector-0}
    	\boldsymbol{\pi}^{(\mathtt{0})}_{\boldsymbol{\pm}}\! = -\rv\hspace{1pt}p^{(\mathtt{0})}_{\boldsymbol{\pm}} + \eta\hspace{1pt}r\frac{\partial}{\partial r}\!\hspace{-1pt}\left[\!\frac{\itbf{V}^{(\mathtt{0})}_{\!{\boldsymbol{\pm}}}}{r}\hspace{-1pt}\right]\!+\frac{\eta}{r}\,\bnabla\!\left[r\rv\cdot\!\itbf{V}^{(\mathtt{0})}_{\!{\boldsymbol{\pm}}}\right]\!\hspace{-1pt}.
    \end{equation}
    \cc{Since the nonconstant pressure harmonics vanish, the corresponding pressure fields reduce to}
    \begin{align}
    	p^{(\mathtt{0})}_{\boldsymbol{-}}
    	&=
    	P^{(\mathtt{0})}_{\boldsymbol{-}}
    	+\sum_{\ell=1}^{\infty}p_\ell^{(\mathtt{0})}
    	=
    	P^{(\mathtt{0})}_{\boldsymbol{-}},
    	\\[6pt]
    	p^{(\mathtt{0})}_{\boldsymbol{+}}
    	&=
    	P^{(\mathtt{0})}_{\boldsymbol{+}}
    	+\sum_{\ell=1}^{\infty}p_{-(\ell+1)}^{(\mathtt{0})}
    	=
    	P^{(\mathtt{0})}_{\boldsymbol{+}}.
    \end{align}
    Here, $P^{(\mathtt{0})}_{\boldsymbol{\pm}}$ are the \cc{zeroth-order constant pressures outside and inside of the organelle, respectively}. \cc{These constants do not contribute to the velocity fields} $\!\itbf{V}^{(\mathtt{0})}_{\!{\boldsymbol{\pm}}}$, \cc{since their gradients vanish}, cf.~Eq.~(\ref{eqn:Lamb-solution}).
    
     \cc{The remaining zeroth-order solid spherical harmonics $\Xi^{(\mathtt{0})}_{\ell}$ and $\Xi^{(\mathtt{0})}_{-(\ell+1)}$ describe the toroidal components of the inner and outer flows, respectively. Since the homogeneous membrane fields generate no tangential force at this order, cf.~Eq.~(\ref{eqn:membrane-equation-tangential}), taking the radial component of the curl of the stress-vector jump obtained from Eq.~(\ref{eqn:radial-stress-vector-0}) gives}
    \begin{equation}
    	{\mathcal{Z}^{(\mathtt{0})}_\ell=0,\qquad \ell\geq1.}
    \end{equation}
    \cc{It then follows from Eqs.~(\ref{eqn:xi-inner}) and (\ref{eqn:xi-outer}) that both $\Xi^{(\mathtt{0})}_{\ell}$ and $\Xi^{(\mathtt{0})}_{-(\ell+1)}$ vanish identically.}
    
    As a result, the {zeroth-order} velocity fields $\!\itbf{V}^{(\mathtt{0})}_{\!{\boldsymbol{\pm}}}\hspace{-1pt}=\boldsymbol{0}$ everywhere within the bulk fluid, including at $r=R$. \cc{The interfacial condition, together with $j_{\scriptscriptstyle V}^{(\mathtt{0})}=0$ and $\itbf{v}_{\scriptscriptstyle S}^{(\mathtt{0})}=\boldsymbol{0}$, then gives $\itbf{v}^{(\mathtt{0})}=\itbf{v}_{\mathfrak{m}}^{(\mathtt{0})}=\boldsymbol{0}$.} By means of Eq.~(\ref{eqn:radial-stress-vector-0}), we find that the {zeroth-order} stress vectors $\boldsymbol{\pi}^{(\mathtt{0})}_{\boldsymbol{\pm}}=-\rv\hspace{1pt}P^{(\mathtt{0})}_{\boldsymbol{\pm}}$, whereas the stress tensors are given by
    \begin{equation}
    	\boxed{\boldsymbol{\mathsf{\Omega}}^{(\mathtt{0})}_{\boldsymbol{\pm}} = -P^{(\mathtt{0})}_{\hspace{0.5pt}{\boldsymbol{\pm}}}\itbf{I}.}
    \end{equation}
    Hence, the {zeroth-order} stress boundary condition in Eq.~(\ref{eqn:force-stress-vector-eq}) {reduces to}
    \begin{equation}
    	\label{eqn:body-forces-0}
    	\itbf{f}^{(\mathtt{0})} = - \rv\left[P^{(\mathtt{0})}_{\boldsymbol{+}}\!-P^{(\mathtt{0})}_{\boldsymbol{-}}\right] =  - \rv\hspace{0.75pt}P^{(\mathtt{0})}_{\hspace{-0.5pt}\scriptscriptstyle\Delta},
    \end{equation}
    where \cc{$P^{(\mathtt{0})}_{\hspace{-0.5pt}\scriptscriptstyle\Delta}=P^{(\mathtt{0})}_{\boldsymbol{+}}-P^{(\mathtt{0})}_{\boldsymbol{-}}$ denotes the zeroth-order hydrostatic pressure difference across the membrane}. Thus, by decomposing the \cc{hydrodynamic} force $\itbf{f}^{(\mathtt{0})}$ onto the local membrane basis, we have
    \begin{equation}
    	\label{eqn:body-forces-comp-0}
    	\boxed{f^{(\mathtt{0})} = -P^{(\mathtt{0})}_{\hspace{-0.5pt}\scriptscriptstyle\Delta},\quad\mathrm{and}\quad \itbf{f}_{\!\scriptscriptstyle\mathcal{S}}^{(\mathtt{0})}=  \ed{\alpha}^{(\mathtt{0})}\left[f^\alpha\right]^{(\mathtt{0})}\!= \boldsymbol{0},}
    \end{equation}
    where $f^{(\mathtt{0})}$ and $\left[f^\alpha\right]^{(\mathtt{0})}$ are the {zeroth-order} normal and tangential components of the \cc{hydrodynamic} force, respectively, and $\itbf{f}_{\!\scriptscriptstyle\mathcal{S}}^{(\mathtt{0})}$ \cc{denotes the tangential hydrodynamic-force vector}.

	\subsubsection{First order perturbation}
    
	At {first order} in the perturbation, the position vector of the membrane is given by $\itbf{R}^{(\mathtt{1})}\hspace{-1pt}=R\hspace{1.25pt}u\!\left(\theta,\varphi\right)\rv$, and the corresponding membrane velocity $\itbf{v}^{(\mathtt{1})}$ is {given by}
	\begin{align}
		\itbf{v}^{(\mathtt{1})}\!=\!\frac{\partial\hspace{-1pt}\itbf{R}^{(\mathtt{1})}\!\!}{\partial t}\!+
		\left[v^{\alpha}\right]^{(\mathtt{1})}\!\hspace{-1pt}\itbf{R}^{(\mathtt{0})}_{\hspace{0.25pt},\alpha}\!=\! \,v^{(\mathtt{1})}\hspace{-1pt}\nv^{(\mathtt{0})}\! + \left[v^{\alpha}\right]^{(\mathtt{1})}\!\hspace{-2pt}\ed{\alpha}^{(\mathtt{0})}\!,
	\end{align} which {follows from} $v^{(\mathtt{0})}=0$ and $\left[v^{\alpha}\right]^{(\mathtt{0})}=0$. As before, we can rewrite the above expression as
	\begin{equation}
		\label{eqn:velocity-first-order-def}
		\itbf{v}^{(\mathtt{1})} = \hat{v}^{(\mathtt{1})}_r\hspace{0.6pt}\rv + \hat{v}^{(\mathtt{1})}_\theta\hspace{0.6pt}\thetav\, + \,\hat{v}^{(\mathtt{1})}_\varphi\hspace{0.6pt}\phiv,
	\end{equation} \cc{where we use the corresponding unit tangent vectors to define the tangential velocity components}
	\begin{equation}
		\label{eqn:radial-velocity-first-order}
		\hat{v}^{(\mathtt{1})}_\theta= R\left[v^{\theta}\hspace{0.5pt}\right]^{\!(\mathtt{1})}\! \quad\mathrm{and}\quad\hat{v}^{(\mathtt{1})}_\varphi= R\sin\theta\left[v^{\varphi}\hspace{0.5pt}\right]^{(\mathtt{1})}\!\!,
	\end{equation} while the radial velocity of $\itbf{v}^{(\mathtt{1})}$ is found to be 
	\begin{equation}
		\label{eqn:m-radial-velocity-1st}
		\hat{v}^{(\mathtt{1})}_r= R\,\frac{\partial u}{\partial t} = R\,\sum_{\ell=0}^{\infty}\frac{\partial u_\ell}{\partial t}.
	\end{equation}
	
    {Since the zeroth-order fluid velocities} $\!\itbf{V}^{(\mathtt{0})}_{\!{\boldsymbol{\pm}}}=\boldsymbol{0}$, the boundary condition in Eq.~(\ref{eqn:velocity-1-BC}) reduces to
    \begin{equation}
    	\label{eqn:boundary-condition-1st}
    	\itbf{v}^{(\mathtt{1})}
    	+\itbf{v}^{(\mathtt{1})}_{\!\scriptscriptstyle S}
    	\cc{-j_{\scriptscriptstyle V}^{(\mathtt{1})}\rv}
    	=
    	\left[\itbf{V}_{{\boldsymbol{\pm}}}^{(\mathtt{1})}\right]_{r=R}.
    \end{equation} which allows us to {apply} Lamb's method to obtain both the inner and outer fluid flows. By employing Eq.~(\ref{eqn:X-ell}), {the degree-$\ell$ surface harmonics for $\ell\geq1$ are}
    \begin{equation}
    	\label{eqn:X-ell-1}
    	\boxed{\mathcal{X}^{(\mathtt{1})}_\ell
    	=R\,\frac{\partial u_\ell}{\partial t}
    	-\mathbb{V}_{\hspace{-1pt}\ell}^{(\mathtt{1})},
    	\qquad \ell\geq1,}
    \end{equation} where $u_\ell$ and $\mathbb{V}_{\hspace{-1pt}\ell}^{(\mathtt{1})}$ are the \cc{degree-$\ell$ components of the shape field and the first-order active volume flux, respectively}. \cc{For $\ell\geq1$, the passive osmotic contribution to $j_{\scriptscriptstyle V}^{(\mathtt{1})}$ vanishes, so that $j_{{\scriptscriptstyle V},\ell}^{(\mathtt{1})}=\mathbb{V}_{\ell}^{(\mathtt{1})}$.}
	
	From Eq.~(\ref{eqn:divergence-law-perturbation}), we have $\Div\itbf{v}^{(\mathtt{1})}=\mathbb{M}^{(\mathtt{1})}$, and since $R\,\Div\itbf{v}^{(\mathtt{1})}_{\mathfrak{m}}=r\bnabla\cdot\itbf{v}^{(\mathtt{1})}_{\mathfrak{m}}$, Eq.~(\ref{eqn:Y-ell}) gives
    \begin{equation}
    	\label{eqn:Y-ell-1}
    	\boxed{\mathcal{Y}^{(\mathtt{1})}_\ell
    	=-R\,\mathbb{M}^{(\mathtt{1})}_\ell
    	+2\hspace{0.5pt}\mathbb{V}^{(\mathtt{1})}_{\hspace{-1pt}\ell}
    	+\frac{\ell(\ell+1)}{2\eta R}\,\mathbb{Q}^{(\mathtt{1})}_\ell.}
    \end{equation}
    where $\mathbb{M}^{(\mathtt{1})}_\ell$ is the \cc{degree-$\ell$ component of the first-order active membrane-area flux}. \cc{The remaining terms follow from}
    \begin{equation}
    	\cc{
    	\Div\itbf{v}^{(\mathtt{1})}_{\mathfrak{m}}
    	=
    	\Div\!\left(\itbf{v}^{(\mathtt{1})}+\itbf{v}^{(\mathtt{1})}_{\scriptscriptstyle S}\right)
    	+2H^{(\mathtt{0})}j_{\scriptscriptstyle V}^{(\mathtt{1})},
    	}
    \end{equation}
    \cc{with $H^{(\mathtt{0})}=-1/R$ and $j_{{\scriptscriptstyle V},\ell}^{(\mathtt{1})}=\mathbb{V}^{(\mathtt{1})}_\ell$ for $\ell\geq1$. The first-order slip velocity is}
    \begin{equation}
    	\label{eqn:slip-velocity-1st}
    	\boxed{\itbf{v}^{(\mathtt{1})}_{\scriptscriptstyle S}
    	=
    	\frac{1}{2\eta}[\eu{\alpha}]^{(\mathtt{0})}\,
        \cc{\nabla_{\!\alpha}}\mathbb{Q}^{(\mathtt{1})}.}
    \end{equation}
    \cc{Hence, $\Div\itbf{v}^{(\mathtt{1})}_{\scriptscriptstyle S}
    =\Delta^{(\mathtt{0})}\mathbb{Q}^{(\mathtt{1})}/(2\eta)$.} Using
    \begin{equation}
    	\label{eqn:laplacian-spherical-harmonics}
    	R^2\Delta^{\!(\mathtt{0})}{Y}_{\ell,m}\hspace{-1pt}\left(\theta,\varphi\right)
    	=
    	-\ell(\ell+1){Y}_{\ell,m}\hspace{-1pt}\left(\theta,\varphi\right),
\end{equation}
    \cc{gives the final term in Eq.~(\ref{eqn:Y-ell-1}), where $\mathbb{Q}^{(\mathtt{1})}_\ell$ is the degree-$\ell$ component of the first-order contribution to the second active force moment $\mathbb{Q}$.}
	
	As both $\mathcal{X}^{(\mathtt{1})}_0$ and $\mathcal{Y}^{(\mathtt{1})}_0$ are zero, their associated equations reduce to
	\begin{equation}
		\label{eqn:volume-area-dynamics-first-order}
		\boxed{R\hspace{1pt}\frac{\partial u_0}{\partial t} = \frac{1}{2}R\hspace{1.5pt}\mathbb{M}^{(\mathtt{1})}_0\! = \mathbb{V}^{(\mathtt{1})}_0\!+w_0\hspace{-1pt}\left[P^{(\mathtt{1})}_{\hspace{-0.25pt}\scriptscriptstyle\Delta}-\Pi^{(\mathtt{1})}_{\scriptscriptstyle\Delta}\right]\!,}
	\end{equation} where $P^{(\mathtt{1})}_{\hspace{-0.25pt}\scriptscriptstyle\Delta}$ and $\Pi^{(\mathtt{1})}_{\scriptscriptstyle\Delta}$ are the {first-order hydrostatic pressure difference and osmotic pressure}, respectively.
	
	Therefore, the surface harmonics $\mathcal{X}^{(\mathtt{1})}_\ell$ and $\mathcal{Y}^{(\mathtt{1})}_\ell$ allows us to obtain both the functions $p^{(\mathtt{1})}_\ell$ and $\Upsilon^{(\mathtt{1})}_\ell$ which are associated with the interior fluid flow $\!\itbf{V}^{(\mathtt{1})}_{\!{\boldsymbol{-}}}$, that is,
	\begin{align}
		\label{eqn:p-inner-1}
		p^{(\mathtt{1})}_{\ell} &= \frac{\eta\left(2\ell+3\right)\hspace{-0.75pt}r^{\hspace{0.25pt}\ell}}{\ell\hspace{0.5pt}R^{\hspace{0.5pt}\ell+1}}
		\hspace{-1pt}\left[\hspace{1pt}\mathcal{Y}^{(\mathtt{1})}_\ell - \left(\ell-1\right)\mathcal{X}^{(\mathtt{1})}_\ell\hspace{1pt}\right]\hspace{-2pt},\\[4pt]
		\label{eqn:upsilon-inner-1}
		\Upsilon^{(\mathtt{1})}_{\ell} &= \frac{r^{\hspace{0.25pt}\ell}}{2\hspace{0.5pt}\ell\hspace{0.5pt}R^{\hspace{0.5pt}\ell-1}}\left[\left(\ell+1\right)\mathcal{X}^{(\mathtt{1})}_\ell-\mathcal{Y}^{(\mathtt{1})}_\ell\hspace{1pt}\right]\hspace{-2pt},
	\end{align} with $\ell\geq1$, and the functions  $p^{(\mathtt{1})}_{-(\ell+1)}$ and $\Upsilon^{(\mathtt{1})}_{-(\ell+1)}$ that correspond to the outer fluid flow $\!\itbf{V}^{(\mathtt{1})}_{\!{\boldsymbol{+}}}$, namely
	\begin{align}
		\label{eqn:p-outer-1}
		p^{(\mathtt{1})}_{-(\ell+1)} &= \frac{\eta\left(2\ell-1\right)\hspace{-1pt}R^{\hspace{0.25pt}\ell}}{\left(\ell+1\right) r^{\hspace{0.5pt}\ell+1}}\left[\hspace{1pt}\mathcal{Y}^{(\mathtt{1})}_\ell+\left(\ell+2\right)\mathcal{X}^{(\mathtt{1})}_\ell\right]\hspace{-2pt},\\[4pt]
		\label{eqn:upsilon-outer-1}
		\Upsilon^{(\mathtt{1})}_{-(\ell+1)} &= \frac{R^{\hspace{0.25pt}\ell+2}}{2\hspace{-0.75pt}\left(\ell+1\right)r^{\hspace{0.5pt}\ell+1}}\left[\hspace{1pt}\mathcal{Y}^{(\mathtt{1})}_\ell+\ell\hspace{0.25pt}\mathcal{X}^{(\mathtt{1})}_\ell\hspace{1pt}\right]\hspace{-2pt},
	\end{align} where $\ell\geq1$. On the other hand, $\hspace{1pt}\Xi^{(\mathtt{1})}_{-(\ell+1)}$ and\vspace{-1pt} $\hspace{1pt}\Xi^{(\mathtt{1})}_{\hspace{0.25pt}\ell}$ can be found as before from the stress boundary condition.
	
	The {traction} vectors $\boldsymbol{\pi}^{(\mathtt{1})}_{\boldsymbol{\pm}}\!$ at first order in the perturbation expansion may be expressed in the form
	\begin{equation}
		\label{eqn:radial-stress-vector-1}
		\boldsymbol{\pi}^{(\mathtt{1})}_{\boldsymbol{\pm}}\! = \eta\hspace{1pt}r\frac{\partial}{\partial r}\!\hspace{-1pt}\left[\!\frac{\itbf{V}^{(\mathtt{1})}_{\!{\boldsymbol{\pm}}}}{r}\hspace{-1pt}\right]\!+\frac{\eta}{r}\,\bnabla\!\left[r\rv\cdot\!\itbf{V}^{(\mathtt{1})}_{\!{\boldsymbol{\pm}}}\right]\!\hspace{-1.5pt}-\hspace{-1pt}\nv^{(\mathtt{1})}\hspace{-1pt} P^{(\mathtt{0})}_{\boldsymbol{\pm}}\!-\rv\hspace{1pt}p^{(\mathtt{1})}_{\boldsymbol{\pm}}\!,
	\end{equation} where the normal vector $\nv^{(\mathtt{1})}$ is found to be
	\begin{equation}
		\label{eqn:normal-1}
		\qquad\quad\nv^{(\mathtt{1})}\! = -\frac{\partial u}{\partial \theta}\,\thetav\,-\frac{1}{\sin\theta}\frac{\partial u}{\partial \varphi}\,\phiv\,  = -r\bnabla u,
	\end{equation} where the last equality follows from the definition of the three-dimensional gradient operator $\bnabla$ written in spherical coordinates. As before, the pressure inside the closed membrane is given by
	\begin{equation}
		p^{(\mathtt{1})}_{\boldsymbol{-}} = P^{(\mathtt{1})}_{\boldsymbol{-}} + \sum\limits_{\ell=1}\,p_\ell^{(\mathtt{1})}\!,
	\end{equation} whilst the outside pressure is
	\begin{equation}
		\qquad\qquad\qquad\quad\;\, p^{(\mathtt{1})}_{\boldsymbol{+}} = P^{(\mathtt{1})}_{\boldsymbol{+}} + \sum\limits_{\ell=1}\,p_{-(\ell+1)}^{(\mathtt{1})}\hspace{1pt},
	\end{equation} where the scalars $P^{(\mathtt{1})}_{\boldsymbol{\pm}}\!$ are the \cc{spatially uniform ($\ell=0$) contributions to the first-order pressure fields}.
    
	Since $\bnabla\times\rv=\boldsymbol{0}$ and $\bnabla\times\nv^{(\mathtt{1})}\!=-\rv\times\bnabla u$, the radial {components} of the curl of the stress vectors $\boldsymbol{\pi}^{(\mathtt{1})}_{\boldsymbol{\pm}}\!$ corresponding to the inner and outer flows are given by
    \begin{equation}
    	\rv\hspace{-1pt}\cdot\hspace{-1.5pt}\left[\bnabla\hspace{-2pt}\times\hspace{-1.5pt}\boldsymbol{\pi}^{(\mathtt{1})}_{\boldsymbol{-}}\right]_{\hspace{-1pt}r=R}
    	=
    	\sum^{\infty}_{\ell=1}
    	\frac{\eta\hspace{1pt}\ell(\ell-1)(\ell+1)\,\Xi^{(\mathtt{1})}_{\hspace{0.5pt}\ell}}{R},
    \end{equation}
    and
    \begin{equation}
    	\rv\hspace{-1pt}\cdot\hspace{-2pt}\left[\bnabla\hspace{-2pt}\times\hspace{-1.5pt}\boldsymbol{\pi}^{(\mathtt{1})}_{\boldsymbol{+}}\right]_{\hspace{-1pt}r=R}
    	=
    	-\sum^{\infty}_{\ell=1}
    	\frac{\eta\hspace{1pt}\ell(\ell+1)(\ell+2)\,\Xi^{(\mathtt{1})}_{-(\ell+1)}}{R},
    \end{equation}
    respectively. \cc{Taking their difference and using Eqs.~(\ref{eqn:xi-inner}) and (\ref{eqn:xi-outer}) gives}
    \begin{equation}
    	\rv\cdot
    	\left[
    	\bnabla\times\boldsymbol{\pi}^{(\mathtt{1})}_{\scriptscriptstyle\Delta}
    	\right]_{r=R}
    	=
    	-\sum_{\ell=1}^{\infty}
    	\frac{\eta(2\ell+1)}{R}\,
    	\mathcal{Z}^{(\mathtt{1})}_\ell.
    \end{equation}
    \cc{At first order, the tangential membrane force is a surface gradient and therefore has no toroidal component; equivalently, its radial curl vanishes. Hence, by orthogonality,}
    \begin{equation}
    	\mathcal{Z}^{(\mathtt{1})}_\ell=0,\qquad \ell\geq1.
    \end{equation}
    \cc{It then follows from Eqs.~(\ref{eqn:xi-inner}) and (\ref{eqn:xi-outer}) that both $\Xi^{(\mathtt{1})}_{\hspace{0.25pt}\ell}$ and $\Xi^{(\mathtt{1})}_{-(\ell+1)}$ vanish identically.}
	
	By employing Eqs.~(\ref{eqn:body-forces-0}) and (\ref{eqn:radial-stress-vector-1}), the normal component of the \cc{hydrodynamic} force $\itbf{f}^{(\mathtt{1})}$ is given by
	\begin{equation}
		\label{body-forces-1}
		f^{(\mathtt{1})}\! = 
		\nv^{(\mathtt{0})}\!\cdot\itbf{f}^{(\mathtt{1})}+
		\nv^{(\mathtt{1})}\!\cdot\itbf{f}^{(\mathtt{0})} = \rv\cdot\hspace{-1pt}\left[\hspace{0.5pt}\boldsymbol{\pi}^{(\mathtt{1})}_{\scriptscriptstyle\Delta}\right]_{\hspace{-1pt}r=R},
	\end{equation} where the identity $\rv\,\cdot\,\nv^{(\mathtt{1})}=0$ and Eq.~(\ref{eqn:stress-1-BC}) are used to derive the last equation. To find $f^{(\mathtt{1})}\!$, we need to compute the radial part of the stress vectors. On the inner side of the closed membrane, we have
	\begin{align}
		\label{eqn:r-stress-vector-inner-1}
		\rv\cdot\hspace{-1pt}\left[\hspace{0.5pt}\boldsymbol{\pi}^{(\mathtt{1})}_{\boldsymbol{-}}\right]_{\hspace{-1pt}r=R} = -P^{(\mathtt{1})}_{\boldsymbol{-}}\!\!&\,+\,\sum^{\infty}_{\ell=1}\hspace{-1pt}\left[\frac{\left(\ell^2-\ell-3\right)\hspace{-1pt}R^\ell}{\left(2\ell+3\right)\hspace{-1pt}r^\ell}\hspace{1pt} p^{(\mathtt{1})}_\ell\right.\notag\\[3pt]
		&\hspace{1pt}\left.+\,\frac{2\hspace{0.5pt}\eta\hspace{1pt}\ell\hspace{-0.5pt}\left(\ell-1\right)\hspace{-0.5pt}R^{\ell-2}}{r^{\ell}}\hspace{1pt}\Upsilon^{(\mathtt{1})}_\ell\right]\!\hspace{-1pt},
	\end{align} whereas, on the outer side of the membrane, we get
	\begin{align}
		\label{eqn:r-stress-vector-outer-1}
		\rv\cdot\hspace{-1pt}\left[\hspace{0.5pt}\boldsymbol{\pi}^{(\mathtt{1})}_{\boldsymbol{+}}\right]_{\hspace{-1pt}r=R} = &-P^{(\mathtt{1})}_{\boldsymbol{+}}-\,\sum^{\infty}_{\ell=1}\hspace{-1pt}\left[\frac{\left(\ell^2+3\ell-1\right)\hspace{-1pt}r^{\ell+1}}{\left(2\ell-1\right)\hspace{-1pt}R^{\ell+1}}\hspace{1pt} p^{(\mathtt{1})}_{-(\ell+1)}\! \right.\notag\\[3pt]
		&\hspace{-10pt}\left.-\frac{2\hspace{0.5pt}\eta\hspace{-1pt}\left(\ell+2\right)\!\left(\ell+1\right)\hspace{-0.5pt}r^{\ell+1}}{R^{\ell+2}}\hspace{1pt}\Upsilon^{(\mathtt{1})}_{-(\ell+1)}\,\right]\!\hspace{-1pt}.
	\end{align} Hence, the normal \cc{hydrodynamic} force $f^{(\mathtt{1})}$ in Eq.~(\ref{body-forces-1}) can be written in terms of the surface harmonics $\mathcal{X}^{(\mathtt{1})}_\ell$ and $\mathcal{Y}^{(\mathtt{1})}_\ell$, by substituting Eq.~(\ref{eqn:p-inner-1}) and (\ref{eqn:upsilon-inner-1}) into Eq.~(\ref{eqn:r-stress-vector-inner-1}), and Eq.~(\ref{eqn:p-outer-1}) and (\ref{eqn:upsilon-outer-1}) into Eq.~(\ref{eqn:r-stress-vector-outer-1}), which yields
	\begin{equation}
		\label{eqn:normal-body-force-1}
		\boxed{f^{(\mathtt{1})}\! = -P^{(\mathtt{1})}_{\hspace{-0.5pt}\scriptscriptstyle\Delta}+ \sum^{\infty}_{\ell=1}\,f^{(\mathtt{1})}_\ell,\vspace{-2pt}}
	\end{equation} where $P^{(\mathtt{1})}_{\hspace{-0.5pt}\scriptscriptstyle\Delta}=P^{(\mathtt{1})}_{\boldsymbol{+}}-P^{(\mathtt{1})}_{\boldsymbol{-}}$ \cc{is the first-order hydrostatic pressure difference across the membrane. The degree-$\ell$ components} $f^{(\mathtt{1})}_\ell$ are found to be
	\begin{equation}
		\label{eqn:normal-body-foce-comps}
		\boxed{f^{(\mathtt{1})}_\ell\! = \frac{\eta\hspace{-1pt}\left(2\ell\!+\!1\right)\hspace{-1.5pt}\left\{\hspace{-0.5pt}3\hspace{1pt}\mathcal{Y}^{(\mathtt{1})}_\ell\!- \hspace{-1pt}\left[\hspace{0.5pt}2\ell\hspace{-0.5pt}\left(\ell\!+\!1\right)\!-3\hspace{0.5pt}\right]\hspace{-1pt}\mathcal{X}^{(\mathtt{1})}_\ell\hspace{-1pt}\right\}}{\ell\left(\ell\!+\!1\right)\hspace{-1pt} R}.}
	\end{equation}
	
	The tangential contribution of the \cc{hydrodynamic} force $\itbf{f}^{(\mathtt{1})}$, at first order in $\varepsilon$, has the components
    \begin{equation}
    	\left[f^\alpha\right]^{(\mathtt{1})} = \itbf{f}^{(\mathtt{1})}\cdot\left[\eu{\alpha}\right]^{(\mathtt{0})}+\itbf{f}^{(\mathtt{0})}\cdot\left[\eu{\alpha}\right]^{(\mathtt{1})},
    \end{equation}
    where $\itbf{f}^{(\mathtt{0})}=-\rv\hspace{0.75pt}P^{(\mathtt{0})}_{\hspace{-0.5pt}\scriptscriptstyle\Delta}$ from Eq.~(\ref{eqn:body-forces-0}). \cc{Using the orthogonality of the normal and tangent vectors at first order, the corresponding tangential body force is}
    \begin{equation}
    	\itbf{f}_{\!\scriptscriptstyle\mathcal{S}}^{(\mathtt{1})}
    	=
    	P^{(\mathtt{0})}_{\hspace{-0.5pt}\scriptscriptstyle\Delta}\nv^{(\mathtt{1})}
    	+
    	\left[
    	\boldsymbol{\pi}^{(\mathtt{1})}_{\scriptscriptstyle\Delta}
    	-\rv\left(\rv\cdot\boldsymbol{\pi}^{(\mathtt{1})}_{\scriptscriptstyle\Delta}\right)
    	\right]_{r=R}.
    \end{equation}

\cc{The term within the square brackets is the tangential part of the first-order traction-vector jump. The corresponding tangential traction vectors for the inner and outer flows are}
\begin{align}
	\label{eqn:tangential-stress-balance-series-inner}\!\!\left[\boldsymbol{\pi}^{(\mathtt{1})}_{\boldsymbol{-}}\hspace{-1pt}\!-\rv\left(\rv\cdot\boldsymbol{\pi}^{(\mathtt{1})}_{\boldsymbol{-}}\right)\hspace{-1pt}\right]_{\hspace{-1pt}r=R}\! &=
	\!\sum_{\ell=1}^{\infty}\!\left[\frac{\ell\left(\ell+2\right)\hspace{-1pt}R^\ell}{\left(\ell+1\right)\!\left(2\ell+3\right)r^\ell}\bnabla_{\hspace{-2pt}\scriptscriptstyle\mathcal{S}}\hspace{1pt} p_\ell^{(\mathtt{1})}\right.\notag\\[2pt]
	& \hspace{-65pt}\left.+\frac{2\eta\left(\ell-1\right)\hspace{-1pt}R^{\ell-2}}{r^\ell}\bnabla_{\hspace{-2pt}\scriptscriptstyle\mathcal{S}}\Upsilon^{(\mathtt{1})}_\ell\hspace{0.5pt}\right]\!+ P^{(\mathtt{0})}_{\boldsymbol{-}}\bnabla_{\hspace{-2pt}\scriptscriptstyle\mathcal{S}}\hspace{0.5pt}u,\\[-3pt]
	&\hspace{-108pt}\mathrm{and}\notag\\[-1pt]
	\label{eqn:tangential-stress-balance-series-outer}\!\!\left[\boldsymbol{\pi}^{(\mathtt{1})}_{\boldsymbol{+}}\hspace{-1pt}\!-\rv\left(\rv\cdot\boldsymbol{\pi}^{(\mathtt{1})}_{\boldsymbol{+}}\right)\hspace{-1pt}\right]_{\hspace{-1pt}r=R}\! &=
	\!\sum_{\ell=1}^{\infty}\!\left[\frac{\left(\ell^{\hspace{0.5pt}2}-1\right)\hspace{-1.5pt}r^{\ell+1}}{\ell\left(2\ell-1\right)\hspace{-1pt}R^{\ell+1}}\bnabla_{\hspace{-2pt}\scriptscriptstyle\mathcal{S}}\hspace{1pt} p_{-(\ell+1)}^{(\mathtt{1})}\right.\notag\\[2pt]
	& \hspace{-75pt}\left.-\frac{2\eta\left(\ell+2\right)\hspace{-0.5pt}r^{\ell+1}}{R^{\ell+2}}\bnabla_{\hspace{-2pt}\scriptscriptstyle\mathcal{S}}\Upsilon^{(\mathtt{1})}_{-(\ell+1)}\hspace{0.5pt}\right]\!+ P^{(\mathtt{0})}_{\boldsymbol{+}}\bnabla_{\hspace{-2pt}\scriptscriptstyle\mathcal{S}}\hspace{0.5pt}u,
\end{align}
respectively, where $\bnabla_{\hspace{-2pt}\scriptscriptstyle\mathcal{S}}$ {denotes the surface gradient on the unit sphere},
	\begin{equation}
		\bnabla_{\hspace{-2pt}\scriptscriptstyle\mathcal{S}}\hspace{-1pt}\left(\boldsymbol{\cdot}\right)= \frac{\partial\hspace{-1pt}\left(\boldsymbol{\cdot}\right)}{\partial\theta}\hspace{1pt}\thetav + \frac{1}{\sin\theta}\frac{\partial\hspace{-1pt}\left(\boldsymbol{\cdot}\right)}{\partial\varphi}\hspace{1pt}\phiv.
	\end{equation} The above expressions in Eq.~(\ref{eqn:tangential-stress-balance-series-inner}) and (\ref{eqn:tangential-stress-balance-series-outer}) are derived by employing the identities:
	\begin{align}
		\;&\;\hspace{1pt}\bnabla h _\ell = \frac{\ell\hspace{1pt}h_\ell\hspace{0.5pt}\rv}{r} + \frac{\bnabla_{\hspace{-2pt}\scriptscriptstyle\mathcal{S}}\hspace{0.5pt}h_\ell}{r},\quad\mathrm{and}\\[3pt]
		\;&\left[\bnabla_{\hspace{-2pt}\scriptscriptstyle\mathcal{S}}\hspace{0.5pt}h_\ell\hspace{0.5pt}\right]_{r=R} = \left[\frac{r}{R}\hspace{0.5pt}\right]^{-\ell}\hspace{-1pt}
		\bnabla_{\hspace{-2pt}\scriptscriptstyle\mathcal{S}}\hspace{0.5pt}h_\ell,
	\end{align} where $h_\ell$ is a solid spherical harmonic of degree $\ell$. Hence, by substituting Eq.~(\ref{eqn:p-inner-1}) and (\ref{eqn:upsilon-inner-1}) into Eq.~(\ref{eqn:tangential-stress-balance-series-inner}), and Eq.~(\ref{eqn:p-outer-1}) and (\ref{eqn:upsilon-outer-1}) into Eq.~(\ref{eqn:tangential-stress-balance-series-outer}), we find
	\begin{equation}
		\label{eqn:tangential-part-body-force-1}
		\boxed{\itbf{f}_{\!\scriptscriptstyle\mathcal{S}}^{(\mathtt{1})}\!\hspace{-1pt} =  -\hspace{-0.65pt}\sum_{\ell=1}^{\infty}\bnabla_{\hspace{-2pt}\scriptscriptstyle\mathcal{S}}\!\hspace{-1pt}\left[\frac{\eta\hspace{-1pt}\left(2\ell\hspace{-0.5pt}+\hspace{-0.5pt}1\right)}{\ell\hspace{-1pt}\left(\ell\hspace{-0.5pt}+\hspace{-0.5pt}1\right)\! R}\!\hspace{-0.5pt}\left(\hspace{-1pt}\mathcal{X}^{(\mathtt{1})}_\ell\!\!\hspace{-0.5pt}+\hspace{-0.5pt}2\hspace{0.25pt}\mathcal{Y}^{(\mathtt{1})}_\ell\hspace{-1pt}\right)\!\right]\!,}
	\end{equation} \cc{which shows that the first-order tangential body force is a surface gradient and has no toroidal component.}

	\subsection{Dynamics of the excess areal fractions}
	
	Here, the {steady-state areal fractions} corresponding to Eq.~(\ref{eqn:phi-k-dynamics}) are denoted by $\bar{\Phi}_k$, and they must satisfy
    \begin{equation}
    	\label{eqn:steady-state-condition}
    	\boxed{\mathbb{M}^{(\mathtt{0})}=\sum_k\,\bar{\Phi}_k\,\bMk = 0.}
    \end{equation}
    On the other hand, the steady-state {areal fraction} $\bar{\Phi}_0$ that is associated with Eq.~(\ref{eqn:phi-0-dynamics}) is obtained via the {areal-fraction constraint}, $\textnormal{\ie}1-\bar{\Phi}_0 = \bar{\Phi}_1 + \bar{\Phi}_2$. By using this constraint, $\bar{\Phi}_k$ can be expressed as follows:
	\begin{equation}
		\label{eqn:steady-state-value-k}
		\boxed{\bar{\Phi}_k = \frac{\displaystyle{\mathbb{M}^{\hspace{0.5pt}k'}_{\CircleArrow}}}{\displaystyle{\mathbb{M}^{\hspace{0.5pt}k'}_{\CircleArrow}}-{\mathbb{M}^{\hspace{0.5pt}k}_{\CircleArrow}}}\left(1-\bar{\Phi}_0\right)\!,\quad\mathrm{with}\,\;k\hspace{-1pt}\neq k',}
	\end{equation} where both the indices $k,\,k'\!\in\{1,2\}$. In deriving this homogeneous steady-state, we assume that the shape is also fixed, corresponding to the undeformed sphere of radius $R$ at zeroth order in the perturbation expansion.
	
	The system is subsequently perturbed about this nonequilibrium steady-state by \cc{small variations in the membrane areal fractions}; namely, we consider that
    \begin{equation}
    	\Phi_0\hspace{-1pt}\left(\xi^{\alpha}\!,\hspace{0.5pt}t\right) = \bar{\Phi}_0 - \varepsilon\,\Psi_0\hspace{-1pt}\left(\xi^{\alpha}\!,\hspace{0.5pt}t\right)\,+ \;\mathcal{O}[\varepsilon^2],
    \end{equation}
    where $\varepsilon$ is the perturbation parameter and $\Psi_0$ is {the corresponding perturbation field of the background areal fraction}. This linear perturbation also leads to a linear response in the other {areal fractions},
    \begin{equation}
    	\label{eqn:linear-response-k}
    	\Phi_k\hspace{-1pt}\left(\xi^{\alpha}\!,\hspace{0.5pt}t\right) = \bar{\Phi}_k + \varepsilon\, \Psi_k\hspace{-1.5pt}\left(\xi^{\alpha}\!,\hspace{0.5pt}t\right)\,+ \;\mathcal{O}[\varepsilon^2],
    \end{equation}
    where $\Psi_k$ is the \cc{excess areal fraction of active membrane components associated with trafficking cycle $k$}. Due to the {constraint} in Eq.~(\ref{eqn:mass-fraction-condition}), $\Psi_k$ must satisfy
	\begin{equation}
		\label{eqn:mass-cons-psi-s}
		\Psi_1+\Psi_2=\Psi_0.
	\end{equation}
	
	Moreover, Eq.~(\ref{eqn:linear-response-k}) implies the following perturbation expansion of the {active membrane-area flux} $\mathbb{M}$ from Eq.~(\ref{eqn:total-mass-flux-rate}):
    \begin{equation}
    	\mathbb{M} = \mathbb{M}^{(\mathtt{0})}+\varepsilon\,\mathbb{M}^{(\mathtt{1})}\, + \; \mathcal{O}[\varepsilon^2],
    \end{equation}
    which is truncated to linear order in $\varepsilon$. Hence, the {first-order contribution} is
    \begin{equation}
    	\mathbb{M}^{(\mathtt{1})} = \sum_k\Psi_k\,\bMk.\vspace{-5pt}
    \end{equation}
	
	Thus, the dynamical equation {for} $\Psi_k$ associated with the {first-order} perturbation of Eq.~(\ref{eqn:phi-k-dynamics-expanded}) is given by
	\begin{equation}
		\label{eqn:psi-k-dynamics}
		\dot{\Psi}_{k}\! =\! 2\hspace{1pt}\Omega_k\hspace{-1pt}\left[\Delta H\hspace{0.5pt}\right]^{(\mathtt{1})}\!-\bar{\Phi}_{k}\hspace{1pt}\mathbb{M}^{(\mathtt{1})}\! + \hspace{1pt}\gamma_{k}\hspace{1pt}\Delta^{\!(\mathtt{0})}\!\!\left[\hspace{-0.2pt}\frac{\Psi_0}{\,\bar{\Phi}^{\phantom{k'}}_0\!\!}+\frac{\Psi_k}{\bar{\Phi}_{k}}\right]\!\!,
	\end{equation} where $\left[\Delta H\hspace{0.5pt}\right]^{(\mathtt{1})}\hspace{-1pt} = \Delta^{\!(\mathtt{0})}H^{(\mathtt{1})}\hspace{-1pt} + \Delta^{\!(\mathtt{1})}H^{(\mathtt{0})}\hspace{-1pt}$ is the {first-order} contribution to the Laplacian of the mean curvature $H$. From the shape parameterization in Eq.~(\ref{eqn:shape-parameterization}), we have  that $H^{(\mathtt{0})}=-1/R$, and thus $\Delta^{\!(\mathtt{1})}H^{(\mathtt{0})}\hspace{-1pt}=0$. Also, we find
	\begin{equation}
		H^{(\mathtt{1})}\hspace{-1.5pt}=\hspace{-0.5pt}\frac{u\!\left(\theta,\hspace{-0.5pt}\varphi\right)}{R}+\frac{{\partial_\theta}\!\left[\hspace{1pt}\sin\theta\hspace{2pt}\partial_\theta\hspace{1pt}u\!\left(\theta,\hspace{-0.5pt}\varphi\right)\right]}{2 R\sin\theta}+\frac{\partial^2_\varphi\hspace{1pt}u\!\left(\theta,\hspace{-0.5pt}\varphi\right)}{2R\sin^2\hspace{-0.75pt}\theta},
	\end{equation} where $\partial^2_\alpha\!\left(\boldsymbol{\cdot}\right)=\partial_\alpha\!\left[\hspace{1pt}\partial_\alpha\!\left(\boldsymbol{\cdot}\right)\hspace{0.5pt}\right]$ is the second derivative with respect to the local surface coordinates.  As the Laplacian operator $\Delta^{\!(\mathtt{0})}\!$ in spherical coordinates is given by
	\begin{equation}
		\label{eqn:Laplacian-zero-order}
		\Delta^{\!(\mathtt{0})}\!\left(\boldsymbol{\cdot}\right) = \frac{{\partial_\theta}\!\left[\hspace{1pt}\sin\theta\hspace{2pt}\partial_\theta\hspace{0.5pt}\!\left(\boldsymbol{\cdot}\right)\right]}{R^2\sin\theta}+\frac{\partial^2_\varphi\hspace{0.5pt}\!\left(\boldsymbol{\cdot}\right)}{R^2\sin^2\hspace{-0.5pt}\theta},
	\end{equation} we can express the term $\left[\Delta H\hspace{0.5pt}\right]^{(\mathtt{1})}\!= \Delta^{\!(\mathtt{0})}H^{(\mathtt{1})}\hspace{-1pt}$ as follows:
	\begin{equation}
		\left[\Delta H\hspace{0.5pt}\right]^{(\mathtt{1})}\! = \frac{\Delta^{\!(\mathtt{0})}u}{R} + \frac{R\hspace{1pt}\Delta^{\!(\mathtt{0})}\!\Delta^{\!(\mathtt{0})}u}{2}.
	\end{equation}
	
	Since $\sum_k\Psi_k=\Psi_0$, the last term in Eq.~(\ref{eqn:psi-k-dynamics}) can be written in terms of the variables $\Psi_1$ and $\Psi_2$, \ie
	\begin{equation*}
		\frac{\Delta^{\!(\mathtt{0})}\Psi_k}{\;\;\bar{\Phi}_{k}^{\phantom{k'}}}+\frac{\Delta^{\!(\mathtt{0})}\Psi_0}{\;\;\bar{\Phi}^{\phantom{k'}}_0} = \Delta^{\!(\mathtt{0})}\Psi_k\!\left[\hspace{-1pt}\frac{1}{\,\bar{\Phi}^{\phantom{k'}}_0\!\!}\!+\!\frac{1}{\,\bar{\Phi}_{k}^{\phantom{k'}}\!}\hspace{-1pt}\right] + \frac{\Delta^{\!(\mathtt{0})}\Psi_{k'}\!}{\,\bar{\Phi}^{\phantom{k'}}_0\!\!\!},
	\end{equation*} with $k'\hspace{-1pt}\neq k$. Also, the total derivative of $\Psi_k$ reduces to
	\begin{equation}
		\dot{\Psi}_{k} = \frac{\partial\Psi_k}{\partial t} + \left[v^\alpha\right]^{(\mathtt{0})}\!\Psi_{k\hspace{0.5pt},\alpha}+ \left[v^\alpha\right]^{(\mathtt{1})}\!\bar{\Phi}_{k\hspace{0.5pt},\alpha}\! =\frac{\partial\Psi_k}{\partial t},
	\end{equation} as the zeroth-order membrane velocity $\itbf{v}^{(\mathtt{0})}=\boldsymbol{0}$ \cc{and the base-state areal fractions $\bar{\Phi}_k$ are spatially uniform}. Thus, Eq.~(\ref{eqn:psi-k-dynamics}) can be expressed in a matrix form as follows:
	\begin{equation}
		\label{eqn:psi-k-dynamics-vector}
		\boxed{\frac{\partial\boldsymbol{\Psi}}{\partial\hspace{0.25pt}t} = -\mathsf{M}\hspace{1pt}\boldsymbol{\Psi}+\mathsf{D}\hspace{1pt}\Delta^{\!(\mathtt{0})}\boldsymbol{\Psi}+2\hspace{1pt}\boldsymbol{\Omega}\left[\Delta H\hspace{0.5pt}\right]^{(\mathtt{1})},}
	\end{equation} by defining the vectors $\boldsymbol{\Psi}=\left(\Psi_1,\Psi_2\right)^{\mathsf{T}}\hspace{-1.5pt}$, $\boldsymbol{\Omega}=\left(\Omega_1,\Omega_2\right)^{\mathsf{T}}\hspace{-1.5pt}$, where the superscript~$\scriptstyle\mathsf{T}$ indicates a vector transpose. On the other hand, $\mathsf{M}$ is a matrix which is defined by
	\begin{equation}
		\mathsf{M} = \!\begin{bmatrix}
			\,\bar{\Phi}_{1}\hspace{1pt}{\mathbb{M}^{\hspace{0.5pt}1}_{\CircleArrow}} &
			\bar{\Phi}_{1}\hspace{1pt}{\mathbb{M}^{\hspace{0.5pt}2}_{\CircleArrow}}\, \\[5pt]
			\,\bar{\Phi}_{2}\hspace{1pt}{\mathbb{M}^{\hspace{0.5pt}1}_{\CircleArrow}} &
			\bar{\Phi}_{2}\hspace{1pt}{\mathbb{M}^{\hspace{0.5pt}2}_{\CircleArrow}}\, \end{bmatrix}\!\hspace{-1pt},
	\end{equation} whilst the diffusion matrix $\mathsf{D}$ is given by
	\begin{equation}
        \label{eqn:composition-diffusion-matrix}
		\mathsf{D} = \!\begin{bmatrix}
			\,\gamma_1\!\left(\hspace{-0.25pt}\left.1\middle/\bar{\Phi}_0\right.+\left.1\middle/\bar{\Phi}_1\right.\hspace{-1pt}\right) &
			\,\left.\gamma_1\hspace{0.25pt}\middle/\bar{\Phi}_0\right. \\[5pt]
			\,\left.\gamma_2\hspace{0.25pt}\middle/\bar{\Phi}_0\right. &
			\,\gamma_2\!\left(\hspace{-0.25pt}\left.1\middle/\hspace{0.25pt}\bar{\Phi}_0\right.+\left.1\middle/\bar{\Phi}_2\right.\hspace{-1pt}\right)
		\end{bmatrix}\!\hspace{-1pt}.
	\end{equation}
	
	By employing the quasi--spherical parametrization in Eq.~(\ref{eqn:shape-parameterization}), and assuming that $\Psi_k(\theta,\varphi)$ are well-behaved functions of the angular variables, we could then expand, as done previously for $u(\theta,\varphi)$, the excess areal fractions $\Psi_k$ in a series of surface spherical harmonics, that is,
    \begin{equation}
    	\Psi_k(\theta,\varphi) = \sum_{\ell=0}^{\infty}\,\Psi_{k,\ell}(\theta,\varphi).
    \end{equation}
    Here, the surface harmonics can also be rewritten as
    \begin{equation}
    	\label{eqn:psi-harmonic-def}
    	\Psi_{k,\ell}\hspace{-1pt}\left(\theta,\varphi\right)
    	=
    	\sum^{\ell}\limits_{m=-\ell}\!
    	\Psi_{k,\ell,m}\,
    	{Y}_{\ell,m}\!\left(\theta,\varphi\right)\!,\vspace{-2pt}
    \end{equation}
    where $\Psi_{k,\ell,m}$ is the amplitude associated with \cc{the individual spherical harmonic} ${Y}_{\ell,m}$. 
    
    Hence, the governing equation in (\ref{eqn:psi-k-dynamics-vector}) can be re-expressed as follows:
    \begin{equation}
    	\label{eqn:psi-k-dynamics-vector-harmonics}
    	\boxed{\frac{\partial\boldsymbol{\Psi}_{\hspace{-0.5pt}\ell}}{\partial\hspace{0.25pt}t}
    	=
    	-\mathsf{M}\hspace{0.5pt}\boldsymbol{\Psi}_{\hspace{-0.5pt}\ell}
    	-\frac{\ell(\ell+1)}{R^2}\mathsf{D}\hspace{0.5pt}\boldsymbol{\Psi}_{\hspace{-0.5pt}\ell}
    	+\frac{\ell(\ell+2)(\ell^2\!-1)}{R^3}\boldsymbol{\Omega}\hspace{1pt}u_\ell,}
    \end{equation} where we define the vectors $\boldsymbol{\Psi}_{\hspace{-0.5pt}\ell}\hspace{-1pt}
    =\hspace{-1pt}\left(\Psi_{1,\ell},\Psi_{2,\ell}\hspace{1pt}\right)^{\!\mathsf{T}}\hspace{-1.5pt}$, \cc{and use Eq.~(\ref{eqn:laplacian-spherical-harmonics}) to evaluate the Laplacian terms. In particular,}
    \begin{equation}
    	\label{eqn:laplacian-mean-curvature}
    	\left[\Delta H\hspace{0.5pt}\right]^{(\mathtt{1})}\! = \sum_{\ell=0}^{\infty}\,\frac{\ell\left(\ell+2\right)\!(\ell^2-1)\hspace{1pt}u_\ell}{2R^3},
    \end{equation}
    which shows that the zeroth and first harmonics of the Laplacian of the mean curvature vanish identically.

	\subsection{Area and volume dynamics}

	The {hydrostatic pressure difference} $P_{\hspace{-0.5pt}\scriptscriptstyle\Delta}$ between the two sides of the membrane depends on the osmotic pressure, as well as {the active volume exchange associated with the trafficking cycles}, as prescribed by Eq.~(\ref{eqn:volume-integral-condition}). At zeroth order in the perturbation expansion, this becomes
	\begin{equation}
		\label{eqn:volume-integral-condition-0}
		\int\limits_{\mathcal{M}\;\;}\!\!\mathrm{d}S^{(\mathtt{0})}\hspace{1pt}\bigg\{\hspace{-1pt}v^{(\mathtt{0})}\!+w_0\hspace{-1pt}\left[\Pi^{(\mathtt{0})}_{\scriptscriptstyle\Delta}\!-P^{(\mathtt{0})}_{\hspace{-0.25pt}\scriptscriptstyle\Delta}\right]\!-\mathbb{V}^{(\mathtt{0})}\hspace{-1pt}\bigg\}\hspace{-1pt}=0.
	\end{equation} Note that the area element $\mathrm{d}S$ also depends on the perturbation parameter $\varepsilon$; namely, we have that
	\begin{equation}
		\mathrm{d}S^{(\mathtt{0})}\! = \left[\sqrt{g}\hspace{1pt}\right]^{(\mathtt{0})}\mathrm{d}\varphi\hspace{1.5pt}\mathrm{d}\theta\, = R^2\sin\theta\;\mathrm{d}\varphi\hspace{1.5pt}\mathrm{d}\theta,
	\end{equation} for the shape parameterization in Eq.~(\ref{eqn:shape-parameterization}). The {zeroth-order active volume flux} is given by
	\begin{equation}
		\label{eqn:volume-rate-zeroth}
		\mathbb{V}^{(\mathtt{0})} = \sum_k\bar{\Phi}_k\hspace{1pt}\bVk.
	\end{equation} Moreover, the osmotic pressure is found to be
	\begin{equation}
		\label{eqn:pressure-drop-zeroth}
		\Pi^{(\mathtt{0})}_{\scriptscriptstyle\Delta}\! =\, k_B T\hspace{1pt}\bigg[\hspace{-0.5pt} c_{\boldsymbol{+}}-\frac{\mathfrak{n}_{\scriptscriptstyle\mathcal{S}}}{V^{(\mathtt{0})}}\bigg] =\, k_B T\left[c_{\boldsymbol{+}} - c^{(\mathtt{0})}_{\boldsymbol{-}}\right]\!,
	\end{equation} where $\mathfrak{n}_{\scriptscriptstyle\mathcal{S}}$ is the number of solutes {inside the organelle}, the volume $V^{(\mathtt{0})}=\frac{4\pi}{3}R^3$, and $c^{(\mathtt{0})}_{\boldsymbol{-}}={\mathfrak{n}_{\scriptscriptstyle \mathcal{S}}}/{\,V^{(\mathtt{0})}}\hspace{1pt}$.
	
	As the normal velocity $v^{(\mathtt{0})}=0$, the integral equation in (\ref{eqn:volume-integral-condition-0}) gives us an expression for the {zeroth-order hydrostatic pressure difference},
	\begin{equation}
		\label{eqn:Laplace-pressure-zeroth-unpacked}
		\boxed{P^{(\mathtt{0})}_{\hspace{-0.25pt}\scriptscriptstyle\Delta}\! = \hspace{1pt} \Pi^{(\mathtt{0})}_{\scriptscriptstyle\Delta}-w_0^{-1}\,\mathbb{V}^{(\mathtt{0})},}
	\end{equation} which in terms of Eq.~(\ref{eqn:volume-rate-zeroth}) and (\ref{eqn:pressure-drop-zeroth}) now reads
	\begin{equation}
		\label{eqn:Laplace-pressure-zeroth}
		P^{(\mathtt{0})}_{\hspace{-0.25pt}\scriptscriptstyle\Delta}\! =\, k_B T\left[c_{\boldsymbol{+}} - c^{(\mathtt{0})}_{\boldsymbol{-}}\right]\hspace{-1pt}-w_0^{-1}\sum_k\bar{\Phi}_k\hspace{1pt}\bVk.
	\end{equation} This is equivalent to the {zeroth-order normal volume flux vanishing}, $j_{\hspace{-1pt}\scriptscriptstyle V}^{(\mathtt{0})}\!=0$, cf.~Eq.~(\ref{eqn:volume-flux-zeroth-order}). \cc{The active contribution in Eq.~(\ref{eqn:Laplace-pressure-zeroth}) may therefore be interpreted as an effective shift of the osmotic pressure difference.}

    If the characteristic time associated with passive leakage is much longer than \cc{those associated with the active volume-transfer rates $\bVk$} (that is, \cc{$w_0\to0$}), then \cc{maintaining a finite zeroth-order pressure difference requires} $\mathbb{V}^{(\mathtt{0})}=\sum_k\bar{\Phi}_k\hspace{1pt}\bVk=0$. {In this limit,} Eq.~(\ref{eqn:Laplace-pressure-zeroth}) {reduces to the {\it van't Hoff relation}}, cf.~Eq.~(\ref{eqn:vantHoff}), {so that $P^{(\mathtt{0})}_{\scriptscriptstyle\Delta}=\Pi^{(\mathtt{0})}_{\scriptscriptstyle\Delta}$}. Together with Eq.~(\ref{eqn:steady-state-condition}), {this condition implies that} the active rates $\bMk$ and $\bVk$ cannot be independent {if a homogeneous steady state with $\bar{\Phi}_k\neq0$ is to exist, which gives the linear relationship} ${\mathbb{V}^{\hspace{0.5pt}1}_{\hspace{-0.85pt}\CircleArrow}}/{\mathbb{M}^{\hspace{0.5pt}1}_{\CircleArrow}} = {\mathbb{V}^{\hspace{0.5pt}2}_{\hspace{-0.85pt}\CircleArrow}}/{\mathbb{M}^{\hspace{0.5pt}2}_{\CircleArrow}}$ \cc{as $w_0\to0$}.
	
	The osmotic pressure at first order in $\varepsilon$ can be determined by expanding the volume in Eq.~(\ref{eqn:osmotic-pressure-concentrations}), namely
	\begin{equation}
		\Pi^{(\mathtt{1})}_{\scriptscriptstyle\Delta} = c^{(\mathtt{0})}_{\boldsymbol{-}}\,k_B T\,\frac{V^{(\mathtt{1})}}{V^{(\mathtt{0})}}\hspace{1pt},
	\end{equation} where the {first-order perturbation of the enclosed volume} can be computed as follows:
	\begin{align}
		V^{(\mathtt{1})}\! = \frac{1}{3}\iint\!\mathrm{d}\varphi\hspace{1.5pt}\mathrm{d}\theta&\,\left\{\left[\sqrt{g}\hspace{1pt}\right]^{(\mathtt{0})}\!\nv^{(\mathtt{1})}\!\cdot\hspace{-0.5pt}\itbf{R}^{(\mathtt{0})}+\left[\sqrt{g}\hspace{1pt}\right]^{(\mathtt{0})}\!\nv^{(\mathtt{0})}\!\cdot\hspace{-0.5pt}\itbf{R}^{(\mathtt{1})}
		\right.\notag\\
		&\left.\qquad+\,\left[\sqrt{g}\hspace{1pt}\right]^{(\mathtt{1})}\!\nv^{(\mathtt{0})}\!\cdot\hspace{-0.5pt}\itbf{R}^{(\mathtt{0})}\right\}\hspace{-1pt},
	\end{align} where the Jacobian term at first order in $\varepsilon$ is given by
	\begin{equation}
		\label{eqn:jacobian-1}
		\left[\sqrt{g}\hspace{1pt}\right]^{(\mathtt{1})}\! =\hspace{1pt} 2\hspace{0.5pt}R^2\hspace{1pt}u(\theta,\varphi)\hspace{0.5pt}\sin\theta.
	\end{equation} Since the lowest perturbation fields of the position vector are given by $\itbf{R}^{(\mathtt{0})}\!=R\hspace{1pt}\rv$ and $\itbf{R}^{(\mathtt{1})}\!=R\,u(\theta,\varphi)\hspace{1pt}\rv$, and the normal vector $\nv^{(\mathtt{0})}$ is the radial unit vector $\rv$, the excess volume $V^{(\mathtt{1})}\!$ can be further simplified to
    \begin{equation}
    	\label{eqn:volume-1}
    	V^{(\mathtt{1})}\! = R^3\!\iint\!\mathrm{d}\varphi\hspace{1.5pt}\mathrm{d}\theta\;u(\theta,\varphi)\hspace{1pt}\sin\theta = 4\pi R^3\hspace{1pt}u_0,
    \end{equation}
    by also using that $\nv^{(\mathtt{1})}$ is orthogonal to $\rv$, cf.~Eq.~(\ref{eqn:normal-1}). The latter equality in Eq.~(\ref{eqn:volume-1}) {follows by expanding} $u(\theta,\varphi)$ in terms of the spherical harmonics $u_\ell(\theta,\varphi)$ {and using their orthogonality}. Since \cc{$Y_{0,0}=1/\sqrt{4\pi}$ and $u_0=u_{0,0}Y_{0,0}$, only the zeroth harmonic contributes to the surface integral}. Thus, the osmotic pressure at first order can be rewritten as
    \begin{equation}
    	\label{eqn:osmotic-pressure-first-order}
    	\Pi^{(\mathtt{1})}_{\scriptscriptstyle\Delta} = 3\hspace{1pt}u_0\hspace{0.75pt} c^{(\mathtt{0})}_{\boldsymbol{-}}\hspace{0.25pt}k_B T.
    \end{equation}
	
	The {first-order active volume flux} is obtained by substituting Eq.~(\ref{eqn:linear-response-k}) into Eq.~(\ref{eqn:total-volume-flux-rate}), which yields
	\begin{equation}
		\mathbb{V}^{(\mathtt{1})}= \sum_k\Psi_k\,\bVk.
	\end{equation}
	
	The last two equations {give the first-order form of the integral condition} in Eq.~(\ref{eqn:volume-integral-condition}),
	\begin{equation}
		\label{eqn:volume-integral-condition-1}
		\int\limits_{\mathcal{M}\;\;}\!\!\mathrm{d}S^{(\mathtt{0})}\hspace{1pt}\bigg\{\hspace{-1pt}v^{(\mathtt{1})}\!+w_0\hspace{-1pt}\left[\Pi^{(\mathtt{1})}_{\scriptscriptstyle\Delta}\!-P^{(\mathtt{1})}_{\hspace{-0.25pt}\scriptscriptstyle\Delta}\right]\!-\mathbb{V}^{(\mathtt{1})}\hspace{-1pt}\bigg\}\hspace{-1pt}=0,\vspace{-3pt}
	\end{equation} where the term pre-multiplying the area element $\mathrm{d}S^{(\mathtt{1})}$ vanishes identically due to the result in Eq.~(\ref{eqn:Laplace-pressure-zeroth-unpacked}). As the normal membrane velocity $v^{(\mathtt{1})}\hspace{-2pt} = \hat{v}^{(\mathtt{1})}_r\hspace{-2pt} = R\,\frac{\partial}{\partial t}u(\theta,\varphi)$, {from Eq.~(\ref{eqn:m-radial-velocity-1st})}, we get
    \begin{equation}
    	\label{eqn:Laplace-pressure-first-unpacked}
    	P^{(\mathtt{1})}_{\hspace{-0.25pt}\scriptscriptstyle\Delta}
    	=
    	\Pi^{(\mathtt{1})}_{\scriptscriptstyle\Delta}
    	+w_0^{-1}\!\left(
    	R\hspace{1.25pt}\frac{\partial u_0}{\partial t}
    	-\sum_k\Psi_{k,0}\,\bVk
    	\right)\!.
    \end{equation}
    \cc{This result follows from the orthogonality of the spherical harmonics, since only the degree-zero components of $u(\theta,\varphi)$ and $\Psi_k(\theta,\varphi)$ contribute to the surface integral. Eq.~(\ref{eqn:Laplace-pressure-first-unpacked}) agrees with Eq.~(\ref{eqn:volume-area-dynamics-first-order}), obtained independently from Lamb's solution, which also gives}
	\begin{equation}
    	\label{eqn:u-0-dynamics}
    	\boxed{\frac{\partial u_0}{\partial t} = \frac{1}{2}\hspace{1pt}\mathbb{M}_0^{(\mathtt{1})}\! = \frac{1}{2}\sum_k\Psi_{k,0}\,\bMk\,.}
    \end{equation} \cc{This relation connects the uniform shape change to the degree-zero active membrane-area flux.} Thus, we can rewrite the {first-order hydrostatic pressure difference as}
	\begin{equation}
    	\label{eqn:Laplace-pressure-1}
    	\boxed{P^{(\mathtt{1})}_{\hspace{-0.25pt}\scriptscriptstyle\Delta}\! = \Pi^{(\mathtt{1})}_{\scriptscriptstyle\Delta}-w_0^{-1}\sum_k \Psi_{k,0}\left(\bVk-\frac{1}{2}R\hspace{1.25pt}\bMk\right)\!.}
    \end{equation}
	
	The dynamics of the excess areal fractions $\Psi_k$ in terms of the surface spherical harmonics is given by Eq.~(\ref{eqn:psi-k-dynamics-vector-harmonics}). For $\ell=0$, this reduces to $\frac{\partial}{\partial t}\Psi_{k,0} = -\bar{\Phi}_k\hspace{1pt}\mathbb{M}_0^{(\mathtt{1})}$, which gives
    \begin{equation}
    	\frac{\partial\hspace{0.5pt}\mathbb{M}_0^{(\mathtt{1})}}{\partial t}
    	=
    	-\mathbb{M}^{(\mathtt{0})}\hspace{0.5pt}\mathbb{M}_0^{(\mathtt{1})}
    	=0.
    \end{equation}
    \cc{Thus, the degree-zero active membrane-area flux is constant in time and set by the initial perturbation, so that}
    \begin{equation}
    	\boxed{\mathbb{M}_0^{(\mathtt{1})}\!\left(t\right)=\tilde{\mathbb{M}}_0^{(\mathtt{1})}\!\quad\!\mathrm{and}\quad\!u_0(t)=\tilde{u}_0+\frac{t}{2}\,\tilde{\mathbb{M}}_0^{(\mathtt{1})},}
    \end{equation}
    where $\tilde{u}_0$ and $\tilde{\mathbb{M}}_0^{(\mathtt{1})}\!$ are the initial values at $t=0$ for their corresponding functions. Note that the second equation is simply obtained by integrating Eq.~(\ref{eqn:u-0-dynamics}). Thus, $u_0$ \cc{changes linearly with time} if $\mathbb{M}_0^{(\mathtt{1})}\!\neq0$. \cc{To avoid this secular change of the uniform mode, membrane-area addition and removal must balance globally, so that} $\mathbb{M}_0^{(\mathtt{1})}\!=0$.
	
	Since the volume of the homogeneous steady-state is given by $V^{(\mathtt{0})}\!=\frac{4\pi}{3}R^3$, we find from Eq.~(\ref{eqn:volume-1}) that
    \begin{equation}
    	\label{eqn:change-volume-1}
    	\boxed{\frac{\partial V^{(\mathtt{1})}\!}{\partial t} = \frac{3}{2}\,\mathbb{M}^{(\mathtt{1})}_0 V^{(\mathtt{0})}.}
    \end{equation}
    Similarly, the total area at zeroth order is
    $\mathcal A^{(\mathtt{0})}=4\pi R^2$, and its {first-order}
    contribution is given by
    \begin{equation}
        \label{eqn:area-1}
        \mathcal A^{(\mathtt{1})}
        =
        2R^2\!\iint\!\mathrm{d}\varphi\hspace{1.5pt}\mathrm{d}\theta\;
        u(\theta,\varphi)\hspace{1pt}\sin\theta
        =
        8\pi R^2\hspace{1pt}u_0,
    \end{equation}
    where the former equality follows from Eq.~(\ref{eqn:jacobian-1}), and the latter is obtained by employing the orthogonality condition of the surface spherical harmonics. Hence, the rate of change {of} the total area at first order is
    \begin{equation}
        \label{eqn:change-area-1}
        \boxed{
        \frac{\partial\hspace{-0.5pt}\mathcal A^{(\mathtt{1})}}{\partial t}
        =
        \mathbb{M}^{(\mathtt{1})}_0
        \mathcal A^{(\mathtt{0})}.
        }
    \end{equation}
    {Combining these results,} the rate of change {of} $u_0$ is related to variations in both the area and the volume:
    \begin{equation}
    	\frac{\partial u_0}{\partial t}
        =
        \frac{1}{2\mathcal A^{(\mathtt{0})}}
        \frac{\partial\mathcal A^{(\mathtt{1})}}{\partial t}
        =
        \frac{1}{3V^{(\mathtt{0})}}
        \frac{\partial V^{(\mathtt{1})}}{\partial t}.
    \end{equation}
    
    \cc{If $\mathbb{M}_0^{(\mathtt{1})}=0$, the first-order corrections to both the area and volume remain constant in time and are set by the initial perturbation $\tilde{u}_0$. When $\mathbb{M}_0^{(\mathtt{1})}\neq0$, both instead vary linearly with time at this order, increasing or decreasing according to the sign of $\mathbb{M}_0^{(\mathtt{1})}$.}
    
	\subsection{Membrane shape equations}

    The surface tension $\Sigma(\xi^\alpha)$ acts as a Lagrange multiplier {enforcing the local membrane-area balance} in Eq.~(\ref{eqn:divergence-law-eq}). \cc{Accordingly, we expand the surface tension order by order in} the perturbation parameter $\varepsilon$,
    \begin{equation}
    	\Sigma(\theta,\varphi)=\sum_{\mathtt{i}=\hspace{0.65pt}0}^{\infty}\varepsilon^{\mathtt{i}}\hspace{0.5pt}\Sigma^{(\mathtt{i})}(\theta,\varphi).
    \end{equation}
    \cc{At each order, $\Sigma^{(\mathtt{i})}$ is determined self-consistently from the tangential membrane traction balance together with the local membrane-area balance.}
	
	\subsubsection{At zeroth order}

    \begin{figure}[t]\includegraphics[width=\columnwidth]{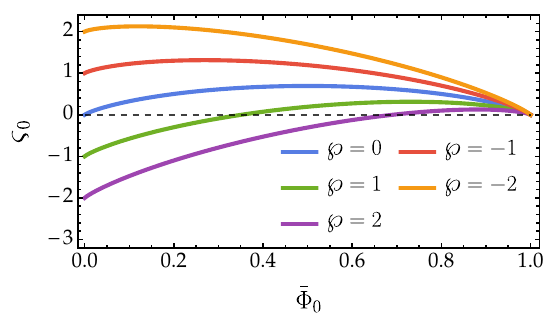}
    	\caption{\label{fig:S7} The reduced surface tension $\bigvarsigma_0$ as a function of the \cc{homogeneous background areal fraction} $\bar{\Phi}_0$ for \cc{several fixed values of the activity parameter} $\bigwp$.}
    \end{figure}

    At zeroth order, the tangential force balance together with Eq.~(\ref{eqn:body-forces-comp-0}) gives
    \begin{equation}
    	\cc{\nabla_{\!\alpha}}\Sigma^{(\mathtt{0})} = 0.
    \end{equation}
    \cc{Thus, the zeroth-order surface tension is spatially uniform.} \cc{The normal force balance in Eq.~(\ref{eqn:membrane-equation-normal}) becomes}
	\begin{align}
		\label{eqn:zero-order-normal-membrane}
		\!\!f^{(\mathtt{0})}\! &= \hspace{-1pt}\sum_{k}\frac{2\hspace{0.5pt}\bar{\Phi}_k}{R}\!\hspace{-1pt}\left[\bPk\hspace{-1pt}+\hspace{-1pt}\frac{k_B T\!}{b_0}\!\left(\!E_k\hspace{-1pt}+\hspace{-1pt}\ln\hspace{-0.5pt}\frac{\bar{\Phi}_k}{\!\!\bar{\Phi}_0^{\phantom{k'}}\!\!\!\!}\right)\hspace{-2.5pt}-\hspace{-0.5pt}\frac{2\hspace{0.5pt}\kappa\hspace{0.25pt}\hspace{0.5pt}C_k\hspace{-1pt}+\hspace{-1.5pt}3\hspace{0.5pt}\bQk}{2R}\right]\notag\\[5pt]
		&\qquad+\frac{2}{R}\hspace{-1pt}\left[\Sigma^{(\mathtt{0})}+\frac{k_B T}{b_0}(\mathcal{E}_0+\ln\bar{\Phi}_0)- \frac{\kappa\hspace{0.25pt}\hspace{1pt}\mathcal{C}_0}{R}\right]\!\hspace{-0.5pt}.
	\end{align} Here, the {zeroth-order normal fluid traction} is given by $f^{(\mathtt{0})}=-P^{(\mathtt{0})}_{\hspace{-0.5pt}\scriptscriptstyle\Delta}$. Using Eq.~(\ref{eqn:Laplace-pressure-zeroth-unpacked}), Eq.~(\ref{eqn:zero-order-normal-membrane}) gives
	\begin{align}
		\label{eqn:normal-balance-zeroth-order}
		\!\hspace{-1pt}\Pi^{(\mathtt{0})}_{\scriptscriptstyle\Delta}\!\hspace{-1pt} &+ \hspace{-0.5pt}\sum_{k}\frac{2\hspace{0.5pt}\bar{\Phi}_k}{R}\!\hspace{-1pt}\left[\bPk\hspace{-1pt}+\hspace{-1pt}\frac{k_B T\!}{b_0}\!\left(\!E_k\hspace{-1pt}+\hspace{-1pt}\ln\hspace{-0.5pt}\frac{\bar{\Phi}_k}{\!\!\bar{\Phi}_0^{\phantom{k'}}\!\!\!\!}\right)\hspace{-2.5pt}-\hspace{-0.5pt}\frac{2\hspace{0.5pt}\kappa\hspace{0.25pt}\hspace{0.5pt}C_k\hspace{-1pt}+\hspace{-1.5pt}3\hspace{0.5pt}\bQk}{2R}\right]\notag\\[6pt]
		&\hspace{-15pt}+\frac{2}{R}\hspace{-1pt}\left(\bar{\Sigma}^{(\mathtt{0})}+\frac{k_B T}{b_0}\ln\bar{\Phi}_0- \frac{\kappa\hspace{0.25pt}\hspace{1pt}\mathcal{C}_0}{R}\right)\!=\sum_k\frac{\bar{\Phi}_k\hspace{1pt}\bVk}{w_0},
	\end{align} where we define $\bar{\Sigma}^{(\mathtt{0})}\!=\Sigma^{(\mathtt{0})}+\frac{k_B T}{b_0}\mathcal{E}_0$, and $\Pi^{(\mathtt{0})}_{\scriptscriptstyle\Delta}$ is the osmotic pressure, whose expression is found in Eq.~(\ref{eqn:pressure-drop-zeroth}).

    \begin{figure}[t]\includegraphics[width=\columnwidth]{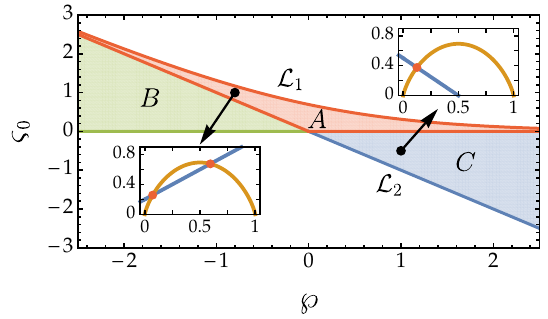}
        \caption{\label{fig:S3} \cc{Diagram showing the regions in $(\bigvarsigma_0,\bigwp)$ parameter space where Eq.~(\ref{eqn:mass-fraction-zero-condition}) admits two, one, or no real solutions for $\bar{\Phi}_0$.} The boundary $\mathcal{L}_1$ is given by $\bigvarsigma_0=\ln\!\left(1+e^{-\bigwp}\right)$, while $\mathcal{L}_2$ is defined by $\bigvarsigma_0=-\bigwp$. In region $A$, \cc{two distinct real solutions exist, as illustrated by the inset at} $(\bigvarsigma_0,\bigwp)=(1,-0.8)$, while \cc{regions $B$ and $C$ each contain a single solution, with the inset at} $(\bigvarsigma_0,\bigwp)=(-0.5,1)$ \cc{illustrating the case in region $C$}. The insets show the left-hand side (in blue) and the right-hand side (in orange) of Eq.~(\ref{eqn:mass-fraction-zero-condition}) as \cc{functions} of $\bar{\Phi}_0$, with the \cc{red intersections marking the corresponding solutions}.}
    \end{figure}

    The {homogeneous steady-state areal fractions} $\bar{\Phi}_k$ are given by Eq.~(\ref{eqn:steady-state-value-k}), and thus $\bar{\Phi}_k\hspace{-1pt}\propto\!\left(1-\bar{\Phi}_0\right)\hspace{-1pt}$. Hence, Eq.~(\ref{eqn:normal-balance-zeroth-order}) {can be written as the relation}
	\begin{equation}
		\label{eqn:mass-fraction-zero-condition}
		\boxed{\bigvarsigma_0 + \bigwp\hspace{1pt}(1\!-\hspace{-1pt}\bar{\Phi}_0) = -\bar{\Phi}_0\ln\bar{\Phi}_0-(1\!-\hspace{-1pt}\bar{\Phi}_0)\ln(1\!-\hspace{-1pt}\bar{\Phi}_0)}
	\end{equation} where the parameters $\bigvarsigma_0$ and $\bigwp$ are given by
	\begin{equation}
		\bigvarsigma_0= \frac{b_0}{k_B T}
		\left(\bar{\Sigma}^{(\mathtt{0})} + \frac{R}{2}\hspace{0.5pt}\Pi^{(\mathtt{0})}_{\scriptscriptstyle\Delta}- \frac{\kappa\hspace{0.25pt}\hspace{1pt}\mathcal{C}_0}{R}\right)\!,
	\end{equation} and $\bigwp=\bigwp_{1}+\bigwp_{2}$, with
	\begin{align}
		\bigwp_{k}&=\frac{|\bMkp|}{|\bMk|+|\bMkp|}\left[\,E_k+\ln\hspace{-1pt}\left(\frac{|\bMkp|}{|\bMk|+|\bMkp|}\right)\right.\notag\\[5pt]
		&\;+\hspace{-1pt}\frac{b_0}{k_B T}\!\left(\bPk\hspace{-1pt}-\frac{R\hspace{1pt}\bVk\hspace{-1pt}}{2w_0}-\frac{2\hspace{0.5pt}\kappa\hspace{0.25pt}\hspace{0.5pt}{C}_k+3\hspace{0.5pt}\bQk}{2R}\right)\!\hspace{-0.5pt}\Bigg]\!,
	\end{align} {where $k'\neq k$.} The right-hand side of Eq.~(\ref{eqn:mass-fraction-zero-condition}) is positive for $\bar{\Phi}_0\in(0,1)$. At thermodynamic equilibrium, {the active-component areal fractions vanish}, $\bar{\Phi}_k=0$, and hence $\bar{\Phi}_0=1$. In this limit, Eq.~(\ref{eqn:mass-fraction-zero-condition}) gives $\bigvarsigma_0=0$, {recovering the {\it Young--Laplace} relation:}
	\begin{equation}
		\Pi^{(\mathtt{0})}_{\scriptscriptstyle\Delta}\! = 2 H^{(\mathtt{0})}\bar{\Sigma}^{(\mathtt{0})}\! + 2\kappa\hspace{1pt}\mathcal{C}_0 K^{(\mathtt{0})},
	\end{equation} \cc{At equilibrium, the hydrostatic pressure difference equals the osmotic pressure,} $P^{(\mathtt{0})}_{\hspace{-0.5pt}\scriptscriptstyle\Delta}\!=\Pi^{(\mathtt{0})}_{\scriptscriptstyle\Delta}$, where $H^{(\mathtt{0})}$ and $K^{(\mathtt{0})}$ \cc{are the zeroth-order mean and Gaussian curvatures, respectively}. Also, $\bigwp=\bigwp_1=\bigwp_2=0$ at equilibrium, \cc{since the cycle rates $\bRk$ vanish}.

    \cc{We first consider the simpler case in which $\bar{\Phi}_0$ is prescribed. Eq.~(\ref{eqn:mass-fraction-zero-condition}) then directly determines the corresponding reduced surface tension $\bigvarsigma_0$.} The dependence of the reduced surface tension $\bigvarsigma_0$ on $\bar{\Phi}_0$ is shown in Fig.~\ref{fig:S7}, \cc{with $\bigvarsigma_0\to0$ as $\bar{\Phi}_0\to1$ in the equilibrium limit.}

    \cc{We next consider the case in which the zeroth-order surface tension is fixed and $\bar{\Phi}_0$ is allowed to adjust. Eq.~(\ref{eqn:mass-fraction-zero-condition}) must then be solved for $\bar{\Phi}_0$ and may admit more than one solution depending on $(\bigvarsigma_0,\bigwp)$. In Fig.~\ref{fig:S3}, we show the corresponding solution regions in} $\left(\bigvarsigma_0,\bigwp\right)$. \cc{Since the right-hand side is strictly concave and the left-hand side is linear, the upper boundary $\mathcal{L}_1$ is defined by their tangency at} $\bar{\Phi}_0^\star=1/(1+e^{-\bigwp})$. \cc{Substitution into Eq.~(\ref{eqn:mass-fraction-zero-condition}) gives}
    $\mathcal{L}_{1}=\left\{\left(\bigvarsigma_0,\bigwp\right)\,\middle|\,\bigvarsigma_0=\ln\left(1+e^{-\bigwp}\right)\right\}$. \cc{This gives the largest value of $\bigvarsigma_0$ for which a solution exists:}
    \begin{equation}
    	\label{eqn:upper-boundary-condition}
    	\bigvarsigma_0\leq\ln\hspace{-1pt}\left(1+e^{-\bigwp}\hspace{0.75pt}\right),
    \end{equation}
    for any real $\bigwp$. \cc{The lower bound is determined by the endpoint values
    $\bigvarsigma_0=-\bigwp$ at $\bar{\Phi}_0=0$
    and $\bigvarsigma_0=0$ at $\bar{\Phi}_0=1$:}
    \begin{equation}
        \label{eqn:lower-boundary-condition}
        \bigvarsigma_0\geq\min(0,-\bigwp).
    \end{equation}
    
    \cc{Two distinct solutions exist in region $A$ when $\bigvarsigma_0$ lies above both endpoint values but below the tangency value:}
    \begin{equation}
    	A=
    	\left\{
    	\left(\bigvarsigma_0,\bigwp\right)
    	\,\middle|\,
    	\max\left(0,-\bigwp\right)
    	<
    	\bigvarsigma_0
    	<
    	\ln\!\left(1\!+\!e^{-\bigwp}\right)
    	\right\}.
    \end{equation}
    \cc{On $\mathcal{L}_1$, the two roots merge at $\bar{\Phi}_0^\star$. Regions $B$ and $C$ each contain a single solution,}
    \begin{align}
    	B&=\left\{\left(\bigvarsigma_0,\bigwp\right)\,\middle|\,0<\bigvarsigma_0<-\bigwp,\;\bigwp<0\right\},\\[4pt]
    	C&=\left\{\left(\bigvarsigma_0,\bigwp\right)\,\middle|\,-\bigwp<\bigvarsigma_0<0,\;\bigwp>0\right\}.
    \end{align}
    \cc{The two branches are shown in Fig.~\ref{fig:S4}. The branch extending through $A\cup B$ follows the larger root in $A$ and approaches $\bar{\Phi}_0=1$ at equilibrium, whereas the branch extending through $A\cup C$ follows the smaller root and does not. Thus, the former is the physical homogeneous branch, which exists for}
    \begin{equation}
    	\label{eqn:homogeneous-condition-0}
    	\boxed{0\leq\bigvarsigma_0\leq\ln\hspace{-1pt}\left(1+e^{-\bigwp}\hspace{0.75pt}\right)\!,}
    \end{equation}
    for any real value of the {activity parameter} $\bigwp$. {Within this domain,} Eq.~(\ref{eqn:mass-fraction-zero-condition}) {determines a unique physical value of $\bar{\Phi}_0$}, which in turn determines the {homogeneous areal fractions} $\bar{\Phi}_k$ via Eq.~(\ref{eqn:steady-state-value-k}).
     
    \begin{widetext}
    
    \subsubsection{At first order}
    
    \cc{Linearizing the tangential membrane force balance in Eq.~(\ref{eqn:membrane-equation-tangential}) about the homogeneous spherical base state gives}
    \begin{equation}
        \label{eqn:tangential-membrane-eqns-1}
        \left[f^\alpha\right]^{(\mathtt{1})}
        =-\nabla^\alpha\Sigma^{(\mathtt{1})}
        -\sum_k\bar{\Phi}_k\bQk\nabla^\alpha\!\left(2H^{(\mathtt{1})}\right)
        -\sum_k\left(\nabla^\alpha\Psi_k\right)
        \left[
        \bPk+\frac{k_B T}{b_0}\left(E_k+\ln\frac{\bar{\Phi}_k}{\bar{\Phi}_0}\right)
        -\frac{2\kappa C_k+\bQk}{R}
        \right].
    \end{equation}
    \cc{Since $\left[f^\alpha\right]^{(\mathtt{0})}=0$, the first-order tangential contribution to the hydrodynamic force is
    $\itbf{f}_{\!\scriptscriptstyle\mathcal{S}}^{(\mathtt{1})}
    =\ed{\alpha}^{(\mathtt{0})}\left[f^\alpha\right]^{(\mathtt{1})}$. Hence,}
    \begin{equation}
    	\label{eqn:surface-body-force-1}
    	\itbf{f}_{\!\scriptscriptstyle\mathcal{S}}^{(\mathtt{1})}
    	=\sum_{\ell=1}^{\infty}\bnabla_{\!\scriptscriptstyle\mathcal{S}}
    	\left\{
    	-\frac{\Sigma^{(\mathtt{1})}_\ell}{R}
    	+\sum_k\bar{\Phi}_k\bQk\frac{(\ell-1)(\ell+2)u_\ell}{R^2}
    	-\sum_k\frac{\Psi_{k,\ell}}{R}
    	\!\left[
    	\bPk+\frac{k_B T}{b_0}\left(E_k+\ln\frac{\bar{\Phi}_k}{\bar{\Phi}_0}\right)
    	-\frac{2\kappa C_k+\bQk}{R}
    	\right]
    	\!\right\}
    \end{equation}
    {where} $\Sigma^{(\mathtt{1})}$ and $H^{(\mathtt{1})}$ {are expanded} in the basis of surface spherical harmonics. The result in Eq.~(\ref{eqn:surface-body-force-1}) {shows} that the body force $\itbf{f}_{\!\scriptscriptstyle\mathcal{S}}^{(\mathtt{1})}$ can be expressed as a surface gradient of a scalar {field defined on} the membrane. {Likewise}, $\itbf{f}_{\!\scriptscriptstyle\mathcal{S}}^{(\mathtt{1})}$ can be written in terms of the ambient fluid stresses acting across the membrane surface, as derived in Eq.~(\ref{eqn:tangential-part-body-force-1}) from Lamb's solution. \cc{Since this expression is also a surface gradient, the orthogonality of spherical harmonics allows the corresponding surface modes to be equated. Thus, for each $\ell\geq1$, the tangential balance gives}
    \begin{align}
    	\label{eqn:tangential-balance-1}
    	&\qquad\qquad
    	\frac{\eta(2\ell+1)}{\ell(\ell+1)}
    	\left[
    	\frac{\partial u_\ell}{\partial t}
    	-\sum_k\Psi_{k,\ell}
    	\left(
    	2\bMk-\frac{3\bVk}{R}
    	-\frac{\ell(\ell+1)\bQk}{\eta R^2}
    	\right)
    	\right]
    	-\frac{\Sigma^{(\mathtt{1})}_\ell}{R}
    	\notag\\[5pt]
    	&+u_\ell\sum_k\bar{\Phi}_k
    	\frac{(\ell-1)(\ell+2)\bQk}{R^2}
    	-\sum_k\frac{\Psi_{k,\ell}}{R}
    	\left[
    	\bPk+\frac{k_B T}{b_0}
    	\left(
    	E_k+\ln\frac{\bar{\Phi}_k}{\bar{\Phi}_0}
    	\right)
    	-\frac{2\kappa C_k+\bQk}{R}
    	\right]
    	=0,
    \end{align}
    {where we have used the explicit forms} $\mathcal{X}^{(\mathtt{1})}_\ell=R\frac{\partial u_\ell}{\partial t}-\sum_k\Psi_{k,\ell}\bVk$ and $\mathcal{Y}^{(\mathtt{1})}_\ell=\sum_k\Psi_{k,\ell}\left[2\bVk-R\bMk+\frac{\ell(\ell+1)\bQk}{2\eta R}\right]$.

    \begin{figure*}[t!]
        \includegraphics[width=\textwidth]{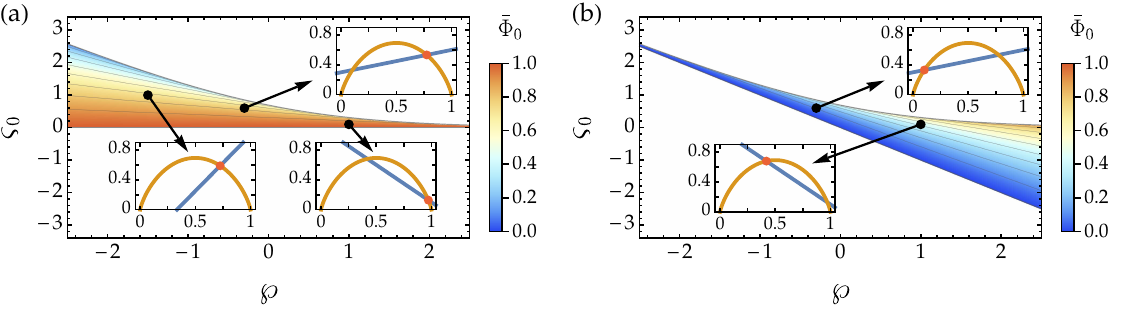}
        \caption{\label{fig:S4}
        (a) Density plot of the background \cc{areal} fraction $\bar{\Phi}_0$ as a function of the parameters $\bigvarsigma_0$ and $\bigwp$, showing \cc{the physical branch} of Eq.~(\ref{eqn:mass-fraction-zero-condition}) \cc{on $A\cup B$, with the regions $A$, $B$, and $C$ defined in Fig.~\ref{fig:S3}}. \cc{In region $A$, this branch follows the larger, right-most of the two solutions.} The \cc{black points mark representative parameter choices:} $(\bigvarsigma_0,\bigwp)=(1,-1.5)$ corresponds to a single solution, while \cc{$(0.6,-0.3)$ and $(0.1,1)$ each correspond to two solutions, with the corresponding roots illustrated in the insets as in Fig.~\ref{fig:S3}.}
        (b) Density plot of $\bar{\Phi}_0$ in terms of $\bigvarsigma_0$ and $\bigwp$, \cc{showing the second branch on $A\cup C$}. \cc{In region $A$, this branch follows the smaller, left-most of the two solutions. The black points at $(\bigvarsigma_0,\bigwp)=(0.6,-0.3)$ and $(0.1,1)$ correspond to the same representative parameter choices as in panel (a), with the corresponding roots illustrated in the insets.} This root is unphysical as it does not respect the requirement that $\bar{\Phi}_0\to1$ as we approach \cc{thermodynamic} equilibrium, \ie $\bigvarsigma_0\to0$ and $\bigwp\to0$.
        }
    \end{figure*}
    
    The equation of motion along the normal direction at first order in $\varepsilon$ {follows} from Eq.~(\ref{eqn:membrane-equation-normal}). {Accordingly,} the normal body force on the membrane, $f^{(\mathtt{1})}$, can be computed as
    \begin{align}
    	&f^{(\mathtt{1})}=\frac{2\Sigma^{(\mathtt{1})}}{R}-2H^{(\mathtt{1})}\left\{\left(\bar{\Sigma}^{(\mathtt{0})}-\frac{2\kappa\mathcal{C}_0}{R}+\frac{k_B T}{b_0}\ln\bar{\Phi}_0\right)+\sum_k\bar{\Phi}_k\left[\bPk+\frac{k_B T}{b_0}\left(E_k+\ln\frac{\bar{\Phi}_k}{\bar{\Phi}_0}\right)-\frac{2\kappa C_k+3\bQk}{R}\right]\right\}\notag\\[5pt]
    	&\qquad+\sum_k\frac{2\Psi_k}{R}\left[\bPk+\frac{k_B T}{b_0}\left(E_k+\ln\frac{\bar{\Phi}_k}{\bar{\Phi}_0}\right)-\frac{2\kappa C_k+3\bQk}{2R}\right]+\sum_k\Delta^{(\mathtt{0})}\Psi_k\left(\kappa C_k-\frac{\bQk}{2}\right)+2\kappa\Delta^{(\mathtt{0})}H^{(\mathtt{1})}.
    \end{align}
    Here, we use that $H^2-K=\mathcal{O}(\varepsilon^2)$, which gives the Gaussian curvature at first order as $K^{(\mathtt{1})}=2H^{(\mathtt{0})}H^{(\mathtt{1})}$. Moreover, since $H^{(\mathtt{0})}$ is spatially uniform, the first-order expansion of the curvature-Laplacian term is $[\Delta H]^{(\mathtt{1})}=\Delta^{(\mathtt{0})}H^{(\mathtt{1})}$. The term in the curly brackets can be {simplified using} Eq.~(\ref{eqn:zero-order-normal-membrane}), {giving}
    \begin{align}
    	&f^{(\mathtt{1})}=2H^{(\mathtt{1})}\left[\frac{R}{2}P^{(\mathtt{0})}_{\scriptscriptstyle\Delta}+\frac{\kappa\mathcal{C}_0}{R}+\sum_k\frac{\bar{\Phi}_k(2\kappa C_k+3\bQk)}{2R}\right]+\sum_k\frac{2\Psi_k}{R}\left[\bPk+\frac{k_B T}{b_0}\left(E_k+\ln\frac{\bar{\Phi}_k}{\bar{\Phi}_0}\right)-\frac{2\kappa C_k+3\bQk}{2R}\right]\notag\\[5pt]
    	&\qquad+\sum_k\Delta^{(\mathtt{0})}\Psi_k\left(\kappa C_k-\frac{\bQk}{2}\right)+2\kappa\Delta^{(\mathtt{0})}H^{(\mathtt{1})}+\frac{2\Sigma^{(\mathtt{1})}}{R},
    \end{align}
    where $P^{(\mathtt{0})}_{\scriptscriptstyle\Delta}$ is the {hydrostatic pressure difference} at zeroth order in $\varepsilon$. Expanding in the basis of spherical harmonics, with $H^{(\mathtt{1})}_\ell=-(\ell-1)(\ell+2)u_\ell/(2R)$ and $\Delta^{(\mathtt{0})}Y_\ell=-\ell(\ell+1)Y_\ell/R^2$, gives $\Delta^{(\mathtt{0})}H^{(\mathtt{1})}_\ell=\ell(\ell-1)(\ell+1)(\ell+2)u_\ell/(2R^3)$. Thus,
    \begin{align}
    	\label{eqn:normal-balance-order-1-expansion}
    	&f^{(\mathtt{1})}
    	=\sum_{\ell=0}^{\infty}
    	\Bigg\{
    	\frac{2\Sigma_\ell^{(\mathtt{1})}}{R}
    	+\frac{\kappa\ell(\ell-1)(\ell+1)(\ell+2)u_\ell}{R^3}
    	-(\ell-1)(\ell+2)u_\ell
    	\left[
    	\frac{P^{(\mathtt{0})}_{\scriptscriptstyle\Delta}}{2}
    	+\frac{\kappa\mathcal{C}_0}{R^2}
    	+\sum_k\frac{\bar{\Phi}_k(2\kappa C_k+3\bQk)}{2R^2}
    	\right]
    	\notag\\[5pt]
    	&\qquad
    	+\sum_k\frac{2\Psi_{k,\ell}}{R}
    	\left[
    	\bPk+\frac{k_B T}{b_0}
    	\left(E_k+\ln\frac{\bar{\Phi}_k}{\bar{\Phi}_0}\right)
    	-\frac{2\kappa C_k+3\bQk}{2R}
    	\right]
    	-\sum_k\frac{\ell(\ell+1)\Psi_{k,\ell}(2\kappa C_k-\bQk)}{2R^2}
    	\Bigg\}.
    \end{align}
    
    \cc{To avoid secular evolution of the isotropic mode,} cf.~Eqs.~(\ref{eqn:change-volume-1}) and (\ref{eqn:change-area-1}), we require that the $\ell=0$ contribution to the {active membrane-area flux vanish}: $\mathbb{M}^{(\mathtt{1})}_0=\sum_k\Psi_{k,0}\bMk=0$. Hence, Eq.~(\ref{eqn:psi-k-dynamics-vector-harmonics}) for $\ell=0$ becomes $\dot{\Psi}_{1,0}=\dot{\Psi}_{2,0}=0$. This implies that the excess areal fractions $\Psi_{k,0}$ are constants independent of time, set by the initial perturbation. Since we perturb about the homogeneous {steady-state areal fractions} $\bar{\Phi}_k$, we {choose} $\Psi_{1,0}=\Psi_{2,0}=0$, \cc{so that the perturbations contain no uniform shift in the areal fractions}. {Likewise, we choose} $u_0=0$, cf.~Eq.~(\ref{eqn:volume-area-dynamics-first-order}). Using Eq.~(\ref{eqn:normal-body-force-1}) and Eq.~(\ref{eqn:Laplace-pressure-1}), we find that the zeroth mode of Eq.~(\ref{eqn:normal-balance-order-1-expansion}) reduces to $\Sigma^{(\mathtt{1})}_0=0$, as $P^{(\mathtt{1})}_{\scriptscriptstyle\Delta}=\Pi^{(\mathtt{1})}_{\scriptscriptstyle\Delta}=0$. Hence, {for this choice of perturbations,} the sum in Eq.~(\ref{eqn:normal-balance-order-1-expansion}) may start from the first {spherical-harmonic} mode ($\ell=1$). \cc{More generally, if the $\ell=0$ perturbations are retained,} $\Sigma^{(\mathtt{1})}_0$ is linearly related to the initial value of $u_0$ and \cc{one independent compositional zero-mode amplitude, since $\Psi_{1,0}$ and $\Psi_{2,0}$ are constrained by} $\mathbb{M}^{(\mathtt{1})}_0=0$.
    
    \cc{Equating the degree-$\ell$ components of Eq.~(\ref{eqn:normal-balance-order-1-expansion}) with the fluid-side normal traction in Eq.~(\ref{eqn:normal-body-foce-comps}), we obtain for each mode $\ell\geq1$}
    \begin{align}
    	&\frac{\eta(2\ell+1)}{\ell(\ell+1)}
    	\left\{
    	[3-2\ell(\ell+1)]\frac{\partial u_\ell}{\partial t}
    	+\frac{3+2\ell(\ell+1)}{R}\sum_k\Psi_{k,\ell}\bVk
    	-3\sum_k\Psi_{k,\ell}
    	\left(
    	\bMk-\frac{\ell(\ell+1)\bQk}{2\eta R^2}
    	\right)
    	\right\}
    	\notag\\[5pt]
    	&-\frac{2\Sigma_\ell^{(\mathtt{1})}}{R}
    	+(\ell-1)(\ell+2)u_\ell
    	\left[
    	\frac{P^{(\mathtt{0})}_{\scriptscriptstyle\Delta}}{2}
    	+\frac{\kappa\mathcal{C}_0}{R^2}
    	+\sum_k\frac{\bar{\Phi}_k(2\kappa C_k+3\bQk)}{2R^2}
    	\right]
    	-\frac{\kappa\ell(\ell-1)(\ell+1)(\ell+2)u_\ell}{R^3}
    	\notag\\[5pt]
    	&\qquad
    	-\sum_k\frac{2\Psi_{k,\ell}}{R}
    	\left[
    	\bPk+\frac{k_B T}{b_0}
    	\left(E_k+\ln\frac{\bar{\Phi}_k}{\bar{\Phi}_0}\right)
    	-\frac{2[\ell(\ell+1)+2]\kappa C_k-(\ell-2)(\ell+3)\bQk}{4R}
    	\right]
    	=0.
    \end{align}
    
    \cc{Solving the tangential balance in Eq.~(\ref{eqn:tangential-balance-1}) for $\Sigma_\ell^{(\mathtt{1})}$ and substituting the result into the normal balance gives the evolution equation for the degree-$\ell$ shape component $u_\ell$:}
    \begin{align}
    	\frac{\partial u_\ell}{\partial t}
    	&=\sum_k\Psi_{k,\ell}
    	\left[
    	\frac{
    	R\bMk+[2\ell(\ell+1)-3]\bVk
    	-\frac{\ell(\ell+1)}{2\eta R}\bQk
    	}{
    	R[2\ell(\ell+1)-1]
    	}
    	\right]
    	-\sum_k
    	\frac{
    	\ell(\ell-1)(\ell+1)(\ell+2)
    	(\Psi_{k,\ell}+u_\ell\bar{\Phi}_k)
    	(\bQk-2\kappa C_k)
    	}{
    	2\eta R^2(2\ell+1)[2\ell(\ell+1)-1]
    	}
    	\notag\\[5pt]
    	&\quad
    	+\frac{
    	\kappa u_\ell\ell(\ell-1)(\ell+1)(\ell+2)
    	[R\mathcal{C}_0-\ell(\ell+1)]
    	}{
    	\eta R^3(2\ell+1)[2\ell(\ell+1)-1]
    	}
    	+\frac{
    	\ell(\ell-1)(\ell+1)(\ell+2)u_\ell
    	}{
    	2\eta(2\ell+1)[2\ell(\ell+1)-1]
    	}
    	\left[
    	\Pi^{(\mathtt{0})}_{\scriptscriptstyle\Delta}
    	-\frac{1}{w_0}\sum_k\bar{\Phi}_k\bVk
    	\right].
    	\label{eqn:shape-governing-eq}
    \end{align}
    \cc{Substituting the resulting shape dynamics back into the tangential balance gives the corresponding degree-$\ell$ component of the surface tension,}
    \begin{align}
    	\label{eqn:surface-tension-eq}
    	\Sigma_\ell^{(\mathtt{1})}
    	&=\sum_k\frac{\bQk\!\left[(8\ell^3+15\ell^2+\ell-3)\Psi_{k,\ell}+(\ell-1)(\ell+2)(2\ell-1)(2\ell+3)u_\ell\bar{\Phi}_k\right]}{2R[2\ell(\ell+1)-1]}
    	-\sum_k\Psi_{k,\ell}\!\left[\bPk+\frac{k_B T}{b_0}\left(E_k+\ln\frac{\bar{\Phi}_k}{\bar{\Phi}_0}\right)\!\right]
    	\notag\\[5pt]
    	&\quad
    	+\sum_k\frac{\kappa C_k\left\{[5\ell(\ell+1)-4]\Psi_{k,\ell}+(\ell-1)(\ell+2)u_\ell\bar{\Phi}_k\right\}}{R[2\ell(\ell+1)-1]}
    	-\sum_k\frac{\eta(2\ell-1)(2\ell+1)(2\ell+3)\Psi_{k,\ell}(R\bMk-2\bVk)}{\ell(\ell+1)[2\ell(\ell+1)-1]}
    	\notag\\[5pt]
    	&\quad
    	+\frac{\kappa u_\ell(\ell-1)(\ell+2)[R\mathcal{C}_0-\ell(\ell+1)]}{R^2[2\ell(\ell+1)-1]}
    	+\frac{(\ell-1)(\ell+2)Ru_\ell}{2[2\ell(\ell+1)-1]}
    	\left[
    	\Pi^{(\mathtt{0})}_{\scriptscriptstyle\Delta}
    	-\frac{1}{w_0}\sum_k\bar{\Phi}_k\bVk
    	\right].
    \end{align}
    
    \cc{The surface-tension component $\Sigma_\ell^{(\mathtt{1})}$ obtained above is the degree-$\ell$ contribution to $\Sigma$, which acts as a Lagrange multiplier enforcing local membrane-area balance. To connect this result with the activity-renormalized surface tension $\Sigma_A$ defined in Eq.~(\ref{eqn:renormalized-surface-tension}), we expand $\Sigma_A$ to first order, giving}
    \begin{equation}
        \label{eqn:renormalized-surface-tension-amplitude}
        \cc{
        \Sigma_{A,\ell}^{(\mathtt{1})}
        =
        \Sigma_\ell^{(\mathtt{1})}
        +
        \sum_k\Psi_{k,\ell}
        \left[
        \bPk+
        \frac{k_B T}{b_0}
        \left(
        E_k+\ln\frac{\bar{\Phi}_k}{\bar{\Phi}_0}
        \right)
        \right].
        }
    \end{equation}
    \cc{Substituting Eq.~(\ref{eqn:surface-tension-eq}) shows that the explicit first-moment and thermodynamic contributions cancel from $\Sigma_{A,\ell}^{(\mathtt{1})}$ through the self-consistent adjustment of the surface-tension Lagrange multiplier. Physically, the first active force moment contributes directly to the homogeneous activity-renormalized surface tension in Eq.~(\ref{eqn:renormalized-surface-tension}), whereas its spatially varying first-order contribution, proportional to $\Psi_{k,\ell}\,\bPk$, is compensated by the local adjustment of $\Sigma$ required by membrane-area balance. Thus, $\bPk$ remains part of the homogeneous state but does not contribute independently to the linear nonuniform response, which also explains its absence from the linear shape equation.}
    
    \cc{The $\ell=1$ mode is of particular interest because it corresponds to translation of the compartment and is associated with the drift instability discussed below. For this mode, Eqs.~(\ref{eqn:surface-tension-eq}) and (\ref{eqn:renormalized-surface-tension-amplitude}) reduce to}
    \begin{equation}
        \label{eqn:renormalized-surface-tension-amplitude-1}
        \cc{
        \Sigma_{A,1}^{(\mathtt{1})}
        =
        \sum_k\Psi_{k,1}
        \!\left[
        \frac{7\bQk+4\kappa C_k}{2R}
        -\frac{5\eta}{2}\!\left(R\bMk-2\bVk\right)
        \right]\!.
        }
    \end{equation}

\end{widetext}
    
    \cc{For the two-point force model, an extensile force dipole during fission and a contractile one during fusion give $\mathrm{sgn}(\bQk)=(-1)^{k+1}$, while $\mathrm{sgn}(\bMk)=(-1)^k$, so $\bQk$ and $\bMk$ have opposite signs for each trafficking cycle. Using Eqs.~(\ref{eqn:renormalized-surface-tension-amplitude}) and (\ref{eqn:surface-tension-eq}), their composition-dependent contributions are seen to enter with opposite prefactors for every $\ell\geq1$ and therefore reinforce one another. For $\Psi_{k,\ell}>0$, both contributions raise the activity-renormalized surface tension for fission and lower it for fusion. This change has the same direction as the direct first-moment contribution discussed in Appendix~A following Eq.~(\ref{eqn:renormalized-surface-tension}), but its origin is different: in the linear nonuniform response it arises from the second active force moment and membrane-area transfer rather than from an explicit first-moment contribution.}

    \subsection{Linear stability analysis}
    
    The dynamics of the excess areal fractions $\Psi_{k,\ell}$ is given by Eq.~(\ref{eqn:psi-k-dynamics-vector-harmonics}), {and the shape dynamics by} Eq.~(\ref{eqn:shape-governing-eq}). {Together, these equations form a closed linear system whose stability about the homogeneous steady state can be analyzed for each spherical-harmonic degree} $\ell\geq1$ {by writing it in matrix form:}
    \begin{equation}
    	\label{eqn:linear-set-eqs}
    	\frac{\partial}{\partial t}
    	\big[\begin{matrix}
    	u_\ell&\Psi_{1,\ell}&\Psi_{2,\ell}
    	\end{matrix}\,\big]^{\mathsf{T}}
    	=
    	\mathsf{S}_\ell
    	\big[\begin{matrix}
    	u_\ell&\Psi_{1,\ell}&\Psi_{2,\ell}
    	\end{matrix}\,\big]^{\mathsf{T}}\!,
    \end{equation}
    where $\scriptstyle\mathsf{T}$ denotes a transpose, and $\mathsf{S}_\ell$ is the {corresponding stability matrix}.
    {For $\ell=1$, Eq.~(\ref{eqn:shape-governing-eq}) reduces to}
    \begin{equation}
    	\label{eqn:shape-eq-1}
    	\frac{\partial u_1}{\partial t}
    	=
    	\sum_k\Psi_{k,1}
    	\!\left[
    	\frac{R\bMk+\bVk}{3R}
    	-\frac{\bQk}{3\eta R^2}
    	\right],
    \end{equation}
    whereas the membrane compositional equation, written in the vector form
    $\boldsymbol{\Psi}_\ell=\left(\Psi_{1,\ell},\Psi_{2,\ell}\right)^{\!\mathsf{T}}\hspace{-1.5pt}$,
    reduces to
    \begin{equation}
    	\label{eqn:composition-eq-1}
    	\frac{\partial\boldsymbol{\Psi}_{\hspace{-0.5pt}1}}{\partial\hspace{0.25pt}t}
    	=
    	-\mathsf{M}\hspace{0.5pt}\boldsymbol{\Psi}_{\hspace{-0.5pt}1}
    	-\frac{2}{R^2}\,\mathsf{D}\hspace{0.5pt}\boldsymbol{\Psi}_{\hspace{-0.5pt}1}.
    \end{equation}

\begin{table}
    \begin{ruledtabular}
        \begin{tabular}{c|cc|c}
            $A$ & Dimensionless fission rate & &
            {$0.1$--$500$}\\[2pt]\hline

            $\zeta$ & Ratio of fusion and fission rates & &
            $0.25$--$4$\\[2pt]\hline

            {$\bar{\Phi}_0$} & {Background areal fraction} & &
            {$3/5$}\\[2pt]\hline

            {$\gamma$} & {Surface diffusion coefficient} & &
            {$0.002$--$0.1~\mu\mathrm{m}^2/\mathrm{s}$}\\[2pt]\hline

            {$\mathcal{K}$} & {Dimensionless bending rigidity} & &
            {$0.2$--$200$}\\[2pt]\hline

            {$|Q_k|$} & {Dimensionless second moment} & &
            {$5$--$35000$}\\[2pt]\hline

            {$|\bMk|$} & {Membrane-area transfer rate} & &
            {$0.01$--$1.2~\mathrm{s}^{-1}$}\\[2pt]\hline

            {$|\bPk|$} & Active force first moment & &
            $0.01$--$0.2~\mathrm{mN}/\mathrm{m}$\\[2pt]\hline

            {$|\bQk|$} & Active force second moment & &
            {$0.5$--$7~\mathrm{pN}$}\\[2pt]
        \end{tabular}
    \end{ruledtabular}

    \caption{\label{tab:scaled-eqs-2}
    Estimated ranges of physical and dimensionless parameters. \cc{To estimate $\gamma$, we associate the effective surface diffusivity $D$ with the slower compositional diffusion mode, corresponding to the smaller eigenvalue of the diffusion matrix $\mathsf{D}$ in Eq.~(\ref{eqn:composition-diffusion-matrix}). For $\gamma_1=\gamma_2=\gamma$ and $\zeta=1$, the slower eigenvalue is $D=2\gamma/(1-\bar{\Phi}_0)$. The membrane-area transfer rate is estimated as $|\bMk|\sim\nu_k r^2/(R^2\bar{\Phi}_k)$, where $\nu_k$ is the frequency of vesicle events associated with a cisterna. Taking $\nu_k\sim1/T_a$ and $\bar{\Phi}_k\sim1/5$ gives the estimated range of $|\bMk|$, from which $A=R^2|\bMone|/\gamma$ follows. We take $\zeta$ to be of order unity,
    allowing both weaker and stronger fusion relative to fission.} The estimates of $\bPk$ and $\bQk$ are obtained
    by considering \cc{a characteristic mechanical energy scale $E_a\sim5$--$50\,k_BT$ acting over a normal distance of order the vesicle radius $r\approx30$--$40$\,nm, giving $|\bPk|\sim E_a/r^2$ and $|\bQk|\sim E_a/r$. The dimensionless second moment and bending rigidity follow from $|Q_k|=|\bQk|/(\gamma\eta)$ and
    $\mathcal{K}=\kappa/(R\gamma\eta)$, respectively. The energy scale $E_a$ represents the combined energetic cost of elastic membrane deformation and hydrodynamic dissipation associated with displacement of the ambient fluid.}}
\end{table}

    We denote the eigenvalues of $\mathsf{S}_\ell$ by $\lambda_\ell$, which satisfy $\det(-\lambda_\ell\hspace{0.5pt}\mathsf{I}+\mathsf{S}_\ell)=0$, where $\mathsf{I}$ is the identity matrix. \cc{For $\ell=1$, one eigenvalue is identically zero, corresponding to the translational mode associated with the first spherical harmonics. The remaining two eigenvalues are determined by the compositional dynamics and are independent of the first-row elements of $\mathsf{S}_1$.}

    Although we could obtain {the eigenvalues} in exact form, it is more instructive to work with {a reduced set of} dimensionless parameters. We choose the surface diffusivities $\gamma_k$ to be the same for both the fissogens and fusogens, namely $\gamma_k=\gamma$. We also {nondimensionalize the membrane-area transfer rates} $\bMk$ using the characteristic diffusion time $R^2/\gamma$:
    \begin{equation}
    	\label{eqn:zeta-m-definitions}
    	{\mathbb{M}^{\hspace{0.5pt}1}_{\CircleArrow}}= -\frac{\gamma A}{R^2},\quad\mathrm{and}\quad{\mathbb{M}^{\hspace{0.5pt}2}_{\CircleArrow}}= \zeta\,\frac{\gamma A}{R^2},
    \end{equation}
    where $A$ is a {dimensionless fission rate}, while $\zeta$ is the ratio of the magnitude of the {fusion and fission membrane-area transfer rates}; namely, $\zeta=-{\mathbb{M}^{\hspace{0.5pt}2}_{\CircleArrow}}/{\mathbb{M}^{\hspace{0.5pt}1}_{\CircleArrow}}$. Estimates of $A$ and $\zeta$ are shown in Table II. 
    
    \cc{To estimate the membrane-area transfer rate $\bMk$, we consider a cisterna with an effective area $4\pi R^2$ and nearly uniform composition. From Eq.~(\ref{eqn:total-mass-flux-rate}), the corresponding area-transfer rate for cycle $k$ is $4\pi R^2\bar{\Phi}_k|\bMk|$. Since each fission or fusion event transfers a membrane area $4\pi r^2$, events occurring at a frequency $\nu_k$ produce an area-transfer rate $4\pi r^2\nu_k$. Equating these rates gives $|\bMk|\sim\nu_k r^2/(R^2\bar{\Phi}_k)$. For the fission cycle, we take $\nu_1\sim1/T_a$, assuming that the local event frequency is comparable to the Golgi-wide estimate in Table I, although turnover may be faster or slower at individual cisternae. Combining this estimate with the definition of $A$ in Eq.~(\ref{eqn:zeta-m-definitions}), we obtain $A\sim r^2/(\gamma\bar{\Phi}_1 T_a)$.}

    The {homogeneous areal fractions} $\bar{\Phi}_k$ from Eq.~(\ref{eqn:steady-state-value-k}) can be rewritten as
    \begin{equation}
    	\bar{\Phi}_1 = \zeta\,\frac{1-\bar{\Phi}_0}{1+\zeta},\quad\mathrm{and}\quad\bar{\Phi}_2 = \frac{1-\bar{\Phi}_0}{1+\zeta},
    \end{equation}
    where $\bar{\Phi}_0$ is {the homogeneous background areal fraction}. This shows that the {areal fractions} $\bar{\Phi}_k$ are independent of $A$. For simplicity, we also neglect the {curvature couplings and volume-transfer rates} by setting $C_k=0$ and $\bVk=0$. {We therefore focus on membrane-area exchange and active force transmission.} {Among the active force moments, only the second moments} $\bQk$ {appear explicitly in} the shape equation in Eq.~(\ref{eqn:shape-governing-eq}), {so we define}
    \begin{equation}
    	\label{eqn:Q-definition}
    	{\mathbb{Q}^{\hspace{0.5pt}1}_{\CircleArrow}}=\gamma\eta\,Q_1,\quad\mathrm{and}\quad
    	{\mathbb{Q}^{\hspace{0.5pt}2}_{\CircleArrow}}=\gamma\eta\,Q_2,
    \end{equation}
    rescaling them by the characteristic force $\gamma\eta$, with $Q_k$ dimensionless. \cc{The signs of the second active force moments are not fixed in general, so $Q_1$ and $Q_2$ may take either sign independently. We therefore consider two representative cases. First, we take $Q_2=-Q_1=Q$, corresponding to second moments of equal magnitude and opposite sign. For the illustrative two-point force model in Appendix~A, $Q<0$ corresponds to an extensile force dipole for fission and a contractile one for fusion, whereas $Q>0$ reverses these force dipoles. Second, we take $Q_1=Q_2=Q$, corresponding to second moments of equal magnitude and the same sign.} Estimates of the {magnitude of} $Q$ {are} shown in Table II.
	
	Using this rescaling of parameters, the nonzero eigenvalues of $\mathsf{S}_1$ are found to be {(with growth rates measured in units of $\gamma/R^2$)}:
	\begin{widetext}
        \begin{equation}
            \label{eqn:eigenvalues-1}
            \lambda^\pm_1 =
            \frac{
            -2\zeta-(1+\zeta^2)\bar{\Phi}_0
            \pm
            \sqrt{
            2\zeta^2(1-\bar{\Phi}_0)^2
            \left[2+A(\zeta-1)\bar{\Phi}_0\right]
            +(1-\zeta^2)^2\bar{\Phi}_0^2
            }
            }{
            \zeta\bar{\Phi}_0(1-\bar{\Phi}_0)
            }.
        \end{equation}
	\end{widetext} \noindent \cc{This shows that $\operatorname{Re}\lambda^{-}_1<0$ for all parameter values, whereas $\lambda^{+}_1$ can acquire a positive real part. The onset of this instability occurs when $\lambda^{+}_1$ crosses zero, giving}
	\begin{equation}
		\label{eqn:instability-condition-1}
		A\geq A^{\star}_{\ell=1}=\frac{2(1+\zeta)^2}{\zeta(\zeta-1)(1-\bar{\Phi}_0)^2},\;\;\mathrm{with}\;\;\zeta>1,
	\end{equation} where $A^{\star}_{\ell=1}(\zeta)$ defines the instability boundary at which the composition fields become spontaneously unstable. By using Eq.~(\ref{eqn:shape-eq-1}), we can see that this also leads to an instability in the shape. For the first spherical harmonics, the instability of these modes ($\ell=1$)  corresponds to a spontaneous drift of the membrane compartment.

    Herein, the drift velocity $\!\itbf{V}_{\!\!\mathrm{cm}}$ of the center of mass of the {organelle} is defined by
    \begin{equation}
    	\!\itbf{V}_{\!\!\mathrm{cm}} = \frac{\mathrm{d}}{\mathrm{d}t}\itbf{R}_\mathrm{cm},
    \end{equation}
    where $\itbf{R}_\mathrm{cm}$ is its center-of-mass position, {given by}
    \begin{equation}
    	\itbf{R}_\mathrm{cm} = \frac{1}{V}\iiint\mathrm{d}V\,\itbf{r},
    \end{equation}
    {where the integration is over the enclosed volume $V$, and $\itbf{r}$ is the position vector.} We can express the center-of-mass position $\itbf{R}_\mathrm{cm}$, as well as $\!\itbf{V}_{\!\!\mathrm{cm}}$, {as power series in} the perturbation parameter $\varepsilon$; that is,
    $\itbf{R}_\mathrm{cm} = \sum_{\mathtt{i}=\hspace{0.65pt}0}^{\infty}\varepsilon^{\mathtt{i}}\itbf{R}_\mathrm{cm}^{(\mathtt{i})}\!$
    and
    $\!\itbf{V}_{\!\!\mathrm{cm}}= \sum_{\mathtt{i}=\hspace{0.65pt}0}^{\infty}\varepsilon^{\mathtt{i}} \itbf{V}_{\!\!\mathrm{cm}}^{(\mathtt{i})}\!$. At zeroth order in $\varepsilon$, we find
    	\begin{equation}
    		\itbf{R}_\mathrm{cm}^{(\mathtt{0})} = \boldsymbol{0},\quad\mathrm{and}\quad\!\itbf{V}_{\!\!\mathrm{cm}}^{(\mathtt{0})}= \boldsymbol{0}.
    	\end{equation} At first order in $\varepsilon$, the center-of-mass position is given by $\itbf{R}_\mathrm{cm}^{(\mathtt{1})} = \frac{3R}{4\pi}\iint\mathrm{d}\varphi\,\mathrm{d}\theta\sin\theta\,\rv\,u(\theta,\varphi)$, and therefore the center-of-mass velocity is found to be
    	\begin{equation}
    		\label{eqn:com-velocity}
    		\itbf{V}_{\!\!\mathrm{cm}}^{(\mathtt{1})} = \frac{3R}{4\pi}\iint\mathrm{d}\varphi\,\mathrm{d}\theta\sin\theta\,\rv\,\frac{\partial}{\partial t}u(\theta,\varphi).
    	\end{equation} {Expanding} $u(\theta,\varphi)$ in spherical harmonics, {we note that the components of the radial unit vector $\rv$ are themselves combinations of $\ell=1$ spherical harmonics. By orthogonality,} only {the $\ell=1$ component $u_1$} contributes to the integral in Eq.~(\ref{eqn:com-velocity}). {For the $m=0$ component, we have}
    $\iint\mathrm{d}\varphi\,\mathrm{d}\theta\sin\theta\,\rv\,Y_{1,0}(\theta,\varphi)
    =\sqrt{{4\pi}/{3}}\,\boldsymbol{e}_z$.

        \begin{figure*}[t]
        \includegraphics[width=\textwidth]{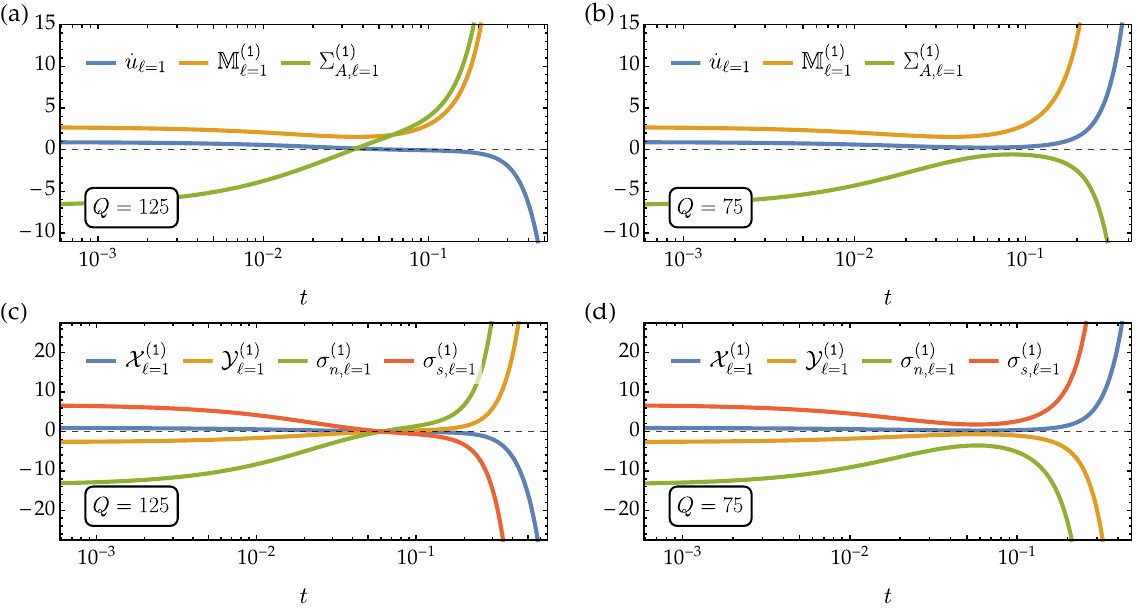}
        \caption{\label{fig:S9}
        \cc{Temporal evolution of the spherical-harmonic amplitudes for mode $\ell=1$. Panels (a) and (b) show the amplitude of the membrane-area flux $\mathbb{M}^{(\mathtt{1})}_{\ell=1}$ and the amplitude of the shape rate $\dot{u}_{\ell=1}$, together with the amplitude of the activity-renormalized surface tension $\Sigma^{(\mathtt{1})}_{A,\ell=1}$. Panels (c) and (d) show the amplitudes $\mathcal{X}^{(\mathtt{1})}_{\ell=1}$ and $\mathcal{Y}^{(\mathtt{1})}_{\ell=1}$ of the surface-harmonic modes entering Lamb's solution, rescaled by $R/\gamma$, together with the amplitudes of the normal and tangential tractions $\sigma^{(\mathtt{1})}_{\!n,\,\ell=1}$ and $\sigma^{(\mathtt{1})}_{\!s,\,\ell=1}$, rescaled by $R^2/(\eta\gamma)$. The membrane-area flux and shape rate are rescaled by $R^2/\gamma$, while the activity-renormalized surface tension is rescaled by $R/(\eta\gamma)$. Time is in units of $R^2/\gamma$. Here, $\zeta= 2.5$, $A= 180$, $\bar{\Phi}_0=3/5$, and $\mathcal{K}=10$, with $u_{\ell=1}(0)=0$ and $\Psi_{k,\ell=1}(0)=0.01$ for both $k$. With $Q_2=-Q_1=Q$, we use $Q=125$ in panels (a) and (c) for anterograde motion, directed away from the region of net membrane addition, and $Q=75$ in panels (b) and (d) for retrograde motion, directed towards it.}
        }
    \end{figure*}
    
    Thus, for the $m=0$ {component of the $\ell=1$ mode, the} center-of-mass velocity can be written as
    \begin{equation}
    	\label{eqn:com-velocity-exp}
    	\itbf{V}_{\!\!\mathrm{cm}}^{(\mathtt{1})}
    	=
    	\boldsymbol{e}_z\hspace{1pt}R\,\frac{\partial u_{1,0}}{\partial t}
    	\sqrt{\frac{3}{4\pi}}.
    \end{equation}
    \cc{Here, $m=0$ corresponds to translation along the $z$-axis, while combinations of the $m=\pm1$ components correspond to translations along the $x$- and $y$-directions.}

	As a result, Eq.~(\ref{eqn:com-velocity-exp}) {and its generalizations show} that $\!\itbf{V}_{\!\!\mathrm{cm}}^{(\mathtt{1})}$ depends linearly on the rate of change of $u_1$, which in turn {is coupled through} Eq.~(\ref{eqn:shape-eq-1}) to the compositional fields $\boldsymbol{\Psi}_1$. {These fields become} linearly unstable if the condition in Eq.~(\ref{eqn:instability-condition-1}) is satisfied, resulting in {spontaneous drift of the organelle}.
	
	The condition for linear stability is given by $A<A^{\star}_{\ell=1}$ if $\zeta\!>\!1$, and for every $A\!>\!0$ if $\zeta\in(0,1)$. In the latter domain, we also find a region in which the solutions show damped oscillations. The boundary of this regime is found by setting the {radicand} in Eq.~(\ref{eqn:eigenvalues-1}) to zero; within the oscillatory regime it is negative, giving a complex-conjugate pair of eigenvalues with negative real part. With $\zeta\in(0,1)$, we find
    \begin{equation}
    	A\geq A^{\textsl{o}}_{\ell=1}
    	=
    	\frac{
    	4\zeta^2\left(1-\bar{\Phi}_0\right)^2
    	+\left(1-\zeta^2\right)^2\bar{\Phi}_0^2
    	}{
    	2\bar{\Phi}_0\left(1-\bar{\Phi}_0\right)^2
    	\left(1-\zeta\right)\zeta^2
    	}.
    \end{equation} 
    
	For higher {spherical} harmonics ($\ell>1$), we also choose to neglect the curvature couplings to the compositional fields by setting $\Omega_k=0$. Thus, the dynamics of $\Psi_{k,\ell}$ is decoupled from the shape, \cc{and the two eigenvalues associated with the compositional fields, denoted by $\lambda_\ell^{\pm}$, are independent of the shape dynamics. We find that $\operatorname{Re}\lambda^{-}_\ell<0$ for all $\ell>1$, {while} $\operatorname{Re}\lambda^{+}_\ell>0$ gives the instability condition for the higher spherical-harmonic modes:}
    \begin{equation}
    	\label{eqn:instability-condition-other}
    	\boxed{A\geq A^{\star}_{\ell}
    	=\frac{\ell(\ell+1)(1+\zeta)^2}
    	{\zeta(\zeta-1)(1-\bar{\Phi}_0)^2},
    	\;\;\mathrm{with}\;\;\zeta>1,}
    \end{equation}
    where $A^{\star}_{\ell}(\zeta)$ marks the onset of the $\ell$-th compositional instability. As for the $\ell=1$ mode, the boundary of the stable damped-oscillation regime is obtained by setting the eigenvalue discriminant to zero, giving
    \begin{equation}
    	A\geq A^{\textsl{o}}_{\ell}
    	=
    	\frac{
    	\ell(\ell+1)
    	\!\left[
    	4\zeta^2(1-\bar{\Phi}_0)^2
    	+(1-\zeta^2)^2\bar{\Phi}_0^2
    	\right]
    	}{
    	4\bar{\Phi}_0(1-\bar{\Phi}_0)^2
    	(1-\zeta)\zeta^2
    	},
    \end{equation} with $\zeta\in(0,1)$. \cc{Note that this instability in the composition also drives in turn unstable growth of the corresponding shape modes through Eq.~(\ref{eqn:shape-governing-eq}).}
    
    \begin{figure*}[t!]
        \includegraphics[width=\textwidth]{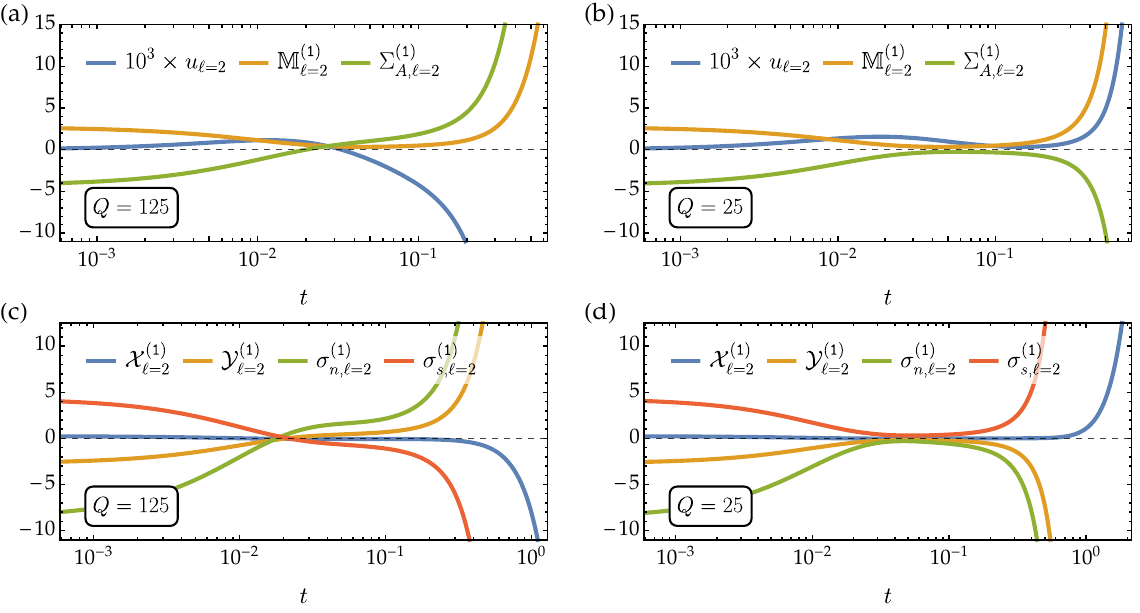}
        \caption{\label{fig:S13}
        \cc{Temporal evolution of the spherical-harmonic amplitudes for mode $\ell=2$. Panels (a) and (b) show the amplitude of the membrane-area flux $\mathbb{M}^{(\mathtt{1})}_{\ell=2}$ \cc{and the amplitude of the shape deformation} $u_{\ell=2}$, together with the amplitude of the activity-renormalized surface tension $\Sigma^{(\mathtt{1})}_{A,\ell=2}$. Panels (c) and (d) show the amplitudes $\mathcal{X}^{(\mathtt{1})}_{\ell=2}$ and $\mathcal{Y}^{(\mathtt{1})}_{\ell=2}$, together with the normal and tangential traction amplitudes $\sigma^{(\mathtt{1})}_{\!n,\,\ell=2}$ and $\sigma^{(\mathtt{1})}_{\!s,\,\ell=2}$. The membrane-area flux is rescaled by $R^2/\gamma$, the shape amplitude $u_{\ell=2}$ is scaled by $10^3$ for clarity, the activity-renormalized surface tension by $R/(\eta\gamma)$, the surface harmonics are rescaled by $R/\gamma$, and the traction amplitudes by $R^2/(\eta\gamma)$. Time is in units of $R^2/\gamma$. Here, $\zeta=2.5$, $A=180$, $\bar{\Phi}_0=3/5$, and $\mathcal{K}=10$, with $u_{\ell=2}(0)=0$ and $\Psi_{k,\ell=2}(0)=0.01$ for both $k$. For these initial conditions, $\mathbb{M}^{(\mathtt{1})}_{\ell=2}$ remains positive, so the long-time deformation is prolate when $u_{\ell=2}$ and $\mathbb{M}^{(\mathtt{1})}_{\ell=2}$ have the same sign and oblate when their signs are opposite. With $Q_2=-Q_1=Q$, we use $Q=125$ in panels (a) and (c) for the oblate case and $Q=25$ in panels (b) and (d) for the prolate case.}
        }
    \end{figure*}
    
    Another instability can arise directly from the shape equation, where the corresponding eigenvalues 
    \begin{equation}
    	\lambda^u_\ell\! =\!- \frac{\ell^2(\ell\!+\!1)^2(\ell\!-\!1)(\ell\!+\!2)}{(4\ell^3+6\ell^2-1)}\!\left[\mathcal{K}\!+\!\displaystyle\frac{(Q_2+\zeta Q_1\hspace{-1pt})\left(1\!-\bar{\Phi}_0\hspace{-1pt}\right)}{2\ell(\ell+1)(1+\zeta)} \right]
    \end{equation}
    where we assume an isotonic solution, with the osmotic pressure $\Pi_{\hspace{-0.5pt}\scriptscriptstyle\Delta}=0$, and $\mathcal{K}=\kappa/(\gamma\eta R)$ is a rescaled bending rigidity, non-dimensionalized by the energy scale $\gamma\eta R$. Hence, a band of unstable modes can be obtained (for which $\lambda^{u}_\ell>0$) when the following condition holds:
    \begin{equation}
    	Q_2+\zeta Q_1<-\frac{2\ell(\ell+1)(1+\zeta)\mathcal{K}}{1-\bar{\Phi}_0}.
    \end{equation}

    \begin{figure*}[t!]
        \includegraphics[width=\textwidth]{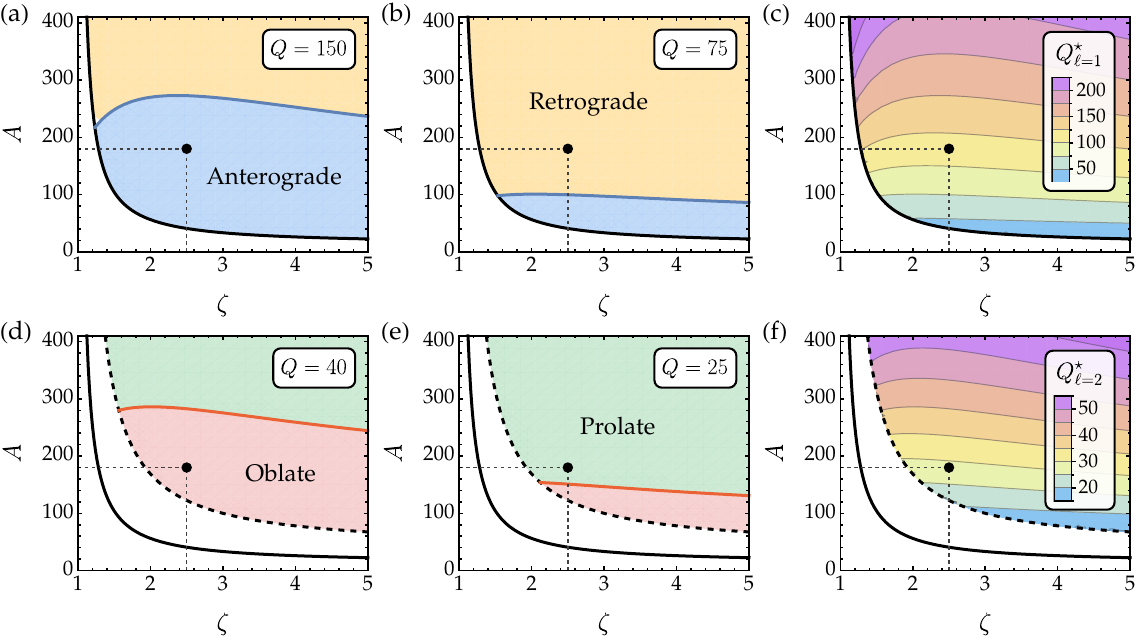}
        \caption{\label{fig:S11}
        \cc{Panels (a) and (b) show the long-time retrograde (yellow) and anterograde (blue) drift regimes in the $A$--$\zeta$ plane for fixed $Q=150$ and $Q=75$, respectively. The colored curves mark the boundary $Q=Q^\star_{\ell=1}(A,\zeta)$ separating the two regimes, whose dependence on $A$ and $\zeta$ is shown in (c). Panels (d) and (e) show the corresponding prolate (green) and oblate (red) deformation regimes for fixed $Q=40$ and $Q=25$, respectively. The colored curves mark the boundary $Q=Q^\star_{\ell=2}(A,\zeta)$, shown as a function of $A$ and $\zeta$ in (f). These deformation regimes correspond to the asymptotic linear response to the chosen initial perturbations, which produce a positive quadrupolar membrane-area flux amplitude. The solid black curve marks the $\ell=1$ instability threshold, while the dashed black curve in panels (d)--(f) marks the $\ell=2$ instability threshold. Black points indicate $(\zeta,A)=(2.5,180)$, as used in Figs.~\ref{fig:S9} and \ref{fig:S13}. Here, $Q_2=-Q_1=Q$, $\bar{\Phi}_0=3/5$, and $\mathcal{K}=10$.}
        }
    \end{figure*}
    
    If we consider \cc{the equal-magnitude, opposite-sign case $Q_1=-Q$ and $Q_2=Q$}, \cc{where the signed parameter $Q$ may take any real value, then $Q<0$ gives the second-moment signs of the illustrative two-point force model discussed in Appendix~A.7, for which the fission second moment is positive and the fusion second moment is negative, while $Q>0$ corresponds to the opposite orientation of the two second moments.} The condition for shape instability can be further simplified as follows:
    \begin{equation}
    	\left\{\!\begin{array}{cc}
    	\displaystyle Q<-\frac{2\ell(\ell+1)(1+\zeta)\mathcal{K}}{(1-\zeta)(1-\bar{\Phi}_0)}, & \mathrm{if}\;\;\zeta<1, \\[14pt]
    	\displaystyle Q>\frac{2\ell(\ell+1)(1+\zeta)\mathcal{K}}{(\zeta-1)(1-\bar{\Phi}_0)}, & \mathrm{if}\;\;\zeta>1.
    	\end{array}\right.
    \end{equation}
    {The corresponding} phase regions are plotted in Fig.~\ref{fig:fig2}(c). If we consider the other case, with equal magnitudes and {the same sign} ($Q_1=Q_2=Q$), we find that
    \begin{equation}
    	Q<-\frac{2\ell(\ell+1)\mathcal{K}}{1-\bar{\Phi}_0}
    \end{equation}
    corresponds to the unstable region, which is independent of the parameter $\zeta$. However, in what follows we consider only \cc{the case $Q_1=-Q$ and $Q_2=Q$}.

	\subsection{Solutions to membrane shape and composition}

    \cc{The governing equations at first order, Eq.~(\ref{eqn:linear-set-eqs}), can be solved exactly for prescribed initial conditions to obtain the time evolution of the shape and composition amplitudes $u_\ell(t)$ and $\Psi_{k,\ell}(t)$. In the unstable regime, the component along an unstable eigenmode grows exponentially. These solutions also determine the surface tension from Eq.~(\ref{eqn:surface-tension-eq}) and the activity-renormalized surface tension from Eq.~(\ref{eqn:renormalized-surface-tension-amplitude}). We first consider the $\ell=1$ mode, governed by Eqs.~(\ref{eqn:shape-eq-1}) and (\ref{eqn:composition-eq-1}), with $u_{\ell=1}(0)=0$ and an initial perturbation in the composition fields. The quantities of interest are the shape rate $\dot{u}_{\ell=1}$, which determines the center-of-mass velocity through Eq.~(\ref{eqn:com-velocity-exp}), the membrane-area flux amplitude $\mathbb{M}^{(\mathtt{1})}_{\ell=1}$, and the activity-renormalized surface-tension amplitude $\Sigma^{(\mathtt{1})}_{A,\ell=1}$, as shown in Fig.~\ref{fig:S9}. We choose $A$ and $\zeta$ in the unstable regime defined by Eq.~(\ref{eqn:instability-condition-1}).}
    \\
    \indent
    \cc{Since this instability condition is independent of the second-moment parameter $Q$, the membrane-area flux grows exponentially at a rate independent of $Q$. The translational response, however, depends on $Q$, and the direction of drift follows from the sign of $\dot{u}_{\ell=1}$ relative to the membrane-area flux amplitude. We call the motion retrograde when $\dot{u}_{\ell=1}$ and $\mathbb{M}^{(\mathtt{1})}_{\ell=1}$ have the same sign, corresponding to motion towards the region of net membrane addition, and anterograde when they have opposite signs, corresponding to motion away from this region. Thus, varying $Q$ can reverse the direction of drift relative to the membrane-area flux.}

    \begin{widetext}
    \cc{The corresponding surface harmonics $\mathcal{X}^{(\mathtt{1})}_{\ell=1}$ and $\mathcal{Y}^{(\mathtt{1})}_{\ell=1}$ determine the surrounding Stokesian flows and are shown together with the hydrodynamic traction amplitudes in Fig.~\ref{fig:S9}(c) and (d). The tangential and normal traction amplitudes, obtained from Eqs.~(\ref{eqn:tangential-part-body-force-1}) and (\ref{eqn:normal-body-foce-comps}), are given by}
    \begin{equation}
        \label{eqn:membrane-tractions}
        \begin{aligned}
            \sigma^{(\mathtt{1})}_{\!s,\,\ell}
            &= -\frac{\eta(2\ell+1)}{\ell(\ell+1)R}
            \left[\mathcal{X}^{(\mathtt{1})}_\ell+2\mathcal{Y}^{(\mathtt{1})}_\ell\right],
            \qquad
            \sigma^{(\mathtt{1})}_{\!n,\,\ell}
            &= \frac{\eta(2\ell+1)}{\ell(\ell+1)R}
            \left[3\mathcal{Y}^{(\mathtt{1})}_\ell
            -(2\ell^2+2\ell-3)\mathcal{X}^{(\mathtt{1})}_\ell\right].
        \end{aligned}
    \end{equation}
    
    \cc{The long-time anterograde and retrograde drift regimes are shown in Figs.~\ref{fig:S11}(a) and (b) as $\zeta$ and $A$ vary within the unstable domain defined by Eq.~(\ref{eqn:instability-condition-1}). The colored curves mark the boundary $Q=Q^\star_{\ell=1}(A,\zeta)$, separating retrograde motion for $Q<Q^\star_{\ell=1}$ from anterograde motion for $Q>Q^\star_{\ell=1}$. This boundary is given by}
        \begin{equation}
            Q^\star_{\ell=1}
            =
            \frac{2A(\zeta+1)}
            {2+\bar{\Phi}_0\left(\zeta+\zeta^{-1}\right)
            +
            \sqrt{
            2A\bar{\Phi}_0(1-\bar{\Phi}_0)^2(\zeta-1)
            +\bar{\Phi}_0^2\left(\zeta+\zeta^{-1}\right)^2
            -8\bar{\Phi}_0+4
            }}.
        \end{equation} 
    \end{widetext}
    \cc{In Fig.~\ref{fig:S11}(c), we show the dependence of $Q^\star_{\ell=1}$ on $\zeta$ and $A$ in the unstable region. In the large-$\zeta$ limit, $Q^\star_{\ell=1}$ approaches $A/\bar{\Phi}_0>0$, so anterograde drift requires $Q>0$ in this limit. The boundary is determined by requiring the leading contribution of the unstable linear mode to the center-of-mass velocity to vanish. Nonlinear mode coupling, however, modifies the finite-amplitude composition and can shift the anterograde--retrograde boundary relative to the linear prediction; see Appendix~C.}
    \\ 
    \indent
    \cc{We next consider the $\ell=2$ mode, with an initially undeformed shape, $u_{\ell=2}(0)=0$, and prescribed perturbations of the composition fields $\Psi_{k,\ell=2}$. The resulting evolution of $u_{\ell=2}$ is shown in Fig.~\ref{fig:S13} for two values of $Q$ with distinct long-time responses. For these initial perturbations, the quadrupolar membrane-area flux amplitude remains positive, so a prolate deformation corresponds to $u_2>0$, whereas an oblate deformation has $u_2<0$. Reversing the initial quadrupolar composition perturbations reverses both the flux amplitude and $u_2$, interchanging the prolate and oblate shapes while preserving their relative sign.}
    \\ 
    \indent
    \cc{These solutions also determine the amplitude of the activity-renormalized surface-tension $\Sigma^{(\mathtt{1})}_{A,\ell=2}$ and the surface harmonics $\mathcal{X}^{(\mathtt{1})}_{\ell=2}$ and $\mathcal{Y}^{(\mathtt{1})}_{\ell=2}$, from which the tangential and normal traction amplitudes $\sigma^{(\mathtt{1})}_{\!s,\,\ell=2}$ and $\sigma^{(\mathtt{1})}_{\!n,\,\ell=2}$ are obtained, as shown in Fig.~\ref{fig:S13}. The long-time prolate and oblate deformation regimes are shown in Figs.~\ref{fig:S11}(d) and (e) as $\zeta$ and $A$ vary within the $\ell=2$ unstable domain. The colored curves mark the boundary $Q=Q^\star_{\ell=2}(A,\zeta)$ separating the two regimes, whose dependence on $\zeta$ and $A$ is shown in panel (f). For the chosen initial perturbations, the oblate response occurs only for $Q>0$, consistent with $Q^\star_{\ell=2}(A,\zeta)>0$.}

	\subsection{Solutions to the surrounding fluid flows}
    
\begin{figure*}[t!]
        \includegraphics[width=\textwidth]{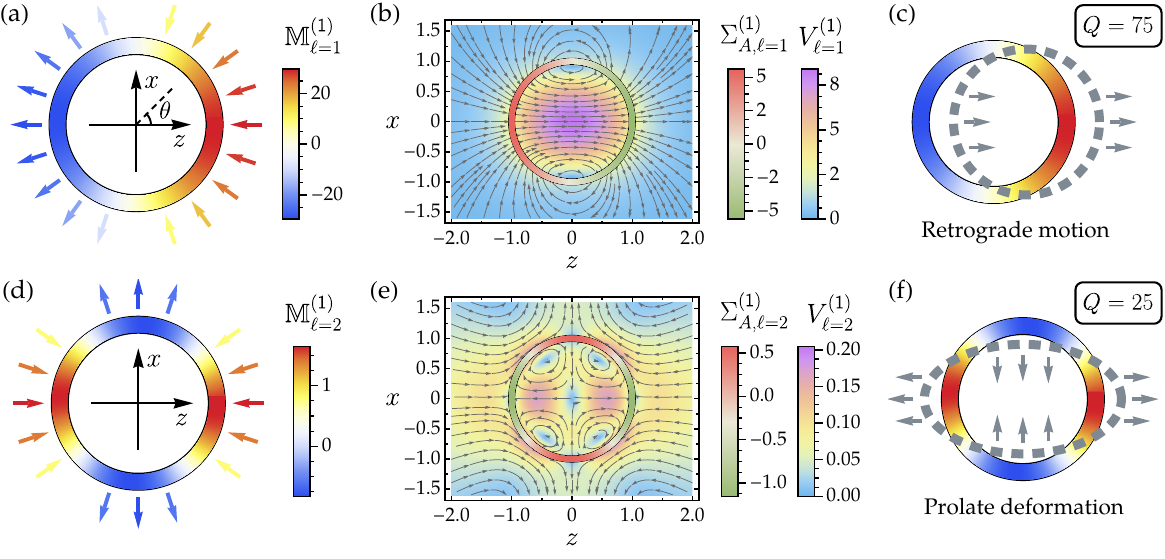}
        \caption{\label{fig:S15}
        \cc{Illustration of the retrograde motion and prolate deformation regimes for $\zeta=2.5$, $A=180$, $\bar{\Phi}_0=3/5$, and $\mathcal{K}=10$, at $t=0.3\,R^2/\gamma$. We take $Q_2=-Q_1=Q$, with $Q=75$ in panels (a)--(c) and $Q=25$ in panels (d)--(f), corresponding to the black points in Fig.~\ref{fig:S11}(b) and (e), respectively. The fields are axisymmetric ($m=0$) and shown in the $x$-$z$ plane through the sphere center, with $Y_{\ell,m}$ denoting the spherical harmonic of degree $\ell$ and order $m$. Panels (a) and (d) show the membrane-area flux profiles} 
        \cc{$\mathbb{M}^{(\mathtt{1})}_{\ell}(\theta)
        =\sum_k\bMk\Psi_{k,\ell,0}Y_{\ell,0}(\theta)$,
        for $\ell=1$ and $2$, respectively,}
        \cc{with arrows indicating membrane addition or removal. Panels (b) and (e) show the corresponding inner and outer fluid flows. In each case, the density plot and its color bar show the velocity magnitude $V^{(\mathtt{1})}_{\ell}(r,\theta)$, the gray streamlines show the flow direction, and the interfacial color and its color bar show the activity-renormalized surface tension} \cc{$\Sigma^{(\mathtt{1})}_{A,\ell}(\theta)
        =\Sigma^{(\mathtt{1})}_{A,\ell,0}Y_{\ell,0}(\theta)$}\cc{. Panels (c) and (f) show the corresponding outcomes: retrograde motion towards the region of net membrane addition for $\ell=1$, and a prolate deformation along the $z$-axis for $\ell=2$. The membrane-area flux is rescaled by $R^2/\gamma$, the fluid velocity by $R/\gamma$, and the activity-renormalized surface tension by $R/(\eta\gamma)$.}
        }
    \end{figure*}

    \cc{The surface harmonics $\mathcal{X}^{(\mathtt{1})}_{\ell}$ and $\mathcal{Y}^{(\mathtt{1})}_{\ell}$, determined by $u_\ell$ and $\Psi_{k,\ell}$, give the surrounding fluid flows through Lamb's solution in Eq.~(\ref{eqn:Lamb-solution}). To obtain the explicit spatial flows, we restore the azimuthal index $m$ and use the individual amplitudes $\mathcal{X}^{(\mathtt{1})}_{\ell,m}$ and $\mathcal{Y}^{(\mathtt{1})}_{\ell,m}$. The inner-fluid velocity for each mode $\ell>0$, with $|m|\leq\ell$, is given by}
    \begin{align}
        \itbf{V}^{(\mathtt{1})}_{\!\!-,\ell,m} &= \frac{\rv Y_{\ell,m}\hspace{1pt}r^{\ell-1}}{2R^{\ell-1}}\left[\mathcal{X}^{(\mathtt{1})}_{\ell,m}\!\left(1+\ell-(\ell-1)\frac{r^2}{R^2}\right)\right.\notag\\[5pt]
        &\hspace{-20pt}-\left.\!\mathcal{Y}^{(\mathtt{1})}_{\ell,m}\!\left(\!1\!-\!\frac{r^2}{R^2}\right)\hspace{-1pt}\right]-\frac{\boldsymbol{\Theta}_{\ell,m}\,r^{\ell-1}}{2\ell R^{\ell-1}}\left[\mathcal{Y}^{(\mathtt{1})}_{\ell,m}\!\left(\!1\!-\!\frac{(\ell\!+\!3)\hspace{1pt}r^2}{(\ell\!+\!1)\hspace{1pt}R^2}\right)\right.\notag\\[5pt]
        &\hspace{-20pt}\left.-\mathcal{X}^{(\mathtt{1})}_{\ell,m}\left(1\!+\!\ell-\frac{(\ell-1)(\ell+3)\hspace{1pt}r^2}{(\ell+1)\hspace{1pt}R^2}\right)\right]\!,
    \end{align}
    \cc{where $\boldsymbol{\Theta}_{\ell,m}=r\boldsymbol{\nabla}Y_{\ell,m}(\theta,\varphi)$ is the tangential vector spherical harmonic. The inner-fluid velocity follows by substituting $p^{(\mathtt{1})}_\ell$ and $\Upsilon^{(\mathtt{1})}_\ell$ from Eqs.~(\ref{eqn:p-inner-1}) and (\ref{eqn:upsilon-inner-1}), respectively, into Lamb's solution, Eq.~(\ref{eqn:Lamb-solution}). Similarly, the outer-fluid velocity is obtained using $p^{(\mathtt{1})}_{-(\ell+1)}$ and $\Upsilon^{(\mathtt{1})}_{-(\ell+1)}$ from Eqs.~(\ref{eqn:p-outer-1}) and (\ref{eqn:upsilon-outer-1}), respectively, giving}
    \begin{align}
        \label{eqn:outside-fluid-flows}
        \itbf{V}^{(\mathtt{1})}_{\!\!+,\ell,m} &= \frac{\rv Y_{\ell,m}R^{\ell+2}}{2\hspace{1pt}r^{\ell+2}}\left[\mathcal{X}^{(\mathtt{1})}_{\ell,m}\!\left(\!-\ell+(\ell+2)\frac{r^2}{R^2}\right)\right.\notag\\[5pt]
        &\hspace{-20pt}-\left.\!\mathcal{Y}^{(\mathtt{1})}_{\ell,m}\!\left(\!1\!-\!\frac{r^2}{R^2}\right)\hspace{-1pt}\right]+\frac{\boldsymbol{\Theta}_{\ell,m}\hspace{1pt}R^{\ell+2}}{2(\ell+1)\hspace{1pt}r^{\ell+2}}\left[\mathcal{Y}^{(\mathtt{1})}_{\ell,m}\!\left(\!1\!-\!\frac{(\ell\!-\!2)\hspace{1pt}r^2}{\ell\hspace{1pt}R^2}\right)\right.\notag\\[5pt]
        &\hspace{-20pt}\left.+\mathcal{X}^{(\mathtt{1})}_{\ell,m}\left(\ell-\frac{(\ell+2)(\ell-2)\hspace{1pt}r^2}{\ell\hspace{1pt}R^2}\right)\right]\!.
    \end{align}
    The fluid velocity $\itbf{v}_\mathfrak{m}^{(\mathtt{1})}$ at the membrane surface ($r=R$) for each surface harmonic mode $\ell$ and $m$ is found to be
    \begin{equation}
        \label{eqn:velocity-on-sphere-1st}
        {\itbf{v}^{(\mathtt{1})}_{\mathfrak{m},\ell,m} = \mathcal{X}^{(\mathtt{1})}_{\ell,m}\,\rv\hspace{1pt}Y_{\ell,m} + \frac{2\mathcal{X}^{(\mathtt{1})}_{\ell,m}+\mathcal{Y}^{(\mathtt{1})}_{\ell,m}}{\ell(\ell+1)}\,\boldsymbol{\Theta}_{\ell,m}.}
    \end{equation}
    \cc{The axisymmetric ($m=0$) inner and outer fluid flows are illustrated in Fig.~\ref{fig:S15}(b) for the $\ell=1$ mode and in Fig.~\ref{fig:S15}(e) for the $\ell=2$ mode, corresponding to the retrograde drift and prolate deformation regimes, respectively. The complementary anterograde $\ell=1$ and oblate $\ell=2$ responses are shown in Fig.~\ref{fig:fig3}.}

    \cc{From Eq.~(\ref{eqn:outside-fluid-flows}), the leading far-field terms of the complete outer fluid velocity are}
    \begin{align}
        \itbf{V}^{(\mathtt{1})}_{\!\!+} =& \sum_{|m|\leq1}\frac{3\mathcal{X}^{(\mathtt{1})}_{1,m}+\mathcal{Y}^{(\mathtt{1})}_{1,m}}{4(r/R)}\,(2\rv\hspace{1pt}Y_{1,m}+\boldsymbol{\Theta}_{1,m})\notag\\[5pt]
        &+\sum_{|m|\leq2}\frac{4\mathcal{X}^{(\mathtt{1})}_{2,m}+\mathcal{Y}^{(\mathtt{1})}_{2,m}}{2(r/R)^2}\,\rv\hspace{1pt}Y_{2,m}+\mathcal{O}[r^{-3}].
    \end{align}
    \cc{Using Eqs.~(\ref{eqn:X-ell-1}) and (\ref{eqn:Y-ell-1}) for $\ell=1$, together with the membrane shape equation in Eq.~(\ref{eqn:shape-eq-1}), we find that the $1/r$ term vanishes identically, since}
    \begin{equation}
        3\mathcal{X}^{(\mathtt{1})}_{1,m}+\mathcal{Y}^{(\mathtt{1})}_{1,m}=0,\qquad |m|\leq1.
    \end{equation}
    \cc{Thus, the far-field flow of the outer fluid becomes}
    \begin{equation}
        \itbf{V}^{(\mathtt{1})}_{\!\!+} = \sum_{|m|\leq2}\frac{4\mathcal{X}^{(\mathtt{1})}_{2,m}+\mathcal{Y}^{(\mathtt{1})}_{2,m}}{2(r/R)^2}\,\rv\hspace{1pt}Y_{2,m}+\mathcal{O}[r^{-3}].
    \end{equation}
    \cc{The leading $\ell=1$ contribution is contained in the $\mathcal{O}(r^{-3})$ term. Retaining it explicitly gives}
    \begin{equation}
        \itbf{V}^{(\mathtt{1})}_{\!\!+,\,\ell=1}
        =
        \sum_{|m|\leq1}
        \frac{\mathcal{X}^{(\mathtt{1})}_{1,m}+\mathcal{Y}^{(\mathtt{1})}_{1,m}}
        {4(r/R)^3}
        (\boldsymbol{\Theta}_{1,m}-2\rv\hspace{1pt}Y_{1,m}).
    \end{equation}
    \cc{Hence, the force-free $\ell=1$ flow decays as $1/r^3$, whereas the leading $\ell=2$ flow decays as $1/r^2$. The amplitude of this leading $\ell=2$ far-field flow is controlled by $4\mathcal{X}^{(\mathtt{1})}_{2}+\mathcal{Y}^{(\mathtt{1})}_{2}$ and can change sign as the balance between membrane-area exchange, the second active force moment, and the shape response is varied.}
    
    \cc{The net hydrodynamic force on the organelle can also be obtained directly by integrating the traction over the membrane surface. The normal contribution is}
    \begin{equation}
        \itbf{F}_{\! n} = \iint\mathrm{d}\varphi\,\mathrm{d}\theta\,R^2\sin\theta\,\sum_{\ell,m}\sigma^{(\mathtt{1})}_{\! n,\,\ell,m}\,\rv\hspace{1pt}Y_{\ell,m}(\theta,\varphi).
    \end{equation}
    \cc{Applying Eq.~(\ref{eqn:membrane-tractions}) to each azimuthal component, we find that only the $\ell=1$ modes contribute to the integral. For the axisymmetric $m=0$ component,}
    \begin{equation}
        \label{eqn:normal-drag-force-n}
        \itbf{F}_{\! n}\! = 10\sqrt{\frac{\pi}{3}}\boldsymbol{e}_z\left(\mathbb{Q}^{(\mathtt{1})}_{1,0}-\eta R^2\mathbb{M}^{(\mathtt{1})}_{1,0}\right)+\cdots
    \end{equation}
    \cc{where $\mathbb{M}^{(\mathtt{1})}_{1,0}=\sum_k\bMk\Psi_{k,1,0}$ and $\mathbb{Q}^{(\mathtt{1})}_{1,0}=\sum_k\bQk\Psi_{k,1,0}$. The remaining terms correspond to $m=\pm1$ and give the equivalent force components along the other spatial directions. Similarly, the tangential contribution to the net hydrodynamic force is}
    \begin{equation}
        \itbf{F}_{\! s} = \iint\mathrm{d}\varphi\,\mathrm{d}\theta\,R^2\sin\theta\,\sum_{\ell,m}\sigma^{(\mathtt{1})}_{\! s,\,\ell,m}\,\boldsymbol{\Theta}_{\ell,m}(\theta,\varphi),
    \end{equation}
    which gives
    \begin{equation}
        \label{eqn:surface-drag-force-n}
        \itbf{F}_{\! s}\! = -10\sqrt{\frac{\pi}{3}}\boldsymbol{e}_z\left(\mathbb{Q}^{(\mathtt{1})}_{1,0}-\eta R^2\mathbb{M}^{(\mathtt{1})}_{1,0}\right)+\cdots
    \end{equation}
    \cc{with the $m=\pm1$ components canceling in the same way. Hence,} $\itbf{F}_{\! n}\!+\itbf{F}_{\! s}\!=\boldsymbol{0}$, \cc{showing that the organelle behaves as a force-free swimmer, moving either retrograde or anterograde relative to the membrane-area flux.}

    \section{Nonlinear Perturbative Analysis}
    
	The linear perturbation analysis indicates that the homogeneous steady state of the membrane compartment becomes unstable, with the compositional instability driving shape distortions (for modes $\ell>1$) and spontaneous drift (for $\ell=1$). \cc{We therefore ask whether nonlinearities in the membrane composition can saturate these instabilities and give rise to stable stationary or steadily drifting states.}

    \subsection{Weakly nonlinear compositional dynamics}
    
    The compositional instability brings the system into a nonlinear regime, where higher-order terms in the perturbations are no longer negligible.
    \cc{As in Appendix~B, we expand the fissogen and fusogen areal fractions and the passive background fraction about the homogeneous spherical state as $\Phi_k=\bar{\Phi}_k+\Psi_k$ and $\Phi_0=\bar{\Phi}_0-\Psi_0$, suppressing the explicit bookkeeping parameter $\varepsilon$ throughout Appendix~C. Here, the homogeneous areal fractions $\bar{\Phi}_k$ are given in Eq.~(\ref{eqn:steady-state-value-k}), while $\bar{\Phi}_0$ is the homogeneous areal fraction of the passive background membrane. The fields $\Psi_k$ denote the perturbations of the fissogen and fusogen areal fractions, while the passive background perturbation is not independent, since $\Psi_0=\Psi_1+\Psi_2$, cf.~Eq.~(\ref{eqn:mass-cons-psi-s}). The composition perturbations $\Psi_k$ and $\Psi_0$, as well as the shape perturbation $u$, are understood to be small, with order tracked by the total degree in these fields; for instance, linear, quadratic, and cubic terms contain one, two, and three powers of the perturbations, respectively.}
    
    \cc{As shown in Appendix~B, at linear order the $\ell=1$ component of the shape perturbation represents rigid translation of the membrane compartment in the laboratory frame. Beyond linear order, translating a deformed membrane mixes spherical-harmonic modes in its radial representation. To separate this motion from intrinsic shape deformations, we work in a frame comoving with the organelle. Denoting its center-of-mass position by $\itbf{X}(t)$, the membrane position in the laboratory frame is}
    \begin{equation}
        \label{eqn:C-comoving-shape}
        \itbf{R}_{\mathrm{lab}}(\theta,\varphi,t)
        =
        \itbf{X}(t)+R\big[1+u(\theta,\varphi,t)\big]\rv .
    \end{equation}
    \cc{The shape deformation field $u$ is defined relative to the moving center, with rigid translation represented by $\itbf{X}(t)$ rather than by an $\ell=1$ shape component. This comoving description removes the rigid translation of the organelle as a whole, but membrane material can still move tangentially relative to the moving center. Such tangential motion changes the angular positions of membrane constituents and therefore advects the composition fields even in the comoving frame.}
    
    \cc{We make use of the same nondimensionalization as in Appendix~B, with length scaled by $R$ and time by $R^2/\gamma$. We also take the fissogens and fusogens to have the same surface diffusivity, $\gamma_1=\gamma_2=\gamma$, and neglect the curvature--composition couplings and active volume fluxes. With $A$ and $\zeta$ defined in Eq.~(\ref{eqn:zeta-m-definitions}), the dimensionless active membrane-area flux is given by}
    \begin{equation}
        \label{eqn:C-mu-field}
        \frac{R^2}{\gamma}\mathbb{M}
        =
        A\big(\zeta\Phi_2-\Phi_1\big)
        =
        \bar{\mu}+\mu(\theta,\varphi,t).
    \end{equation}
    \cc{Here, $\bar{\mu}=A(\zeta\bar{\Phi}_2-\bar{\Phi}_1)$ is dimensionless and represents the homogeneous steady-state contribution to the active membrane-area flux, which vanishes, $\bar{\mu}=0$, upon using the areal fractions in Eq.~(\ref{eqn:steady-state-value-k}). The dimensionless contribution from the composition perturbations is}
    \begin{equation}
        \mu(\theta,\varphi,t)
        =
        A\big(\zeta\Psi_2-\Psi_1\big).
    \end{equation}
    
    \cc{We denote the dimensionless first-order tangential membrane velocity relative to the moving center by $\boldsymbol{w}^{(\mathtt{1})}$. The material derivative of the composition perturbations is then}
    \begin{equation}
        \label{eqn:C-material-derivative}
        \dot{\Psi}_k
        \simeq
        \frac{\partial\Psi_k}{\partial t}
        +\boldsymbol{w}^{(\mathtt{1})}\!\cdot
        \nabla_{\!\Omega}\Psi_k,
    \end{equation}
    \cc{where
    $\nabla_{\!\Omega}
    =\thetav\,\partial_\theta
    +\phiv\,(\sin\theta)^{-1}\partial_\varphi$
    is the surface gradient on the unit sphere, with $\thetav$ and $\phiv$ the corresponding unit tangent vectors. The partial time derivative is evaluated at fixed angular coordinates in the comoving frame.}
    
    \cc{Substituting these expansions into the governing equation for the areal fractions, Eq.~(\ref{eqn:phi-k-dynamics-expanded}), and using the material derivative in Eq.~(\ref{eqn:C-material-derivative}), we obtain}
    \begin{align}
        \label{eqn:psi-eqn-nonlinear}
        \frac{\partial\Psi_k}{\partial t}
        ={}&
        -\mu\left(\bar{\Phi}_k+\Psi_k\right)
        +\Delta\ln\!\left(
        \frac{\bar{\Phi}_k+\Psi_k}
             {\bar{\Phi}_0-\Psi_0}
        \right)\notag\\
        &-\boldsymbol{w}^{(\mathtt{1})}\!\cdot
        \nabla_{\!\Omega}\Psi_k,
    \end{align}
    \cc{where $\Delta$ is the dimensionless Laplace--Beltrami operator on the deformed membrane surface. The three terms on the right-hand side describe active membrane-area exchange, surface diffusion, and tangential advection, respectively. We retain the nonlinear surface-diffusion contributions to the composition equations through cubic order, while the separate mechanical shape equation and tangential membrane velocity are retained at linear order. For the logarithmic term, neglecting terms of fourth and higher order gives}
    \begin{equation}
        \label{eqn:C-log-series}
        \ln\!\left(
        \frac{\bar{\Phi}_k+\Psi_k}
             {\bar{\Phi}_0-\Psi_0}
        \right)
        \simeq
        \ln\!\left(\frac{\bar{\Phi}_k}{\bar{\Phi}_0}\right)
        +\mathcal{L}^{(\mathtt{1})}_k
        +\mathcal{L}^{(\mathtt{2})}_k
        +\mathcal{L}^{(\mathtt{3})}_k .
    \end{equation}
    \cc{The first term is a constant and does not contribute under the Laplace--Beltrami operator, while the remaining terms are given by}
    \begin{align}
        \label{eqn:C-log-expansion}
        \mathcal{L}^{(\mathtt{1})}_k
        &=
        \frac{\Psi_0}{\bar{\Phi}_0}
        +\frac{\Psi_k}{\bar{\Phi}_k},
        \\[4pt]
        \mathcal{L}^{(\mathtt{2})}_k
        &=
        \frac{1}{2}
        \left(
        \frac{\Psi_0^2}{\bar{\Phi}_0^2}
        -\frac{\Psi_k^2}{\bar{\Phi}_k^2}
        \right),
        \\[4pt]
        \mathcal{L}^{(\mathtt{3})}_k
        &=
        \frac{1}{3}
        \left(
        \frac{\Psi_0^3}{\bar{\Phi}_0^3}
        +\frac{\Psi_k^3}{\bar{\Phi}_k^3}
        \right).
    \end{align}
    
    \cc{Because the spatially varying part of the logarithm begins at linear order in the composition perturbations, the Laplace--Beltrami operator must be expanded through quadratic order in $u$ to retain the corresponding cubic surface-diffusion terms. For an arbitrary scalar field $f$ on the membrane, this gives}
    \begin{equation}
        \label{eqn:C-laplace-expansion}
        \Delta f
        \simeq
        \Delta_{\Omega}f
        -2u\Delta_{\Omega}f
        +\mathcal{D}_2[u]f,
    \end{equation}
    \cc{where $\Delta_{\Omega}=\nabla_{\!\Omega}\!\cdot\nabla_{\!\Omega}$ is the Laplace--Beltrami operator on the unit reference sphere. Terms of third and higher order in $u$ are neglected, and the quadratic correction is}
    \begin{align}
        \label{eqn:C-D2-operator}
        \mathcal{D}_2[u]f={}&
        \nabla_{\!\Omega} \cdot \left[
        \frac{1}{2}|\nabla_{\!\Omega}u|^2\,\nabla_{\!\Omega}f
        -\nabla_{\!\Omega}u
        \left(\nabla_{\!\Omega}u\!\cdot\nabla_{\!\Omega}f\right)
        \right]\notag\\[4pt]
        &\quad+
        \left(3u^2-\frac{1}{2}|\nabla_{\!\Omega}u|^2\right)
        \Delta_{\Omega}f.
    \end{align}
    
    \cc{Substituting the logarithmic and Laplace--Beltrami expansions into Eq.~(\ref{eqn:psi-eqn-nonlinear}) gives}
    \begin{align}
        \label{eqn:psi-dynamics-nonlinear}
        \frac{\partial\Psi_k}{\partial t}
        ={}\!&
        -\bar{\Phi}_k\mu
        +\Delta_{\Omega}\mathcal{L}^{(\mathtt{1})}_k
        \notag\\[1pt]
        &+\left[
        -\Psi_k\mu
        +\Delta_{\Omega}\mathcal{L}^{(\mathtt{2})}_k
        -2u\Delta_{\Omega}\mathcal{L}^{(\mathtt{1})}_k
        -\boldsymbol{w}^{(\mathtt{1})}\!\cdot
        \nabla_{\!\Omega}\Psi_k
        \right]
        \notag\\[6pt]
        &+\left[
        \Delta_{\Omega}\mathcal{L}^{(\mathtt{3})}_k
        -2u\Delta_{\Omega}\mathcal{L}^{(\mathtt{2})}_k
        +\mathcal{D}_2[u]\mathcal{L}^{(\mathtt{1})}_k
        \right].
    \end{align}
    \cc{The three lines on the right-hand side contain the retained linear, quadratic, and cubic contributions to the composition dynamics, respectively. Since $\mu$ is linear in the composition perturbations, active membrane-area exchange contributes only at linear and quadratic order. The cubic terms retained here arise from nonlinear surface diffusion and its coupling to the shape perturbation.}

    \subsection{Galerkin projection and truncation method}
    
    \cc{We restrict to axisymmetric perturbations about the $z$-axis and retain only the $m=0$ spherical harmonics. For the harmonic basis, we write $Y_\ell\equiv Y_{\ell,0}$, with}
    \begin{equation}
        Y_\ell(\theta)
        =
        \sqrt{\frac{2\ell+1}{4\pi}}\,
        P_\ell(\cos\theta),
        \quad
        \int Y_\ell Y_{\ell'}\,\mathrm{d}\Omega
        =
        \delta_{\ell\ell'}.
    \end{equation}
    \cc{Here, $\mathrm{d}\Omega=\sin\theta\,\mathrm{d}\theta\,\mathrm{d}\varphi$ is the solid-angle element and $P_\ell$ is the Legendre polynomial of degree $\ell$.}
    
    \cc{Likewise, we suppress the azimuthal index in the composition and shape amplitudes and write}
    \begin{equation}
        \label{eqn:C-axisymmetric-amplitudes}
        \Psi_{k,\ell}\equiv\Psi_{k,\ell,0},
        \qquad
        u_\ell\equiv u_{\ell,0}.
    \end{equation}
    \cc{Retaining the axisymmetric modes with $\ell=0,1,2$, the perturbations of the fissogen and fusogen areal fractions and the intrinsic shape perturbation are}
    \begin{align}
        \label{eqn:C-retained-fields}
        \Psi_1(\theta,t)
        &=
        \Psi_{1,0}(t)Y_0
        +\Psi_{1,1}(t)Y_1
        +\Psi_{1,2}(t)Y_2,
        \\[4pt]
        \Psi_2(\theta,t)
        &=
        \Psi_{2,0}(t)Y_0
        +\Psi_{2,1}(t)Y_1
        +\Psi_{2,2}(t)Y_2,
        \\[4pt]
        u(\theta,t)
        &=
        u_0(t)Y_0+u_2(t)Y_2.
    \end{align}
    \cc{The composition amplitudes $\Psi_{1,\ell}$ and $\Psi_{2,\ell}$ describe the fissogen and fusogen areal-fraction perturbations at harmonic degree $\ell$. The monopolar ($\ell=0$) mode gives a spatially uniform perturbation of each areal fraction, the dipolar ($\ell=1$) modes describe the fore--aft asymmetry associated with translation, and the quadrupolar ($\ell=2$) modes describe the leading fore--aft-symmetric variation.}
    
    \cc{The evolution of each retained composition amplitude is obtained by projecting Eq.~(\ref{eqn:psi-dynamics-nonlinear}) onto the corresponding spherical harmonic}
    \begin{equation}
        \label{eqn:mmode-dynamics-lm}
        \frac{\partial\Psi_{k,\ell}}{\partial t}
        =
        \int Y_\ell\,
        \frac{\partial\Psi_k}{\partial t}\,
        \mathrm{d}\Omega.
    \end{equation}
    \cc{where $\mathrm{d}\Omega=\sin\theta\,\mathrm{d}\theta\,\mathrm{d}\varphi$, which after integration over the azimuthal angle $\varphi$ reduces to $2\pi\sin\theta\,\mathrm{d}\theta$ for the axisymmetric fields considered here. For compactness, we denote the harmonic projection of a scalar field $f(\theta)$ by}
    \begin{equation}
        \label{eqn:first-nonlinear-term}
        f_\ell
        \equiv
        \int Y_\ell\,f\,\mathrm{d}\Omega.
    \end{equation}
    \cc{Unlike the linear equations, in which each harmonic degree evolves independently, the nonlinear terms couple different degrees. Only the modes $\ell=0,1,2$ are retained as independent amplitudes. Nonlinear products can nevertheless generate higher harmonics as intermediate contributions, which must be retained until the product is projected back onto the modes $\ell=0,1,2$. For example, $Y_1Y_2$ contains both $\ell=1$ and $\ell=3$ contributions, and at cubic order the latter can combine with another $Y_1$ to generate an $\ell=2$ contribution. These projections require integrals of products of three or four spherical harmonics over the unit sphere. For example, projecting $Y_1^2$ onto $Y_0$ gives $G_0=\int Y_0Y_1^2\,\mathrm{d}\Omega$. The required angular coupling coefficients are summarized in Table~\ref{tab:C-angular-overlaps}, with the fourth-order integrals expressed in terms of the five independent coefficients $G_0$, $G_1$, $G_2$, $H_1$, and $H_2$.}
    
    \cc{For the fields in Eq.~(\ref{eqn:C-retained-fields}), the active membrane-area flux defined in Eq.~(\ref{eqn:C-mu-field}) takes the form}
    \begin{equation}
        \label{eqn:C-mu-harmonics}
        \mu(\theta,t)
        =
        \mu_0Y_0+\mu_1Y_1+\mu_2Y_2,
    \end{equation}
    \cc{with each harmonic amplitude given by}
    \begin{equation}
        \mu_\ell
        =
        A\left(\zeta\Psi_{2,\ell}-\Psi_{1,\ell}\right).
    \end{equation}
    
    \cc{Similarly, using the dimensionless active second moments $Q_1$ and $Q_2$ defined in Eq.~(\ref{eqn:Q-definition}), the total dimensionless second moment becomes}
    \begin{equation}
        \label{eqn:C-q-field}
        \frac{\mathbb{Q}}{\eta\gamma}
        =
        Q_1\Phi_1+Q_2\Phi_2
        =
        \bar q+q_0Y_0+q_1Y_1+q_2Y_2,
    \end{equation}
    \cc{where $\bar q=Q_1\bar{\Phi}_1+Q_2\bar{\Phi}_2$ is the homogeneous contribution, with each perturbation harmonic amplitude given by}
    \begin{equation}
        q_\ell
        =
        Q_1\Psi_{1,\ell}+Q_2\Psi_{2,\ell}.
    \end{equation}

    \begin{table*}[t!]
        \setlength{\tabcolsep}{5pt}
        \renewcommand{\arraystretch}{1.35}
        \begin{ruledtabular}
            \begin{tabular}{c|ccc|cccccc}
                & \multicolumn{3}{c|}{Quadratic product} & \multicolumn{6}{c}{Cubic product} \\[2pt]
                & $Y_1^2$ & $Y_1Y_2$ & $Y_2^2$ & $Y_0^3$ & $Y_0Y_1^2$ & $Y_0Y_2^2$ & $Y_1^3$ & $Y_1^2Y_2$ & $Y_2^3$ \\
                \hline
                $Y_0$ & $G_0$ & $0$ & $G_0$ & $G_0^2$ & $G_0^2$ & $G_0^2$ & $0$ & $G_0G_1$ & $G_0G_2$ \\
                $Y_1$ & $0$ & $G_1$ & $0$ & $0$ & $0$ & $0$ & $G_0^2+G_1^2$ & $0$ & $0$ \\
                $Y_2$ & $G_1$ & $0$ & $G_2$ & $0$ & $G_0G_1$ & $G_0G_2$ & $0$ & $H_1$ & $H_2$ \\[1pt]
                \hline
                \multicolumn{10}{c}{\rule{0pt}{4.5ex}$\displaystyle G_0=\frac{1}{2\sqrt{\pi}},\qquad G_1=\frac{1}{\sqrt{5\pi}},\qquad G_2=\frac{\sqrt5}{7\sqrt{\pi}},\qquad H_1=\frac{11}{28\pi},\qquad H_2=\frac{15}{28\pi}$} \\[2ex]
            \end{tabular}
        \end{ruledtabular}
        \caption{\label{tab:C-angular-overlaps}
        \cc{Angular coupling coefficients generated by the nonlinear Galerkin projection. The row gives the spherical harmonic used for the projection and the column gives the quadratic or cubic harmonic product being projected. Each entry is the corresponding integral over the unit sphere. For example, the entry $G_0$ in row $Y_0$ and column $Y_1^2$ denotes $\int Y_0Y_1^2\,\mathrm{d}\Omega=G_0$, while the entry $H_1$ in row $Y_2$ and column $Y_1^2Y_2$ denotes $\int Y_1^2Y_2^2\,\mathrm{d}\Omega=H_1$. Zero entries follow from orthogonality or parity. The final row gives the values of the five independent nonzero coefficients.}}
    \end{table*}
    
    \cc{In the comoving formulation, the linear $\ell=1$ shape response of Appendix~B is represented by the center-of-mass position $\itbf{X}(t)$ rather than by an intrinsic shape mode. We denote the corresponding dimensionless translational rate by $U_1$, while $U_2=\dot u_2$ denotes the intrinsic quadrupolar shape rate. For the retained axisymmetric modes, the linear shape equation, Eq.~(\ref{eqn:shape-governing-eq}), gives}
    \begin{align}
        \label{eqn:C-U1-U2}
        U_1 &= \frac{\mu_1-q_1}{3},\\[2pt]
        \label{eqn:C-U2}
        U_2 &= \frac{\mu_2}{11}
        -\frac{27q_2}{55}
        -\frac{144\mathcal K+12\bar q}{55}\,u_2.
    \end{align}
    \cc{where $\mathcal K=\kappa/(R\eta\gamma)$ is the dimensionless bending rigidity. The center-of-mass relation in Eq.~(\ref{eqn:com-velocity-exp}), expressed in dimensionless form, gives}
    \begin{equation}
        \label{eqn:C-Vcm}
        V_{\mathrm{cm}}
        =
        \sqrt{\frac{3}{4\pi}}\,U_1
        =
        \frac{\mu_1-q_1}{\sqrt{12\pi}}.
    \end{equation}
    
    \cc{For the tangential flow retained here, the velocity in Eq.~(\ref{eqn:C-material-derivative}) is along the $\theta$ direction and can be written in terms of a scalar potential as}
    \begin{equation}
        \label{eqn:C-angular-flow}
        \boldsymbol{w}^{(\mathtt{1})}
        =
        \nabla_{\!\Omega}\chi^{(\mathtt{1})},
        \qquad
        \chi^{(\mathtt{1})}
        =
        B_1Y_1+{B_2}Y_2/3,
    \end{equation}
    \cc{with $\nabla_{\!\Omega}=\thetav\,\partial_\theta$ for the axisymmetric fields. The coefficients $B_1$ and $B_2$ are determined by the local membrane-area balance in Eq.~(\ref{eqn:divergence-law-eq}), which gives}
    \begin{equation}
        \label{eqn:C-area-balance-nonuniform}
        \nabla_{\!\Omega}\!\cdot\boldsymbol{w}^{(\mathtt{1})}
        +2U_2Y_2
        =
        \mu_1Y_1+\mu_2Y_2.
    \end{equation}
    \cc{Using the identity $\Delta_\Omega Y_\ell=-\ell(\ell+1)Y_\ell$ in Eq.~(\ref{eqn:C-angular-flow}) gives $\nabla_{\!\Omega}\!\cdot\boldsymbol{w}^{(\mathtt{1})}=\Delta_\Omega\chi^{(\mathtt{1})}=-2B_1Y_1-2B_2Y_2$. Substitution into Eq.~(\ref{eqn:C-area-balance-nonuniform}) and matching the coefficients of the $Y_1$ and $Y_2$ spherical harmonics gives}
    \begin{equation}
        \label{eqn:C-B-amplitudes}
        B_1=-{\mu_1}/{2},
        \qquad
        B_2=U_2-{\mu_2}/{2}.
    \end{equation}
    
    \cc{To write the projected composition equations compactly, we denote the harmonic amplitudes of the linear logarithmic term $\mathcal L_k^{(\mathtt{1})}$ in Eq.~(\ref{eqn:C-log-expansion}) by}
    \begin{equation}
        \label{eqn:C-Lij}
        L_{k\ell}
        \equiv
        \left(\mathcal L_k^{(\mathtt{1})}\right)_\ell
        =
        \frac{\Psi_{0,\ell}}{\bar\Phi_0}
        +
        \frac{\Psi_{k,\ell}}{\bar\Phi_k}.
    \end{equation}
    \cc{where $\Psi_{0,\ell}=\Psi_{1,\ell}+\Psi_{2,\ell}$ follows from Eq.~(\ref{eqn:mass-cons-psi-s}). Thus, for the retained dipolar and quadrupolar modes, the projected composition dynamics can be separated into linear, quadratic, and cubic contributions as}
    \begin{equation}
        \label{eqn:C-composition-orders}
        \dot\Psi_{k,\ell}
        =
        F_{k,\ell}^{(\mathtt{1})}
        +F_{k,\ell}^{(\mathtt{2})}
        +F_{k,\ell}^{(\mathtt{3})}.
    \end{equation} 
    
    \cc{The linear contributions recover the corresponding composition dynamics of Appendix~B and are given by}
    \begin{align}
        \label{eqn:C-F-linear}
        F_{k,\ell}^{(\mathtt{1})}&=-\bar\Phi_k\mu_\ell-\ell(\ell+1)L_{k\ell},
    \end{align}
    
    \cc{The corresponding quadratic contributions, using the tangential-flow coefficients in Eq.~(\ref{eqn:C-B-amplitudes}), are given by}
    \begin{widetext}
    \begin{align}
        \label{eqn:C-F1-quadratic}
        F_{k,1}^{(\mathtt{2})}={}&
        -G_0\!\left(\Psi_{k,0}\mu_1+\Psi_{k,1}\mu_0\right)
        +G_1\!\left[
        {\mu_1}\Psi_{k,2}/2
        -\left(U_2+{\mu_2}/{2}\right)\Psi_{k,1}
        \right]
        +4\left(G_0u_0+G_1u_2\right)L_{k1}
        \notag\\[3pt]
        &-2G_0\!\left(
        {\Psi_{0,0}\Psi_{0,1}}/{\bar\Phi_0^2}
        -{\Psi_{k,0}\Psi_{k,1}}/{\bar\Phi_k^2}
        \right)
        -2G_1\!\left(
        {\Psi_{0,1}\Psi_{0,2}}/{\bar\Phi_0^2}
        -{\Psi_{k,1}\Psi_{k,2}}/{\bar\Phi_k^2}
        \right),
        \\[7pt]
        \label{eqn:C-F2-quadratic}
        F_{k,2}^{(\mathtt{2})}={}&
        -G_0\!\left(\Psi_{k,0}\mu_2+\Psi_{k,2}\mu_0\right)
        -{3}G_1\mu_1\Psi_{k,1}/2
        -G_2\!\left(U_2+{\mu_2}/{2}\right)\Psi_{k,2}
        +12\left(G_0u_0+G_2u_2\right)L_{k2}
        \notag\\[4pt]
        &-6G_0\!\left(
        {\Psi_{0,0}\Psi_{0,2}}/{\bar\Phi_0^2}
        -{\Psi_{k,0}\Psi_{k,2}}/{\bar\Phi_k^2}
        \right)
        -3G_1\!\left(
        {\Psi_{0,1}^2}/{\bar\Phi_0^2}
        -{\Psi_{k,1}^2}/{\bar\Phi_k^2}
        \right)
        -3G_2\left(
        {\Psi_{0,2}^2}/{\bar\Phi_0^2}
        -{\Psi_{k,2}^2}/{\bar\Phi_k^2}
        \right).
    \end{align}
    
    \cc{The cubic contribution to the dipolar mode is}
    \begin{align}
        \label{eqn:C-F1-cubic}
        F_{k,1}^{(\mathtt{3})}={}&
        -{2\Psi_{0,1}}\!
        \left(
        G_0^2\Psi_{0,0}^2
        +2G_0G_1\Psi_{0,0}\Psi_{0,2}
        +H_1\Psi_{0,2}^2
        +{(G_0^2+G_1^2)}\Psi_{0,1}^2/3
        \right)/{\,\bar\Phi_0^3}
        \notag\\[4pt]
        &-{2\Psi_{k,1}}\!
        \left(
        G_0^2\Psi_{k,0}^2
        +2G_0G_1\Psi_{k,0}\Psi_{k,2}
        +H_1\Psi_{k,2}^2
        +(G_0^2+G_1^2)\Psi_{k,1}^2/3
        \right)/{\,\bar\Phi_k^3}
        \notag\\[4pt]
        &+\left(
        4G_0G_1u_0+7H_1u_2+5H_2u_2
        \right)
        \left(
        {\Psi_{0,1}\Psi_{0,2}}/{\bar\Phi_0^2}
        -{\Psi_{k,1}\Psi_{k,2}}/{\bar\Phi_k^2}
        \right)
        \notag\\[4pt]
        &+4\left(G_0^2u_0+G_0G_1u_2\right)
        \left(
        {\Psi_{0,0}\Psi_{0,1}}/{\bar\Phi_0^2}
        -{\Psi_{k,0}\Psi_{k,1}}/{\bar\Phi_k^2}
        \right)\notag\\[4pt]
        &-6\left(
        G_0^2u_0^2
        +2G_0G_1u_0u_2
        +H_1u_2^2-H_2u_2^2
        \right)L_{k1}.
    \end{align}
    
    \cc{The corresponding quadrupolar contribution is}
    \begin{align}
        \label{eqn:C-F2-cubic}
        F_{k,2}^{(\mathtt{3})}={}
        &12G_0\left(G_0u_0+G_2u_2\right)
        \left(
        {\Psi_{0,0}\Psi_{0,2}}/{\bar\Phi_0^2}
        -{\Psi_{k,0}\Psi_{k,2}}/{\bar\Phi_k^2}
        \right)
        +6\left[
        G_0G_1u_0+(H_2-H_1)u_2
        \right]
        \left(
        {\Psi_{0,1}^2}/{\bar\Phi_0^2}
        -{\Psi_{k,1}^2}/{\bar\Phi_k^2}
        \right)
        \notag\\[4pt]
        &+2\left(
        3G_0G_2u_0+4H_2u_2
        \right)
        \left(
        {\Psi_{0,2}^2}/{\bar\Phi_0^2}
        -{\Psi_{k,2}^2}/{\bar\Phi_k^2}
        \right)
        -18G_0u_0\left(G_0u_0+2G_2u_2\right)L_{k2}\notag\\[4pt]
        &-6\left(
        G_0^2\Psi_{0,0}^2\Psi_{0,2}
        +G_0G_1\Psi_{0,0}\Psi_{0,1}^2
        +G_0G_2\Psi_{0,0}\Psi_{0,2}^2
        +H_1\Psi_{0,1}^2\Psi_{0,2}
        +{H_2}\Psi_{0,2}^3/3
        \right)/{\,\bar\Phi_0^3}
        \notag\\[4pt]
        &-6\left(
        G_0^2\Psi_{k,0}^2\Psi_{k,2}
        +G_0G_1\Psi_{k,0}\Psi_{k,1}^2
        +G_0G_2\Psi_{k,0}\Psi_{k,2}^2
        +H_1\Psi_{k,1}^2\Psi_{k,2}
        +{H_2}\Psi_{k,2}^3/3
        \right)/\,\bar\Phi_k^3.
    \end{align}
    \end{widetext}
    
    \cc{These expressions contain the cubic logarithmic contribution and the shape-dependent surface-diffusion corrections retained in Eq.~(\ref{eqn:psi-dynamics-nonlinear}). The nonlinear terms couple the dipolar and quadrupolar composition modes to one another and to the shape perturbation. Together with $\dot u_2=U_2$ from Eq.~(\ref{eqn:C-U2}), they give five differential equations for $\Psi_{1,1}$, $\Psi_{2,1}$, $\Psi_{1,2}$, $\Psi_{2,2}$, and $u_2$.}
    \cc{The monopolar ($\ell=0$) composition modes behave differently because $\Delta_\Omega Y_0=0$, so surface diffusion provides no restoring contribution. Their linear dynamics is $\dot\Psi_{k,0}=-\bar\Phi_k\mu_0$, with $\mu_0=A(\zeta\Psi_{2,0}-\Psi_{1,0})$, and the steady-state relation $\bar\Phi_1=\zeta\bar\Phi_2$ gives $\dot\mu_0=0$. A nonzero $\mu_0$ would therefore drive a secular linear change in the monopolar amplitudes. We consequently do not evolve $\Psi_{1,0}$ and $\Psi_{2,0}$ independently, but determine them together with $u_0$ from the fixed-volume and component-conservation constraints derived in the following subsection.}
    
    \subsection{Constraints and stability of drifting states}
    
    \cc{The remaining isotropic amplitudes $\Psi_{1,0}$, $\Psi_{2,0}$, and $u_0$ are determined by the global constraints. For the membrane shape in Eq.~(\ref{eqn:C-comoving-shape}), the area element through cubic order in the perturbations is}
    \begin{equation}
        \label{eqn:C-area-element}
        \mathrm{d}S
        \simeq
        R^2
        \!\left(
        1+2u+u^2+\frac{1}{2}|\nabla_{\!\Omega}u|^2
        \right)\mathrm{d}\Omega.
    \end{equation}
    \cc{We define the excess area as $\Delta\mathcal A=\mathcal A-\mathcal A^{(\mathtt{0})}$,
    where $\mathcal A^{(\mathtt{0})}=4\pi R^2$ is the area of the homogeneous spherical state. Using the shape expansion in Eq.~(\ref{eqn:C-retained-fields}), we obtain}
    \begin{equation}
        \label{eqn:C-area-excess}
        \frac{\Delta\mathcal A}{R^2}
        \simeq
        4\sqrt{\pi}\,u_0+u_0^2+4u_2^2.
    \end{equation} \cc{The cubic contributions to the area element cancel identically. The excess area is determined by the isotropic and quadrupolar shape amplitudes and need not vanish at fixed enclosed volume. The volume is given by}
    \begin{equation}
        V=\frac{R^3}{3}\int(1+u)^3\,\mathrm{d}\Omega.
    \end{equation}
    \cc{For the retained shape modes, the angular integrals required to expand the volume through cubic order are}
    \begin{align}
        &\int u\,\mathrm d\Omega
        =2\sqrt{\pi}\,u_0,\quad\int u^2\,\mathrm d\Omega
        =u_0^2+u_2^2,\\[2pt]
        &\int u^3\,\mathrm d\Omega
        =G_0u_0^3+3G_0u_0u_2^2+G_2u_2^3.
    \end{align}
    \cc{Requiring the enclosed volume to remain equal to its homogeneous steady-state value, $V^{(\mathtt{0})}=4\pi R^3/3$, then gives the cubic fixed-volume constraint}
    \begin{align}
        \label{eqn:C-volume-constraint}
        \mathcal C_V={}&
        2\sqrt{\pi}\,u_0+u_0^2+u_2^2 \notag\\
        &+\frac{1}{3}
        \left(
        G_0u_0^3+3G_0u_0u_2^2+G_2u_2^3
        \right)=0.
    \end{align}
    
    \cc{The total amounts of the two active membrane components are separately conserved. Since the total membrane density is constant, we absorb this constant factor into the conserved amount}
    \begin{equation}
        N_k
        \equiv
        \int_{\mathcal M}\Phi_k\,\mathrm{d}S
        =
        4\pi R^2\bar\Phi_k,
    \end{equation}
    \cc{Substituting $\Phi_k=\bar\Phi_k+\Psi_k$ and the area element in Eq.~(\ref{eqn:C-area-element}), and retaining terms through cubic order, gives}
    \begin{equation*}
        \int
        \left[
        (\bar\Phi_k+\Psi_k)
        \!\left(
        1+2u+u^2+\frac{|\nabla_{\!\Omega}u|^2}{2}
        \right)\!
        -\bar\Phi_k
        \right]
        \mathrm{d}\Omega = 0.
    \end{equation*}
    \cc{The dipolar amplitudes $\Psi_{1,1}$ and $\Psi_{2,1}$ do not contribute to the global conservation constraints for the fissogen and fusogen amounts. Since the retained shape contains only the even harmonics $Y_0$ and $Y_2$, the area element is even under $\cos\theta\mapsto-\cos\theta$, whereas $Y_1$ is odd. Any contribution proportional to a dipolar amplitude is odd and integrates to zero over the sphere. Evaluating the remaining angular integrals using Table~\ref{tab:C-angular-overlaps} gives the conservation constraint for each active membrane component:}
    \begin{align}
        \label{eqn:C-component-constraint}
        \mathcal C_k={}&
        \bar\Phi_k\,\frac{\Delta\mathcal A}{R^2}
        +\left[
        2\sqrt{\pi}
        +2u_0
        +G_0\left(u_0^2+4u_2^2\right)
        \right]\Psi_{k,0}
        \notag\\[4pt]
        &+\left(
        2u_2
        +2G_0u_0u_2
        +\frac{5}{2}G_2u_2^2
        \right)\Psi_{k,2}
        =0.
    \end{align}
    \cc{Together with the fixed-volume constraint in Eq.~(\ref{eqn:C-volume-constraint}), these component-conservation constraints determine the isotropic amplitudes $\Psi_{1,0}$, $\Psi_{2,0}$, and $u_0$.}
    
    \begin{figure*}[t!]
        \includegraphics[width=\textwidth]{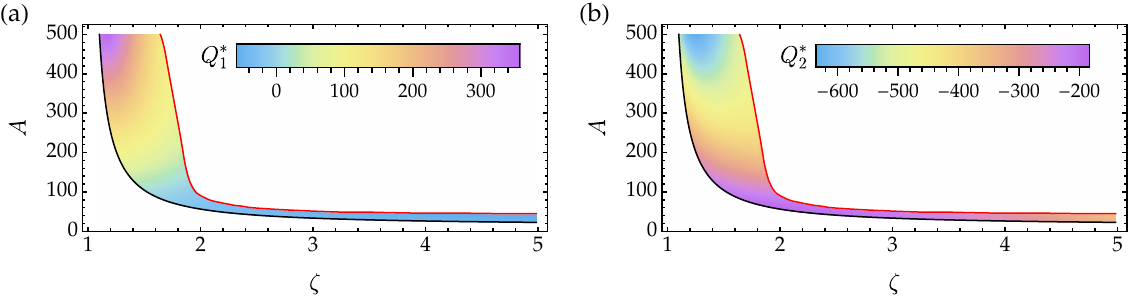}
        \caption{\label{fig:C-critical-moments}
        \cc{Color maps of the critical active second moments (a) $Q_1^{\ast}$ and (b) $Q_2^{\ast}$ as functions of $\zeta$ and $A$. At each point in the colored region, the critical pair $(Q_1^{\ast},Q_2^{\ast})$ is obtained by solving the nonlinear fixed-point equations with $V_{\mathrm{cm}}=u_2=0$, excluding the homogeneous spherical solution. These conditions identify the intersection of the drift-reversal and prolate--oblate boundaries. The black curve marks the linear drift-instability threshold, below which the homogeneous sphere is stable and satisfies both conditions identically. Above the red curve, only unstable fixed points satisfying both conditions were found. The colors indicate the corresponding values of the critical moments.}
        }
    \end{figure*}
    
    \cc{In the weakly nonlinear reduction, the enclosed volume and the total amounts of the active components are conserved, while the mechanical shape dynamics are retained only at linear order. Consequently, the reduced transient dynamics does not enforce the full nonlinear membrane-area balance exactly. At each nonlinear fixed point, however, the intrinsic shape is stationary, $U_2=0$, so the rate of change of the total membrane area vanishes, $\mathrm{d}\mathcal{A}/\mathrm{d}t=0$. The corresponding excess area need not vanish and is given in terms of the isotropic and quadrupolar shape amplitudes by Eq.~(\ref{eqn:C-area-excess}).}
    
    \cc{Combining the retained dynamical equations with the global constraints leads to a reduced dynamical system of differential-algebraic equations. We collect its five dynamical amplitudes and three constrained isotropic amplitudes as follows:}
    \begin{align}
        \label{eqn:C-state-vectors}
        &\boldsymbol{x}
        =
        \left(
        \Psi_{1,1},\Psi_{2,1},\Psi_{1,2},\Psi_{2,2},u_2
        \right)^{\mathsf T},\\[4pt]
        &\boldsymbol{z}
        =
        \left(
        \Psi_{1,0},\Psi_{2,0},u_0
        \right)^{\mathsf T}.
    \end{align}
    \cc{The reduced dynamical system can then be written as}
    \begin{equation}
        \label{eqn:C-DAE}
        \dot{\boldsymbol{x}}
        =
        \boldsymbol{F}(\boldsymbol{x},\boldsymbol{z}),
        \quad
        \boldsymbol{C}(\boldsymbol{x},\boldsymbol{z})
        =
        \boldsymbol{0},
    \end{equation}
    \cc{where the dynamical and constraint functions are}
    \begin{align}
        &\boldsymbol{F}
        =
        \left(
        F_{1,1},F_{2,1},F_{1,2},F_{2,2},U_2
        \right)^{\mathsf T},\\[4pt]
        &\boldsymbol{C}
        =
        \left(
        \mathcal C_V,\mathcal C_1,\mathcal C_2
        \right)^{\mathsf T}.
    \end{align}
    \cc{Thus, the dipolar and quadrupolar composition amplitudes $\Psi_{1,1}$, $\Psi_{2,1}$, $\Psi_{1,2}$, and $\Psi_{2,2}$, together with the shape amplitude $u_2$, evolve dynamically, whereas the monopolar amplitudes $\Psi_{1,0}$, $\Psi_{2,0}$, and $u_0$ are determined by the fixed-volume constraint $\mathcal C_V=0$ and the fissogen and fusogen conservation constraints $\mathcal C_k=0$ for $k=1,2$.}
    
    \cc{A steadily drifting state is stationary in the comoving frame, with constant shape and composition amplitudes and a center-of-mass position $\itbf{X}(t)$ that advances at constant velocity. For the reduced dynamical system in Eq.~(\ref{eqn:C-DAE}), such a fixed point $(\boldsymbol{x}^{\ast},\boldsymbol{z}^{\ast})$ satisfies}
    \begin{equation}
        \label{eqn:fixedpoints-rel}
        \boldsymbol F(\boldsymbol x^{\ast},\boldsymbol z^{\ast})=\boldsymbol0,
        \quad
        \boldsymbol C(\boldsymbol x^{\ast},\boldsymbol z^{\ast})=\boldsymbol0.
    \end{equation}
    \cc{The condition $U_2=0$ makes the intrinsic quadrupolar shape stationary in the comoving frame. Since all amplitudes are constant at a fixed point, Eq.~(\ref{eqn:C-Vcm}) also gives a constant center-of-mass velocity. Denoting quantities evaluated at the fixed point by a superscript $\ast$, the corresponding fixed point represents a drifting state when}
    \begin{equation}
        V_{\mathrm{cm}}^{\ast}
        =
        \frac{\mu_1^{\ast}-q_1^{\ast}}{\sqrt{12\pi}}
        \neq0.
    \end{equation}
    
    \cc{To determine the linear stability of a fixed point, we consider small perturbations of the dynamical amplitudes $\boldsymbol x$. The constrained amplitudes $\boldsymbol z$ must then adjust so that $\boldsymbol C=\boldsymbol 0$ remains satisfied. We write}
    \begin{align}
        \boldsymbol{x}
        &=
        \boldsymbol{x}^{\ast}+\delta\boldsymbol{x},
        \\[3pt]
        \boldsymbol{z}
        &=
        \boldsymbol{z}^{\ast}+\delta\boldsymbol{z}.
    \end{align}
    
    \cc{Linearizing the differential equations and constraints about the fixed point gives}
    \begin{align}
        \label{eqn:C-linearized-DAE}
        \delta\dot{\boldsymbol x}
        &=
        F_x\,\delta\boldsymbol x
        +F_z\,\delta\boldsymbol z,
        \\[3pt]
        \boldsymbol0
        &=
        C_x\,\delta\boldsymbol x
        +C_z\,\delta\boldsymbol z,
    \end{align}
    \cc{where all derivatives are evaluated at $(\boldsymbol{x}^{\ast},\boldsymbol{z}^{\ast})$, with}
    \begin{align}
        F_x&=\left.\frac{\partial\boldsymbol F}{\partial\boldsymbol x}\right|_{(\boldsymbol{x}^{\ast},\boldsymbol{z}^{\ast})},
        &
        F_z&=\left.\frac{\partial\boldsymbol F}{\partial\boldsymbol z}\right|_{(\boldsymbol{x}^{\ast},\boldsymbol{z}^{\ast})},
        \\[4pt]
        C_x&=\left.\frac{\partial\boldsymbol C}{\partial\boldsymbol x}\right|_{(\boldsymbol{x}^{\ast},\boldsymbol{z}^{\ast})},
        &
        C_z&=\left.\frac{\partial\boldsymbol C}{\partial\boldsymbol z}\right|_{(\boldsymbol{x}^{\ast},\boldsymbol{z}^{\ast})}.
    \end{align}
    
    \cc{If $C_z$ is nonsingular, the linearized constraints determine the response of the constrained amplitudes to a perturbation $\delta\boldsymbol x$ as}
    \begin{equation}
        \delta\boldsymbol z
        =
        -C_z^{-1}C_x\,\delta\boldsymbol x.
    \end{equation}
    \cc{Substituting this into the differential equations gives}
    \begin{align}
        \label{eqn:C-constrained-Jacobian}
        \delta\dot{\boldsymbol x}
        =
        J_{\mathrm{con}}\,\delta\boldsymbol x,
        \quad
        J_{\mathrm{con}}
        =
        F_x-F_z\,C_z^{-1}C_x.
    \end{align}
    
    \begin{figure*}[t!]
        \includegraphics[width=\textwidth]{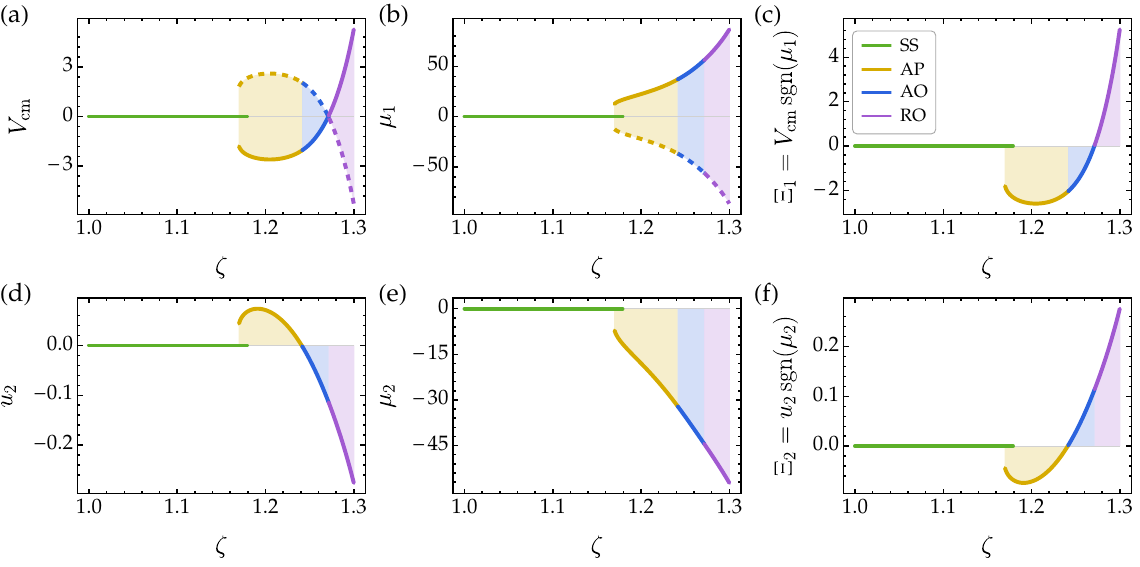}
        \caption{\label{fig:C-bifurcation-A280}
        \cc{Panels (a)--(f) show, respectively, the center-of-mass velocity $V_{\mathrm{cm}}$, dipolar membrane-area flux amplitude $\mu_1$, drift order parameter $\Xi_1$, quadrupolar shape amplitude $u_2$, quadrupolar membrane-area flux amplitude $\mu_2$, and shape order parameter $\Xi_2$, as functions of $\zeta$ at $A=280$, $Q_1=250$, $Q_2=-575$, $\mathcal K=10$, and $\bar\Phi_0=3/5$. Green, yellow, blue, and purple indicate the stationary sphere (SS), anterograde prolate (AP), anterograde oblate (AO), and retrograde oblate (RO) states, respectively. In panels (a) and (b), solid and colored dashed curves represent symmetry-related solutions with opposite signs of $V_{\mathrm{cm}}$ and $\mu_1$, while all other quantities remain unchanged. The AP--AO shape transition precedes the AO--RO drift reversal. For these moving states, $\mu_2<0$, giving $\Xi_2=-u_2$. The velocity $V_{\mathrm{cm}}$ and drift order parameter $\Xi_1$ are scaled by $\gamma/R$, while the flux amplitudes $\mu_1$ and $\mu_2$ are scaled by $\gamma/R^2$; $u_2$ and $\Xi_2$ are dimensionless.}
        }
    \end{figure*}
    
    \cc{The drifting fixed point is locally stable when all eigenvalues $\lambda_i$ of $J_{\mathrm{con}}$ satisfy $\operatorname{Re}\lambda_i<0$. A positive real part identifies an unstable direction. Since $V_{\mathrm{cm}}$ depends on the dipolar composition amplitudes, the effect of each perturbation direction on the drift velocity follows from}
    \begin{equation}
        \label{eqn:C-delta-Vcm}
        \delta V_{\mathrm{cm}}
        =
        \frac{(A\zeta-Q_2)\,\delta\Psi_{2,1} - (A+Q_1)\,\delta\Psi_{1,1}}{\sqrt{12\pi}}.
    \end{equation}
    \cc{Thus, the eigenvectors of $J_{\mathrm{con}}$ determine both the perturbation directions and their first-order effect on the drift velocity. Numerically, the action of $C_z^{-1}$ is obtained by solving the corresponding linear system; a singular $C_z$ prevents this reduction, while a small singular value indicates poor conditioning.}
    
    \begin{figure*}[t!]
        \includegraphics[width=\textwidth]{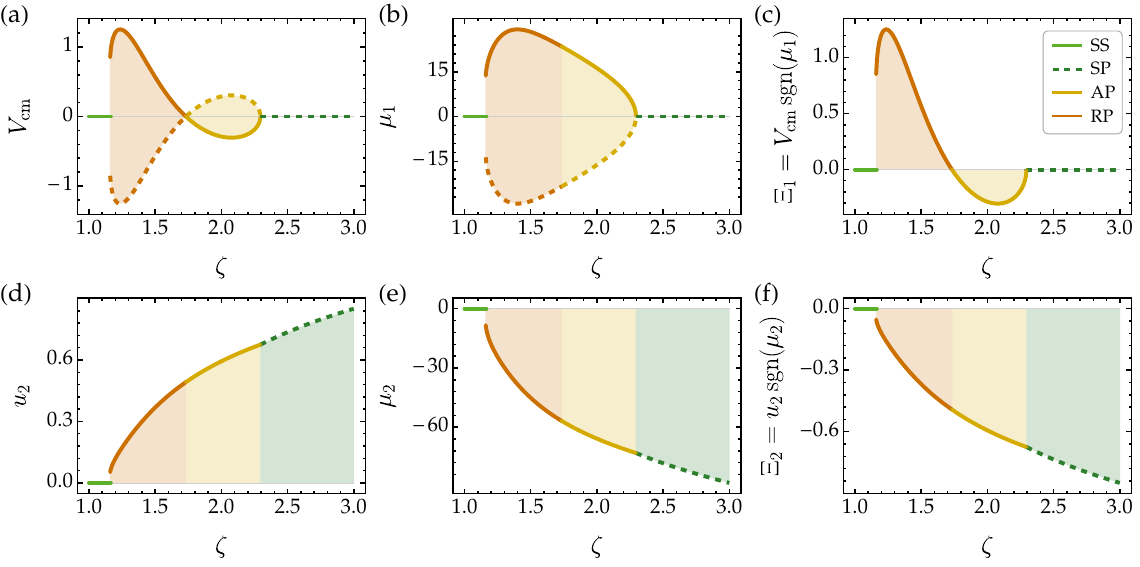}
        \caption{\label{fig:C-bifurcation-A300}
        \cc{Panels (a)--(f) show, respectively, the center-of-mass velocity $V_{\mathrm{cm}}$, dipolar membrane-area flux amplitude $\mu_1$, drift order parameter $\Xi_1$, quadrupolar shape amplitude $u_2$, quadrupolar membrane-area flux amplitude $\mu_2$, and shape order parameter $\Xi_2$, as functions of $\zeta$ at $A=300$, $Q_1=-125$, $Q_2=125$, $\mathcal K=10$, and $\bar\Phi_0=3/5$. Light green, orange, yellow, and dark green indicate the stationary sphere (SS), retrograde prolate (RP), anterograde prolate (AP), and stationary prolate (SP) states, respectively. The RP--AP drift reversal occurs while the shape remains prolate ($u_2>0$). At the subsequent AP--SP transition, $V_{\mathrm{cm}}$ and $\mu_1$ vanish, while the prolate deformation and quadrupolar membrane-area flux remain finite. In panels (a) and (b), solid and colored dashed curves represent symmetry-related drifting solutions with opposite signs of $V_{\mathrm{cm}}$ and $\mu_1$, while the other quantities remain unchanged. For the RP, AP, and SP states, $\mu_2<0$, giving $\Xi_2=-u_2$. The velocity $V_{\mathrm{cm}}$ and drift order parameter $\Xi_1$ are scaled by $\gamma/R$, while the flux amplitudes $\mu_1$ and $\mu_2$ are scaled by $\gamma/R^2$; $u_2$ and $\Xi_2$ are dimensionless.}
        }
    \end{figure*}

    \subsection{Fixed points and phase boundaries}
    
    \cc{We numerically determine the nonlinear fixed points of Eq.~(\ref{eqn:fixedpoints-rel}) using a Newton--Raphson method, repeating the search from randomly sampled initial guesses at several amplitude scales and retaining only distinct solutions. Their stability is assessed using the constrained Jacobian in Eq.~(\ref{eqn:C-constrained-Jacobian}). Repeating this analysis across parameter space, we identify stable stationary and drifting states with spherical, prolate, or oblate shapes. To characterize the direction of drift and shape deformation relative to their respective membrane-area flux amplitudes, we define two signed order parameters:}
    \begin{align}
        \label{eqn:C-signed-indicators}
        \Xi_1 &= V_{\mathrm{cm}}\operatorname{sgn}(\mu_1),\\[4pt]
        \label{eqn:C-shape-order-parameter}
        \Xi_2 &= u_2\operatorname{sgn}(\mu_2).
    \end{align} \cc{Note that the drift order parameter $\Xi_1$ is denoted simply by $\Xi$ in the main text; see Fig.~\ref{fig:fig4}(c). Negative and positive values of $\Xi_1$ correspond to anterograde and retrograde motion, respectively, while $u_2>0$ and $u_2<0$ distinguish prolate and oblate shapes. By reflection symmetry, stable drifting solutions occur in pairs with opposite $V_{\mathrm{cm}}$ and $\mu_1$, but unchanged $u_2$ and $\mu_2$. Consequently, both order parameters $\Xi_1$ and $\Xi_2$ remain unchanged, and the paired solutions belong to the same phase.}
    
    \cc{To find the precise locations of the drift-reversal and prolate--oblate boundaries, we seek parameter values for which the nonlinear system admits fixed points with $V_{\mathrm{cm}}^{\ast}=0$ and $u_2^{\ast}=0$, respectively. Using Eqs.~(\ref{eqn:C-Vcm}) and~(\ref{eqn:C-U2}), we find that}
    \begin{align}
        \label{eqn:C-drift-boundary}
        V_{\mathrm{cm}}^{\ast}=0
        &\quad\Longrightarrow\quad \mu_1^{\ast}=q_1^{\ast},\\[4pt]
        \label{eqn:C-shape-boundary}
        u_2^{\ast}=0
        &\quad\Longrightarrow\quad \mu_2^{\ast}=27q_2^{\ast}/5.
    \end{align}
    
    \cc{We next determine where the drift-reversal and prolate--oblate boundaries intersect as both active second moments $Q_1$ and $Q_2$ vary. At fixed $A$ and $\zeta$, this intersection defines a critical pair $(Q_1^{\ast},Q_2^{\ast})$. Since $q_1$ and $q_2$ depend linearly on $Q_1$ and $Q_2$, the preceding relations allow us to express both moments in terms of the composition amplitudes. We then obtain the critical pair and the corresponding amplitudes self-consistently by solving the nonlinear fixed-point equations, Eq.~(\ref{eqn:fixedpoints-rel}), subject to the simultaneous conditions $V_{\mathrm{cm}}^{\ast}=0$ and $u_2^{\ast}=0$.}
    
    \cc{We exclude the homogeneous spherical solution, which satisfies both zero conditions for arbitrary active second moments. The critical moments corresponding to stable nontrivial fixed points are shown in Fig.~\ref{fig:C-critical-moments}. These solutions occupy the region between the black curve, marking the linear $\ell=1$ instability threshold of the homogeneous sphere, and the red curve, beyond which only linearly unstable critical fixed points are found. Throughout this region, $Q_2^{\ast}<0$, while $Q_1^{\ast}$ is predominantly positive at moderate $\zeta$ and becomes negative at larger $\zeta$.}
    
    \cc{Each critical pair provides a reference point for exploring the nearby phase structure. By varying $Q_1$ and $Q_2$ slightly around $(Q_1^\ast,Q_2^\ast)$ and evaluating stable fixed points on a local stencil, we identify four neighbouring phases combining anterograde or retrograde motion with prolate or oblate morphology. These phases occupy regions separated by the intersecting drift-reversal and prolate--oblate boundaries. In particular, the anterograde oblate (AO) phase forms a wedge between these boundaries, with an opening angle and orientation that vary with $A$ and $\zeta$. This reveals conditions on the active second moments for obtaining stable AO motion, which we find predominantly for $Q_1>0$ and $Q_2<0$. Since a phase diagram at fixed $(Q_1,Q_2)$ represents only a two-dimensional cross section of the four-dimensional parameter space, a narrow AO region in such a diagram need not imply that the phase is similarly restricted in the full parameter space. The right-hand diagram in Fig.~\ref{fig:fig4}(c), with $(Q_1,Q_2)=(250,-575)$, illustrates one such choice supporting stable AO states.}
    
    \cc{However, the active second moments need not be close to these critical values for the system to exhibit stable stationary or drifting states. Different choices of $Q_1$ and $Q_2$ can produce other combinations of motion and shape. For example, the left-hand diagram in Fig.~\ref{fig:fig4}(c), with $(Q_1,Q_2)=(-125,125)$, exhibits stationary sphere (SS), retrograde prolate (RP), anterograde prolate (AP), and stationary prolate (SP) phases, without an AO region in the parameter range shown.}
    
    \cc{To examine the transitions between these phases, we follow the nonlinear fixed-point branches as $\zeta$ varies at fixed $A$, $Q_1$, and $Q_2$. The bifurcation curves in Fig.~\ref{fig:C-bifurcation-A280} reveal a region where both the SS state and finite-amplitude drifting states are stable near the linear drift-instability threshold, consistent with a subcritical onset of drift. Along the stable drifting branches, the AP--AO shape transition precedes the AO--RO drift reversal, so the organelle changes from prolate to oblate while moving anterogradely before reversing its direction of motion.}
    
    \cc{The bifurcation curves in Fig.~\ref{fig:C-bifurcation-A300} likewise show a discontinuous onset of drift, with a finite jump from SS to the retrograde prolate (RP) state. This is followed by an RP--AP drift reversal and a continuous transition to the stationary prolate (SP) phase. At the latter transition, both $V_{\mathrm{cm}}$ and $\mu_1$ vanish, while $u_2>0$ and $\mu_2\neq0$, so the prolate deformation and quadrupolar membrane-area flux persist even when translation ceases.}
    
    \cc{To estimate the center-of-mass speeds in physical units, we take $R=1~\mu\mathrm{m}$, $\gamma=0.002~\mu\mathrm{m}^2/\mathrm{s}$, $\eta=1~\mathrm{Pa}\cdot\mathrm{s}$, and $\kappa=2\times10^{-20}~\mathrm{J}$, consistent with the ranges in Tables I and II. These values give $\mathcal{K}=10$ and a characteristic velocity $\gamma/R=120~\mathrm{nm/min}$. In the right-hand phase diagram of Fig.~\ref{fig:fig4}(c), the active second moments $Q_1=250$ and $Q_2=-575$ correspond to $0.5~\mathrm{pN}$ and $-1.15~\mathrm{pN}$, respectively. The numerically sampled stable AO states are found to be approximately within $A=203$--$424$ and $\zeta=1.18$--$1.33$, corresponding to membrane-area transfer rates $|\bMone|\sim0.41$--$0.85~\mathrm{s}^{-1}$ and $\bMtwo\sim0.48$--$1.13~\mathrm{s}^{-1}$. For these states, the average dimensionless center-of-mass speed is approximately $1.1$, with a maximum of about $2$, corresponding to physical speeds of approximately $130~\mathrm{nm/min}$ and $250~\mathrm{nm/min}$, respectively.}
    
    \cc{For comparison, experimental measurements of effective cis-to-trans cargo progression through the Golgi give velocities of order $10~\mathrm{nm/min}$~\cite{Tie2025}, roughly an order of magnitude smaller than our predicted center-of-mass speeds. However, we expect cisternal motion to be hindered by interactions with the surrounding cytoskeleton and neighbouring compartments, which are not included in our model. These additional mechanical constraints could substantially reduce the resulting drift speed, potentially bringing it closer to the measured cargo progression rate. Such a comparison remains indirect, however, since cargo progression also reflects vesicular exchange and biochemical maturation rather than physical cisternal motion alone.}

    \section{Mathematical Prerequisites}
	\subsection{Differential Geometry of Surfaces}
	
	We briefly outline the mathematical language of differential geometry of two-dimensional surfaces, concentrating on the practical aspects of the subject rather than mathematical rigor. This brief review assumes that the reader is acquainted with some basic knowledge of vector and tensor calculus. It is noteworthy to mention that an extensive literature exists on differential geometry, with emphasis on both pure mathematics (e.g.~Ref.~\cite{Spivak1999}) and applications in physics (e.g.~Refs.~\cite{Deserno2015} and \cite{Kamien2002}).
    
	\subsubsection{Surface parametrization}
    
    Each point on a two-dimensional surface, say $\mathcal{S}$, embedded in the three-dimensional space $\mathbb{R}^3$ is specified by a position vector $\boldsymbol{R} = \left(X,\,Y,\,Z\right)^{\!\mathsf{T}}\!$, where $X$, $Y$, and $Z$ are the Cartesian coordinates, and $\mathsf{T}$ is a transpose~\cite{Kreyszig1991}. However, these coordinates are constrained by the geometry of the surface and are thus not all independent.

    \begin{figure}[t]\includegraphics[width=0.9\columnwidth]{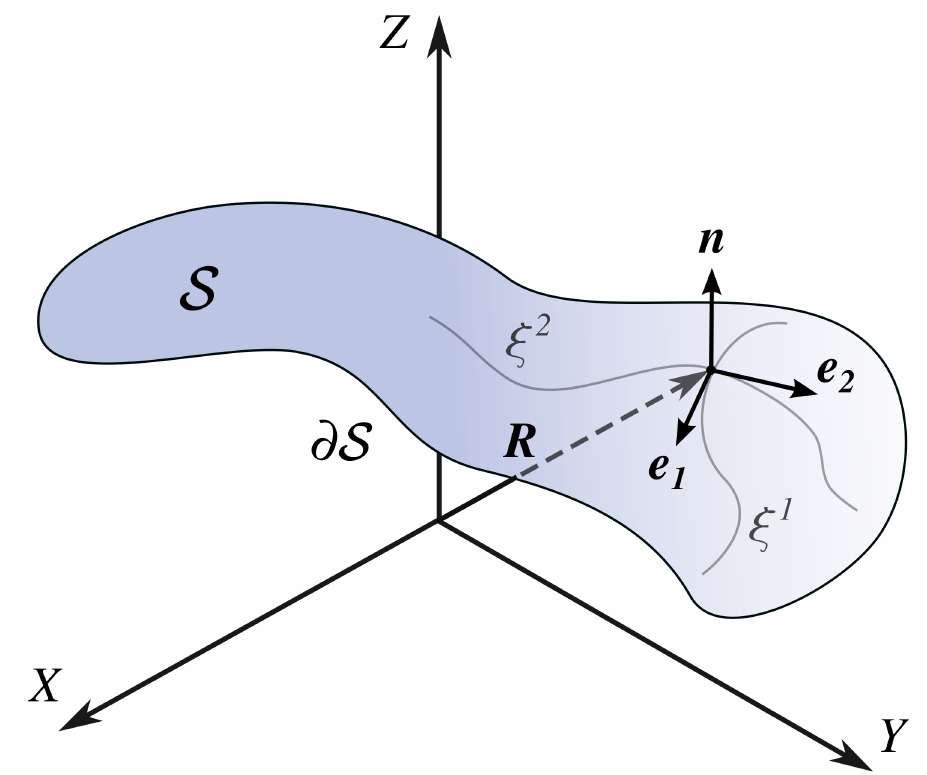}
		\caption{\label{fig:A1} Schematic diagram of a two-dimensional surface $\mathcal{S}$, whose boundary is denoted by $\partial\mathcal{S}$. Every point on the surface can be identified by a three-dimensional vector $\itbf{R}$ (described by the Cartesian coordinates $X$, $Y$ and $Z$). Also, for each vector $\itbf{R}$, we define the tangent vectors $\itbf{e}_1$ and $\itbf{e}_2$, corresponding to the internal coordinates $\xi^1$ and $\xi^2$, respectively, as well as the unit vector $\nv$ normal to the tangent plane.
	}\end{figure}
	
	Therefore, the embedding surface function $\boldsymbol{R}$ can be described locally by a pair of two variables, denoted by~$(\xi^{1},\xi^{2})$, which represent the internal {\it curvilinear} coordinates associated with a parametrization of the surface. At each point on the surface, a basis can be formed by constructing the tangent vectors and their corresponding normal vector, as shown in Fig.~\ref{fig:A1}. The tangent vector associated with the internal coordinate $\xi^\alpha$ is defined by
	\begin{equation}
		\label{eqn:tangent-vector}
		\ed{\alpha} = \frac{\partial\hspace{-0.5pt}\itbf{R}}{\partial\xi^\alpha}\equiv\partial_\alpha\boldsymbol{R},
	\end{equation} with index $\alpha=1,2$, whilst the unit vector $\nv$ normal to the surface (\ie $\nv\cdot\ed{\alpha}=0$ and $\nv\cdot\nv=1$) is given by
	\begin{equation}
		\label{eqn:normal-vector}
		\nv=\left(\ed{1}\times\ed{2}\right)/\left\|\ed{1}\times\ed{2}\right\|,
	\end{equation} where $\times$ is the three-dimensional cross product, and $\left\|\cdot\right\|$ gives the norm of the vector. The vectors $\ed{1}$ and $\ed{2}$ are not orthonormal in general, namely the four elements of
	\begin{equation}
		g_{\alpha\beta} = \ed{\alpha}\cdot\ed{\beta},
	\end{equation} known as the {\it metric tensor}, when written in matrix form, typically give $[g_{\alpha\beta}] \neq \Big[\!\begin{array}{cc}1 & 0 \\[-2pt] 0 & 1 \end{array}\!\Big]$. Instead, we introduce the corresponding dual basis vectors, $\itbf{e}^1$ and $\itbf{e}^2$, which satisfy
	\begin{equation}
		\ed{\alpha}\cdot\eu{\beta} =\delta_\alpha^\beta,
	\end{equation} where $\delta_\alpha^\beta$ is the Kronecker delta, \ie $\left[\delta_\alpha^\beta\right]=\Big[\!\begin{array}{cc}1 & 0 \\[-2pt] 0 & 1 \end{array}\!\Big]$. Note that metric tensor is a symmetric tensor, \ie $g_{\alpha\beta}=g_{\beta\alpha}$. Also, the inverse of the metric tensor, $g^{\alpha\beta}$, satisfies
	\begin{equation}
		\label{eqn:gamma-invgamma}
		g^{\alpha\beta}g_{\beta\gamma}=\delta^\alpha_\gamma,
	\end{equation} where Einstein convention is employed, \ie summation over repeated super- and sub-script indices is  implied. It follows that $g_{\alpha\beta}$ and $g^{\alpha\beta}$ can be used to lower and raise indices, respectively; for example,
	\begin{equation}
		\label{eqn:lower-raise}
		\eu{\alpha} = g^{\alpha\beta}\ed{\beta},\quad\text{and}\quad\ed{\alpha} = g_{\alpha\beta}\eu{\beta},
	\end{equation} 
	
	The metric tensor captures the distance information on the surface, being an intrinsic geometrical property of that surface, and defines the scalar product on $\mathcal{S}$; namely, for two vectors $\itbf{v}=v^\alpha\ed{\alpha}$ and $\itbf{w}=w^\alpha\ed{\alpha}$, we have that $\itbf{v}\cdot\itbf{w}=\left(v^\alpha\ed{\alpha}\right)\cdot\left(w^\beta\ed{\beta}\right)=g_{\alpha\beta}\,v^\alpha\,w^\beta = v_\alpha w^\alpha = v^\alpha w_\alpha$.
	
	\subsubsection{Surface decomposition}
    
    Both triads $\{\ed{1},\ed{2},\nv\}$ and $\{\itbf{e}^1,\itbf{e}^2,\nv\}$ can be used as bases to decompose any three-dimensional vector $\itbf{v}$ into its normal and in-plane vectorial parts, namely $\itbf{v}=\itbf{v}_{\nv}+\itbf{v}_0$, where $\itbf{v}_{\nv} = v\,\nv$ and $\itbf{v}_0 = v^\alpha\ed{\alpha} = v_\alpha\eu{\alpha}$, in which we define the normal vector component by $v = \itbf{v}\cdot\nv$, the co- and contra-variant vector components by $v_\alpha = \itbf{v}\cdot\ed{\alpha}$ and $v^\alpha = \itbf{v}\cdot\eu{\alpha}$, respectively. From (\ref{eqn:gamma-invgamma}) and (\ref{eqn:lower-raise}), we find that the metric tensor and its inverse can be used to raise and lower the index of vector components, \ie $v^\alpha = g^{\alpha\beta}\,v_\beta$ and $v_\alpha = g_{\alpha\beta}\,v^\beta$.
	
	Furthermore, we define the surface identity tensor,
	\begin{equation}
		\label{eqn:surface-identity}
		\itbf{I}_{\!\scriptscriptstyle\mathcal{S}}=\ed{\alpha}\otimes\eu{\alpha}=\eu{\alpha}\otimes\ed{\alpha},
	\end{equation} where $\otimes$ is the tensor product. This second-order tensor acts as projection operator onto the tangent plane of the surface, extracting the in-plane vector $\itbf{v}_0$ from $\itbf{v}$ (that is, $\itbf{v}_0 = \itbf{I}_{\!\scriptscriptstyle\mathcal{S}}\cdot\itbf{v} = \itbf{v}\cdot\itbf{I}_{\!\scriptscriptstyle\mathcal{S}}$). Similarly, we define the three-dimensional identity tensor $\itbf{I}$ as follows:
	\begin{equation}
		\label{eqn:full-identity}
		\itbf{I}=\itbf{I}_{\!\scriptscriptstyle\mathcal{S}}+\nv\otimes\nv,
	\end{equation} which satisfies that $\itbf{I}\cdot\itbf{v} = \itbf{v}\cdot\itbf{I} = \itbf{v}$ for any vector $\itbf{v}$.
	
	The surface identity can also be used to extract the tangential part of any second-order tensor $\itbf{T}\in\mathbb{R}^3\times\mathbb{R}^3$, namely
    \begin{equation}
        \itbf{T}_{\!\scriptscriptstyle\mathcal{S}} = \itbf{I}_{\!\scriptscriptstyle\mathcal{S}}\cdot\itbf{T}\cdot\itbf{I}_{\!\scriptscriptstyle\mathcal{S}} = T^{\alpha\beta}\,\ed{\alpha}\otimes\ed{\beta},
    \end{equation} where 
    \begin{equation}
        T^{\alpha\beta} = \left(\eu{\alpha}\cdot\itbf{T}\right)\cdot\eu{\beta}.
    \end{equation} In a similar fashion, the normal component of the tensor $\itbf{T}$ can be written as
	\begin{equation}
		\itbf{T}_{\!\nv} = T\,\nv\otimes\nv,\quad\,T=\left(\nv\cdot\itbf{T}\right)\cdot\nv.
	\end{equation} In addition to these, a second-order tensor may also have in general components along $\nv\otimes\ed{\alpha}$ and $\ed{\alpha}\otimes\nv$.
	
	\subsubsection{Surface differentiation} 
    
    We denote ordinary differentiation with respect to the surface coordinates $\xi^\alpha$ by $\partial_\alpha$. Differentiating Eq.~(\ref{eqn:tangent-vector}) gives $\partial_\beta\ed{\alpha}=\partial_\beta\partial_\alpha\boldsymbol{R}$, which generally has both tangential and normal components. We denote covariant differentiation by $\nabla_\beta$. For the tangent basis, it removes the tangential connection term, leaving the normal component, \ie
	\begin{equation}
		\nabla_\beta\ed{\alpha}=\left(\nv\otimes\nv\right)\partial_\beta\ed{\alpha}.
	\end{equation} Using Eqs.~(\ref{eqn:surface-identity}) and (\ref{eqn:full-identity}), this can also be expressed in terms of the Christoffel symbols as
	\begin{equation}
		\label{eqn:covariant-derivative-basis}
		\nabla_\beta\ed{\alpha}=\partial_\beta\ed{\alpha}-\Gamma^{\gamma}_{\alpha\beta}\itbf{e}_\gamma,
	\end{equation} where $\Gamma^{\gamma}_{\alpha\beta}=\itbf{e}^{\gamma}\cdot\partial_\beta\ed{\alpha}$ are the {\it Christoffel symbols} of the second kind, which can be also rewritten in terms of the derivatives of the metric tensor, namely
	\begin{equation}
		\label{eqn:Christoffel}
		\Gamma^{\gamma}_{\alpha\beta} = \frac{g^{\gamma\delta}}{2}\left(\partial_\beta g_{\alpha\delta}+\partial_\alpha g_{\delta\beta}-\partial_\delta g_{\alpha\beta}\right)\!,
	\end{equation} which shows the symmetry property that $\Gamma^{\gamma}_{\alpha\beta}=\Gamma^{\gamma}_{\beta\alpha}$.
	
	For the basis vector $\eu{\alpha}$, its covariant derivative is $\nabla_\beta\eu{\alpha}=\left(\nv\otimes\nv\right)\partial_\beta\eu{\alpha}$. From (\ref{eqn:surface-identity}) and (\ref{eqn:full-identity}), we find
	\begin{equation}
		\nabla_\beta\eu{\alpha}=\partial_\beta\eu{\alpha}+\Gamma^{\alpha}_{\beta\gamma}\itbf{e}^\gamma.
	\end{equation}
	
	For a scalar $w\in\mathbb{R}$, covariant and ordinary differentiation coincide, $\nabla_\alpha w=\partial_\alpha w$. For a three-dimensional vector field $\itbf{v}$ without surface indices, the derivative of the full vector similarly satisfies $\nabla_\beta\itbf{v}=\partial_\beta\itbf{v}$, whereas the covariant derivatives of its tangential components include the Christoffel symbols:
	\begin{align}
		\nabla_\beta v_\alpha &= \partial_\beta v_\alpha-\Gamma^\gamma_{\alpha\beta}v_\gamma,\\[4pt]
        \nabla_\beta v^\alpha &= \partial_\beta v^\alpha+\Gamma^\alpha_{\beta\gamma}v^\gamma.
	\end{align} A useful property of the covariant derivative is the fact that the metric tensor and its inverse is constant with respect to the covariant derivative, \ie
	\begin{equation}
		\label{eqn:metrinilic-property}
		\nabla_\gamma g_{\alpha\beta}=\nabla_\gamma g^{\alpha\beta}=0,
	\end{equation} which expresses the metric compatibility of the surface connection. In practice, this means that the process of raising and lowering of indices commutes with the covariant differentiation.
	
	Although covariant differentiation gives a tensorial derivative of the tangent basis, successive covariant derivatives do not generally commute, namely
	\begin{equation}
		(\nabla_\beta\nabla_\alpha-\nabla_\alpha\nabla_\beta)v_\gamma = -v^\delta R_{\alpha\beta\gamma\delta},
	\end{equation} where $R_{\alpha\beta\gamma\delta}$ is the {\it Riemann curvature tensor}. This is, in general, a nonzero tensor, vanishing for intrinsically flat surfaces, and its components are given by
	\begin{equation}
		R_{\alpha\beta\gamma\delta} = \partial_\gamma\Gamma_{\alpha\beta\delta}-\partial_\delta\Gamma_{\alpha\beta\gamma}-\Gamma^\omega_{\alpha\gamma}\Gamma_{\omega\beta\delta}+\Gamma^\omega_{\alpha\delta}\Gamma_{\omega\beta\gamma},
	\end{equation} where $\Gamma_{\gamma\alpha\beta} = g_{\gamma\delta}\Gamma^{\delta}_{\phantom{\delta}\alpha\beta}$ are the Christoffel symbols of the first kind. Since $R_{\alpha\beta\gamma\delta}$ depends only on the metric tensor, it is then an intrinsic quantity of the surface.
	
	With the help of the covariant derivative, we can now define the surface analogous of the regular differential operators, encountered in vector calculus: gradients, divergences, and Laplacian operators. The surface gradient of a scalar function $\phi$ that lives on $\mathcal{S}$ is defined by
	\begin{equation}
		\label{eqn:surface-gradient}
		\nabla\phi=\itbf{I}_{\!\scriptscriptstyle\mathcal{S}}\cdot\bnabla\phi = \partial_\alpha\phi\,\itbf{e}^{\alpha}=\nabla_\alpha\phi\,\itbf{e}^{\alpha},
	\end{equation} where $\bnabla$ is the usual gradient operator in $\mathbb{R}^3$. Similarly, the gradient of a vector function $\bPhi$ can be written as follows $\nabla\bPhi=(\bnabla\bPhi)\cdot\itbf{I}_{\!\scriptscriptstyle\mathcal{S}} = (\partial_\alpha\bPhi)\otimes\eu{\alpha}$. This allows us to define the surface divergence of $\bPhi$, namely
	\begin{equation}
		\label{eqn:surface-divergence}
		\Div\bPhi=\nabla\bPhi\hspace{0.5pt}\cddot\itbf{I} = (\partial_\alpha\bPhi)\cdot\eu{\alpha}=(\nabla_\alpha\bPhi)\cdot\eu{\alpha},
	\end{equation} where $\,\cddot\,$ denotes the double dot product. Let $\bPhi$ be solely defined on $\mathcal{S}$, with $\bPhi=\phi^\alpha\ed{\alpha}$, then Eq.~(\ref{eqn:surface-divergence}) can be written in component form as follows:
	\begin{equation}
		\label{eqn:divergence-Sfield}
		\Div\bPhi=\nabla_\alpha\phi^\alpha = \partial_\alpha\!\left(\sqrt{g}\hspace{1pt}\phi^\alpha\right)/\sqrt{g}\,,
	\end{equation} where the former follows from (\ref{eqn:covariant-derivative-basis}) and (\ref{eqn:Christoffel}), while the latter is a result known as the {\it Voss--Weyl formula}, in which we define that 
	\begin{equation}
		\label{eqn:g-def}
		g=\det[g_{\alpha\beta}].
	\end{equation} Metric compatibility also implies $\partial_\alpha\sqrt{g}=\sqrt{g}\,\Gamma^\beta_{\beta\alpha}$. 
    
    We can define the surface Laplacian for a scalar function $\phi$ as follows:
	\begin{equation}
		\label{eqn:laplacian}
		\Delta\phi=\Div\nabla\phi = g^{\alpha\beta}\nabla_\alpha\nabla_\beta\phi,
	\end{equation} where the latter is given by
	\begin{equation}
		\nabla_\alpha\nabla_\beta\phi = \partial_\alpha\partial_\beta\phi-\Gamma^{\gamma}_{\alpha\beta}\,\partial_\gamma\phi.
	\end{equation} Lastly, we note that the most effective way to compute the surface Laplacian of a scalar function is via the Voss--Weyl formula, which allows us to rewrite (\ref{eqn:laplacian}) as
	\begin{equation}
		\Delta\phi=\partial_\alpha\!\left(\sqrt{g}\hspace{1pt}g^{\alpha\beta}\partial_\beta\phi\right)/\sqrt{g}.
	\end{equation}
	
	\subsubsection{Surface curvature} 
    
    As previously mentioned, the curvature of a surface can be described via the normal component of the second (parametric) derivatives of $\boldsymbol{R}$. Consequently, a second-order tensor can be constructed as $\itbf{B} = b_{\alpha\beta}\,\eu{\alpha}\otimes\eu{\beta}$, which is known as the curvature tensor, and its covariant elements $b_{\alpha\beta}$ are defined by
	\begin{equation}
		\label{eqn:curvature-tensor}
		b_{\alpha\beta}=\nv\cdot\partial_\beta\partial_\alpha\boldsymbol{R} = \nv\cdot\partial_\beta\ed{\alpha}=\nv\cdot\nabla_\beta\ed{\alpha}
	\end{equation} which is a symmetric tensor, and the sign of $b_{\alpha\beta}$ depends on the choice of the normal. The mixed components of $\itbf{B}$ are thus given by $b_{\alpha}^{\hspace{0.5pt}\beta} = b_{\alpha\gamma}\hspace{1pt}g^{\gamma\beta}$, whereas its contravariant elements are found to be $b^{\alpha\beta} = b^{\alpha}_{\gamma}\hspace{1pt}g^{\gamma\beta}$. This tensor measures the extrinsic curvature, \ie the bending of the two-dimensional surface in the three-dimensional space. The connection between intrinsic and extrinsic geometry is captured by the {\it Gauss--Weingarten equations}, \ie
	\begin{equation}
		\label{eqn:gauss-weingarten}
		\nabla_\beta\ed{\alpha} = b_{\alpha\beta}\,\nv,\quad\text{and}\quad\,\nabla_\alpha\nv = -b^{\hspace{0.5pt}\beta}_{\alpha}\,\ed{\beta},
	\end{equation} where the former equation follows from the definition of curvature tensor and covariant differentiation, while the latter is derived by differentiating the identities $\nv\cdot\ed{\alpha}\hspace{-1.5pt}=0$ and $\nv\cdot\nv=1$, and then solving for the derivative $\nabla_\alpha\nv$.
	
	The eigenvalues of the matrix $[b^{\hspace{0.5pt}\alpha}_{\beta}]$ are commonly called the {\it principal curvatures}, and their associated eigenvectors yield the {\it principal directions}. Instead of working with the principal curvatures, we use two scalar invariants: the mean curvature $H$, and the Gaussian curvature $K$, which are defined by
	\begin{align}
		\label{eqn:mean-gauss}
		&H=\frac{1}{2}\Tr[b^{\hspace{0.5pt}\alpha}_{\beta}]=\frac{1}{2}b^{\alpha}_\alpha,\quad\text{and}\\[5pt]
		&K=\det[b^{\hspace{0.5pt}\alpha}_{\beta}],
	\end{align} respectively. The Gaussian curvature can also be written as $K=\det[b_{\alpha\beta}]\,/\det[g_{\alpha\beta}]$, or in index notation,
	\begin{equation}
		\label{eqn:gaussian-index}
		K=\frac{1}{2}\varepsilon^{\alpha\beta}\varepsilon^{\gamma\delta}b_{\alpha\gamma}\hspace{1pt}b_{\beta\delta},
	\end{equation} where the permutation tensor $\varepsilon^{\alpha\beta}$ is defined by vanishing diagonal terms, \ie $\varepsilon^{11}=\varepsilon^{22}=0$, while its off-diagonal components are given by $\varepsilon^{12}=-\varepsilon^{21}=1/{\hspace{-1pt}\sqrt{g}}$. From Eq.~(\ref{eqn:gaussian-index}), we can see that the tensor $\varepsilon^{\alpha\beta}\varepsilon^{\gamma\delta}b_{\beta\delta}=\bar{b}^{\alpha\gamma}$ acts as the cofactor of the curvature tensor, which can be rewritten as follows:
	\begin{equation}
		\label{eqn:cofactor-curvature}
		\bar{b}^{\alpha\beta}=\varepsilon^{\alpha\gamma}\varepsilon^{\beta\delta}b_{\gamma\delta} = 2 H g^{\alpha\beta}-b^{\alpha\beta},
	\end{equation} resulting in $\bar{b}^{\alpha\gamma}b_{\gamma\beta} = K\delta^\alpha_\beta$. This gives us yet another~way to write the Gaussian curvature, \ie $K = 2H^2-\frac{1}{2}b^{\hspace{0.5pt}\alpha}_{\beta}b^{\hspace{0.5pt}\beta}_{\alpha}$, which leads to the following useful formula:
	\begin{equation}
		\label{eqn:cofactor-curvature-2}
		b_{\alpha\beta}\hspace{1pt}b^{\alpha\beta} = b^{\hspace{0.5pt}\alpha}_{\beta}\hspace{1pt}b^{\hspace{0.5pt}\beta}_{\alpha} = 4H^2 - 2 K.
	\end{equation}
	
	In Eq.~(\ref{eqn:divergence-Sfield}), the domain of definition of the vector field $\bPhi$ has been restricted to the tangent plane of the surface. However, Eqs.~(\ref{eqn:gauss-weingarten}) and (\ref{eqn:mean-gauss}) provide now the tools to extend this to vector fields in $\mathbb{R}^3$. By using the definition of the surface divergence, as in (\ref{eqn:surface-divergence}), for a vector $\itbf{v}=v^\alpha\ed{\alpha} + v\,\nv$, then $\Div\hspace{-1pt}(\itbf{v})$ is found to be
	\begin{equation}
		\label{eqn:div-surface-3D}
		\Div\hspace{-1pt}(\itbf{v}) = \nabla_\alpha v^\alpha - 2Hv.
	\end{equation}
	
	We note that the tensors $g_{\alpha\beta}$ and $b_{\alpha\beta}$ are not independent (in other words, not all choices of tensor fields $g_{\alpha\beta}$ and $b_{\alpha\beta}$ can describe a surface). This can be seen as an integrability condition on the Gauss--Weingarten equations, namely one needs to make sure that the identity
	\begin{equation}
		\label{eqn:integrability-condition}
		\partial_\beta\partial_\alpha\ed{\gamma} = \partial_\alpha\partial_\beta\ed{\gamma}
	\end{equation} is satisfied. Thus, the normal and tangential components of (\ref{eqn:integrability-condition}) leads to {\it Gauss--Codazzi--Mainardi equations}:
	\begin{align}
		\label{eqn:Gauss-Codazzi-Mainardi-1}
		&\nabla_\alpha b_{\beta\gamma}-\nabla_\beta b_{\alpha\gamma}=0,\quad\text{and}\\[2.5pt]
		\label{eqn:Gauss-Codazzi-Mainardi-2}
		&b_{\alpha\gamma}b_{\beta\delta}-b_{\alpha\delta}b_{\beta\gamma}=R_{\alpha\beta\gamma\delta}.
	\end{align} The latter equation tells us that the combination given by  $b_{\alpha\gamma}b_{\beta\delta}-b_{\alpha\delta}b_{\beta\gamma}$ is an intrinsic quantity of the surface, which can be purely computed from the metric (as $R_{\alpha\beta\gamma\delta}$ depends only on the metric tensor), despite the fact that the curvature tensor $b_{\alpha\beta}$ itself is an extrinsic quantity that requires an embedding. Moreover, by contracting twice the Riemann tensor, we obtain a scalar quantity known as the Ricci scalar, \ie $R= g^{\alpha\gamma}g^{\beta\delta}R_{\alpha\beta\gamma\delta}$. Using Eq.~(\ref{eqn:Gauss-Codazzi-Mainardi-2}), we have $R = 4H^2-b^{\hspace{0.5pt}\alpha}_{\beta}b^{\hspace{0.5pt}\beta}_{\alpha} = 2K$, where the last equality follows from Eq.~(\ref{eqn:cofactor-curvature-2}), and thus the Gaussian curvature is an intrinsic geometrical quantity. In fact, for two-dimensional surfaces, the Riemann tensor is completely determined by the Gaussian curvature $K$:
	\begin{equation}
		R_{\alpha\beta\gamma\delta} =  K\left(g_{\alpha\gamma}\,g_{\beta\delta}-g_{\alpha\delta}\,g_{\beta\gamma}\right)\!.
	\end{equation} On the other hand, Eq.~(\ref{eqn:Gauss-Codazzi-Mainardi-1}), which is often called the {\it Codazzi equation}, states that $\nabla_\beta b_{\alpha\gamma}$ must be symmetric with respect to the indices $\alpha$ and $\beta$. Since the curvature tensor $b_{\alpha\beta}$ is also symmetric, then we find that $\nabla_\beta b_{\alpha\gamma}$ is fully symmetric with respect to all its indices.	
	
	\subsection{Calculus of Moving Surfaces}
    
    The formulation of kinetic and kinematic relationships on a deformable membrane is much easier achieved by parameterizing the surface through a convected coordinate system $\varrho^\alpha$ with respect to a fixed reference system~$\xi^\alpha$, where the latter is depicted in Fig.~\ref{fig:A1}. Specifically, we identify by $\hspace{1pt}\hat{\!\itbf{R}}(\varrho^\alpha\hspace{-1pt},t)$ the current position at time $t$ of a material point that was located at
    \begin{equation}
        \hspace{1pt}\hat{\!\itbf{R}}(\varrho^\alpha\hspace{-1pt},\hspace{1pt}t_0)\hspace{-1pt} = \itbf{R}(\xi^\alpha),
    \end{equation}
    with the initial (reference) time $t_0\hspace{-1pt}\leq t$. In other words, we specify the fixed coordinates $\xi^\alpha$ as functions which depend on both the time and the convected coordinates $\varrho^\alpha$, subject to the condition that
    \begin{equation}
        \varrho^\alpha = \xi^\alpha(\varrho^\beta\hspace{-1pt},\hspace{1pt}t_0).
    \end{equation}
    In addition, we assume that these mappings between~$\varrho^\alpha$ and $\xi^\alpha$ are invertible, reflecting the notion that, at any fixed time $t$, a material point that is identified by constant $\varrho^\alpha$ can be uniquely associated with the local surface coordinates $\xi^\alpha$ at that time. Thus, we write
    \begin{equation}
        \label{eqn:convected-fixed}
        \hat{\!\itbf{R}}(\varrho^\alpha\hspace{-1pt},\hspace{1pt}t)\hspace{-1pt} = \itbf{R}\!\left(\hspace{1pt}\xi^\beta\hspace{-1pt}(\varrho^\alpha\hspace{-1pt},t),\hspace{1pt}t\hspace{1pt}\right)\!.
    \end{equation}
    
    The velocity $\itbf{v}$ of a material point that lives solely on the surface is defined by the rate of change of the position vector $\hat{\!\itbf{R}}(\varrho^\alpha\hspace{-1pt},\hspace{-1pt}t)$ at fixed $\varrho^\alpha$, namely
    \begin{equation}
        \itbf{v}=\left.\frac{\partial\hat{\!\hspace{1pt}\itbf{R}}}{\partial t}\right|_{\{\varrho^\beta\}} \!\!= \,\left.\frac{\partial\!\hspace{1pt}\itbf{R}}{\partial t}\right|_{\{\xi^\beta\}} \!+ \,\ed{\alpha}\!\left.\frac{\partial\xi^\alpha}{\partial t}\right|_{\{\varrho^\beta\}},
    \end{equation}
    where the latter equality follows via the chain rule. Hereinafter, we define the tangential velocity components $v^\alpha$ as the rate of change of the fixed coordinates $\xi^\alpha$ for a given material point of constant $\varrho^\alpha$, that is,
    \begin{equation}
        v^\alpha =\,\left.\frac{\partial\xi^\alpha}{\partial t}\right|_{\{\varrho^\beta\}}.
    \end{equation}
    On the other hand, the normal velocity is given by
    \begin{equation}
        v = \nv\cdot\hspace{-1pt}\left.\frac{\partial\!\hspace{1pt}\itbf{R}}{\partial t}\right|_{\{\xi^\alpha\}},
    \end{equation}
    so that the surface points of constant $\xi^\alpha$ change only along the normal direction to the surface, without any loss of generality. Thus, the velocity $\itbf{v}$ can be written as
    \begin{equation}
        \label{eqn:velocity-decomposition}
        \itbf{v} = v \nv +v^\alpha\hspace{-1pt}\ed{\alpha}.
    \end{equation}
    
    Any scalar function $\phi(\xi^\alpha\hspace{-1pt},\hspace{1pt}t)$ that lives on the surface can also be written as a function in terms of the convected coordinates, say $\hat{\phi}(\varrho^\alpha\hspace{-1pt},\hspace{1pt}t)$, where we have that
    \begin{equation}
        \hat{\phi}(\varrho^\alpha\hspace{-1pt},\hspace{1pt}t) = \phi\!\left(\xi^\beta\hspace{-1pt}(\varrho^\alpha\hspace{-1pt},t),\hspace{1pt}t\right)\!.
    \end{equation}
    The material derivative of $\phi$ is simply the rate of change of $\hat{\phi}$ for a given material point of constant $\varrho^\alpha$, that is,
    \begin{equation}
        \dot{\phi}=\frac{\mathrm{d}\phi}{\mathrm{d}t}=\left.\frac{\partial\hat{\phi}}{\partial t}\right|_{\{\varrho^\alpha\}},
    \end{equation}
    where the {\sl overdot} symbol denotes the material derivative of a function. By utilizing the chain rule, this leads to
    \begin{equation}
        \dot{\phi} = \,\left.\frac{\partial\phi}{\partial t}\right|_{\{\xi^\beta\}} \!+ \,\partial_\alpha\phi\!\left.\frac{\partial\xi^\alpha}{\partial t}\right|_{\{\varrho^\beta\}} =\; \frac{\partial\phi}{\partial t} +v^\alpha\hspace{1pt}\partial_\alpha\phi.
    \end{equation}
    
    Similarly, we introduce the tangent basis in convected coordinates, $\hat{\itbf{e}}_\alpha=\partial\hat{\itbf{R}}/\partial\varrho^\alpha=(\partial\xi^\beta/\partial\varrho^\alpha)\ed{\beta}$. For quantities carrying surface indices, the overdot denotes their material rate in convected coordinates, expressed in the fixed-coordinate basis. In particular, the material rate of the tangent basis includes the change in the coordinate mapping, giving
    \begin{equation}
        \dot{\itbf{e}}_{\alpha} = \left(\frac{\partial}{\partial t}+v^\beta\partial_\beta\right)\ed{\alpha}+\partial_\alpha v^\beta\ed{\beta} = \partial_\alpha\itbf{v}.
    \end{equation}
    This can be decomposed into the local surface basis,
    \begin{equation}
        \label{eqn:dot-basis}
        \dot{\!\itbf{e}}_{\alpha}= \partial_\alpha\itbf{v} = \left(\nabla_\alpha v^\beta-v\hspace{1pt} b_\alpha^\beta\right)\!\ed{\beta} + \left(v^\beta b_{\alpha\beta} + \partial_\alpha v\right)\!\nv,
    \end{equation}
    by employing the form of the velocity field in Eq.~(\ref{eqn:velocity-decomposition}), and by also using the Gauss--Weingarten equations.
    
    By applying the material derivative to the orthogonality condition $\nv\cdot\ed{\alpha}=0$, we have $\ed{\alpha}\cdot\dot{\nv}=-\nv\cdot\dot{\itbf{e}}_\alpha$. Since $\nv\cdot\dot{\nv}=0$, the material derivative of the normal unit vector acts only within the tangent plane of the surface, and thus from Eq.~(\ref{eqn:dot-basis}) we obtain
    \begin{equation}
        \label{eqn:dot-normal}
        \dot{\nv} = -\left(v^\beta\hspace{0.25pt}b^{\alpha}_\beta + \nabla^\alpha v\right)\!\ed{\alpha},
    \end{equation} where $\nabla^\alpha=g^{\alpha\beta}\nabla_\beta$. Eqs.~(\ref{eqn:dot-basis}) and (\ref{eqn:dot-normal}) then allow us to find that
    \begin{align}
        \dot{\itbf{v}}\; =\; &\left[\frac{\partial v^\alpha}{\partial t}-\left(v^\beta\hspace{0.25pt}b^{\alpha}_\beta + \nabla^\alpha v\right)\hspace{-1pt}v + \left(\nabla_\beta v^\alpha-v\hspace{1pt} b_{\beta}^{\alpha}\right)\hspace{-1pt}v^\beta\right]\!\ed{\alpha}\notag\\[4pt]
        &\;+\;\left[\frac{\partial v}{\partial t} + 2v^\alpha \partial_\alpha v + v^\alpha v^\beta b_{\alpha\beta}\right]\!\nv.
    \end{align}
    
    For the metric and curvature tensors, the overdot likewise denotes their convected material rates, which include the change in their surface-coordinate components. Explicitly, these rates are defined by
    \begin{align}
        \dot{g}_{\alpha\beta} &= \frac{\partial g_{\alpha\beta}}{\partial t}+v^\gamma\partial_\gamma g_{\alpha\beta}+g_{\gamma\beta}\partial_\alpha v^\gamma+g_{\alpha\gamma}\partial_\beta v^\gamma,\\[4pt]
        \dot{b}_{\alpha\beta} &= \frac{\partial b_{\alpha\beta}}{\partial t}+v^\gamma\partial_\gamma b_{\alpha\beta}+b_{\gamma\beta}\partial_\alpha v^\gamma+b_{\alpha\gamma}\partial_\beta v^\gamma.
    \end{align}
    Using the evolution of the tangent basis and normal vector, the corresponding expressions become
    \begin{align}
        \label{eqn:g-dot}
        \dot{g}_{\alpha\beta} &= \nabla_\alpha v_\beta+\nabla_\beta v_\alpha - 2v\hspace{1pt}b_{\alpha\beta},\quad\mathrm{and}\\[4pt]
        \label{eqn:b-dot}
        \dot{b}_{\alpha\beta} &=b_{\gamma\beta}\left(\nabla_\alpha v^\gamma - v\hspace{1pt}b^{\gamma}_\alpha\right)+\nabla_\beta\!\left(v^\gamma\hspace{0.5pt}b_{\gamma\alpha}+\partial_\alpha v\right),
    \end{align}
    respectively. The above expressions follow by applying the product rule to the convected material rates and expanding the definitions of $g_{\alpha\beta}$ and $b_{\alpha\beta}$ in terms of the local tangent basis and the normal. For instance, we have $\dot{g}_{\alpha\beta}=\ed{\alpha}\cdot\dot{\!\itbf{e}}_\beta+\dot{\!\itbf{e}}_\alpha\cdot\ed{\beta}$, which together with Eq.~(\ref{eqn:dot-basis}) gives Eq.~(\ref{eqn:g-dot}). Similarly, from Eq.~(\ref{eqn:curvature-tensor}), we obtain $\dot{b}_{\alpha\beta}=\nv\cdot\partial_\beta\dot{\itbf{e}}_\alpha+\dot{\nv}\cdot\partial_\beta\ed{\alpha}=-\partial_\beta\nv\cdot\dot{\itbf{e}}_\alpha-\partial_\beta\dot{\nv}\cdot\ed{\alpha}$, which yields Eq.~(\ref{eqn:b-dot}) upon substituting Eqs.~(\ref{eqn:dot-basis}) and (\ref{eqn:dot-normal}).
	
	\subsection{Integral and Transport Theorems}
	
	The area element $\mathrm{d}S$ of a surface that is parameterized by the fixed curvilinear coordinates $(\xi^1,\xi^2)$ is given by
    \begin{equation}
        \mathrm{d}S = \mathrm{d}\xi^1\mathrm{d}\xi^2\sqrt{g},
    \end{equation} where $g=\det[g_{\alpha\beta}]$, as previously defined in Eq.~(\ref{eqn:g-def}). This allows us to compute the total area of a surface $\mathcal{M}$ by integrating over the fixed internal coordinates $\xi^\alpha$, \ie
    \begin{equation}
        \mathcal A = \!\int_{\mathcal{M}\;\;}\!\mathrm{d}S\, = \iint\!\mathrm{d}\xi^1\mathrm{d}\xi^2\sqrt{g}\,.
    \end{equation} To determine the rate of change of area, we consider the convected rate of the surface area density. In the fixed-coordinate description, this is given by $\frac{\mathrm{d}}{\mathrm{d}t}\sqrt{g}=\partial_t\sqrt{g}+\partial_\alpha(\sqrt{g}\,v^\alpha)$, where the second term accounts for tangential transport. Applying Jacobi's formula, we obtain
    \begin{equation}
        \frac{1}{\sqrt{g}}\frac{\mathrm{d}(\sqrt{g})}{\mathrm{d}t} = \frac{1}{2}\,g^{\alpha\beta}\dot{g}_{\alpha\beta} = \nabla_\alpha v^\alpha - 2vH = \Div\hspace{-1pt}(\itbf{v}),
    \end{equation} where the last two equalities follow from Eqs.~(\ref{eqn:g-dot}) and (\ref{eqn:div-surface-3D}), respectively. Hence,
    \begin{equation}
        \label{eqn:derivative-sqrt-g}
        \frac{\mathrm{d}}{\mathrm{d}t}\sqrt{g} = \Div\hspace{-1pt}(\itbf{v})\hspace{1pt}\sqrt{g},
    \end{equation} which implies that the rate of change of area of the closed surface is given by
    \begin{equation}
        \label{eqn:rate-change-area}
        \frac{\mathrm{d}\mathcal A}{\mathrm{d}t} = \iint\!\mathrm{d}\xi^1\mathrm{d}\xi^2\,\frac{\mathrm{d}}{\mathrm{d}t}\sqrt{g} = \!\int_{\mathcal{M}\;\;}\!\Div\hspace{-1pt}(\itbf{v})\;\mathrm{d}S.
    \end{equation}
    
    The identity in Eq.~(\ref{eqn:derivative-sqrt-g}) can also be used to derive the {\it Reynolds transport theorem}, that is,
    \begin{equation}
        \frac{\mathrm{d}}{\mathrm{d}t}\hspace{1pt}\int_{\mathcal{S}\;\;}\!\itbf{T}\,\mathrm{d}S = \!\int_{\mathcal{S}\;\;}\!\mathrm{d}S\left[\itbf{T}\,\Div\hspace{-1pt}(\itbf{v})+\frac{\mathrm{d}\itbf{T}}{\mathrm{d}t}\hspace{1pt}\right]\!,
    \end{equation} where $\mathcal{S}$ is a material surface patch moving with the membrane, and $\itbf{T}$ can be any tensor field defined on $\mathcal{S}$.
	
	Furthermore, the volume $V$ enclosed by an arbitrary closed (orientable) surface, say $\mathcal{M}$, can be obtained from the position vector $\itbf{R}$ as follows:
	\begin{equation}
		V=\!\iiint\!\mathrm{d}V\! = \frac{1}{3}\int_{\mathcal{M}\,\;}\!\left(\nv\cdot\itbf{R}\right)\mathrm{d}S,
	\end{equation} where the latter is found by employing the vector calculus identity, $\bnabla\cdot\itbf{r} = 3$, with $\itbf{r}$ being any position vector, and lastly making use of the divergence theorem (by also noting that boundary of the volume $\partial V$ is trivially the same as the closed surface of the membrane~$\mathcal{M}$). 
	
	To find the rate of change of volume, we can simply apply the Reynolds transport theorem, which yields
	\begin{equation}
		\frac{\mathrm{d}V}{\mathrm{d}t} = \frac{1}{3}\int_{\mathcal{M}\,\;}\!\Big\{ v + \itbf{R}\cdot\left[\dot{\nv} + \nv\hspace{1pt}\Div\hspace{-1pt}(\itbf{v})\right]\!\Big\}\,\mathrm{d}S,
	\end{equation} where the term in the square brackets can be written as: $\dot{\nv} + \nv\hspace{1pt}\Div\hspace{-1pt}(\itbf{v}) = -\big(v^\beta\hspace{0.25pt}b^{\alpha}_\beta + \nabla^\alpha v\big)\hspace{-0.5pt}\ed{\alpha} + \big(\nabla_\alpha v^\alpha -2Hv\big)\nv$. Notice that the term $\itbf{R}\cdot(\nabla^\alpha v\,\ed{\alpha})$ can be expressed in terms of a total derivative, namely
	\begin{equation}
		\itbf{R}\cdot(\nabla^\alpha v\,\ed{\alpha}) = \nabla^\alpha\!\left[v\hspace{1pt}(\ed{\alpha}\cdot\itbf{R})\right] - 2\hspace{0.5pt}v\hspace{1pt}(1+ H \nv\cdot\itbf{R}),
	\end{equation} where Eq.~(\ref{eqn:curvature-tensor}) is used to derive the latter. Since $\mathcal{M}$ is a closed surface, we obtain that
	\begin{equation}
		\label{eqn:volume-eq-appex}
		\frac{\mathrm{d}V}{\mathrm{d}t} = \frac{1}{3}\int_{\mathcal{M}\,\;}\!\Big\{3v + \itbf{R}\cdot\left[\nv\hspace{1pt}\nabla_\alpha v^\alpha -v^\beta b^{\alpha}_\beta\ed{\alpha}\right]\!\Big\}\,\mathrm{d}S.
	\end{equation} Using the Gauss--Weingarten equation, $\nabla_\beta\nv = -b^{\alpha}_{\beta}\ed{\alpha}$, the second term in the integrand of Eq.~(\ref{eqn:volume-eq-appex}) simply becomes $\itbf{R}\cdot\big[\nv\hspace{1pt}\nabla_\alpha v^\alpha\hspace{-0.5pt}+v^\alpha\nabla_\alpha\nv\big]\! =\hspace{-1pt} \nabla_\alpha\!\left[(\nv\cdot\itbf{R})v^\alpha\right]$. Its integral vanishes on the closed surface by the surface divergence theorem. Therefore, this leads to the following simple (and geometrically intuitive) result:
	\begin{equation}
		\frac{\mathrm{d}V}{\mathrm{d}t} = \int_{\mathcal{M}\,\;}\!v\;\mathrm{d}S,
	\end{equation} that is, the rate of change of the volume $V$ is identical to the integrated value of the normal velocity over the total area of the membrane surface.

    \bibliographystyle{apsrev4-2}

%
	
\end{document}